\documentstyle[epsfig]{livrev}
\begin{document}

\title{Numerical hydrodynamics in general relativity}

\author{Jos\'e~A.~Font \\
Departamento de Astronom\'{\i}a y Astrof\'{\i}sica \\
Edificio de Investigaci\'on ``Jeroni Mu\~noz" \\
Universidad de Valencia \\
Dr. Moliner 50 \\
E-46100 Burjassot (Valencia), Spain}

\date{}

\maketitle

\begin{center}
{\bf Abstract}
\end{center}
The current status of numerical solutions for the equations of ideal
general relativistic hydrodynamics is reviewed. With respect to an
earlier version of the article the present update provides additional 
information on numerical schemes and extends the discussion of 
astrophysical simulations in general relativistic hydrodynamics. Different 
formulations of the equations are presented, with special mention of 
conservative and hyperbolic formulations well-adapted to advanced numerical 
methods. A large sample of available numerical schemes is discussed,
paying particular attention to solution procedures based on schemes 
exploiting the characteristic structure of the equations through 
linearized Riemann solvers. A comprehensive summary of astrophysical 
simulations in strong gravitational fields is presented. These include 
gravitational collapse, accretion onto black holes and hydrodynamical
evolutions of neutron stars. The material contained in these sections 
highlights the numerical challenges of various representative simulations. 
It also follows, to some extent, the chronological development of the field, 
concerning advances on the formulation of the gravitational field and
hydrodynamic equations and the numerical methodology designed to solve them. 


\section{Introduction}
\label{intro}

The description of important areas of modern astronomy, such as high-energy 
astrophysics or gravitational wave astronomy, requires General Relativity.  
High energy radiation is often emitted by highly relativistic events in regions 
of strong gravitational fields near compact objects such as neutron stars or 
black holes. The production of relativistic radio jets in active galactic nuclei, 
explained by pure hydrodynamical effects as in the twin-exhaust model
\cite{blandford74}, by hydromagnetic centrifugal acceleration as in the 
Blandford-Payne mechanism~\cite{blandford82}, or by electromagnetic extraction
of energy as in the Blandford-Znajek mechanism~\cite{blandford77}, involves an 
accretion disk around a rotating supermassive black hole. The discovery of kHz 
quasi-periodic oscillations in low-mass X-ray binaries extended the frequency 
range over which these oscillations occur into timescales associated with the 
relativistic, innermost regions of accretion disks (see, e.g.~\cite{vanderklis98}).
A relativistic description is also necessary in scenarios involving explosive 
collapse of very massive stars ($\sim 30 M_{\odot}$) to a black hole (in 
the so-called collapsar and hypernova models), or during the last phases of 
the coalescence of neutron star binaries. These catastrophic events are
believed to exist at the central engine of highly energetic $\gamma$-ray bursts 
(GRBs) \cite{paczynski86,narayan92,woosley93,paczynski98}. In addition, non 
spherical gravitational collapse leading to black hole formation or to a supernova 
explosion, and neutron star binary coalescence are among the most promising 
sources of detectable gravitational radiation. Such astrophysical scenarios 
constitute one of the main targets for the new generation of ground-based laser 
interferometers, just starting their gravitational wave search (LIGO, VIRGO, 
GEO600, TAMA)~\cite{thorne96,new02}.

A powerful way to improve our understanding of the above scenarios is
through accurate, large scale, three-dimensional numerical simulations. 
Nowadays, computational general relativistic astrophysics is an increasingly 
important field of research. In addition to the large amount of observational 
data by high-energy X- and $\gamma$-ray satellites such as Chandra, 
XMM-Newton or INTEGRAL, and the new generation of gravitational wave detectors,
the rapid increase in computing power through parallel supercomputers 
and the associated advance in software technologies is making possible
large scale numerical simulations in the framework of general relativity. 
However, the computational astrophysicist and the numerical relativist
face a daunting task. In the most general case, the equations governing 
the dynamics of relativistic astrophysical systems are an intricate, coupled 
system of time-dependent partial differential equations, comprising the 
(general) relativistic (magneto-) hydrodynamic (MHD) equations and the 
Einstein gravitational field equations. In many cases, the number of 
equations must be augmented to account for non-adiabatic processes, e.g., 
radiative transfer or sophisticated microphysics (realistic equations of 
state for nuclear matter, nuclear physics, magnetic fields, etc.).

Nevertheless, in some astrophysical situations of interest, e.g., accretion of 
matter onto compact objects or oscillations of relativistic stars, the `test-fluid' 
approximation is enough to get an accurate description of the underlying
dynamics. In this approximation the fluid self-gravity is neglected in comparison 
to the {\it background} gravitational field. This is best exemplified in accretion 
problems where the mass of the accreting fluid is usually much smaller than 
the mass of the compact object. Additionally, a description employing ideal 
hydrodynamics (i.e., with the stress-energy tensor being that of a perfect fluid), 
is also a fairly standard choice in numerical astrophysics.

The main purpose of this review is to summarize the existing efforts 
to solve numerically the equations of (ideal) general relativistic 
hydrodynamics. To this aim, the most important numerical schemes
will be first presented in some detail. Prominence will be given
to the so-called Godunov-type schemes written in conservation form.
Since~\cite{marti91} it has been gradually demonstrated~(\cite{font94,
eulderink95,romero96,falle96,banyuls97,wen97,pons98}) that conservative
methods exploiting the hyperbolic character of the relativistic hydrodynamic 
equations are optimally suited for accurate numerical integrations, even well 
inside the ultrarelativistic regime. The explicit knowledge of the characteristic
speeds (eigenvalues) of the equations, together with the corresponding
eigenvectors, provides the mathematical (and physical) framework for such
integrations, by means of either exact or approximate Riemann solvers.

The article includes, furthermore,
a comprehensive description of `relevant' numerical applications in
relativistic astrophysics, including gravitational collapse, accretion 
onto compact objects and hydrodynamical evolution of neutron stars.
Numerical simulations of strong-field scenarios employing Newtonian gravity 
and hydrodynamics, as well as possible post-Newtonian extensions, have received 
considerable attention in the literature and will not be covered in the review,
which focuses in relativistic simulations. Nevertheless, we must emphasize that 
most of what is known about hydrodynamics near compact objects, in particular 
in black hole astrophysics, has been accurately described using Newtonian models. 
Probably the best known example is the use of a pseudo-Newtonian potential for 
non-rotating black holes which mimics the existence of an event horizon at the 
Schwarzschild gravitational radius~\cite{paczynski80}, which has allowed accurate 
interpretations of observational phenomena. 

The organization of the article is as follows: Section 2 presents the
equations of general relativistic hydrodynamics, summarizing the most 
relevant theoretical formulations which, to some extent, have helped
to drive the development of numerical algorithms for their solution. 
Section 3 is mainly devoted to describing numerical schemes specifically
designed to solve non-linear hyperbolic systems of conservation laws. Hence, 
particular emphasis will be paid on conservative high-resolution shock-capturing 
(HRSC) upwind methods based on linearized Riemann solvers. Alternative schemes 
such as Smoothed Particle Hydrodynamics (SPH), (pseudo-) spectral methods 
and others will be briefly discussed as well. Section 4 summarizes a 
comprehensive sample of hydrodynamical simulations in strong-field general 
relativistic astrophysics. Finally, in Section 5 we provide additional 
technical information needed to build up upwind HRSC schemes for the 
general relativistic hydrodynamics equations. Geometrized units ($G=c=1$) 
are used throughout the paper except where explicitly indicated, as well 
as the metric conventions of~\cite{MTW}. Greek (Latin) indices run from 0 
to 3 (1 to 3).

\section{Equations of general relativistic hydrodynamics}
\label{equations}

The general relativistic hydrodynamic equations consist of the local
conservation laws of the stress-energy tensor, $T^{\mu \nu}$ (the Bianchi 
identities) and of the matter current density, $J^{\mu}$ (the continuity
equation)
\begin{equation}
{\nabla}_{\mu} T^{\mu \nu} = 0,
\label{eq:stressenergycons}
\end{equation}
\begin{equation}
{\nabla}_{\mu} J^{\mu} = 0.
\label{eq:masscons}
\end{equation}

As usual ${\nabla}_{\mu}$ stands for the covariant derivative associated with
the four-dimensional spacetime metric $g_{\mu\nu}$.
The density current is given by $J^{\mu} = \rho u^{\mu}$, $u^{\mu}$
representing the fluid 4-velocity and $\rho$ the rest-mass density in a
locally inertial reference frame.

The stress-energy tensor for a non-perfect fluid is defined as
\begin{equation}
T^{\mu \nu} = \rho(1+\varepsilon) u^{\mu} u^{\nu} +
(p-\zeta\theta)h^{\mu\nu} - 2\eta\sigma^{\mu\nu} +
q^{\mu}u^{\nu} + q^{\nu}u^{\mu}
\end{equation}
\noindent
where $\varepsilon$ is the rest frame specific internal energy
density of the fluid, $p$ is the pressure and $h^{\mu\nu}$ is the 
spatial projection tensor $h^{\mu\nu}=u^{\mu}u^{\nu} + g^{\mu\nu}$. 
In addition, $\eta$ and $\zeta$ are the shear and bulk viscosities. 
The expansion $\theta$, describing the divergence or convergence of 
the fluid world lines is defined as $\theta=\nabla_{\mu}u^{\mu}$. 
The symmetric, trace-free, and spatial shear tensor $\sigma^{\mu\nu}$, 
is defined by
\begin{equation}
\sigma^{\mu\nu}=\frac{1}{2}
(\nabla_{\alpha}u^{\mu} h^{\alpha\nu}  +
 \nabla_{\alpha}u^{\nu} h^{\alpha\mu}) -
\frac{1}{3}\theta h^{\mu\nu},
\end{equation}
and, finally, $q^{\mu}$ is the energy flux vector.

In the following we will neglect non-adiabatic effects, such as viscosity
or heat transfer, assuming the stress-energy tensor to be that of a perfect
fluid
\begin{equation}
T^{\mu \nu} = \rho h u^{\mu} u^{\nu} + p g^{\mu \nu},
\label{perf_fluid}
\end{equation}
where we have introduced the relativistic specific enthalpy, $h$, 
defined by 
\begin{equation}
h = 1 + \varepsilon + \frac{p}{\rho}.
\end{equation}

Introducing an explicit coordinate chart $(x^{0},x^{i})$ the previous
conservation equations read
\begin{eqnarray}
\label{initial1}
\frac{\partial}{\partial x^{\mu}} \sqrt{-g} J^{\mu} & = & 0 \, , \\
\frac{\partial}{\partial x^{\mu}} \sqrt{-g} T^{\mu\nu} & = & - \sqrt{-g}
\Gamma^{\nu}_{\mu\lambda} T^{\mu\lambda} \, ,
\label{initial2}
\end{eqnarray}
where the scalar $x^{0}$ represents a foliation of the spacetime with
hypersurfaces (coordinatized by $x^{i}$). Additionally, $\sqrt{-g}$ is 
the volume element associated with the 4-metric, with $g=\det(g_{\mu\nu})$, 
and $\Gamma^{\nu}_{\mu\lambda}$ are the 4-dimensional Christoffel symbols.

In order to close the system, the equations of motion~(\ref{eq:stressenergycons}) 
and the continuity equation~(\ref{eq:masscons}) must be supplemented with 
an equation of state (EOS) relating some fundamental thermodynamical 
quantities. In general, the EOS takes the form
\begin{equation}
p=p(\rho,\varepsilon).
\label{eos}
\end{equation}

Due to their simplicity, the most widely employed EOS in numerical simulations 
are the {\it ideal fluid} EOS, $p=(\Gamma -1)\rho\varepsilon$, where $\Gamma$ is 
the adiabatic index, and the {\it polytropic} EOS (e.g. to build equilibrium 
stellar models), $p=K\rho^{\Gamma}\equiv K\rho^{1+1/N}$, $K$ being the polytropic 
constant and $N$ the polytropic index.

In the `test-fluid' approximation, where the fluid self-gravity is neglected,
the dynamics of the system is completely governed by 
Eqs.~(\ref{eq:stressenergycons}) and (\ref{eq:masscons}), together
with the EOS (\ref{eos}). In those situations where such approximation does
not hold, the previous equations must be solved in conjunction with
the Einstein gravitational field equations,
\begin{equation}
G^{\mu\nu}=8\pi T^{\mu\nu},
\label{einstein}
\end{equation}
\noindent
which describe the evolution of the geometry in a dynamical spacetime. A
detailed description of the various numerical approaches to solve the
Einstein equations is beyond the scope of the present article (see, e.g.
Lehner~\cite{lehner} for a recent review). We only mention that the 
formulation of the Einstein equations as an initial value (Cauchy) problem, 
in the presence of matter fields, adopting the so-called 3+1 decomposition 
of the spacetime~\cite{arnowitt62} can be found in, e.g.,~\cite{york79}. 
Given a choice of gauge, the Einstein equations in the 3+1 
formalism~\cite{arnowitt62} split into evolution equations for the 3--metric 
$\gamma_{ij}$ and the extrinsic curvature $K_{ij}$ (the second fundamental
form), and constraint equations, the Hamiltonian and momentum constraints,
that must be satisfied at every time slice. Long-term stable evolutions of 
the Einstein equations have recently been accomplished using various 
reformulations of the original 3+1 system (see, 
e.g.~\cite{baumgarte98,shibata99,alcubierre99,font02a} for simulations 
involving matter sources, and~\cite{alcubierre02a} and references therein 
for (vacuum) black hole evolutions). Alternatively, a characteristic initial value 
problem formulation of the Einstein equations was developed in the 1960s by 
Bondi, van der Burg and Metzner~\cite{bondi62}, and Sachs~\cite{sachs62}. 
This approach has gradually advanced to a state where long-term stable 
evolutions of caustic-free spacetimes in multidimensions are possible, even 
including matter fields (see~\cite{lehner} and references therein).
A recent review of the characteristic formulation is presented in a
{\it Living Reviews} article by Winicour~\cite{winicour98}. Examples of
this formulation in general relativistic hydrodynamics are discussed in
various sections of the present article.

Traditionally, most of the approaches for numerical integrations of the
general relativistic hydrodynamic equations have adopted spacelike 
foliations of the spacetime, within the 3+1 formulation. Recently, 
however, covariant forms of these equations, well suited for advanced 
numerical methods, have also been developed. This is reviewed next in a 
chronological way.

\subsection{Spacelike (3+1) approaches}

In the 3+1 (ADM) formulation~\cite{arnowitt62}, spacetime is foliated into 
a set of non-intersecting spacelike hypersurfaces.  There are two kinematic
variables describing the evolution between these surfaces: the
lapse function $\alpha$, which describes the rate of advance of time
along a timelike unit vector $n^\mu$ normal to a surface, and the
spacelike shift vector $\beta^i$ that describes the motion of
coordinates within a surface.

The line element is written as
\begin{equation}
 ds^2 = -(\alpha^{2} -\beta _{i}\beta ^{i}) dx^0 dx^0 +
 2 \beta _{i} dx^{i} dx^0 +\gamma_{ij} dx^{i} dx^{j},
\label{metric}
\end{equation}
where $\gamma_{ij}$ is the 3--metric induced on each spacelike slice.

\subsubsection{1+1 Lagrangian formulation (May and White)}

The pioneering numerical work in general relativistic hydrodynamics
dates back to the one-dimensional gravitational collapse code of May and 
White~\cite{may66,may67}. Building on theoretical work by Misner and
Sharp~\cite{misner64}, May and White developed a time-dependent numerical 
code to solve the evolution equations describing adiabatic spherical collapse 
in general relativity. This code was based on a Lagrangian finite difference 
scheme (see Section~\ref{fds}), in which the coordinates are co-moving with 
the fluid. Artificial viscosity terms were included in the equations to damp 
the spurious numerical oscillations caused by the presence of shock waves in 
the flow solution. May and White's formulation became the starting point of 
a large number of numerical investigations in subsequent years and, hence, 
it is worth describing its main features in some detail.

For a spherically symmetric spacetime the line element can be written as
\begin{equation}
ds^2=-a^2(m,t)dt^2+b^2(m,t)dm^2+R^2(m,t)(d\theta^2+\sin^2\theta d\phi^2),
\end{equation}
\noindent
$m$ being a radial (Lagrangian) coordinate, indicating the total rest-mass 
enclosed inside the circumference $2\pi R(m,t)$.

The co-moving character of the coordinates leads, for a perfect fluid, to
a stress-energy tensor of the form:
\begin{eqnarray}
T^{1}_1=T^2_2=T^3_3=-p,
\\
T^0_0=(1+\varepsilon)\rho,
\\
T^{\mu}_{\nu}=0 \hspace{0.1in} \mbox{if} \hspace{0.1in} \mu \ne \nu.
\end{eqnarray}
\noindent
In these coordinates the local conservation equation for the baryonic mass,
Eq.~(\ref{eq:masscons}), can be easily integrated to yield the metric potential
$b$:
\begin{equation}
b=\frac{1}{4\pi\rho R^2}.
\end{equation}

The gravitational field equations, Eq.~(\ref{einstein}), and the
equations of motion, Eq.~(\ref{eq:stressenergycons}), reduce to the
following quasi-linear system of partial differential equations (see
also~\cite{misner64}):
\begin{equation}
\frac{\partial p}{\partial m} + \frac{1}{a}\frac{\partial a}{\partial
m}\rho h = 0,
\label{mw1}
\end{equation}
\begin{equation}
\frac{\partial \varepsilon}{\partial t} + p\frac{\partial}{\partial
t}\left(\frac{1}{\rho}\right) = 0,
\label{mw2}
\end{equation}
\begin{equation}
\frac{\partial u}{\partial t} = 
-a \left(4 \pi \frac{\partial p}{\partial m}R^{2}
\frac{\Gamma}{h} + \frac{M}{R^{2}} + 4 \pi pR \right),
\label{mw3}
\end{equation}
\begin{equation}
\frac{1}{\rho R^2}\frac{\partial \rho R^2}{\partial t} = -a\frac{\partial
u/\partial m}{\partial R/\partial m},
\label{mw4}
\end{equation}
with the definitions $u=\frac{1}{a}\frac{\partial R}{\partial t}$ and
$\Gamma=\frac{1}{b}\frac{\partial R}{\partial m}$, satisfying
$\Gamma^{2}=1-u^{2}-\frac{2M}{R}$. Additionally, 
\begin{equation}
M=\int_{0}^{m}4\pi R^2 \rho (1+\varepsilon) \frac{\partial R}{\partial m} dm,
\label{35}
\end{equation}
represents the total mass interior to radius $m$ at time $t$.
The final system, Eqs.~(\ref{mw1})-(\ref{mw4}), is closed with an EOS 
of the form given by Eq.~(\ref{eos}).

Hydrodynamics codes based on the original formulation of May and White and 
on later versions (e.g.,~\cite{vanriper79}) have been used in many non-linear
simulations of supernova and neutron star collapse (see, e.g.,~\cite{miralles91,
swesty94} and references therein), as well as in perturbative computations of 
spherically symmetric gravitational collapse within the framework of the linearized 
Einstein equations~\cite{seidel87,seidel88}. In Section 4.1.1 below some of these 
simulations are discussed in detail. An interesting analysis of the above 
formulation in the context of gravitational collapse is provided by Miller and 
Sciama~\cite{misci}. By comparing the Newtonian and relativistic equations, 
these authors showed that the net acceleration of the infalling mass shells is
larger in general relativity than in Newtonian gravity. The Lagrangian character 
of May and White's formulation, together with other theoretical considerations 
concerning the particular coordinate gauge, has prevented its extension to 
multidimensional calculations. However, for one-dimensional problems, the 
Lagrangian approach adopted by May and White has considerable advantages with 
respect to an Eulerian approach with spatially fixed coordinates, most notably 
the lack of numerical diffusion. 

\subsubsection{3+1 Eulerian formulation (Wilson)}
\label{wilson}

The use of Eulerian coordinates in multidimensional numerical relativistic 
hydrodynamics started with the pioneering work by Wilson~\cite{wilson72}. 
Introducing the basic dynamical variables $D$, $S_{\mu}$ and $E$, representing
the relativistic density, momenta and energy, respectively, defined as
\begin{equation}
D=\rho u^{0}, \hspace{0.5cm}
S_{\mu}=\rho h u_{\mu}u^{0}, \hspace{0.5cm}
E=\rho \varepsilon u^{0},
\label{wilson_vars}
\end{equation}
\noindent
the equations of motion in Wilson's formulation~\cite{wilson72,wilson79} 
are:
\begin{equation}
\frac{1}{\sqrt{-g}}\frac{\partial}{\partial x^0}
(D\sqrt{-g})+\frac{1}{\sqrt{-g}}\frac{\partial}{\partial
x^{i}}(DV^{i}\sqrt{-g})=0,
\label{36}
\end{equation}
\begin{equation}
\frac{1}{\sqrt{-g}}\frac{\partial}{\partial x^0}
(S_{\mu}\sqrt{-g})+\frac{1}{\sqrt{-g}}\frac{\partial}{\partial
x^{i}}(S_{\mu}V^{i}\sqrt{-g})+\frac{\partial p}{\partial
x^{\mu}}+\frac{1}{2}\frac{\partial g^{\alpha \beta}}{\partial
x^{\mu}}\frac{S_{\alpha}S_{\beta}}{S^{0}}=0,
\label{40}
\end{equation}
\begin{equation}
\frac{\partial}{\partial x^0}(E\sqrt{-g})+\frac{\partial}{\partial
x^{i}}(EV^{i}\sqrt{-g})+p\frac{\partial}{\partial
x^{\mu}}(u^{0}V^{\mu}\sqrt{-g})=0,
\label{42}
\end{equation}
\noindent
with the ``transport velocity" given by $V^{\mu}=u^{\mu}/u^{0}$. We note
that in the original formulation~\cite{wilson79} the momentum density 
equation, Eq.~(\ref{40}), is only solved for the three spatial components, 
$S_i$, and $S_0$ is obtained through the 4-velocity normalization condition 
$u_{\mu}u^{\mu}=-1$.

A direct inspection of the system shows that the equations are written as 
a coupled set of advection equations. In doing so, the terms containing 
derivatives (in space or time) of the pressure are treated as source terms. 
This approach, hence, sidesteps an important guideline for the formulation 
of non-linear hyperbolic systems of equations, namely the preservation of 
their {\it conservation form}. This is a necessary condition to guarantee 
correct evolution in regions of sharp entropy generation (i.e., shocks).  
Furthermore, some amount of numerical dissipation must be used to stabilize 
the solution across discontinuities. In this spirit the first attempt to solve 
the equations of general relativistic hydrodynamics in the original Wilson's 
scheme~\cite{wilson72} used a combination of finite difference upwind techniques 
with artificial viscosity terms. Such terms adapted the classic treatment of 
shock waves introduced by von Neumann and Richtmyer~\cite{vonneumann50} to 
the relativistic regime (see Section 3.1.1).

Wilson's formulation has been widely used in hydrodynamical codes developed 
by a variety of research groups. Many different astrophysical scenarios were 
first investigated with these codes, including axisymmetric stellar 
core-collapse~\cite{naka80,naka81,naka82,bardeen83,stark85,piran86,evans86}, 
accretion onto compact objects~\cite{hawley84a,petrich89}, numerical 
cosmology~\cite{centrella83,centrella84,anninos98} and, more recently, the 
coalescence of neutron star binaries~\cite{wilson95,wilson96,mathews99}.
This formalism has also been employed, in the special relativistic limit, in 
numerical studies of heavy-ion collisions~\cite{wilson89,mcabee94}. We note 
that in most of these investigations, the original formulation of the 
hydrodynamic equations was slightly modified by re-defining the dynamical 
variables, Eq.~(\ref{wilson_vars}), with the addition of a multiplicative 
$\alpha$ factor (the lapse function) and the introduction of the Lorentz 
factor, $W\equiv \alpha u^0$:
\begin{equation}
D=\rho W, \hspace{0.5cm}
S_{\mu}=\rho h W u_{\mu}, \hspace{0.5cm}
E=\rho \varepsilon W.
\label{wilson_vars2}
\end{equation}

As mentioned before, the description of the evolution of self-gravitating
matter fields in general relativity requires a joint integration
of the hydrodynamic equations and the gravitational field equations 
(the Einstein equations). Using Wilson's formulation for the fluid dynamics, 
such coupled simulations were first considered in~\cite{wilson79}, building 
on a vacuum, numerical relativity code specifically developed to investigate 
the head-on collision of two black holes~\cite{smarr75}. The resulting code was
axially symmetric and aimed to integrate the coupled set of equations in
the context of stellar core collapse~\cite{evans86b}.

More recently, Wilson's formulation has been applied to the numerical study 
of the coalescence of binary neutron stars in general 
relativity~\cite{wilson95,wilson96,mathews99} (see Section 4.4.2). These
studies adopted an approximation scheme for the gravitational field, by 
imposing the simplifying condition that the three-geometry (the 3-metric 
$\gamma_{ij}$) is {\it conformally flat}. The line element, Eq.~(\ref{metric}), 
then reads
\begin{equation}
ds^2=-(\alpha^2-\beta_i\beta^i)dx^0 dx^0+
2\beta_i dx^i dx^0 + \phi^4\delta_{ij}dx^i dx^j.
\end{equation}
The curvature of the three metric is then described by a position dependent
conformal factor $\phi^4$ times a flat-space Kronecker delta. Therefore, in this
approximation scheme all radiation degrees of freedom are removed, while the field 
equations reduce to a set of five Poisson-like elliptic equations in flat spacetime 
for the lapse, the shift vector and the conformal factor. While in spherical 
symmetry this approach is no longer an approximation, being identical to Einstein's
theory, beyond spherical symmetry its quality degrades. In~\cite{kley98} it
was shown by means of numerical simulations of extremely relativistic
disks of dust that it has the same accuracy as the first post-Newtonian 
approximation. 

Wilson's formulation showed some limitations in handling situations involving 
ultrarelativistic flows ($W\gg 2$), as first pointed out by Centrella and 
Wilson~\cite{centrella84}. Norman and Winkler~\cite{norman86} performed a 
comprehensive numerical assessment of such formulation by means of special 
relativistic hydrodynamical simulations. Figure 1 reproduces a plot 
from~\cite{norman86} in which the relative error of the density compression 
ratio in the so-called relativistic shock reflection problem -- the heating 
of a cold gas which impacts at relativistic speeds with a solid wall and 
bounces back -- is displayed as a function of the Lorentz factor $W$ of the 
incoming gas. The source of the data is Ref.~\cite{centrella84}. This figure 
shows that for Lorentz factors of about 2 ($v\approx 0.86c$), the threshold of
the ultrarelativistic limit, the relative errors are between 5\% and 7\% 
(depending on the adiabatic exponent of the gas), showing a linear growth 
with $W$.


\begin{figure}[h]
\centerline{\psfig{figure=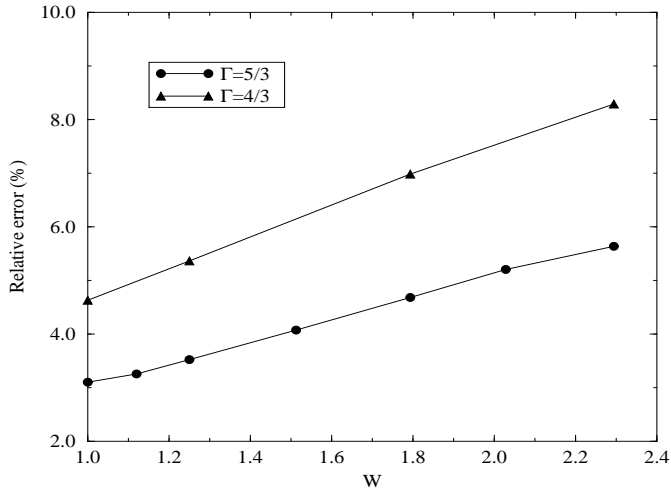,width=4.0in,height=3.0in}}
\caption{{ Results for the shock heating test of a cold,
relativistically inflowing gas against a wall using the explicit
Eulerian techniques of Centrella and Wilson~\cite{centrella84}.
Dependence of the relative errors of the density compression ratio 
versus the Lorentz factor $W$ for two different values of the adiabatic 
index of the flow, $\Gamma=4/3$ (triangles) and $\Gamma=5/3$ (circles) gases. 
The relative error is measured with respect to the average value of the
density over a region in the shocked material. The data are from Centrella 
and Wilson~\cite{centrella84} and the plot reproduces a similar one from 
Norman and Winkler~\cite{norman86}.
}}
\label{fig1}
\end{figure}

Norman and Winkler~\cite{norman86} concluded that those large errors 
were mainly due to the way in which the artificial viscosity terms are 
included in the numerical scheme in Wilson's formulation. These terms, 
commonly called $Q$ in the literature (see Section 3.1.1), are only added 
to the pressure terms in some cases, namely at the pressure gradient in 
the source of the momentum equation, Eq.~(\ref{40}), and at the divergence 
of the velocity in the source of the energy equation, Eq.~(\ref{42}). 
However,~\cite{norman86} proposed to add the $Q$ terms in a relativistically
consistent way, in order to consider the artificial viscosity as a real 
viscosity. Hence, the hydrodynamic equations should be rewritten for a 
modified stress-energy tensor of the following form:
\begin{equation}
T^{\mu \nu} = \rho (1+\varepsilon+(p+Q)/\rho) u^{\mu} u^{\nu} + 
(p+Q) g^{\mu \nu}.
\label{viscous}
\end{equation}
\noindent
In this way, for instance, the momentum equation takes the following form 
(in flat spacetime):
\begin{equation}
 \frac{\partial}{\partial x^0}[(\rho h+Q)W^2V_j]
+\frac{\partial}{\partial x^i}[(\rho h+Q)W^2V_jV^i] +
\frac{\partial(p+Q)}{\partial x^j} = 0.
\label{momdensvis}
\end{equation}
\noindent
In Wilson's original formulation $Q$ is omitted in the two terms containing 
the quantity $\rho h$. In general $Q$ is a non-linear function of the
velocity and, hence, the quantity $QW^2V$ in the momentum density of
Eq.~(\ref{momdensvis}) is a highly non-linear function of the velocity and 
its derivatives. This fact, together with the explicit presence of the Lorentz 
factor in the convective terms of the hydrodynamic equations, as well as the 
pressure in the specific enthalpy, make the relativistic equations much more 
coupled than their Newtonian counterparts. As a result Norman and Winkler 
proposed the use of implicit schemes as a way to describe more accurately 
such coupling. Their code, which in addition incorporates an adaptive grid, 
reproduces very accurate results even for ultra relativistic flows with 
Lorentz factors of about 10 in one-dimensional, flat spacetime, simulations.

Very recently, Anninos and Fragile~\cite{anninos02} have compared 
state-of-the-art artificial viscosity schemes and high-order non-oscillatory 
central schemes (see Section 3.1.3) using Wilson's formulation for the former 
class of schemes and a conservative formulation (similar to the one considered 
in~\cite{papadopoulos99b,papadopoulos99c}; Section 2.2.2) for the latter. 
These authors found, using a three-dimensional Cartesian code, that earlier 
results for artificial viscosity schemes in shock tube tests or shock reflection 
tests are not improved, i.e. the numerical solution becomes increasingly 
unstable for shock velocities greater than about $\sim 0.95c$. On the other 
hand, results for the shock reflection problem with a second-order finite 
difference central scheme show the suitability of such a scheme to handle 
ultrarelativistic flows, the underlying reason being, most likely, the use 
of a conservative formulation of the hydrodynamic equations rather than the 
particular scheme employed (see Section 3.1.3). Tests concerning spherical 
accretion on to a Schwarzschild black hole using both schemes yield the 
maximum relative errors near the event horizon, as large as $\sim 24$\% for 
the central scheme.

\subsubsection{3+1 conservative Eulerian formulation (Ib\'a\~nez and coworkers)}
\label{valencia}

In 1991, Mart\'{\i}, Ib\'a\~nez and Miralles~\cite{marti91} presented a new 
formulation of the (Eulerian) general relativistic hydrodynamic equations. 
This formulation was aimed to take fundamental advantage of the hyperbolic 
and conservative character of the equations, contrary to the one discussed
in the previous Section. From the numerical point of view, the hyperbolic and 
conservative nature of the relativistic Euler equations allows for the use 
of schemes based on the characteristic fields of the system, translating to 
relativistic hydrodynamics existing tools of classical fluid dynamics. This 
procedure departs from earlier approaches, most notably in avoiding the need 
for artificial dissipation terms to handle discontinuous 
solutions~\cite{wilson72,wilson79} as well as implicit schemes as proposed 
in~\cite{norman86}.

If a numerical scheme written in conservation form converges, it automatically 
guarantees the correct Rankine-Hugoniot (jump) conditions across discontinuities 
- the shock-capturing property (see, e.g.~\cite{leveque92}). Writing the relativistic 
hydrodynamic equations as a system of conservation laws, identifying the suitable 
vector of unknowns and building up an approximate Riemann solver permitted the 
extension of state-of-the-art {\it high-resolution shock-capturing} schemes (HRSC 
in the following) from classical fluid dynamics into the realm of 
relativity~\cite{marti91}.

Theoretical advances on the mathematical character of the relativistic 
hydrodynamic equations were first achieved studying the special relativistic
limit. In Minkowski spacetime, the hyperbolic character of relativistic 
hydrodynamics and magneto-hydrodynamics (MHD) was exhaustively
studied by Anile and collaborators (see~\cite{anile89} and references
therein) by applying Friedrichs' definition of
hyperbolicity~\cite{friedrichs74} to a quasi-linear form of the system
of hydrodynamic equations,
\begin{eqnarray}
{\cal A}^{\mu}({\bf w})\frac{\partial {\bf w}}
{\partial x^{\mu}} = 0,
\label{quasilinear}
\end{eqnarray}
where ${\cal A}^{\mu}$ are the Jacobian matrices of the system and 
${\bf w}$ is a suitable set of {\it primitive} (physical) variables (see below).
The system~(\ref{quasilinear}) is hyperbolic in the time direction defined
by the vector field $\xi$ with $\xi_{\mu}\xi^{\mu}=-1$, if the
following two conditions hold: (i) $\det({\cal A}^{\mu}\xi_{\mu})\ne 0$ and
(ii) for any $\zeta$ such that $\zeta_{\mu}\xi^{\mu}=0$, $\zeta_{\mu}
\zeta^{\mu}=1$, the eigenvalue problem ${\cal A}^{\mu}(\zeta_{\mu}-\lambda
\xi_{\mu}){\bf r}=0$ has only real eigenvalues $\{\lambda_n\}_{n=1,\cdots,5}$
and a complete set of right-eigenvectors $\{{\bf r}_n\}_{n=1,\cdots,5}$.
Besides verifying the hyperbolic character of the relativistic hydrodynamic
equations, Anile and collaborators~\cite{anile89} obtained the explicit
expressions for the eigenvalues and eigenvectors in the local rest frame,
characterized by $u^{\mu}=\delta_{0}^{\mu}$. In Font et al.~\cite{font94}
those calculations were extended to an arbitrary reference frame in which the
motion of the fluid is described by the 4-velocity $u^{\mu}=W(1,v^i)$.

The approach followed in~\cite{font94} for the equations of special relativistic 
hydrodynamics was extended to general relativity in~\cite{banyuls97}. The choice 
of evolved variables ({\it conserved quantities}) in the 3+1 Eulerian formulation 
developed by Banyuls et al.~\cite{banyuls97} differs slightly from that of Wilson's 
formulation~\cite{wilson72}. It comprises the rest-mass density ($D$), 
the momentum density  in the $j$-direction ($S_j$), and the total energy density 
($E$), measured by a family of observers which are the natural extension (for 
a generic spacetime) of the Eulerian observers in classical fluid dynamics. 
Interested readers are addressed to~\cite{banyuls97} for their definition and 
geometrical foundations.

In terms of the so-called {\it primitive variables} ${\bf w} = (\rho, v_{i},
\varepsilon)$, the conserved quantities are written as:
\begin{equation}
D =  \rho W,
\label{D}
\end{equation}
\begin{equation}
S_j  =  \rho h W^2 v_j,
\label{S}
\end{equation}
\begin{equation}
E  =  \rho h W^2 - p,
\label{E}
\end{equation}
\noindent
where the contravariant components $v^i = \gamma^{ij} v_j$ of
the three-velocity are defined as
\begin{equation}
v^i= \frac{u^i}{\alpha u^0} + \frac{\beta^i}{\alpha},
\end{equation}
\noindent
and $W$ is the relativistic Lorentz factor $W\equiv\alpha u^0=
(1-{v}^{2})^{-1/2}$ with ${v}^{2}= \gamma_{ij} v^i v^j$.

With this choice of variables the equations can be written in {\it 
conservation} form. Strict conservation is only possible in flat 
spacetime. For curved spacetimes there exist source terms, arising 
from the spacetime geometry. However, these terms do not contain 
derivatives of stress-energy tensor components. More precisely, the 
first-order flux-conservative hyperbolic system, well suited for 
numerical applications, reads:
\begin{equation}
\frac{1}{\sqrt{-g}} \left(
\frac {\partial \sqrt{\gamma}{\bf U}({\bf w})}
{\partial x^{0}} +
\frac {\partial \sqrt{-g}{\bf F}^{i}({\bf w})}
{\partial x^{i}} \right)
 = {\bf S}({\bf w}),
\label{F}
\end{equation}
\noindent
with $g\equiv \det(g_{\mu\nu})$ satisfying
$\sqrt{-g} = \alpha\sqrt{\gamma}$ with $\gamma\equiv \det(\gamma_{ij})$.
The state vector is given by
\begin{equation}
{\bf U}({\bf w})  =   (D, S_j, \tau),
\label{val1}
\end{equation}
\noindent
with $\tau \equiv  E - D$. The vector of fluxes is
\begin{equation}
{\bf F}^{i}({\bf w})  =   \left(D \left(v^{i}-\frac{\beta^i}{\alpha}\right),
 S_j \left(v^{i}-\frac{\beta^i}{\alpha}\right) + p \delta^i_j,
\tau \left(v^{i}-\frac{\beta^i}{\alpha}\right)+ p v^{i} \right),
\label{val2}
\end{equation}
\noindent
and the corresponding sources ${\bf S}({\bf w})$ are
\begin{equation}
{\bf S}({\bf w}) =  \left(0,
T^{\mu \nu} \left(
\frac {\partial g_{\nu j}}{\partial x^{\mu}} -
\Gamma^{\delta}_{\nu \mu} g_{\delta j} \right),
\alpha  \left(T^{\mu 0} \frac {\partial {\rm ln} \alpha}{\partial x^{\mu}} -
T^{\mu \nu} \Gamma^0_{\nu \mu} \right)
                     \right).
\label{val3}
\end{equation}

The local characteristic structure of the previous system of equations 
was presented in~\cite{banyuls97}. The eigenvalues (characteristic speeds) 
of the corresponding Jacobian matrices are all real (but not distinct, one 
showing a threefold degeneracy as a result of the assumed directional
splitting approach) and a complete set of right-eigenvectors
exists. System~(\ref{F}) satisfies, hence, the definition of hyperbolicity.
As it will become apparent in Section 3.1.2 below, the knowledge of the 
spectral information is essential in order to construct HRSC schemes based 
on Riemann solvers. This information can be found in~\cite{banyuls97} (see 
also~\cite{font99a}). 

The range of applications considered so far in general relativity employing 
the above formulation of the hydrodynamic equations, Eq.~(\ref{F})-(\ref{val3}),
is still small and mostly devoted to the study of stellar core collapse and
accretion flows on to black holes (see Sections 4.1.1 and 4.2 below). In the 
special relativistic limit this 
formulation is being successfully applied to simulate the evolution of (ultra-) 
relativistic extragalactic jets, using numerical models of increasing complexity 
(see, e.g.~\cite{marti97,aloy99}). The first applications in general 
relativity were performed, in one spatial dimension, in~\cite{marti91}, using 
a slightly different form of the equations. Preliminary investigations of 
gravitational stellar collapse were attempted by coupling the above formulation 
of the hydrodynamic equations to a hyperbolic formulation of the Einstein 
equations developed by~\cite{bona89}. These results are discussed 
in~\cite{marti91b,bona93}. More recently, successful evolutions of fully 
dynamical spacetimes in the context of adiabatic stellar core-collapse,
both in spherical symmetry and in axisymmetry, have been 
achieved~\cite{ibanez92,romero96,dimmelmeier02c}. These investigations are
considered in Section 4.1.1 below.

An ambitious three-dimensional, Eulerian code which evolves the coupled system 
of Einstein and hydrodynamics equations was developed by Font et al.~\cite{font99a} 
(see Section 3.3.2). The formulation of the hydrodynamic equations in this code 
follows the conservative Eulerian approach discussed in this section. The code 
is constructed for a completely general spacetime metric based on a Cartesian 
coordinate system, with arbitrarily specifiable lapse and shift conditions. 
In~\cite{font99a} the spectral decomposition (eigenvalues and right-eigenvectors) 
of the general relativistic hydrodynamic equations, valid for general spatial 
metrics, was derived, extending earlier results of~\cite{banyuls97} for 
non-diagonal metrics. A complete set of left-eigenvectors was presented by 
Ib\'a\~nez et al.~\cite{ibanez99b}. Due to the paramount importance of the 
characteristic structure of the equations in the design of upwind HRSC schemes 
based upon Riemann solvers, we summarize all necessary information in Section 5.2 
of the article. 

\subsection{Covariant approaches}

General (covariant) conservative formulations of the general relativistic 
hydrodynamic equations for ideal fluids, i.e., not restricted to spacelike 
foliations, have been reported in~\cite{eulderink95} and, more recently, 
in~\cite{papadopoulos99b,papadopoulos99c}. The {\em form invariance} of 
these approaches with respect to the nature of the spacetime foliation 
implies that existing work on highly specialized techniques for fluid 
dynamics (i.e. HRSC schemes) can be adopted straightforwardly. In the next 
two sections we describe the existing covariant formulations in some detail. 

\subsubsection{Eulderink and Mellema}

Eulderink and Mellema~\cite{eulderink93,eulderink95} first derived a covariant
formulation of the general relativistic hydrodynamic equations. As in the 
formulation discussed in Section~\ref{valencia}, these authors took special 
care of the conservative form of the system, with no derivatives of the dependent 
fluid variables appearing in the source terms.  Additionally, this formulation 
is strongly adapted to a particular numerical method based on a generalization of Roe's 
approximate Riemann solver. Such solver was first applied to the non-relativistic
Euler equations in \cite{roe81} and has been widely employed since in simulating
compressible flows in computational fluid dynamics. Furthermore, their procedure 
is specialized for a perfect fluid EOS, $p=(\Gamma-1)\rho\varepsilon$, $\Gamma$ 
being the (constant) adiabatic index of the fluid.

After the appropriate choice of the state vector variables, the conservation
laws, Eqs.~(\ref{initial1}) and~(\ref{initial2}), are re-written in
flux-conservative form. The flow variables are then expressed in terms of
a parameter vector $\omega$ as
\begin{eqnarray}
{\bf F}^{\alpha} = \left( \left[K-\frac{\Gamma}{\Gamma-1}\omega^4\right]
\omega^{\alpha},\omega^{\alpha}\omega^{\beta}+K\omega^4 g^{\alpha\beta}
\right),
\label{em1}
\end{eqnarray}
\noindent
where $\omega^{\alpha}\equiv Ku^{\alpha}$, $\omega^4\equiv K \frac{p}
{\rho h}$ and $K^2\equiv \sqrt{-g}\rho h = -g_{\alpha\beta}\omega^{\alpha}
\omega^{\beta}$. The vector ${\bf F}^0$ represents the state vector
(the unknowns), and each vector ${\bf F}^i$ is the corresponding flux in the
coordinate direction $x^i$.

Eulderink and Mellema computed the exact ``Roe matrix"~\cite{roe81} for
the vector~(\ref{em1}) and obtained the corresponding spectral decomposition.
The characteristic information is used to solve the system numerically
using Roe's generalized approximate Riemann solver. Roe's linearization can 
be expressed in terms of the average state $\tilde{\omega}=\frac{\omega_L + 
\omega_R}{K_L+K_R}$, where $L$ and $R$ denote the left and right states in a 
Riemann problem (see Section 3.1.2).  Further technical details can be found 
in~\cite{eulderink93,eulderink95}. 

The performance of this general relativistic Roe solver was tested in a
number of one-dimensional problems for which exact solutions are known,
including non-relativistic shock tubes, special relativistic shock tubes
and spherical accretion of dust and a perfect fluid onto a (static)
Schwarzschild black hole. In its special relativistic version it has
been used in the study of the confinement properties of relativistic 
jets~\cite{eulderink94}. However, no astrophysical applications in strong-field 
general relativistic flows have yet been attempted with this formulation.

\subsubsection{Papadopoulos and Font}
\label{papafont}

In this formulation~\cite{papadopoulos99b} the spatial velocity components of 
the 4-velocity, $u^{i}$, together with the rest-frame density and internal 
energy, $\rho$ and $\varepsilon$, provide a unique description of the state 
of the fluid at a given time and are taken as the {\em primitive} variables. 
They constitute a vector in a five dimensional space $ {\bf w} = (\rho, u^{i}, 
\varepsilon)$. The initial value problem for equations~(\ref{initial1}) and
(\ref{initial2}) is defined in terms of another vector in the same fluid state 
space, namely the {\em conserved variables}, ${\bf U}$, individually denoted 
$(D,S^{i},E)$:
\begin{eqnarray}
\label{eq:D}
D     & = & {\bf U}^{0} = J^{0}  = \rho u^{0} \, ,  \\
S^{i} & = & {\bf U}^{i} = T^{0i} = \rho h u^{0} u^{i} + p g^{0i} \, ,  \\
E     & = & {\bf U}^{4} = T^{00} = \rho h u^{0} u^{0} + p g^{00} \, .
\label{eq:E}
\end{eqnarray}
Note that the state vector variables slightly differ from previous choices 
(see, e.g., Eqs.~(\ref{wilson_vars}), (\ref{D}), (\ref{S}), (\ref{E}) and 
(\ref{em1})). With those definitions the equations of general relativistic 
hydrodynamics take the standard conservation law form,
\begin{equation}
\frac{\partial(\sqrt{-g} {\bf U}^{A})}{\partial x^0}
+ \frac{\partial(\sqrt{-g} {\bf F}^{j})}{\partial x^j}
= {\bf S} \, ,
\label{eq:cons-law}
\end{equation}
with $A=(0,i,4)$. The flux vectors ${\bf F}^{j}$ and the source terms ${\bf S}$ 
(which depend only on the metric, its derivatives and the undifferentiated 
stress energy tensor), are given by
\begin{eqnarray}
{\bf F}^{j} = (J^{j}, T^{ji}, T^{j0}) = 
(\rho u^{j},  \rho h u^{i} u^{j} + p g^{ij}, 
\rho h u^{0} u^{j} + p g^{0j}),
\label{eq:fluxes}
\end{eqnarray}
and
\begin{eqnarray}
{\bf S} = (0, - \sqrt{-g} \, \Gamma^{i}_{\mu\lambda} T^{\mu\lambda},
- \sqrt{-g} \, \Gamma^{0}_{\mu\lambda} T^{\mu\lambda}).
\label{pf:sources}
\end{eqnarray}

The state of the fluid is uniquely described using either vector of
variables, i.e., either ${\bf U}$ or ${\bf w}$, and each one can be
obtained from the other via the definitions~(\ref{eq:D})-(\ref{eq:E}) 
and the use of the normalization condition for the 4-velocity, $g_{\mu\nu} 
u^{\mu} u^{\nu} = -1$. The local characteristic structure of the above 
system of equations was presented in~\cite{papadopoulos99b}, where the 
formulation proved well suited for the numerical implementation of HRSC 
schemes. The formulation presented in this section has been 
developed for a perfect fluid EOS. Extensions to account for generic EOS
are given in~\cite{papadopoulos99c}. This reference further contains
a comprehensive analysis of general relativistic hydrodynamics
in conservation form. 

A technical remark must be included here: In all conservative formulations
discussed in Sections 2.1.3, 2.2.1 and 2.2.2, the time update of a given
numerical algorithm is applied to the conserved quantities ${\bf U}$. After 
this update the vector of primitive quantities ${\bf w}$ must be reevaluated,
as those are needed in the Riemann solver (see Section 3.1.2). The relation 
between the two sets of variables is, in general, not in closed form and, hence, 
the recovery of the primitive variables is done using a root-finding
procedure, typically a Newton-Raphson algorithm. This feature, distinctive of
the equations of (special and) general relativistic hydrodynamics -- it does
not exist in the Newtonian limit -- may lead in some cases to accuracy losses 
in regions of low density and small speeds, apart from being computationally 
inefficient. Specific details on this issue for each formulation of the 
equations can be found in Refs.~\cite{banyuls97,eulderink95,papadopoulos99b}. 
In particular, for the covariant formulation discussed in Section 2.2.1, there
exists an analytic method to determine the primitive variables, which is,
however, computationally very expensive since it involves many extra variables
and solving a quartic polynomial. Therefore, iterative methods are still
preferred~\cite{eulderink95}. On the other hand, we note that the covariant 
formulation discussed in this section, when applied to null spacetime foliations, 
allows for a simple and explicit recovery of the primitive variables, as a 
consequence of the particular form of the Bondi-Sachs metric.

\paragraph{Lightcone hydrodynamics:}
A comprehensive numerical study of the formulation of the hydrodynamic
equations discussed in this section was presented in~\cite{papadopoulos99b}, 
where it was applied to simulate one-dimensional relativistic flows on null 
(lightlike) spacetime foliations. The various demonstrations performed 
include standard shock tube tests in Minkowski spacetime, perfect fluid 
accretion onto a Schwarzschild black hole using ingoing null 
Eddington-Finkelstein coordinates, dynamical spacetime evolutions of 
relativistic polytropes (i.e., stellar models satisfying the so-called 
Tolman-Oppenheimer-Volkoff equations of hydrostatic equilibrium) sliced along 
the radial null cones, and accretion of self-gravitating matter onto a central 
black hole.

Procedures for integrating various forms of the hydrodynamic equations on null 
hypersurfaces are much less common than on spacelike (3+1) hypersurfaces. They 
have been presented before in~\cite{isaacson83} (see~\cite{bishop99} for a 
recent implementation). This approach is geared towards smooth isentropic flows. 
A Lagrangian method, limited to spherical symmetry, was developed 
by~\cite{miller89}. Recent work in~\cite{dubal98} includes an Eulerian, 
non-conservative, formulation for general fluids in null hypersurfaces and 
spherical symmetry, including their {\em matching} to a spacelike section.

The general formalism laid out in \cite{papadopoulos99b,papadopoulos99c} is 
currently being systematically applied to astrophysical problems of increasing 
complexity. Applications in spherical symmetry include the investigation of 
accreting dynamic black holes, which can be found 
in~\cite{papadopoulos99b,papadopoulos01a}. Studies of the gravitational collapse 
of supermassive stars are discussed in~\cite{linke01a} and studies of the 
interaction of scalar fields with relativistic stars are presented 
in~\cite{siebel02a}. Axisymmetric neutron star spacetimes have been considered in 
\cite{siebel02b}, as part of a broader program aimed at the study of relativistic 
stellar dynamics and gravitational collapse using characteristic numerical 
relativity. We note that there has been already a proof-of-principle demonstration 
of the inclusion of matter fields in three dimensions~\cite{bishop99}.

\subsection{Going further: non-ideal hydrodynamics}

Formulations of the equations of non-ideal hydrodynamics in general
relativity are also available in the literature. The term ``non-ideal"
accounts for additional physics such as viscosity, magnetic fields and 
radiation. These non-adiabatic effects can play a major role in some 
astrophysical systems as, e.g., accretion disks or relativistic stars.

The equations of viscous hydrodynamics, the Navier-Stokes-Fourier
equations, have been formulated in relativity in terms of causal
dissipative relativistic fluids (see the {\it Living Reviews} article
by M\"uller~\cite{mueller99} and references therein). These extended 
fluid theories, however, remain unexplored, numerically, in astrophysical 
systems. The reason may be the lack of an appropriate formulation 
well-suited for numerical studies. Work in this direction was done 
by Peitz and Appl~\cite{peitz99} who provided a 3+1 coordinate-free 
representation of different types of dissipative relativistic fluid 
theories which possess, in principle, the potentiality of being well 
adapted to numerical applications.

The inclusion of magnetic fields and the development of formulations
for the MHD equations, attractive to numerical studies, is still very 
limited in general relativity. Numerical approaches in special relativity 
are presented in~\cite{komissarov99,vanputten98,balsara01}. In particular, 
Komissarov~\cite{komissarov99}, and Balsara~\cite{balsara01} developed two 
different upwind HRSC (or Godunov-type) schemes, providing the characteristic 
information of the corresponding system of equations, and proposed a battery 
of tests to validate numerical MHD codes. 3+1 representations of relativistic MHD 
can be found in~\cite{sloan85,evans88}. In~\cite{yokosawa93} the transport 
of energy and angular momentum in magneto-hydrodynamical accretion onto a 
rotating black hole was studied adopting Wilson's formulation for the 
hydrodynamic equations (conveniently modified to account for the magnetic 
terms), and the magnetic induction equation was solved using the constrained 
transport method of~\cite{evans88}. Recently, Koide et al.~\cite{koide98,koide02a} 
performed the first MHD simulation, in general relativity, of magnetically 
driven relativistic jets from an accretion disk around a Schwarzschild black 
hole (see Section~\ref{jetformation}). These authors used a second-order finite 
difference central scheme with nonlinear dissipation developed by 
Davis~\cite{davis84}. Even though astrophysical applications of Godunov-type 
schemes (see Section 3.1.2) in general relativistic MHD are still absent, it 
is realistic to believe this situation may change in the near future.

The interaction between matter and radiation fields, present in different 
levels of complexity in all astrophysical systems, is described by the 
equations of radiation hydrodynamics. The Newtonian framework is highly 
developed (see, e.g.~\cite{mihalas84}; the special relativistic transfer
equation is also considered in that reference). Pons et al.~\cite{pons00b}
discuss a hyperbolic formulation of the radiative transfer equations, paying
particular attention to the closure relations and to extend HRSC schemes
to those equations. General relativistic formulations of radiative transfer 
in curved spacetimes are considered in, e.g.,~\cite{rezzolla94} 
and~\cite{zampieri96} (see also references therein). 

\section{Numerical schemes}
\label{methods}

We turn now to describe the numerical schemes, mainly those based on
finite differences, specifically designed to solve non-linear hyperbolic
systems of conservation laws. As discussed in the previous section, the
equations of general relativistic hydrodynamics fall in this category,
irrespective of the formulation. Even though we also consider schemes 
based on artificial viscosity techniques, the emphasis is given on the 
so-called high-resolution shock-capturing (HRSC) schemes (or Godunov-type 
methods), based on (either exact or approximate) solutions of local
Riemann problems using the characteristic structure of the equations. 
Such finite difference schemes (or, in general, finite volume schemes) 
have been the subject of diverse review articles and textbooks (see, 
e.g.,~\cite{leveque92,leveque98,toro97,ibanez99}). For this reason only 
the most relevant features will be covered here, addressing the 
reader to the appropriate literature for further details. In particular, 
an excellent introduction on the implementation of HRSC schemes in special 
relativistic hydrodynamics is presented in the {\it Living Reviews} article
by Mart\'{\i} and M\"uller~\cite{marti99}. Alternative techniques to finite 
differences, such as Smoothed Particle Hydrodynamics, (pseudo-) Spectral 
Methods and others, are briefly considered last.

\subsection{Finite difference schemes}
\label{fds}

Any system of equations presented in the previous section can be solved
numerically by replacing the partial derivatives by finite differences on 
a discrete numerical grid and then advancing the solution in time via some 
time-marching algorithm. Hence, specification of the state-vector ${\bf U}$ 
on an initial hypersurface, together with a suitable choice of EOS, followed 
by a recovery of the primitive variables, leads to the computation of the 
fluxes and source terms. Through this procedure the first time derivative of 
the data is obtained, which then leads to the formal propagation of the 
solution forward in time, with a time-step constrained by the 
Courant-Friedrichs-Lewy (CFL) condition.

The hydrodynamic equations (either in Newtonian physics or in general relativity)
constitute a non-linear hyperbolic system and, hence, smooth initial data can 
transform into discontinuous data (cross of characteristics in the case of shocks) 
in a finite time during the evolution. As a consequence, classical finite 
difference schemes (see e.g.~\cite{leveque92,toro97}) present important 
deficiencies when dealing with such systems. Typically, first order accurate 
schemes are much too dissipative across discontinuities (excessive smearing) and 
second order (or higher) schemes produce spurious oscillations near 
discontinuities which do not disappear as the grid is refined. To avoid these 
effects standard finite difference schemes have been conveniently modified in 
various ways which ensure high-order, oscillation-free accurate representations 
of discontinuous solutions, as we discuss next.

\subsubsection{Artificial viscosity approach}

The idea of modifying the hydrodynamic equations by introducing {\it artificial 
viscosity} terms to damp the amplitude of spurious oscillations near 
discontinuities was originally proposed by von Neumann and 
Richtmyer~\cite{vonneumann50} in the context of the (classical) Euler 
equations. The basic idea is to introduce a 
purely artificial dissipative mechanism whose form and strength are such 
that the shock transition becomes smooth, extending over a small number
of intervals $\Delta x$ of the space variable. 
In their original work von Neumann and Richtmyer proposed the following 
expression for the viscosity term:
\[ Q = \left\{ \begin{array}{cl}
                         -\alpha \frac{\partial v}{\partial x}
                         & \mbox{if
                            $\frac{\partial v}{\partial x} <0$ or 
                            $\frac{\partial \rho}{\partial t}>0$},\\
                        0  & \mbox{otherwise},
                        \end{array}
                  \right. \]
with $\alpha=\rho(k\Delta x)^2\frac{\partial v}{\partial x}$, $v$ being the
fluid velocity, $\rho$ the density, $\Delta x$ the spatial interval, and
$k$ a constant parameter whose value is conveniently adjusted in every
numerical experiment. This parameter controls the number of zones in which 
shock waves are spread.

This type of generic recipe, with minor modifications, has been used in
conjuction with standard finite difference schemes in all
numerical simulations employing May and White's formulation, mostly in the 
context of gravitational collapse, as well as Wilson's formulation. So, for 
example, in May and White's original code~\cite{may66} the artificial viscosity 
term, obtained in analogy with the one proposed by von Neumann and Richtmyer
\cite{vonneumann50}, is introduced in the equations accompanying the pressure, 
in the form:
\[ Q = \left\{ \begin{array}{cl}
                        \rho(\frac{a\Delta m}{R^{2}})^{2}
                        \frac{\partial R^{2}u}{\partial m}/\Gamma
                         & \mbox{if $\frac{\partial \rho}{\partial
                         t}>0$},\\
                        0  & \mbox{otherwise}.
                        \end{array}
                  \right. \]

Further examples of similar expressions for the artificial viscosity
terms, in the context of Wilson formulation, can be found in, 
e.g.,~\cite{wilson72,hawley84b}. A state-of-the-art formulation of the
artificial viscosity approach is reported in~\cite{anninos02}.

The main advantage of the artificial viscosity approach is its simplicity, 
which results in high computational efficiency. Experience has shown, however, 
that this procedure is both, problem dependent and inaccurate for ultra 
relativistic flows~\cite{norman86,anninos02}. Furthermore, the artificial 
viscosity approach has the inherent ambiguity of finding the appropriate 
form for $Q$ that introduces the necessary amount of dissipation to reduce
the spurious oscillations and, at the same time, avoids introducing 
excessive smearing in the discontinuities. In many instances both properties 
are difficult to achieve simultaneously. A comprehensive numerical study
of artificial viscosity induced errors in strong shock calculations in
Newtonian hydrodynamics (including also proposed improvements) was presented 
by Noh~\cite{noh87}.

\subsubsection{High-resolution shock-capturing (HRSC) upwind schemes}

In finite difference schemes, convergence properties under grid refinement must 
be enforced to ensure that the numerical results are correct (i.e., if a scheme 
with an order of accuracy $\alpha$ is used, the global error of the numerical 
solution has to tend to zero as ${\cal O}(\Delta x)^{\alpha}$ as the cell 
width $\Delta x$ tends to zero). For hyperbolic systems of conservation laws, 
schemes written in {\it conservation form} are preferred since, according to 
the Lax-Wendroff theorem~\cite{lax60}, they guarantee that the convergence, 
if it exists, is to one of the so-called {\it weak solutions} of the original 
system of equations. Such weak solutions are generalized solutions that 
satisfy the integral form of the conservation system. They are ${\cal C}^1$
classical solutions (continuous and differentiable) in regions where they are 
continuous and have a finite number of discontinuities. 

For the sake of simplicity let us consider in the following an initial value 
problem for a one-dimensional scalar hyperbolic conservation law,
\begin{equation}
\frac{\partial u}{\partial t} + \frac{\partial f(u)}{\partial x} = 0,
\hspace{1cm}
u(x,t=0)=u_0(x),
\label{scalar}
\end{equation}
and let us introduce a discrete numerical grid of space-time points $(x_j,t^n)$.
An explicit algorithm written in conservation form updates the solution from 
time $t^n$ to the next time level $t^{n+1}$ as:
\begin{equation}
u_j^{n+1} = u_j^n - \frac{\Delta t}{\Delta x} (
\hat{f}(u_{j-p}^n,u_{j-p+1}^n,\cdots,u_{j+q}^n) -
\hat{f}(u_{j-p-1}^n,u_{j-p}^n,\cdots,u_{j+q-1}^n) ),
\end{equation}
where $\hat{f}$ is a consistent numerical flux function (i.e.,
$\hat{f}(u,u,\cdots,u)=f(u)$) of $p+q+1$ arguments and $\Delta t$ and 
$\Delta x$ are the time-step and cell width respectively.  Furthermore,
$u_j^n$ is an approximation to the average of $u(x,t)$ within the 
numerical cell $[x_{j-1/2},x_{j+1/2}]$ $(x_{j\pm1/2}=(x_j+x_{j\pm1})/2)$:
\begin{equation}
u_j^n\approx \frac{1}{\Delta x}\displaystyle\int_{x_{j-1/2}}^{x_{j+1/2}}
u(x,t^n) dx.
\end{equation}

The class of all weak solutions is too wide in the sense that there is no 
uniqueness for the initial value problem. The numerical method should, hence, 
guarantee convergence to the {\it physically admissible solution}. This is the 
vanishing-viscosity limit solution, that is, the solution when $\eta\rightarrow 
0$, of the ``viscous version" of the scalar conservation law, Eq.~(\ref{scalar}):
\begin{equation}
\frac{\partial u}{\partial t} + \frac{\partial f(u)}{\partial x} =
\eta\frac{\partial^2u}{\partial x^2}.
\end{equation}
Mathematically, the solution one needs to search for is characterized by the 
so-called {\it entropy condition} (in the language of fluids, the condition that 
the entropy of any fluid element should increase when running into a discontinuity). 
The characterization of these {\it entropy-satisfying solutions} for scalar 
equations was given by Oleinik~\cite{oleinik57}. For hyperbolic systems of 
conservation laws it was developed by Lax~\cite{lax72}.

The Lax-Wendroff theorem~\cite{lax60} cited above does not establish whether the 
method converges. To guarantee convergence, some form of stability is required, as
Lax first proposed for linear problems ({\it Lax equivalence theorem}; see, e.g.,
~\cite{richtmyer67}). Along this direction, the notion of total-variation
stability has proven very successful although powerful results have only
been obtained for scalar conservation laws. The total variation of a solution
at time $t=t^n$, TV$(u^n)$, is defined as
\begin{equation}
\mbox{TV}(u^n)=\displaystyle\sum_{j=0}^{+\infty} |u_{j+1}^n-u_j^n|.
\end{equation}

A numerical scheme is said to be TV-stable if TV$(u^n)$ is bounded for all
$\Delta t$ at any time for each initial data. In the case of non-linear, scalar 
conservation laws it can be proved that TV-stability is a sufficient condition 
for convergence~\cite{leveque92}, as long as the numerical schemes are written
in conservation form and have consistent numerical flux functions. Current 
research has focused on the development of high-resolution numerical schemes 
in conservation form satisfying the condition of TV-stability, e.g., the 
so-called {\it total variation diminishing} (TVD) schemes~\cite{harten84}
(see below). 

Let us now consider the specific system of hydrodynamic equations as formulated in 
Eq.~(\ref{F}) and let us consider a single computational cell of our 
discrete spacetime. Let $\Omega$ be a region (simply connected) of a given 
four-dimensional manifold $\cal M$, bounded by a closed three-dimensional 
surface $\partial \Omega$. We further take the 3-surface $\partial\Omega$ 
as the standard-oriented hyper-parallelepiped made up of two spacelike
surfaces $\{\Sigma_{x^0}, \Sigma_{x^0+\Delta x^0}\}$ plus timelike surfaces
$\{\Sigma_{x^i}, \Sigma_{x^i+\Delta x^i}\}$ that join the two temporal
slices together. By integrating system~(\ref{F}) over a domain $\Omega$ 
of a given spacetime, the variation in time of the state vector ${\bf U}$ 
within $\Omega$ is given -- keeping apart the source terms -- by the fluxes 
${\bf F}^{i}$ through the boundary $\partial \Omega$. The integral form of 
system (\ref{F}) is
\begin{equation}
\int_{\Omega} \frac{1}{\sqrt{-g}} \frac {\partial \sqrt{\gamma}{\bf U}}
{\partial x^{0}} d\Omega +
\int_{\Omega} \frac{1}{\sqrt{-g}}
\frac {\partial \sqrt{-g}{\bf F}^{i}}{\partial x^{i}} d\Omega
 = \int_{\Omega} {\bf S} d\Omega,
\label{int}
\end{equation}
which can be written in the following conservation form, well-adapted
to numerical applications:
\begin{eqnarray}
& &
(\displaystyle{\overline{\bf U}} {\Delta V})_{x^0+\Delta x^0} -
(\displaystyle{\overline{\bf U}} {\Delta V})_{x^0}   =
\nonumber \\ 
&-& \left(
\displaystyle{\int_{\Sigma_{x^1+\Delta x^1}}}\sqrt{-g} {{\bf F}}^{1}
dx^0 dx^2 dx^3 -
\displaystyle{\int_{\Sigma_{x^1}}}\sqrt{-g} {{\bf F}}^{1}
dx^0 dx^2 dx^3 \right) 
\nonumber \\ &
-&\left(
\displaystyle{\int_{\Sigma_{x^2+\Delta x^2}}}\sqrt{-g} {{\bf F}}^{2}
dx^0 dx^1 dx^3 -
\displaystyle{\int_{\Sigma_{x^2}}}\sqrt{-g} {{\bf F}}^{2}
dx^0 dx^1 dx^3 \right) 
\nonumber \\ &
-&\left(
\displaystyle{\int_{\Sigma_{x^3+\Delta x^3}}}\sqrt{-g} {{\bf F}}^{3}
dx^0 dx^1 dx^2 -
\displaystyle{\int_{\Sigma_{x^3}}}\sqrt{-g} {{\bf F}}^{3}
dx^0 dx^1 dx^2 \right) 
\nonumber \\ &
+& \displaystyle{\int_{\Omega}} {\bf S} d\Omega,
\label{integ}
\end{eqnarray}
\noindent
where
\begin{equation}
\displaystyle{\overline{\bf U}} = \frac{1}{\Delta V}
\int_{\Delta V} \sqrt{\gamma} {\bf U} dx^1 dx^2 dx^3,
\end{equation}
\begin{equation}
{\Delta V} = \int^{x^1+\Delta x^1}_{x^1}
\int^{x^2+\Delta x^2}_{x^2}
\int^{x^3+\Delta x^3}_{x^3}
\sqrt{\gamma} dx^1 dx^2 dx^3.
\end{equation}

A numerical scheme written in conservation form ensures that, in the absence 
of sources, the (physically) conserved quantities, according to the partial 
differential equations, are numerically conserved by the finite difference 
equations. 

The computation of the time integrals of the interface fluxes appearing in 
Eq.~(\ref{integ}) is one of the distinctive features of upwind HRSC schemes.
One needs first to define the so-called numerical fluxes, which are recognized 
as approximations to the time-averaged fluxes across the cell-interfaces, which 
depend on the solution at those interfaces, ${\bf U}(x^j+\Delta x^j/2,x^0)$,
during a time step, 
\begin{equation}
\hat{{\bf F}}_{j+\frac{1}{2}}\approx \frac{1}{\Delta t}\int_{t^n}^{t^{n+1}}
{\bf F}({\bf U}(x^{j+1/2},x^0)).
\end{equation}

\begin{figure}[t]
\centerline{\psfig{figure=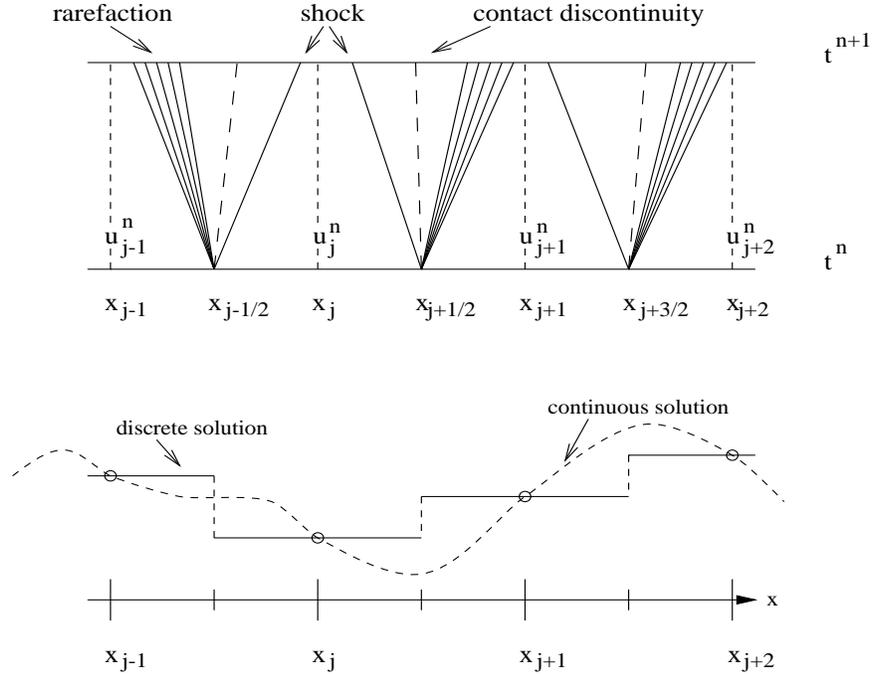,width=4.5in,height=3.5in}}
\caption{{ Godunov's scheme: local solutions of Riemann problems.
At every interface, $x_{j-\frac{1}{2}}$, $x_{j+\frac{1}{2}}$ and
$x_{j+\frac{3}{2}}$, a local Riemann problem is set up as a result of
the discretization process (bottom panel), when approximating the 
numerical solution by piecewise constant data. At time $t^n$ these 
discontinuities decay into three elementary waves which propagate the 
solution forward to the next time level $t^{n+1}$ (top panel). The time 
step of the numerical scheme must satisfy the Courant-Friedrichs-Lewy
condition, being small enough to prevent the waves from advancing
more than $\Delta x/2$ in $\Delta t$.
}}
\label{fig2}
\end{figure}

At the cell-interfaces the flow can be discontinuous 
and, following the seminal idea of Godunov~\cite{godunov59}, the numerical 
fluxes can be obtained by solving a collection of local Riemann problems, as
is depicted in Fig.~\ref{fig2}. This is the approach followed by the
so-called Godunov-type methods~\cite{harten83,einfeldt88}. Fig.~\ref{fig2}
shows how the continuous solution is locally 
averaged on the numerical grid, a process which leads to the appearance of 
discontinuities at the cell-interfaces. Physically, every discontinuity 
decays into three elementary waves: a shock wave, a rarefaction wave and a 
contact discontinuity. The complete structure of the Riemann problem can be 
solved analytically (see~\cite{godunov59} for the solution in Newtonian 
hydrodynamics and~\cite{marti94,pons00} in special relativistic hydrodynamics) 
and, accordingly, used to update the solution forward in time.


For reasons of numerical efficiency and, particularly in multidimensions,
the exact solution of the Riemann problem is frequently avoided and
linearized (approximate) Riemann solvers are preferred. These solvers
are based on the exact solution of Riemann problems corresponding to a 
linearized version of the original system of equations. After
extensive experimentation, the results achieved with approximate
Riemann solvers are comparable to those obtained with the exact solver 
(see~\cite{toro97} for a comprehensive overview of Godunov-type methods,
and~\cite{marti99} for an excellent summary of approximate Riemann solvers 
in special relativistic hydrodynamics).

In the frame of the local characteristic approach the numerical fluxes 
appearing in Eq.~(\ref{integ}) are computed according to some generic 
flux-formula which makes use of the characteristic information of the system. 
For example, in Roe's approximate Riemann solver~\cite{roe81} it adopts the 
following functional form:
\begin{eqnarray}
\hat{\bf F}_{j+\frac{1}{2}} = \frac{1}{2}
\left({\bf F}({\bf w}_{\rm R}) +
 {\bf F}({\bf w}_{\rm L}) - \sum_{n=1}^5 |\widetilde{\lambda}_n |
 \Delta \widetilde{\omega}_n
\widetilde{\bf r}_n \right),
\label{nflux}
\end{eqnarray}

\noindent
where ${\bf w}_{\rm L}$ and ${\bf w}_{\rm R}$ are the values of the
primitive variables at the left and right sides, respectively, of a given
cell interface. They are obtained from the cell-centered quantities after 
a suitable monotone reconstruction procedure. 

The way these variables are computed determines the spatial order of 
accuracy of the numerical algorithm and controls 
the amplitude of the local jumps at every cell interface. If these jumps 
are monotonically reduced, the scheme provides more accurate initial guesses
for the solution of the local Riemann problems (the average values give only
first-order accuracy). A wide variety of cell 
reconstruction procedures is available in the literature. Among the 
slope limiter procedures (see, e.g.,~\cite{toro97,leveque98})
most commonly used for TVD schemes~\cite{harten84} are the second order, 
piecewise linear reconstruction, introduced by van Leer~\cite{vanleer79} in 
the design of the MUSCL scheme (Monotonic Upstream Scheme for Conservation Laws), 
and the third order, piecewise parabolic reconstruction developed by Colella 
and Woodward~\cite{colella84} in their Piecewise Parabolic Method (PPM). 
Since TVD schemes are only first-order accurate at local extrema, alternative
reconstruction procedures where some growth of the total variation is
allowed have also been developed. Among those we mention the {\it total variation
bounded} (TVB) schemes~\cite{shu87} and the {\it essentially non-oscillatory}
(ENO) schemes~\cite{harten87}.

Alternatively, high-order methods for non-linear hyperbolic systems have also
been designed using flux limiters rather than slope limiters, as in
the FCT scheme of Boris and Book~\cite{boris73}. In this approach, 
the numerical flux consists of two pieces, a high-order flux (e.g. the 
Lax-Wendroff flux) for smooth regions of the flow, and a low-order flux 
(e.g. the flux from some monotone method) near discontinuities, $\hat{\bf F} = 
\hat{\bf F}_h - (1-\Phi) (\hat{\bf F}_h-\hat{\bf F}_l)$ with the limiter 
$\Phi\in[0,1]$ (see~\cite{toro97,leveque98} for further details).

The last term in the flux-formula, Eq.~(\ref{nflux}), represents the numerical 
viscosity of the scheme, and it makes explicit use of the characteristic information
of the Jacobian matrices of the system. This information is used to 
provide the appropriate amount of numerical dissipation to obtain accurate
representations of discontinuous solutions without excessive smearing,
avoiding, at the same time, the growth of spurious numerical oscillations 
associated with the Gibbs phenomenon. In Eq.~(\ref{nflux}), $\{\widetilde{\lambda}_n, 
\widetilde{\bf r}_n\}_{n=1\dots 5}$ are, respectively, the eigenvalues and 
right-eigenvectors of the Jacobian matrix of the flux vector. Correspondingly, 
the quantities $\{\Delta \widetilde{\omega}_n\}_{n=1\dots 5}$ are the jumps of 
the so-called characteristic variables across each characteristic field. They are 
obtained by projecting the jumps of the state-vector variables with the 
left-eigenvectors matrix:
\begin{equation}
{\bf U}({\bf w}_{\rm R})-{\bf U}({\bf w}_{\rm L}) =
\sum_{n=1}^5 \Delta \widetilde{\omega}_n \widetilde{\bf r}_n.
\label{charvar}
\end{equation}
\noindent
The ``tilde" in Eqs.~(\ref{nflux}) and~(\ref{charvar}) indicates that the 
corresponding fields are averaged at the cell interfaces from the left and 
right (reconstructed) values.

During the last few years most of the standard Riemann solvers developed 
in Newtonian fluid dynamics have been extended to relativistic hydrodynamics: 
Eulderink~\cite{eulderink95}, as discussed in Section 2.2.1, explicitly derived 
a relativistic Roe's Riemann solver~\cite{roe81}, Schneider et al.~\cite{schneider93}
carried out the extension of Harten, Lax, van Leer and Einfeldt's (HLLE)
method~\cite{harten83,einfeldt88}, Mart\'{\i} and M\"uller~\cite{marti96}
extended the PPM method of Woodward and Colella~\cite{woodward84}, Wen et 
al~\cite{wen97} extended Glimm's exact Riemann solver, Dolezal and
Wong~\cite{dolezal95} put into practice Shu-Osher ENO techniques,
Balsara~\cite{balsara94} extended Colella's two-shock approximation
and Donat et al.~\cite{donat98} extended Marquina's method~\cite{donat96}. 
Recently, much effort has been spent concerning the exact special 
relativistic Riemann solver and its extension to multidimensions
\cite{marti94,pons00,rezzolla00,zanotti02}. The interested reader is addressed 
to~\cite{marti99} and references therein for a comprehensive description of 
such solvers in special relativistic hydrodynamics.

\subsubsection{High-order central schemes}

The use of high-order non-oscillatory symmetric (central) TVD schemes for
solving hyperbolic systems of conservation laws, emerged at the mid 1980s
\cite{davis84,roe84,yee87,tadmor90} (see also \cite{yee89} and \cite{toro97} 
and references therein) as an alternative approach to upwind HRSC schemes.
Symmetric schemes are based on either high-order schemes (e.g. Lax-Wendroff) with 
additional dissipative terms \cite{davis84,roe84,yee87}, or on non-oscillatory 
high-order extensions of first-order central schemes (e.g. Lax-Friedrichs) 
\cite{tadmor90}. One of the nicest properties of central schemes is that they 
exploit the conservation form of the Lax-Wendroff or Lax-Friedrichs schemes. 
Therefore, they yield the correct propagation speeds of all nonlinear waves 
appearing in the solution. Furthermore, central schemes sidestep the use of 
Riemann solvers, which results in enhanced computational efficiency, especially 
in multidimensional problems. Its use is, thus, not limited to those systems 
of equations where the characteristic information is explicitly known or to 
systems where the solution of the Riemann problem is prohibitively expensive 
to compute. This approach has gradually developed during the last decade to 
reach a mature status where a number of straightforward central schemes of 
high order can be applied to any nonlinear hyperbolic system of conservation 
laws. The typical results obtained for the Euler equations show a 
quality comparable to that of upwind HRSC schemes, at the expense of a small loss 
of sharpness of the solution at discontinuities \cite{toro97}. An up-to-date 
summary of the status and applications of this approach is discussed in 
\cite{toro97,kurganov00,tadmorweb}.

In the context of special and general relativistic MHD, Koide et al 
\cite{koide98,koide02a} applied a second-order central scheme with nonlinear 
dissipation developed by \cite{davis84} to the study of black hole accretion 
and formation of relativistic jets. One-dimensional test simulations in 
special relativistic hydrodynamics performed by the author and coworkers 
\cite{font02c} using the SLIC (slope limiter centred) scheme (see \cite{toro97} 
for details) showed its capabilities to yield satisfactory results, comparable 
to those obtained by HRSC schemes based on Riemann solvers, even well inside 
the ultrarelativistic regime. The slopes of the original central scheme were 
limited using high-order reconstruction procedures such as PPM \cite{colella84}, 
which was essential to keep the inherent diffusion of central schemes at 
discontinuities at reasonable levels. Very recently, Del Zanna and 
Bucciantini~\cite{delzanna02a} assessed a third-order convex essentially 
non-oscillatory central scheme in multidimensional special relativistic 
hydrodynamics. Again, these authors obtained results as accurate as those 
of upwind HRSC schemes in standard tests (shock tubes, shock reflection test).
Yet another central scheme has been assessed by \cite{anninos02} in one-dimensional 
special and general relativistic hydrodynamics, where similar results to those 
of \cite{delzanna02a} are reported. These authors also validate their central 
scheme in simulations of spherical accretion on to a Schwarzschild black hole, 
and further provide a comparison with an artificial viscosity based scheme.

It is worth to underline that early pioneer approaches in the field of 
relativistic hydrodynamics \cite{norman86,centrella84} used standard 
finite difference schemes with artificial viscosity terms to stabilize 
the solution at discontinuities. However, as discussed in Section 2.1.2, 
those approaches only succeeded in obtaining accurate results for moderate 
values of the Lorentz factor ($W\sim 2$). A key feature lacking in those 
investigations was the use of a conservative approach for both, the system 
of equations and the numerical schemes. A posteriori, and to the light of 
the results reported by \cite{delzanna02a,anninos02,font02c}, it appears 
that this was the ultimate reason preventing the extension of the early 
computations to the ultrarelativistic regime.

The alternative of using high-order central schemes instead of upwind HRSC schemes 
becomes apparent when the spectral decomposition of the hyperbolic system under 
study is not known. The straightforwardness of a central scheme makes its use very 
appealing, especially in multi-dimensions where computational efficiency is an 
issue. Perhaps the most important example in relativistic astrophysics is the system 
of (general) relativistic MHD equations. Despite some recent progress has been 
accomplished in recent years (see, e.g.,~\cite{balsara01,komissarov99}), much more 
work is needed concerning their solution with HRSC schemes based upon Riemann solvers. 
Meanwhile, an obvious choice is the use of central schemes \cite{koide98,koide02a}. 

\subsubsection{Source terms}

Most ``conservation laws" involve source terms on the right hand side of the
equations. In hydrodynamics, for instance, those terms arise when considering 
external forces such as gravity, which make the right hand side of the
momentum and energy equations no longer zero (see Section 2). Other effects
leading to the appearance of source terms are radiative heat transfer (accounted
for in the energy equation) or ionization (resulting in a collection of
non-homogeneous continuity equations for the mass of each species, which is
not conserved separately). The incorporation of the source terms in the solution
procedure is a common issue to all numerical schemes considered so far. Since
a detailed discussion on the numerical treatment of source terms is beyond
the scope of this article, we simply provide some basic information in
this section, addressing the interested reader to~\cite{toro97,leveque98}
and references therein for further details.

There are, essentially, two basic ways of handling source terms. The first
approach is based on {\it unsplit methods} by which a single finite difference 
formula advances the entire equation over one time step (as in Eq.~(\ref{integ})):
\begin{eqnarray}
{\bf U}_j^{n+1} = {\bf U}_j^n - \frac{\Delta t}{\Delta x}
\left(\hat{\bf F}_{j+\frac{1}{2}} - \hat{\bf F}_{j-\frac{1}{2}}\right) +
\Delta t \,\, {\bf S}_j^n.
\end{eqnarray}

\noindent
The temporal order of this basic algorithm can be improved by introducing successive 
sub-steps to perform the time update (e.g. predictor-corrector, Shu \& Osher's 
conservative high order Runge-Kutta schemes, etc.)

Correspondingly, the second approach is based on
{\it fractional step or splitting methods}. The basic idea is to split the equation 
into different pieces (transport + sources) and apply appropriate methods for
each piece independently. In the first-order Godunov splitting,
${\bf U}^{n+1} = {\cal L}_s^{\Delta t} {\cal L}_f^{\Delta t} {\bf U}^n$,
the operator ${\cal L}_f^{\Delta t}$ solves the homogeneous PDE in the first
step to yield the intermediate value ${\bf U}^*$. Then, in the second step, the
operator ${\cal L}_s^{\Delta t}$ solves the ordinary differential equation
$\partial_t{\bf U}^*={\bf S}({\bf U}^*)$ to yield the final value ${\bf U}^{n+1}$.
In order to achieve second-order accuracy (assuming each independent method is
second order) a common fractional step method is the {\it Strang splitting},
where ${\bf U}^{n+1} = {\cal L}_s^{\Delta t/2} {\cal L}_f^{\Delta t}
{\cal L}_s^{\Delta t/2} {\bf U}^n$. Therefore, this method advances half
time step the solution for the ODE containing the source terms, and a
full time step the conservation law (the PDE) in between each source step.

We note that in some cases the source terms may become {\it stiff}, as in
phenomena which either occur on much faster time scales than the hydrodynamic 
time-step $\Delta t$, or act over much smaller spatial scales than the grid 
resolution $\Delta x$. Stiff source terms may easily lead to numerical 
difficulties. The interested reader is addressed to~\cite{leveque98} (and 
references therein) for further information on various approaches to overcome
the problems of stiff source terms.

\subsection{Other techniques}

Two of the most frequently employed alternatives to finite difference schemes 
in numerical hydrodynamics are Smoothed Particle Hydrodynamics (SPH) and, 
to a lesser extent, (pseudo-) Spectral Methods. This section contains
a brief description of both approaches. In addition, we also mention
the Field-Dependent Variation method and the Finite Element method. We note,
however, that both of these approaches have barely been used so far in 
relativistic hydrodynamics.

\subsubsection{Smoothed particle hydrodynamics}

The Lagrangian particle method SPH, derived independently by Lucy~\cite{lucy77} 
and Gingold and Monaghan~\cite{gingold77}, has shown successful performance
to model fluid flows in astrophysics and cosmology. Most studies
to date consider Newtonian flows and gravity, enhanced with the inclusion 
of the fluid self-gravity. 

In the SPH method a finite set of extended Lagrangian particles replaces
the continuum of hydrodynamical variables, the finite extent of the
particles being determined by a smoothing function (the kernel)
containing a characteristic length scale $h$. The main advantage of
this method is that it does not require a computational grid, avoiding
mesh tangling and distortion. Hence, compared to grid-based finite
volume methods, SPH avoids wasting computational power in multidimensional 
applications, when, e.g., modelling regions containing large voids. Experience 
in Newtonian hydrodynamics shows that SPH produces very accurate
results with a small number of particles ($\approx 10^3$ or even less).
However, if more than $10^4$ particles have to be used for certain problems,
and self-gravity  has to be included, the computational power, which grows
as the square of the number of particles, may exceed the capabilities of 
current supercomputers. Among the limitations of SPH we note the
difficulties in modelling systems with extremely different characteristic lengths
and the fact that boundary conditions usually require a more involved 
treatment than in finite volume schemes.

Reviews of the classical SPH equations are abundant in the literature (see, 
e.g.,~\cite{monaghan92,mueller98} and references therein). The reader is addressed 
to~\cite{mueller98} for a summary of comparisons between SPH and HRSC methods.

Recently, implementations of SPH to handle
(special) relativistic (and even ultra relativistic) flows have been
developed (see, e.g.,~\cite{chow97} and references therein). However,
SPH has been scarcely applied to simulate relativistic flows in curved 
spacetimes. Relevant references include~\cite{kheyfets90,laguna93a,
laguna93b,siegler99}.

Following~\cite{laguna93a} let us describe the implementation of an SPH
scheme in general relativity. Given a function $f({\bf x})$, its mean 
{\it smoothed} value $\langle f({\bf x})\rangle , ({\bf x}=(x,y,z))$ can be 
obtained from
\begin{eqnarray}
\langle f({\bf x})\rangle
\equiv\int W({\bf x},{\bf x'};h) f({\bf x'}) \sqrt{g'} d^3 x',
\end{eqnarray}
where $W$ is the smoothing kernel, $h$ the smoothing length and $\sqrt{g'} d^3 x'$ the
volume element. The kernel must be differentiable at least once, and the derivative should
be continuous to avoid large fluctuations in the force felt by a particle. Additional
considerations for an appropriate election of the smoothing kernel can be
found in~\cite{gingold82}. The kernel is required to satisfy a normalization condition
\begin{eqnarray}
\int W({\bf x},{\bf x'};h) \sqrt{g'} d^3 x' = 1,
\end{eqnarray}
which is assured by choosing $W({\bf x},{\bf x'};h)=\xi({\bf x})\Omega(v)$,
with $v=|{\bf x}-{\bf x'}| /h$, $\xi({\bf x})$ a normalization function and 
$\Omega(v)$ a standard spherical kernel.

The smooth approximation of gradients of scalar functions can be written as
\begin{eqnarray}
\langle \nabla f({\bf x})\rangle = \nabla \langle f({\bf x})\rangle -
\langle f({\bf x})\rangle \nabla \ln\xi({\bf x}),
\end{eqnarray}
and that corresponding to the divergence of a vector reads
\begin{eqnarray}
\langle \nabla \cdot {\bf A}({\bf x}) \rangle =
\nabla \cdot \langle {\bf A}({\bf x}) \rangle -
\langle {\bf A}({\bf x}) \rangle \cdot \ln\xi({\bf x}).
\end{eqnarray}

Discrete representations of these procedures are obtained
after introducing the number density distribution of particles
$\langle n({\bf x}) \rangle \equiv \sum_{a=1}^{N} 
\delta({\bf x}- {\bf x}_a)/\sqrt{g}$, with 
$\{ {\bf x}_a \}_{a=1,\dots ,N}$ the collection of $N$-particles 
where the functions are known. The previous representations then read: 
\begin{eqnarray}
\langle f({\bf x}_a)\rangle &=& \xi({\bf x}_a)
\sum_{b=1}^{N} \frac{f({\bf x}_b)}{\langle n({\bf x}_b) \rangle}
\Omega_{ab},
\\
\langle \nabla f({\bf x}_a)\rangle &=& \xi({\bf x}_a)
\sum_{b=1}^{N} \frac{f({\bf x}_b)}{\langle n({\bf x}_b) \rangle}
\nabla_{{\bf x}_a} \Omega_{ab},
\\
\langle \nabla \cdot {\bf A}({\bf x}_a) \rangle &=&
\xi({\bf x}_a)
\sum_{b=1}^{N} \frac{{\bf A}({\bf x}_b)}{\langle n({\bf x}_b) \rangle}
\cdot \nabla_{{\bf x}_a} \Omega_{ab},
\end{eqnarray}
with $\Omega_{ab}\equiv\Omega({\bf x}_a,{\bf x}_b;h)$. These approximations
can then be used to derive the SPH version of the general relativistic
hydrodynamic equations. Explicit formulae are reported in~\cite{laguna93a}.
The time evolution of the final system of ODEs is performed 
in~\cite{laguna93a} using a second-order Runge-Kutta time integrator
with adaptive time step. As in non-Riemann solver based finite volume
schemes, in SPH simulations involving the presence of shock waves, artificial 
viscosity terms must be introduced as a viscous pressure term~\cite{mann91}.

Recently, Siegler and Riffert~\cite{siegler99} have developed a Lagrangian 
conservation form of the general relativistic hydrodynamic equations for 
perfect fluids, with artificial viscosity, in arbitrary background spacetimes. 
Within that formulation these authors~\cite{siegler99} have built a general 
relativistic SPH code using the {\it standard} SPH formalism as known from 
Newtonian fluid dynamics (in contrast to previous approaches, 
e.g.,~\cite{mann91,kheyfets90,laguna93a}). The conservative character of their 
scheme has allowed the modelling of ultrarelativistic flows including shocks 
with Lorentz factors as large as 1000.

\subsubsection{Spectral methods}

The basic principle underlying spectral methods consists of transforming
the partial differential equations into a system of ordinary differential
equations by means of expanding the solution in a series on a
complete basis. The mathematical theory of these schemes is presented
in~\cite{gottlieb77,canuto88}. Spectral methods are particularly well
suited to the solution of elliptic and parabolic equations. Good results can 
also be obtained for hyperbolic equations as long as no discontinuities appear in
the solution. When a discontinuity is present some amount of artificial
viscosity must be added to avoid spurious oscillations. In the specific case of
relativistic problems, where coupled systems of elliptic equations (i.e.,
the Einstein constraint equations) and hyperbolic equations (i.e., 
hydrodynamics) must be solved, an interesting strategy is to use 
spectral methods to solve the elliptic equations and HRSC schemes for the 
hyperbolic ones. Using such combined methods first results have been obtained 
in one-dimensional supernova collapse simulations, both in the framework of 
a tensor-scalar theory of gravitation~\cite{novak98,novak99} and in General 
Relativity \cite{novak01}. 

Following~\cite{bonazzola99} we illustrate the main ideas of spectral 
methods considering the quasi-linear one-dimensional scalar equation:
\begin{eqnarray}
\frac{\partial u}{\partial t} =
\frac{\partial^2 u}{\partial x^2} + \lambda u 
\frac{\partial u}{\partial x},
\hspace{1cm}
t\ge 0,
\hspace{1cm}
x\in[0,1],
\label{burgers}
\end{eqnarray}
with $u=u(t,x)$ and $\lambda$ a constant parameter. In the linear case
($\lambda=0$), and assuming the function $u$ to be periodic, spectral 
methods expand the function in a Fourier series:
\begin{eqnarray}
u(x,t)=\sum_{k=0}^{\infty}[a_k(t)\cos(2\pi kx) +
                           b_k(t)\sin(2\pi kx)].
\label{sm1}
\end{eqnarray}
From the numerical point of view, the series is truncated for a suitable
value of $k$. Hence, Eq.~(\ref{burgers}), with $\lambda=0$, can be 
rewritten as
\begin{eqnarray}
\frac{da_k}{dt}=-k^2a_k(t),
\hspace{1cm}
\frac{db_k}{dt}=-k^2b_k(t).
\end{eqnarray}
Finding a solution of the original equation is then equivalent to solve an
``infinite" system of ordinary differential equations, where the initial
values of the coefficients $a_k$ and $b_k$ are given by the Fourier expansion
of $u(x,0)$.

In the non-linear case, $\lambda\ne 0$, spectral methods proceed in a more 
convoluted way: first, the derivative of $u$ is computed in the Fourier space. 
Then, a step back to the configuration space is taken through an inverse 
Fourier transform. Finally, after multiplying $\partial u/\partial x$ by 
$u$ in the configuration space the scheme returns again to the Fourier space.

The particular set of trigonometric functions used for the expansion in
Eq.~(\ref{sm1}) is chosen because it automatically fulfills the boundary 
conditions, and because a fast transform algorithm is available (the latter
is no longer an issue for today's computers). If the initial or boundary 
conditions are not periodic, Fourier expansion is no longer useful because 
of the presence of a Gibbs phenomenon at the boundaries of the interval. 
Legendre or Chebyshev polynomials are, in this case, the most common base 
of functions used in the expansions (see~\cite{gottlieb77, canuto88} for 
a discussion on the different expansions).

Extensive numerical applications using (pseudo-) spectral methods have been 
undertaken by the LUTH Relativity Group at the Observatoire de Paris in Meudon.
The group is focused in the study of compact objects, as well as the associated 
violent phenomena, gravitational collapse and supernova explosion. They have 
developed a fully object-oriented library (based on the C++ computer language) 
called LORENE \cite{lorene} to implement (multi domain) spectral methods within 
spherical coordinates. A comprehensive summary of applications in general 
relativistic astrophysics is presented in~\cite{bonazzola99}. The most recent 
ones deal with the computation of quasi-equilibrium configurations of either 
synchronized or irrotational binary neutron stars in General Relativity 
\cite{gourgoulhon01a,gourgoulhon01b,taniguchi02a}. Such initial data are currently 
being used by fully relativistic, finite difference time-dependent codes 
(see Section 3.3.2) to simulate the coalescence of binary neutron stars. 

\subsubsection{Flow field-dependent variation method}

Richardson and Chung \cite{richardson02} have recently proposed the flow 
field-dependent variation (FDV) method as a new approach for general 
relativistic (non ideal) hydrodynamics computations. In the FDV method, 
parameters characteristic of the flow field are computed in order to guide 
the numerical scheme toward a solution. The basic idea is to expand the 
conservation flow variables into a Taylor series in terms of the FDV 
parameters, which are related to variations of physical parameters such as
the Lorentz factor, the relativistic Reynolds number and the relativistic
Froude number.

The general relativistic hydrodynamic equations are expanded in a special form
of the Taylor series:
\begin{eqnarray}
{\bf U}^{n+1}&=&{\bf U}^n + \Delta t \frac{\partial {\bf U}^{n+s_a}}{\partial t}
+ \frac{\Delta t^2}{2}\frac{\partial^2 {\bf U}^{n+s_b}}{\partial t^2} +
{\cal O}(\Delta t^3),
\\
\frac{\partial {\bf U}^{n+s_a}}{\partial t} &=& \frac{\partial {\bf U}^n}{\partial t}
+ s_a \frac{\partial\Delta {\bf U}^{n+1}}{\partial t},
\hspace{0.5cm} 0<s_a<1,
\\
\frac{\partial^2 {\bf U}^{n+s_b}}{\partial t^2} &=&
\frac{\partial^2 {\bf U}^n}{\partial t^2} +
s_b \frac{\partial^2 \Delta {\bf U}^{n+1}}{\partial t^2},
\hspace{0.5cm} 0<s_b<1,
\end{eqnarray}
with $s_a$ and $s_b$ denoting the first-order and second-order variation parameters.
using the above expressions, the time update then reads:
\begin{eqnarray}
{\bf U}^{n+1} = {\bf U}^n &+& \Delta t 
\left(\frac{\partial {\bf U}^n}
{\partial t} + s_a \frac{\partial \Delta {\bf U}^{n+1}}{\partial t} \right)
\nonumber
\\
&+& \frac{\Delta t^2}{2}
\left(\frac{\partial^2 {\bf U}^n}
{\partial t^2} + s_b \frac{\partial^2 \Delta {\bf U}^{n+1}}{\partial t^2} \right)
+ {\cal O}(\Delta t^3).
\end{eqnarray}
Combining the conservation form of the equations and the rearranged Taylor series
expansion allows to rewrite the time update into standard matrix (residual) form,
which can then be discretized using either standard finite difference or finite element
methods \cite{richardson02}.

The physical interpretation of the coefficients $s_a$ and $s_b$ is the foundation 
of the FDV method. The first-order parameter $s_a$ is proportional to ${\bf a}_i 
\partial {\bf U}^{n+1}/ \partial x_i$, where ${\bf a}_i$ is the convection 
Jacobian representing the change of convective motion. If the Lorentz factor 
remains constant in space and time then $s_a=0$. However, if the Lorentz factor 
between adjacent zones is large, $s_a=1$. Similar assessments in terms of the 
Reynolds number can be provided for the diffusion and diffusion gradients, while 
the Froude number is used for the source term Jacobian $\partial {\bf S}/\partial 
{\bf U}$. Correspondingly, the second-order FDV parameters $s_b$ are chosen to be 
exponentially proportional to the first-order ones.

Obviously, the main drawback of the FDV method is the dependence of the solution 
procedure on a large number of problem-dependent parameters, a drawback shared 
to some extent by artificial viscosity schemes. Richardson and Chung 
\cite{richardson02} have implemented the FDV method in a C++ code called GRAFSS 
(General Relativistic Astrophysical Flow and Shock Solver). The test problems 
they report are the special relativistic shock tube (problem 1 in the notation 
of \cite{marti96}) and the Bondi accretion on to a Schwarzschild black hole. 
While their method yields the correct wave propagation, the numerical solution 
near discontinuities has considerably more diffusion than with upwind HRSC schemes. 
Nevertheless, the generality of the approach is worth considering. Applications 
to non ideal hydrodynamics and relativistic MHD are in progress \cite{richardson02}.

\subsubsection{Going further}

The finite element method is the preferred choice to solve multidimensional
structural engineering problems since the late 1960s~\cite{fem}. First steps 
in the development of the finite element method for modeling astrophysical 
flows in general relativity are discussed by Meier~\cite{meier99a}. The method, 
designed in a fully covariant manner, is valid not only for the hydrodynamic 
equations but also for the Einstein and electromagnetic field equations. The 
most unique aspect of the approach presented in~\cite{meier99a} is that the 
grid can be arbitrarily extended into the time dimension. Therefore, the 
standard finite element method generalizes to a four-dimensional covariant
technique on a spacetime grid, with the engineer's ``isoparametric
transformation" becoming the generalized Lorentz transform. At present,
the method has shown its suitability to accurately compute the 
equilibrium stellar structure of Newtonian polytropes, either spherical
or rotating. The main limitation of the finite element method, as Meier 
points out~\cite{meier99a}, is that it is generally fully implicit, which
results in severe computer time and memory limitations for three- and 
four-dimensional problems.

\subsection{State-of-the-art three-dimensional codes}
\label{codes}

The most advanced time-dependent, finite-difference, three-dimensional Cartesian 
codes to solve the system of Einstein and hydrodynamics equations are those 
developed by Shibata \cite{shibata99} and by the Washington University/NCSA/AEI-Golm 
Numerical Relativity collaboration \cite{font99a,font02a}. The main difference 
between both codes lies in the numerical methods used to solve the relativistic 
hydrodynamic equations, artificial viscosity in Shibata's code \cite{shibata99} 
and upwind HRSC schemes in the code of \cite{font99a,font02a}. We note, however, 
that very recently Shibata has upgraded his code to incorporate HRSC schemes 
(in particular, a Roe-type approximate Riemann solver and piecewise parabolic 
cell-reconstruction procedures)~\cite{shibata02c}. Further 3D codes are currently
being developed by a group in the U.S. (Duez et al.~\cite{duez02}) and by a E.U. 
Research Training Network collaboration \cite{whisky}.

\subsubsection{Shibata}
\label{shibata}

In Shibata's code \cite{shibata99} the hydrodynamics equations are formulated 
following Wilson's approach (Section 2.1.2) for a conformal-traceless reformulation 
of the spacetime variables (see below). An important difference with respect to
the original system, Eqs.~(\ref{36})-(\ref{42}), is that an equation for the 
entropy is solved instead of the energy equation. The hydrodynamic equations 
are integrated using van Leer's~\cite{vanleer79} second order finite difference 
scheme with artificial viscosity, following the approach of a precursor code 
developed by \cite{oohara96}.

The ADM Einstein equations are reformulated into a conformal traceless system, 
an idea originally introduced by Shibata and Nakamura~\cite{shibata95} (see
also~\cite{naka87a}) and further developed by Baumgarte and Shapiro 
\cite{baumgarte98}. This ``BSSN" formulation, which shows enhanced long-term 
stability compared to the original ADM system, makes use of a conformal 
decomposition of the 3-metric, $\tilde \gamma_{ij} = e^{- 4 \phi} \gamma_{ij}$ 
and the trace-free part of the extrinsic curvature, $A_{ij} = K_{ij} - \gamma_{ij} K/3$, 
with the conformal factor $\phi$ chosen to satisfy $e^{4 \phi} = \gamma^{1/3} \equiv 
\det(\gamma_{ij})^{1/3}$. In this formulation, as shown by \cite{baumgarte98}, in
addition to the evolution equations for the conformal three--metric $\tilde \gamma_{ij}$ 
and the conformal-traceless extrinsic curvature variables $\tilde A_{ij}$, there 
are evolution equations for the conformal factor $\phi$, the trace of the extrinsic 
curvature $K$ and the ``conformal connection functions'' $\tilde
\Gamma^i$. Further details can be found in \cite{baumgarte98,shibata99}. 

The code uses an approximate maximal slicing condition, which amounts to 
solving a parabolic equation for $\ln\alpha$, and a minimal distortion 
gauge condition for the shift vector. It also admits $\pi$-rotation 
symmetry around the $z$-axis, as well as plane symmetry with respect 
to the $z=0$ plane, allowing computations in a quadrant region. In 
addition, Shibata has also implemented in the code the ``cartoon" method
proposed by the AEI Numerical Relativity group~\cite{alcubierre00b}, which 
makes possible axisymmetric computations with a Cartesian grid. ``Approximate" 
outgoing boundary conditions are used at the outer boundaries, which do not 
completely eliminate numerical errors due to spurious back reflection of 
gravitational waves\footnote{We note however that all codes based on the
3+1 formalism share this problem since the outer boundaries are located at finite 
radii. Further work on the development of more sophisticated boundary conditions
is needed to solve this problem. Alternative solutions are to follow the light-cone
approach developed by Winicour et al~\cite{winicour98} or the conformal
formalism of Friedrich~\cite{friedrich02}}. A staggered Leapfrog method is used 
for the time evolution of the metric quantities. Correspondingly, the hydrodynamic 
equations are updated using a second-order two-step Runge-Kutta scheme. 
In each time step, the staggered metric quantities needed for the hydrodynamics 
update are properly extrapolated to intermediate time levels to reach the desired 
order of accuracy.

The code developed by Shibata \cite{shibata99,shibata02c} has been tested in a 
variety of problems, including spherical collapse of dust to a black 
hole, signalled by the appearance of the apparent horizon (comparing 1D 
and 3D simulations), stability of spherical stars and computation of the 
radial oscillation period, quadrupole oscillations of perturbed spherical 
stars and computation of the associated gravitational radiation, preservation 
of the rotational profile of (approximate) rapidly rotating stars, and 
the preservation of a co-rotating binary neutron star in a quasi-equilibrium 
state (assuming a conformally flat 3-metric) for more than one orbital 
period. Improvements of the hydrodynamical schemes have been considered 
very recently~\cite{shibata02c}, in order to tackle problems in which shocks 
play an important role, e.g. stellar core collapse to a neutron star.
Shibata's code has allowed important breakthroughs in the study of the 
morphology and dynamics of various general relativistic astrophysical 
problems, such as black hole formation resulting from both, the rotational
gravitational collapse of neutron stars and supermassive stars, and, perhaps
most importantly, the coalescence of binary neutron stars, a long-standing 
problem in numerical relativistic hydrodynamics. These applications are 
discussed in Section~\ref{simulations}. The most recent simulations of binary 
neutron star coalescence~\cite{shibata02a} have been performed on a FACOM 
VPP5000 computer with typical grid sizes of (505,505,253) in ($x,y,z$). 

\subsubsection{CACTUS/GR\_ASTRO}
\label{grastro}

A three-dimensional general relativistic hydrodynamics code developed by the 
Washington University/NCSA/AEI-Golm collaboration for the NASA Neutron Star 
Grand Challenge Project~\cite{GR_ASTRO} is discussed in Refs.~\cite{font99a,font02a}. 
The code is built upon the Cactus Computational Toolkit~\cite{Cactusweb}. A
version of this code which passed the milestone requirement of the
NASA Grand Challenge project, was released to the community. This code has 
been benchmarked at over 140 GFlop/sec on a $1024$ node Cray T3E, with a scaling 
efficiency of over 95\%, showing the potential for large scale 3D simulations
of realistic astrophysical systems.

The hydrodynamics part of the code uses the conservative formulation discussed 
in Section \ref{valencia}. A variety of state-of-the-art Riemann solvers are 
implemented, including a Roe-like solver~\cite{roe81} and Marquina's flux 
formula~\cite{donat96}. The code uses slope-limiter methods to construct 
second-order TVD schemes by means of monotonic piecewise linear reconstructions 
of the cell-centered quantities for the computation of the numerical fluxes. 
The standard ``minmod'' limiter and the ``monotonized central-difference" 
limiter~\cite{vanLeer77} are implemented. Both schemes provide the desired 
second-order accuracy for smooth solutions, while still satisfying the TVD 
property. In addition, third-order piecewise parabolic (PPM) reconstruction 
has been recently implemented and used in \cite{stergioulas01a}.

The Einstein equations are solved using different approaches, (i) the standard 
ADM formalism, (ii) a hyperbolic formulation developed by~\cite{bona95}, and (iii) 
the BSSN formulation of \cite{naka87a,shibata95,baumgarte98}. The time evolution of both 
the ADM and the BSSN systems can be performed using several numerical 
schemes~\cite{font99a,alcubierre99,font02a}. Currently, a Leapfrog (non-staggered 
in time), and an iterative Crank-Nicholson scheme have been coupled to the hydrodynamics 
solver. The code is designed to handle arbitrary shift and lapse conditions, which 
can be chosen as appropriate for a given spacetime simulation. The AEI Numerical 
Relativity group \cite{AEIgroup} is currently focused in developing robust 
gauge conditions for (vacuum) black hole spacetimes (see e.g. \cite{alcubierre02a} 
and references therein). Hence, all advances accomplished here can be incorporated 
in future versions of the code for non-vacuum spacetime evolutions. Similarly, being 
a general purpose code, a number of different boundary conditions can be imposed for 
either the spacetime variables or for the hydrodynamical variables. We refer the 
reader to \cite{font99a,alcubierre99,font02a} for additional details. 

The code has been subjected to a series of convergence tests \cite{font99a}, with many 
different combinations of the spacetime and hydrodynamics finite differencing 
schemes, demonstrating the consistency of the discrete equations with the
differential equations. The simulations performed in~\cite{font99a} include,
among others, the evolution of equilibrium configurations of compact stars 
(solutions to the TOV equations), and the evolution of relativistically 
boosted TOV stars ($v=0.87c$) transversing diagonally across the computational 
domain, test for which an exact solution exists. In~\cite{alcubierre99,alcubierre99b} 
the improved stability properties of the BSSN formulation of the Einstein equations 
were compared to the ADM system. In particular, in~\cite{alcubierre99} a number of 
strongly gravitating systems were simulated, including weak and strong gravitational 
waves, black holes, boson stars and relativistic stars. While the error growth-rate 
can be decreased by going to higher grid resolutions, the BSSN formulation requires 
grid resolutions higher than the ones needed in the ADM formulation to achieve the same 
accuracy. Furthermore, it was shown in Ref.~\cite{font02a} that the code successfully
passed stringent long-term evolution tests, such as the evolution of both, spherical 
and rapidly rotating, stationary stellar configurations, and the formation 
of an apparent horizon during the collapse of a relativistic star to a black 
hole. The high accuracy of the hydrodynamical schemes employed 
has allowed the detailed investigation of the frequencies of radial, 
quasi-radial and quadrupolar oscillations of relativistic stellar models, 
and use them as a strong assessment of the code. The frequencies obtained 
have been compared to the frequencies computed with perturbative methods and 
in axisymmetric nonlinear evolutions \cite{font01a}. In all of the cases 
considered, the code reproduces these results with excellent accuracy 
and is able to extract the gravitational waveforms that might be produced
during non-radial stellar pulsations.  


\section{Hydrodynamical simulations in relativistic astrophysics}
\label{simulations}

With the exception of the vacuum two-body problem (i.e. the coalescence 
of two black holes), all realistic astrophysical systems and sources of 
gravitational radiation involve matter. Not surprisingly,
the joint integration of the equations of motion for matter and 
geometry was in the minds of theorists from the very beginning 
of numerical relativity.

Nowadays there is a large body of numerical investigations in the
literature dealing with hydrodynamical integrations in {\it static}
background spacetimes. Most of those are based on Wilson's Eulerian formulation
of the hydrodynamic equations and use schemes based on finite differences
with some amount of artificial viscosity. The use of conservative formulations 
of the equations, and the incorporation of the characteristic information 
in the design of numerical schemes, has started in more recent years.

On the other hand, time-dependent simulations of self-gravitating flows 
in general relativity, evolving the spacetime {\it dynamically} with the 
Einstein equations coupled to a hydrodynamic source, constitute a much 
smaller sample. Although there is much interest in this direction, 
only the spherically symmetric case (1D) has been extensively studied.
In axisymmetry (2D) fewer attempts have been made, with most of them 
devoted to the study of the gravitational collapse of rotating stellar 
cores, black hole formation and the subsequent emission of gravitational
radiation. Three-dimensional simulations have only started more recently.
Much effort is nowadays focused on the study of the coalescence of compact 
neutron star binaries (as well as the vacuum black hole binary counterpart). 
These theoretical investigations are driven by the emerging possibility of 
soon detecting gravitational waves with the different experimental efforts 
currently underway. The waveform catalogues resulting from time-dependent
hydrodynamical simulations may provide some help to data analysis groups,
since the chances for detection may be enhanced through matched-filtering 
techniques.

In the following we review the status of the numerical investigations in three
astrophysical scenarios all involving strong gravitational fields and,
hence, relativistic physics: gravitational collapse, accretion onto
black holes, and hydrodynamical evolution of neutron stars. Relativistic 
cosmology, another area where fundamental advances have been accomplished 
through numerical simulations, is not considered. The interested reader is 
addressed to the {\it Living Reviews} article by Anninos~\cite{anninos97} and 
references therein.


\subsection{Gravitational collapse}

The study of gravitational collapse of massive stars has been largely
pursued numerically over the years. This problem was already the main
motivation when May and White built the first one-dimensional
code~\cite{may66,may67}. Such code was conceived to integrate the 
coupled system of Einstein field equations and relativistic hydrodynamics, 
sidestepping Newtonian approaches.

By browsing through the literature one realizes that the numerical
study of gravitational collapse has been neatly split in two main
directions since the early investigations. First, the {\it computational 
astrophysics community} has traditionally focused on the physics of 
the (type II/Ib/Ic) Supernova mechanism, i.e., collapse of unstable 
stellar iron cores followed by a bounce and explosion. Nowadays, numerical
simulations are highly developed, as they include rotation 
and detailed state-of-the-art microphysics (e.g., EOS and 
sophisticated treatments for neutrino transport). These studies are 
currently performed, routinely, in multidimensions with advanced HRSC 
schemes. Progress in this direction has been achieved, however,
at the expense of necessary approximations in the treatment of
the hydrodynamics and the gravitational field, by using Newtonian physics.
In this approach the emission of gravitational radiation is computed
through the quadrupole formula. For reviews of the current status
in this direction see~\cite{mueller98,janka01} and references therein.

On the other hand, the {\it numerical relativity community} has been more
interested, since the beginning, in highlighting those aspects of the
collapse problem having to do with relativistic gravity, in particular
the emission of gravitational radiation from non-spherical collapse
(see \cite{new02} for a recent {\it Living Reviews} article on gravitational
radiation from gravitational collapse). Much of the effort has focused 
on developing reliable numerical schemes to solve the gravitational 
field equations and much less emphasis, if any, has been given to the 
details of the microphysics of the collapse. In addition, this approach has 
mostly considered gravitational collapse leading to black hole formation, 
employing relativistic hydrodynamics and gravity. Quite surprisingly 
both approaches have barely interacted over the years, except for a 
handful of simulations in spherical symmetry. Nevertheless, numerical 
relativity is steadily reaching a state in which it is not unrealistic to 
expect a much closer interaction in the near future (see, e.g. 
\cite{shibata99b,shibata00b,font02a,shibata02c} and references therein).

\subsubsection{Core collapse supernovae}


\begin{figure}[t]
\centerline{\psfig{figure=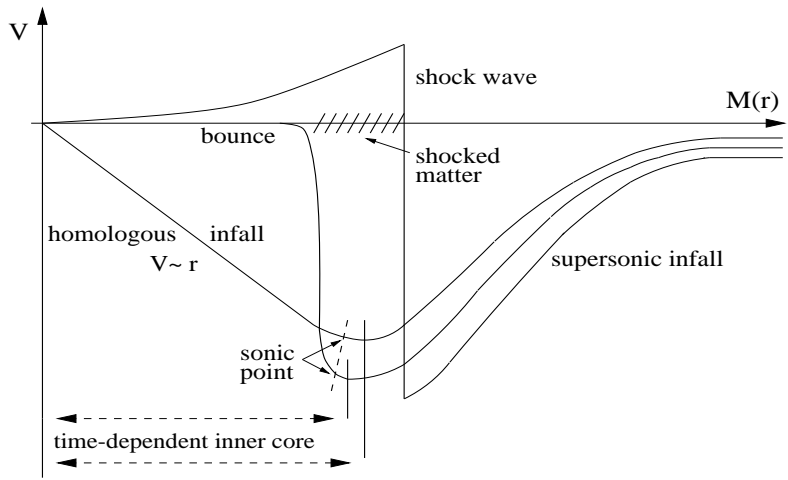,width=4.1in,height=2.5in}}
\caption{{ Schematic profiles of the velocity versus radius at three
different times during core collapse: at the point of ``last good
homology", at bounce and at the time when the shock wave has just
detached from the inner core.
}}
\label{collapse}
\end{figure}

At the end of their thermonuclear evolution, massive stars in the
(Main Sequence) mass range from 9$M_{\odot}$ to 30$M_{\odot}$ develop a
core composed of iron group nuclei which becomes dynamically unstable 
against gravitational collapse. The iron core collapses to a neutron 
star or a black hole releasing gravitational binding energy of the order
$ \sim 3 \times 10^{53} {\rm\ erg} \, (M / M_\odot)^2 (R / 10 {\rm\ km})^{-1}$,
which is sufficient to power a supernova explosion. The standard model 
of type II/Ib/Ic Supernovae involves the presence of a strong shock wave 
formed at the edge between the homologous inner core and the outer core, 
which falls at roughly free-fall speed. A schematic illustration of the 
dynamics of this process is depicted in Fig.~\ref{collapse}. The shock 
is produced after the bounce of the inner core when its density exceeds 
nuclear density. Numerical simulations have tried to elucidate if this 
shock is strong enough to propagate outwards through the star's 
mantle and envelope given certain initial conditions for the material in 
the core (an issue subject to important uncertainties in the nuclear EOS), 
as well as through the outer layers of the star. In the accepted scenario which 
has emerged from numerical simulations, the existence of the shock wave 
together with neutrino heating that re-energizes it (in the so-called 
delayed-mechanism by which the stalled prompt supernova shock is reactivated 
by an increase in the post-shock pressure due to neutrino energy 
deposition~\cite{wilson85,bethe85}), and convective overturn which rapidly 
transports energy into the shocked region~\cite{colgate89,bethe90} (and 
which can lead to large-scale deviations from spherically symmetric 
explosions), are all necessary ingredients for any plausible explosion 
mechanism (see \cite{janka01} for a review).

However, the path to reach such conclusions has been mostly delineated 
from numerical simulations in one spatial dimension. Fully relativistic 
simulations of microphysically detailed core collapse off spherical symmetry 
are still absent and they might well introduce some modifications. The 
broad picture described above has been demonstrated in multidimensions 
using sophisticated Newtonian models, as is documented in~\cite{mueller98}.

\paragraph{Spherically symmetric simulations.}
May and White's formulation and their corresponding one-dimensional code 
formed the basis of most spherically symmetric codes built to study the 
collapse problem.  Wilson~\cite{wilson71} investigated the effect of
neutrino transport on stellar collapse concluding that heat conduction
by neutrinos does not produce the ejection of material, in contrast to
previous results~\cite{colgate66,arnett66}. The code solved the
coupled set of hydrodynamic and Einstein equations, supplemented
with a Boltzmann transport equation to describe the neutrino flow.
Van Riper~\cite{vanriper79} used a spherically symmetric general
relativistic code to study adiabatic collapse of stellar cores, considering
the purely hydrodynamical bounce as the preferred explosion mechanism.
The important role of general relativistic effects to produce collapses
otherwise absent in Newtonian simulations was emphasized. 
Bruenn~\cite{bruenn85} developed a code in which the neutrino transport was 
handled with an approximation, the multigroup flux-limited diffusion.
Baron et al.~\cite{baron85} obtained a successful and very energetic explosion
for a model in which the value of the incompressibility modulus
of symmetric nuclear matter at zero temperature was particularly small, 
$K_0^{sym}=180$ MeV (its precise value, nowadays preferred around 
$240$ MeV~\cite{glendening}, is still a matter of debate). Mayle et 
al.~\cite{mayle87} computed neutrino spectra from stellar collapse to 
stable neutron stars for different collapse models using, as in Ref.~\cite{bruenn85}, 
multigroup flux-limited diffusion for the neutrino transport. Van 
Riper~\cite{vanriper88} and Bruenn~\cite{bruenn89} verified that a 
softer supranuclear EOS, combined with general relativistic hydrodynamics, 
increases the chances for a prompt explosion. In~\cite{schinder88,schinder89} 
and~\cite{mezzacappa89} the neutrino transport was first handled without 
approximation by using a general relativistic Boltzmann equation. Whereas 
in~\cite{schinder88,schinder89} the neutrino transport is described by 
moments of the general relativistic Boltzmann equation in polar slicing 
coordinates~\cite{bardeen83}, \cite{mezzacappa89} used maximal slicing 
coordinates.  Miralles et al.~\cite{miralles91}, employing a realistic EOS 
based upon a Hartree-Fock approximation for Skyrme-type nuclear forces for 
densities above nuclear density, found that the explosion was favoured by
soft EOS, a result qualitatively similar to that of Baron et al.~\cite{baron85},
who used a phenomenological EOS. Swesty et al.~\cite{swesty94} also 
focused on the role of the nuclear EOS in stellar collapse on prompt timescales. 
Contrary to previous results they found that the dynamics of the shock is
almost independent of the nuclear incompressibility once the EOS is not
unphysically softened as in earlier simulations (e.g.~\cite{vanriper79,
baron85,vanriper88,bruenn89,miralles91}). Swesty and coworkers used a 
finite temperature compressible liquid drop model EOS for the
nuclear matter component~\cite{lattimer91}. For the shock to propagate promptly 
to a large radius they found that the EOS must be very soft at densities just 
above nuclear densities, which is inconsistent with the $1.44 M_{\odot}$ neutron 
star mass constraint imposed by observations of the binary pulsar PSR 1913+16.

From the above discussion it is clear that numerical simulations 
show a strong sensitivity of the explosion mechanism to the details
of the post-bounce evolution: general relativity, the nuclear EOS and
the properties of the nascent neutron star, the treatment of the neutrino
transport, and the neutrino-matter interaction. Recently, state-of-the-art 
treatments of the neutrino transport have been achieved, even in the general 
relativistic case, by solving the Boltzmann equation in connection with the 
hydrodynamics, instead of using multigroup flux-limited diffusion
\cite{mezzacappa01,rampp00a,rampp02a,liebendoerfer01}. The results of the 
few spherically symmetric simulations available show, however, that the 
models do not explode, neither in the Newtonian or in the general relativistic 
case. Therefore, computationally expensive multidimensional simulations 
with Boltzmann neutrino transport, able to
account for convective effects, are needed to draw further conclusions
on the viability of the neutrino-driven explosion mechanism.

From the numerical point of view, many of the above investigations used 
artificial viscosity techniques to handle shock waves. Together with a 
detailed description of the microphysics, the correct numerical modeling 
of the shock wave is the major issue in stellar collapse. In this context, 
the use of HRSC schemes, specifically designed to capture discontinuities, is
much more recent, starting in the late 1980s with the Newtonian simulations
of Fryxell et al.~\cite{fryxell89} using an Eulerian second order PPM scheme 
(see~\cite{mueller98} for a review of the present status). There are only a 
few relativistic simulations, so far restricted to spherical symmetry
\cite{ibanez92,romero96,novak99}.

Mart\'{\i} et al.~\cite{marti90} implemented Glaister's approximate Riemann
solver~\cite{glaister88} in a Lagrangian Newtonian hydrodynamical code. 
They performed a comparison of the energetics of a stellar
collapse simulation using this HRSC scheme and a standard May and White
code with artificial viscosity, for the same initial model, grid size and EOS.
They found that the simulation employing a Godunov-type scheme produced
larger kinetic energies than that corresponding to the artificial viscosity
scheme, with a factor of two difference in the maximum of the infalling
velocity. Motivated by this important difference Janka et
al.~\cite{janka93} repeated this computation with a different EOS, using a
PPM second order Godunov-type scheme, disagreeing with Mart\'{\i} et
al.~\cite{marti90}. The state-of-the-art implementation of the tensorial
artificial viscosity terms in~\cite{janka93}, together with the very fine
numerical grids employed (unrealistic for three dimensional simulations),
could be the reason of the discrepancies.


\begin{figure}[t]
\centerline{\psfig{figure=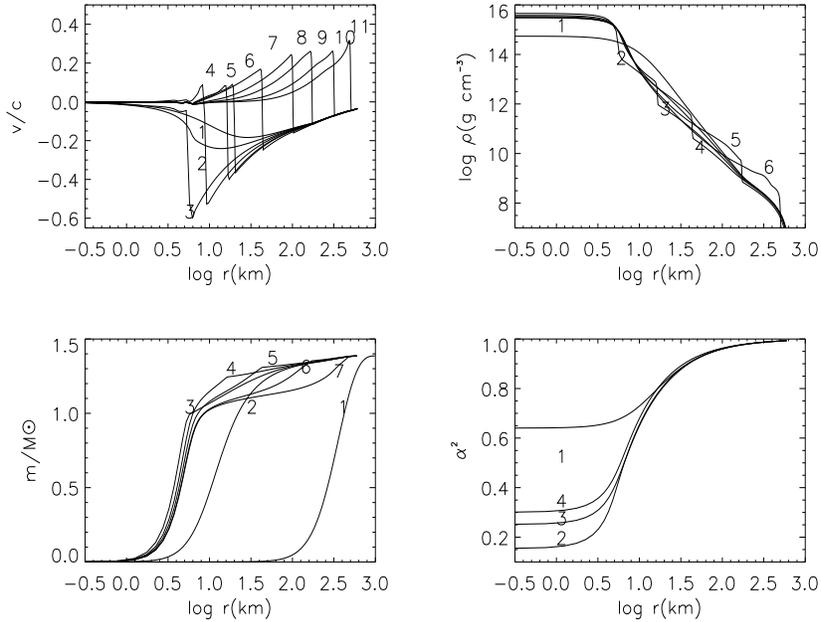,width=4.5in,height=3.5in}}
\caption{{
Stills from a movie showing the animations of a relativistic adiabatic core 
collapse using HRSC schemes ({\it snapshots of the radial profiles of various 
variables are shown at different times}). The simulations are taken from 
Ref.~\cite{romero96}: Velocity (top-left), logarithm of the rest-mass density 
(top-right), gravitational mass (bottom-left), and lapse function squared  
(bottom-right). See text for details of the initial model. Visualization
by Jos\'e V. Romero. 
(To see the movie, please go to the electronic version of
this review article at 
{\tt http://www.livingreviews.org/Articles/Volume3/2000-2font}).
}}
\label{mpeg1}
\end{figure}

As mentioned in Section 2.1.3, Godunov-type methods were first used to solve
the equations of general relativistic hydrodynamics in~\cite{marti91},
where the characteristic fields of the one-dimensional (spherically
symmetric) system were derived. The first astrophysical application
was stellar collapse. In~\cite{bona93} the hydrodynamic equations
were coupled to the Einstein equations in spherical symmetry. The
field equations were formulated as a first-order flux-conservative hyperbolic 
system for a harmonic gauge~\cite{bona89}, somehow ``resembling" the
hydrodynamic equations. HRSC schemes were applied to both systems 
simultaneously (only coupled through the source terms of the equations). 
Results for simple models of adiabatic collapse can be found 
in~\cite{marti91b,bona93,ibanez93}. 

A comprehensive study of adiabatic, one-dimensional, core collapse using
explicit upwind HRSC schemes was presented in~\cite{romero96} (see~\cite{yamada96}
for a similar computation using implicit schemes). In this investigation 
the equations for the hydrodynamics and the geometry are written using 
radial gauge polar slicing (Schwarzschild-type) coordinates. The collapse 
is modeled with an ideal gas EOS with a non-constant adiabatic index, which 
depends on the density as $\Gamma=\Gamma_{min}+\eta(\log\rho-\log\rho_b)$, 
where $\rho_b$ is the bounce density and $\eta=0$ if $\rho<\rho_b$ and 
$\eta>0$ otherwise~\cite{vanriper79}. A set of animations of the simulations 
presented in~\cite{romero96} is included in the four MPEG movies in 
Fig.~\ref{mpeg1}. They correspond to the rather stiff Model B of~\cite{romero96}: 
$\Gamma_{min}=1.33, \eta=5$ and $\rho_b=2.5\times 10^{15}$ g\,cm$^{-3}$. The 
initial model is a white dwarf having a gravitational mass of $1.3862M_{\odot}$. 
The animations show the time evolution of the radial profiles of the following 
fields: velocity (movie 1), logarithm of the rest-mass density (movie 2), 
gravitational mass (movie 3) and the square of the lapse function (movie 4).

The movies show that the shock is sharply resolved and free of spurious
oscillations. The radius of the inner core at the time of maximum compression,
at which the infall velocity is maximum ($v=-0.62c$), is $6.3$ km. The 
gravitational mass of the inner core at the time of maximum compression is 
$\sim 1.0 M_{\odot}$.  The minimum value the quantity $\alpha^2$ reaches is 
$0.14$ which indicates the highly relativistic character of these simulations 
(at the surface of a typical neutron star the value of the lapse function
squared is $\alpha^2\sim 0.75$).




\begin{figure}[t]
\centerline{\psfig{figure=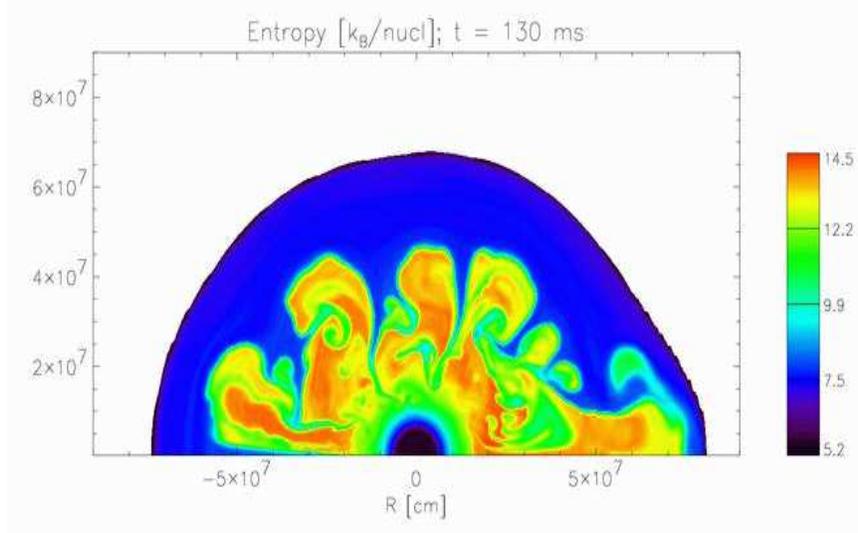,width=4.5in,height=2.8in}}
\caption{{ 
Still from a movie showing the animation of the time evolution of the entropy
in a core collapse supernova explosion~\cite{kifonidis99}. The movie
shows the evolution within the innermost 3000 km of the star and up to 
220 ms after core bounce. See text for explanation. Visualization by
Konstantinos Kifonidis. (To see the movie, please go to the electronic 
version of this review article at 
{\tt http://www.livingreviews.org/Articles/Volume3/2000-2font.})
}}
\label{mpeg3}
\end{figure}

\paragraph{Axisymmetric Newtonian simulations.}
Beyond spherical symmetry most of the existing simulations of core collapse
supernova are Newtonian. Axisymmetric investigations have been performed
by~\cite{mueller82,bodenheimer83a,bonazzola93} using a realistic EOS and 
some treatment of weak interaction processes but neglecting neutrino transport, 
and by~\cite{janka89a,moenchmeyer91a,imshennik92a,fryer00a,fryer02a} 
employing some approximate description of neutrino transport. In addition, 
\cite{finn90a,yamada94a,zwerger97} have performed Newtonian parameter studies 
of the axisymmetric collapse of rotating polytropes. 

Among the more detailed multi-dimensional non-relativistic hydrodynamical 
simulations are those performed by the MPA/Garching group (an on-line sample
can be found at their website~\cite{MPAweb}). As mentioned before these include 
advanced microphysics and employ accurate HRSC integration schemes. To 
illustrate the degree of sophistication achieved in Newtonian simulations we 
present in the MPEG movie in Fig.~\ref{mpeg3} an animation of the early 
evolution of a core collapse supernova explosion up to 220 ms after core 
bounce and shock formation (only an intermediate snapshot at 130 ms is 
depicted in Fig.~\ref{mpeg3}). The movie shows the evolution of the entropy 
within the innermost 3000 km of the star.

The initial data used in these calculations is taken from the
15\,$\rm M_{\odot}$ pre-collapse model of a blue supergiant star 
of~\cite{woosley88}. The computations start 20 ms after core bounce
from a one-dimensional model of~\cite{bruenn93}. This model is
obtained with the one-dimensional general relativistic code mentioned
previously~\cite{bruenn85} which includes a detailed treatment of the 
neutrino physics and a realistic EOS, and which is able to follow core 
collapse, bounce and the associated formation of the supernova shock.
Because of neutrino cooling and energy losses due to the dissociation 
of iron nuclei the shock initially stalls inside the iron core.

The movie shows how the stalling shock is ``revived'' by neutrinos
streaming from the outer layers of the hot, nascent neutron star in
the center.  A negative entropy gradient builds up between the so
called ``gain-radius'', which is the position where
cooling of hot gas by neutrino emission switches into net energy gain
by neutrino absorption, and the shock further out.  Convective
instabilities therefore set in, which are characterized by large
rising blobs of neutrino heated matter and cool, narrow downflows.
The convective energy transport increases the efficiency of
energy deposition in the post-shock region by transporting
heated material near the shock  and cooler matter near the
gain radius where the heating is strongest. The freshly
heated matter then rises again while the shock is distorted
by the upward streaming bubbles. The reader is addressed
to~\cite{kifonidis99} for a more detailed description of
a more energetic initial model.


\paragraph{Axisymmetric relativistic simulations.}
Previous investigations in general relativistic rotational core collapse 
were mainly concerned with the question of black hole formation under 
idealized conditions (see Section \ref{bhformation}), but none of those
studies has really addressed the problem of supernova core collapse
which proceeds from white dwarf densities to neutron star densities,
involves core bounce, shock formation, and shock propagation. 

Wilson \cite{wilson79} first computed neutron star bounces of
$\Gamma = 2$ polytropes, and measured the gravitational wave
emission. The initial configurations were either prolate or slightly
aspherical due to numerical errors of an otherwise spherical
collapse. 

More than twenty years later, and with no other simulations in between,
the first comprehensive numerical study of relativistic rotational 
supernova core collapse in axisymmetry has been performed by Dimmelmeier 
et al.~\cite{dimmelmeier01a,dimmelmeier02a,dimmelmeier02b,dimmelmeier02c}, 
who computed the gravitational radiation emitted in such events. The Einstein
equations were formulated using the so-called conformally flat metric
approximation \cite{wilson96}. Correspondingly, the hydrodynamic equations 
were cast as the first-order flux-conservative hyperbolic system described 
in Section~\ref{valencia}. Details of the methodology and of the numerical 
code are given in \cite{dimmelmeier02b}. 

Dimmelmeier et al.~\cite{dimmelmeier02c} have simulated the evolution of 26 
models in both Newtonian and relativistic gravity. The initial configurations 
are differentially rotating relativistic $4/3$-polytropes in equilibrium, 
which have a central density of $10^{10} {\rm\ g\ cm}^{-3}$. Collapse is 
initiated by decreasing the adiabatic index to some prescribed fixed value 
$\Gamma_1 $ with $ 1.28 \le \Gamma_1 \le 1.325$. The EOS consists of a polytropic 
and a thermal part for a more realistic treatment of shock waves. Any 
microphysics like electron captures and neutrino transport is neglected. The 
simulations show that the three different types of rotational supernova core 
collapse and gravitational waveforms identified in previous Newtonian 
simulations by Zwerger and M\"uller~\cite{zwerger97} (regular collapse, 
multiple bounce collapse, and rapid collapse) are also present in relativistic 
gravity. However, rotational core collapse with multiple bounces is only 
possible in a much narrower parameter range in general relativity. Relativistic 
gravity has, furthermore, a qualitative impact on the dynamics: If the density 
increase due to the deeper relativistic potential is sufficiently large, a 
collapse which is stopped by centrifugal forces at subnuclear densities (and 
thus undergoes multiple bounces) in a Newtonian simulation, becomes a regular, 
single bounce collapse in relativistic gravity. Such collapse type transitions 
have important consequences for the maximum gravitational wave signal amplitude. 
Moreover, in several of the relativistic models discussed in \cite{dimmelmeier02c}, 
the rotation rate of the compact remnant exceeds the critical value where 
MacLaurin spheroids become secularly, and in some cases even dynamically, 
unstable against triaxial perturbations.

\begin{figure}[t]
\centerline{\psfig{figure=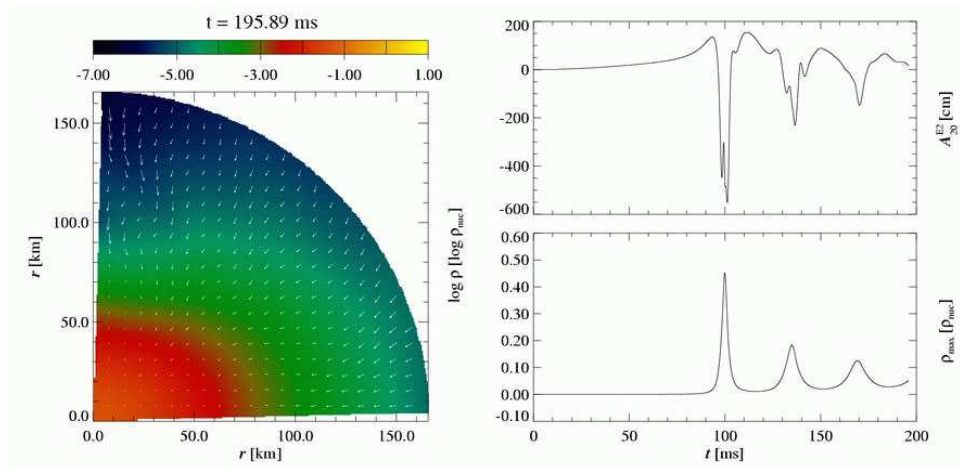,width=5.0in}}
\caption{{ Still from an animation showing the time evolution of a relativistic core
collapse simulation (model A2B4G1 of \cite{dimmelmeier02c}). Left: Velocity field and
isocontours of the density. Right: gravitational waveform (top) and central density 
evolution (bottom). Multiple bounce collapse (fizzler), type II signal. The camera
follows the multiple bounces. Visualization by Harald Dimmelmeier.
(To see the movie, please go to the electronic 
version of this review article at 
{\tt http://www.livingreviews.org/Articles/Volume3/2000-2font.})
}}
\label{cocoa2}
\end{figure}

The MPEG movie contained in Fig.~\ref{cocoa2} shows the time evolution of 
a multiple bounce model (model A2B4G1 in the notation of Ref.~\cite{dimmelmeier02b}), 
with a type II gravitational wave signal. The left panel shows isocontours of 
the logarithm of the density together with the corresponding velocity field 
distribution. The two panels on the right show the time evolution of the 
gravitational wave signal (top panel) and of the central rest-mass density. 
In the animation the ``camera" follows the multiple bounces by zooming 
in and out. The panels on the right show how each burst of gravitational 
radiation coincides with the time of bounce of the collapsing core. The 
oscillations of the nascent proto-neutron star are therefore imprinted on the 
gravitational waveform. Additional animations of the simulations performed 
by \cite{dimmelmeier02c} can be found at the MPA's website \cite{MPAweb}.

The relativistic models analyzed by Dimmelmeier et al.~\cite{dimmelmeier02c}
 cover almost the same range of gravitational wave amplitudes
($ 4 \times 10^{-21} \le h^{\rm TT} \le 3 \times 10^{-20} $ for a
source at a distance of 10~kpc) and frequencies ($ 60 {\rm\ Hz} \le 
\nu \le 1000 {\rm\ Hz} $) as the corresponding Newtonian ones \cite{zwerger97}. 
Averaged over all models, the total energy radiated in form of gravitational 
waves is $ 8.2 \times 10^{-8}\, M_{\odot} c^2 $ in the relativistic case, 
and $ 3.6 \times 10^{-8}\, M_{\odot} c^2$ in the Newtonian case. For all 
collapse models which are of the same type in both Newtonian and relativistic 
gravity, the gravitational wave signal is of lower amplitude. If the collapse
type changes, either weaker or stronger signals are found in the relativistic 
case. Therefore, the prospects for 
detection of gravitational wave signals from axisymmetric supernova rotational 
core collapse do not improve when taking into account relativistic 
gravity. The signals are within the sensitivity range of the first generation 
laser interferometer detectors if the source is located within the Local
Group. An on-line waveform catalogue for all models analyzed by Dimmelmeier
et al.~\cite{dimmelmeier02c} is available at the MPA's website~\cite{MPAweb}.
Finally, we note that a program to extend these simulations to full general
relativity (relaxing the conformally flat metric assumption) has been very
recently started by Shibata~\cite{shibata02c}.


\paragraph{Three-dimensional simulations.}
To date, there are no three-dimensional relativistic simulations of gravitational 
collapse, in the context of supernova core collapse, yet available. All the existing 
computations have employed Newtonian physics. This situation, however, might change 
in the near future, as very recently the first fully relativistic three-dimensional 
simulations of gravitational collapse leading to black hole formation have been 
accomplished, for rapidly-rotating, supermassive neutron stars~\cite{shibata99b} and 
supermassive stars~\cite{shibata02b} (see Section~\ref{bhformation}), for the head-on 
collision of two neutron stars~\cite{miller99}, and for the coalescence of neutron
star binaries~\cite{shibata99,shibata99c,shibata02a} (see Section 4.3).

Concerning Newtonian studies Bonazzola and Marck~\cite{bonazzola93} performed 
pioneer three-dimensional simulations, using pseudo-spectral methods, very 
accurate and free of numerical or intrinsic viscosity. They confirmed the low 
amount of energy radiated in gravitational waves regardless of the initial 
conditions of the collapse: axisymmetric, rotating or tidally deformed (see 
also~\cite{mueller82}). This result only applies to the pre-bounce phase of the 
supernova collapse as the simulations did not consider shock propagation after
bounce.  More recently, \cite{rampp98a,brown01a} have extended the work of 
\cite{zwerger97} studying, with two different PPM hydrodynamics codes, the 
dynamical growth of non-axisymmetric (bar mode) instabilities appearing in 
rotational post-collapse cores. A relativistic extension has been performed 
by \cite{shibata00a} (see Section \ref{pulsations}).

\subsubsection{Black hole formation}
\label{bhformation}

Apart from a few one-dimensional simulations, most numerical studies of general 
relativistic gravitational collapse leading to black hole formation have
used Wilson's formulation of the hydrodynamics equations and finite difference
schemes with artificial viscosity. The use of conservative formulations and
HRSC schemes to study black hole formation, both in two and three dimensions,
has started more recently.

\paragraph{Spherically symmetric simulations.}
Van Riper~\cite{vanriper79}, using a (Lagrangian) May and White code, analyzed
the mass division between the formation of a neutron star or a black hole after 
gravitational collapse. For the typical (cold) EOS used, the critical state was 
found to lie between the collapses of 1.95$M_{\odot}$ and 1.96$M_{\odot}$ cores.

In~\cite{shapiro80} a one-dimensional code based on Wilson's hydrodynamical
formulation was used to simulate a general relativistic (adiabatic)
collapse to a black hole. The fluid equations were finite differenced
in a mixed Eulerian-Lagrangian form with the introduction of an arbitrary 
``grid velocity" to ensure sufficient resolution throughout the entire 
collapse. The Einstein equations were formulated following the ADM 
equations. Isotropic coordinates and a maximal time slicing condition 
were used to avoid the physical singularity once the black hole 
forms. Due to the non-dynamical character of the gravitational field in 
the case of spherical symmetry (i.e., the metric variables can be computed 
at every time step solving elliptic equations), the code developed 
by~\cite{shapiro80} could follow relativistic configurations for many 
collapse time scales $\Delta t\gg GM/c^3$ after the initial appearance 
of an event horizon.

A Lagrangian scheme based on the Misner and Sharp~\cite{misner64}
formulation for spherically symmetric gravitational collapse (as in 
Ref.~\cite{vanriper79}) was developed by Miller and Motta~\cite{miller89} 
and later by Baumgarte et al.~\cite{baumgarte95}. The novelty of these 
codes was the use of an outgoing null coordinate $u$ (an ``observer-time" 
coordinate, as suggested previously by~\cite{hernandez66}) instead of the 
usual Schwarzshild time $t$ appearing in the Misner and Sharp equations.
Outgoing null coordinates are ideally suited to study black hole formation 
as they never cross the black hole event horizon. In these codes the 
Hern\'andez-Misner equations~\cite{hernandez66} (or, alternatively, the 
Misner-Sharp equations) were solved by an explicit finite difference scheme 
similar to the one used by~\cite{vanriper79}. In~\cite{miller89} the 
collapse of an unstable polytrope to a black hole was first achieved using 
null slicing. In~\cite{baumgarte95} the collapse of a 1.4$M_{\odot}$ polytrope
with $\Gamma=4/3$ was compared to the result of \cite{vanriper79}, using 
the Misner-Sharp equations, finding a 10\% agreement. This work showed 
numerically that the use of a retarded time coordinate allows for stable 
evolutions after the black hole has formed. The Lagrangian character of both 
codes has prevented their multidimensional extension.

Linke et al.~\cite{linke01a} analyzed the gravitational collapse of 
supermassive stars in the range $5\times 10^5 M_{\odot}-10^9 M_{\odot}$. 
In the same spirit as in Refs.~\cite{miller89,baumgarte95}, the coupled 
system of Einstein and fluid equations was solved employing coordinates 
adapted to a foliation of the spacetime with outgoing null hypersurfaces. 
From the computed neutrino luminosities, estimates of the energy deposition 
by $\nu\bar{\nu}$-annihilation were obtained. Only a small fraction of 
this energy is deposited near the surface of the star, where it could cause 
the ultrarelativistic flow believed to be responsible for GRBs. However, 
the simulations show that for collapsing supermassive stars with masses 
larger than $5\times 10^5 M_{\odot}$, the energy deposition is at least 
two orders of magnitude too small to explain the energetics of 
observed long-duration bursts at cosmological redshifts. 

The interaction of massless scalar fields with self-gravitating relativistic stars 
was analyzed in \cite{siebel02a} by means of fully dynamic numerical simulations 
of the Einstein-Klein-Gordon perfect fluid system. A sequence of stable, 
self-gravitating, $K=100$, $N=1$ relativistic polytropes, increasing the central 
density from $\rho_c=1.5\times 10^{-3}$ to $3.0\times 10^{-3}$ ($G=c=M_{\odot}=1$)
was built. Using a compactified spacetime foliation with outgoing null cones,
Siebel et al.~\cite{siebel02a} studied the fate of the relativistic stars when 
they are hit by a strong scalar field pulse with a mass of the order of 
$10^{-3} \ M_{\odot}$, finding that the star is either forced to oscillate in 
its radial modes of pulsation or to collapse to a black hole on a dynamical 
timescale. The radiative signals, read off at future null infinity, consist of 
several quasi-normal oscillations and a late time power-law tail, in agreement 
with the results predicted by (linear) perturbation analysis of wave propagation 
in an exterior Schwarzschild geometry.

\paragraph{Axisymmetric simulations.}
Beyond spherical symmetry the investigations of black hole formation
started with the work of Nakamura~\cite{naka81}, who first simulated 
general relativistic rotating stellar collapse. He adopted the (2+1)+1 
formulation of the Einstein equations in cylindrical coordinates and 
introduced regularity conditions to avoid divergence behavior at coordinate 
singularities (the plane $z=0$)~\cite{naka80}. The equations were 
finite differenced using the donor cell scheme plus Friedrichs-Lax type 
viscosity terms. Nakamura used a ``hypergeometric" slicing condition 
(which reduces to maximal slicing in spherical symmetry), which
prevents the grid points from hitting the singularity if a black hole forms.
The simulations could track the evolution of the collapse of a 
$ 10 M_\odot $ ``core'' of a massive star with different amounts of 
rotational energy and an initial central density of $ 3 \times 10^{13} 
{\rm\ g\ cm}^{-3}$, up to the formation of a rotating black hole. However, 
the resolution affordable due to limitations in computer resources
($42\times 42$ grid points) was not high enough to compute the emitted 
gravitational radiation. Note that the energy emitted in gravitational waves 
is very small compared to the total rest mass energy, which makes its 
accurate numerical computation very challenging. In subsequent works,
Nakamura \cite{naka83a} (see also \cite{naka87a}) considered a configuration 
consisting of a neutron star ($M=1.09 M_\odot$, $ \rho_{\rm c} = 10^{15} 
{\rm\ g\ cm}^{-3} $) with an accreted envelope of $0.81 M_\odot$, which 
was thought to mimic mass fall-back in a supernova explosion. Rotation 
and infall velocity were added to such configuration, simulating the 
evolution depending on the prescribed rotation rates and rotation laws. 

Bardeen, Stark and Piran, in a series of 
papers~\cite{bardeen83,stark85,piran86,stark87}, studied the collapse of 
rotating relativistic ($\Gamma=2$) polytropic stars to black 
holes, succeeding in computing the associated gravitational radiation.  
The field and hydrodynamic equations were formulated using the 3+1
formalism, with radial gauge and a mixture of (singularity avoiding) 
polar and maximal slicing. The initial model was a spherically symmetric 
relativistic polytrope in equilibrium of mass $M$, central density 
$ 1.9 \times 10^{15} (M / M_\odot)^{-2} $, and radius $ 6 G M / c^2 = 8.8 
\times 10^5 M / M_\odot {\rm\ cm} $. Rotational collapse was induced by 
lowering the pressure in the initial model by a prescribed fraction, and 
by simultaneously adding an angular momentum distribution approximating 
rigid-body rotation. Their parameter space survey showed how black 
hole formation (of the Kerr class) occurs only for angular momenta less than 
a critical value. The numerical results also demonstrated that
the gravitational wave emission from axisymmetric rotating collapse
to a black hole was $\Delta E/Mc^2<7\times 10^{-4}$, and
that the general waveform shape was nearly independent of the details
of the collapse. 

Evans~\cite{evans86}, building on previous work by Dykema~\cite{dykema80}, 
also studied the axisymmetric gravitational collapse problem for non-rotating 
matter configurations. His numerical scheme to integrate the matter fields was 
more sophisticated than in previous approaches, as it included monotonic 
upwind reconstruction procedures and flux limiters.  Discontinuities 
appearing in the flow solution were stabilized by adding artificial 
viscosity terms in the equations, following Wilson's approach. 

Most of the axisymmetric codes discussed so far adopted spherical polar 
coordinates. Numerical instabilities are encountered at the origin 
($r=0$) and at the polar axis ($\theta=0,\pi$) where some fields diverge 
due to the coordinate singularities. Evans did important contributions 
towards regularizing the gravitational field equations in such 
situations~\cite{evans86}. These coordinate problems have deterred 
researchers from building three-dimensional codes in spherical 
coordinates. In this case Cartesian coordinates are adopted which, 
despite the desired property of being everywhere regular, present the 
important drawback of not being adapted to the topology of astrophysical 
systems. As a result this has important consequences on the grid 
resolution requirements. The only extension of an axisymmetric 2+1 code
in spherical coordinates to three dimensions has been accomplished by
Stark~\cite{stark89}, although no applications in relativistic
astrophysics have yet been presented.

Recently, Alcubierre et al.~\cite{alcubierre00b} proposed a method 
(``cartoon") which does not suffer from stability problems at coordinate 
singularities and where, in essence, Cartesian coordinates 
are used even for axisymmetric systems. Using this method, Shibata
\cite{shibata00b} investigated the effects of rotation on the
criterion for prompt adiabatic collapse of rigidly and differentially
rotating ($\Gamma = 2$) polytropes to a black hole. Collapse of the
initial approximate equilibrium models (computed by assuming a conformally 
flat spatial metric) was induced by a pressure reduction. In 
\cite{shibata00b} it was shown that the criterion for black hole 
formation depends strongly on the amount of angular momentum, but
only weakly on its (initial) distribution. Shibata also studied the
effects of shock heating using a gamma-law EOS, concluding that 
it is important in preventing prompt collapse to black holes in
the case of large rotation rates. Using the same numerical approach, 
Shibata and Shapiro\cite{shibata02b} have recently studied the collapse 
of a uniformly rotating supermassive star in general relativity. The 
simulations show that the star, initially rotating at the mass-shedding 
limit, collapses to form a supermassive Kerr black hole with a spin 
parameter of $\sim 0.75$. Roughly 90\% of the mass of the system is 
contained in the final black hole, while the remaining matter forms 
a disk orbiting around the hole.

Alternatively, existing axisymmetric codes employing the characteristic 
formulation of the Einstein equations~\cite{winicour98}, such as the (vacuum) 
code presented in~\cite{gomez94}, do not share the axis instability problems 
of most 2+1 codes. Hence, axisymmetric characteristic codes, once 
conveniently coupled to hydrodynamics codes, are a promising way of 
studying the axisymmetric collapse problem. First steps in this
direction are reported in \cite{siebel02b}, where an axisymmetric
Einstein-perfect fluid code is presented. This code achieves global
energy conservation for a strongly perturbed neutron star spacetime,
for which the total energy radiated away by gravitational waves 
corresponds to a significant fraction of the Bondi mass. 

\paragraph{Three-dimensional simulations.}
%
Hydrodynamical simulations of the collapse of supermassive ($\Gamma=2$) 
uniformly rotating neutron stars to rotating black holes, using the code 
discussed in Section~\ref{shibata}, are presented in \cite{shibata99b}. 
The simulations show no evidence for massive disk formation or outflows, 
which can be related to the moderate initial compactness of the stellar 
models ($R/M\sim 6$). A proof-of-principle of the capabilities of the code 
discussed in Section~\ref{grastro} to study black hole formation is presented 
in \cite{font02a}, where the gravitational collapse of a spherical unstable 
relativistic polytrope is discussed. Similar tests with differentially
rotating polytropes are given in~\cite{duez02} for a recent artificial 
viscosity-based, three-dimensional general relativistic hydrodynamics code.

\subsubsection{Critical collapse}

Critical phenomena in gravitational collapse were first discovered numerically 
by Choptuik in spherically-symmetric simulations of the massless Klein-Gordon 
field minimally coupled to gravity~\cite{choptuik93}. Since then critical phenomena 
arising at the threshold of black hole formation have been found in a variety of 
physical systems, including the perfect fluid model (see \cite{gundlach} for a review). 

Evans and Coleman \cite{evans93} first observed critical phenomena in spherical
collapse of radiation fluid, i.e. a fluid obeying an EOS $p=(\Gamma-1)\rho(1+
\varepsilon)$ with $\Gamma=4/3$ and $\varepsilon\gg 1$. The threshold of black 
hole formation was found for a critical exponent $\Gamma\sim 0.36$, in close agreement 
with that obtained in scalar field collapse \cite{choptuik93}. Their study used 
Schwarzschild coordinates in radial gauge and polar slicing, and the hydrodynamic 
equations followed Wilson's formulation. Subsequently, Maison \cite{maison96}, using
a linear stability analysis of the critical solution, showed that the critical
exponent varies strongly with the parameter $\kappa\equiv \Gamma-1$ of the EOS. 
More recently, Neilsen and Choptuik \cite{neilsen99b}, using a conservative form 
of the hydrodynamic equations and HRSC schemes, revisited this problem. The 
use of a conservative formulation and numerical schemes well adapted to describe 
ultrarelativistic flows~\cite{neilsen99a}, made it possible to find evidence 
of the existence of critical solutions for large values of the adiabatic index 
$\Gamma$ ($1.89\le\Gamma \le 2$), extending the previous upper limit.



\subsection{Accretion on to black holes}

The study of relativistic accretion and black hole astrophysics is a very 
active field of research, both theoretically and observationally (see, e.g.
\cite{blandford99} and references there in). On the one hand, advances in 
satellite instrumentation, e.g., the Rossi X-Ray Timing Explorer (RXTE), 
and the Advanced Satellite for Cosmology and Astrophysics (ASCA), are 
greatly stimulating and guiding theoretical research on accretion physics. 
The discovery of kHz quasi-periodic oscillations in low-mass X-ray binaries, 
extends the frequency range over which these oscillations occur into timescales 
associated with the innermost regions of the accretion process (for a review 
see~\cite{vanderklis98}). Moreover, in extragalactic sources, spectroscopic 
evidence (broad iron emission lines) increasingly points to (rotating) black 
holes being the accreting central objects~\cite{tanaka95,kormendy95,bromley98}. 
Thick accretion discs (or tori) are probably orbiting the central black holes of many
astrophysical objects such as quasars and other active galactic nuclei (AGNs), 
some X-ray binaries, and microquasars. In addition, they are believed to exist 
at the central engine of cosmic GRBs. 

Disk accretion theory is primarily based on the study of (viscous) stationary
flows and their stability properties through linearized perturbations
thereof. A well-known example is the solution consisting of isentropic
constant angular momentum tori, unstable to a variety of non-axisymmetric
global modes, discovered by Papaloizou and Pringle~\cite{papaloizou84}
(see~\cite{balbus98} for a review of instabilities in astrophysical accretion 
disks). Since the pioneering work by Shakura and Sunyaev~\cite{shakura73}, thin 
disk models, parametrized by the so-called $\alpha-$viscosity, in which the 
gas rotates with Keplerian angular momentum which is transported radially by 
viscous stress, have been applied successfully to many astronomical objects. 
The thin disk model, however, is not valid for high luminosity systems, as 
it is unstable at high mass accretion rates. In this regime Abramowicz et 
al.~\cite{abramowicz88} found the so-called slim disk solution, which is 
stable against viscous and thermal instabilities. More recently, much 
theoretical work has been devoted to the problem of slow accretion, motivated 
by the discovery that many galactic nuclei are under-luminous (e.g., NGC 4258). 
Proposed accretion models involve the existence of advection-dominated 
accretion flows (ADAF solution; see, e.g.~\cite{narayan94,narayan98}) and 
advection-dominated inflow outflow solutions (ADIOS solution~\cite{blandford98}). 
The importance of convection for low values of the viscosity parameter $\alpha$ 
is currently being discussed in the so-called convection-dominated accretion 
flow (CDAF solution; see~\cite{igumenshchev00} and references therein). The 
importance of magnetic fields and its consequences on the stability properties 
of this solution is critically discussed in \cite{balbus01}.

For a wide range of accretion problems, the Newtonian theory of gravity
is adequate for the description of the background gravitational forces 
(see, e.g.,~\cite{frank92}). The extensive experience with Newtonian 
astrophysics has shown that explorations of the relativistic regime 
benefit from the use of model potentials. In particular, the Paczy\'nski-Wiita 
pseudo-Newtonian potential for a Schwarzschild black hole~\cite{paczynski80}, 
gives approximations of general relativistic effects with accuracy of 
$10\%-20\%$ outside the {\it marginally stable} radius, which corresponds 
to three times the Schwarzschild radius. Nevertheless, for comprehensive 
numerical work, a three-dimensional formalism is required, able to cover 
also the maximally rotating hole. In rotating spacetimes the gravitational 
forces cannot be captured fully with scalar potential formalisms. Additionally, 
geometric regions such as the ergo-sphere would be very hard to model without 
a metric description. Whereas the bulk of emission occurs in regions with 
almost Newtonian fields, only the observable features attributed to the 
inner region may crucially depend on the nature of the spacetime.

In the following we present a summary of representative {\it time-dependent} 
accretion simulations in relativistic hydrodynamics. We concentrate on 
multidimensional simulations.  In the limit of spherical accretion,
exact stationary solutions exist for both Newtonian gravity~\cite{bondi52}
and relativistic gravity~\cite{michel72}. Such solutions are
commonly used to calibrate time-dependent hydrodynamical codes,
by analyzing whether stationarity is maintained during a numerical
evolution~\cite{hawley84b,marti91,eulderink95,romero96,banyuls97}.

\subsubsection{Disk accretion}
\label{diskaccretion}

Pioneering numerical efforts in the study of black hole
accretion~\cite{wilson72, hawley84b,hawley86,hawley91} made use
of the the so-called {\it frozen star} paradigm of a black hole.
In this framework, the time ``slicing" of the spacetime is synchronized 
with that of asymptotic observers far from the hole. Within this approach
Wilson~\cite{wilson72} first investigated numerically the time-dependent
accretion of inviscid matter onto a rotating black hole. This was
the first problem to which his formulation of the hydrodynamic equations, 
as presented in Section 2.1.2, was applied. Wilson used an axisymmetric 
hydrodynamical code in cylindrical coordinates to study the formation of 
shock waves and the X-ray emission in the strong-field regions close to 
the black hole horizon, being able to follow the formation of thick accretion 
disks during the simulations.

Wilson's formulation has been extensively used in time-dependent
numerical simulations of thick disk accretion. In a system formed
by a black hole surrounded by a thick disk, the gas flows in an effective
(gravitational plus centrifugal)  potential, whose structure is comparable
with that of a close binary. The Roche torus surrounding the black hole
has a cusp-like inner edge located at the Lagrange point $\mathrm{L}_{1}$
where mass transfer driven by the radial pressure gradient is possible
\cite{abramowicz78}. In~\cite{hawley84b} (see also~\cite{hawley86}) Hawley 
and collaborators studied, in the test-fluid approximation and axisymmetry, 
the evolution and development of non-linear instabilities in pressure-supported
accretion disks formed as a consequence of the spiraling infall of fluid 
with some amount of angular momentum. The code used explicit second-order 
finite difference schemes with a variety of choices to integrate the 
transport terms of the equations (i.e. those involving changes in the 
state of the fluid at a specific volume in space). The code also used
a staggered grid (with scalars located at the cell centers and
vectors at the cell boundaries) for its suitability to difference
the transport equations. Discontinuous solutions were stabilized with
artificial viscosity terms.

With a three-dimensional extension of the axisymmetric code of Hawley et 
al.~\cite{hawley84a,hawley84b} Hawley~\cite{hawley91} studied the
global hydrodynamic non-axisymmetric instabilities in thick, constant
angular momentum accretion gas tori orbiting around a Schwarzschild
black hole. Such simulations showed that the Papaloizu-Pringle instability 
saturates in a strong spiral pressure wave, not in turbulence. In addition, the
simulations confirmed that accretion flows through the torus could
reduce and even halt the growth of the global instability. Extensions to 
Kerr spacetimes have been recently reported in \cite{devilliers02a}.

Yokosawa~\cite{yokosawa93,yokosawa95}, also using Wilson's formulation,
studied the structure and dynamics of relativistic accretion disks
and the transport of energy and angular momentum in magneto-hydrodynamical
accretion on to a rotating black hole. In his code the hydrodynamic equations
are solved using the Flux-Corrected-Transport (FCT) scheme~\cite{boris73}
(a second-order flux-limiter method which avoids oscillations near 
discontinuities by reducing the magnitude of the numerical flux), and 
the magnetic induction equation is solved using the constrained transport 
method~\cite{evans88}. The code contains a parametrized viscosity based 
on the $\alpha$-model~\cite{shakura73}. The simulations revealed different 
flow patterns, inside the marginally stable orbit, depending on the thickness, 
$h$, of the accretion disk. For thick disks with $h\ge 4r_h$, $r_h$ being 
the radius of the event horizon, the flow pattern becomes turbulent.

Igumenshchev and Beloborodov~\cite{igumenshchev97} have performed 
two-dimensional relativistic hydrodynamical simulations of inviscid
transonic disk accretion on to a Kerr black hole. The hydrodynamical 
equations follow Wilson's formulation but the code avoids the use of 
artificial viscosity. The advection terms are evaluated with an 
upwind algorithm which incorporates the PPM scheme~\cite{colella84} 
to compute the fluxes. Their numerical work confirms analytic
expectations: (i) the structure of the innermost disk region 
strongly depends on the black hole spin, and (ii) the mass 
accretion rate goes as $\dot{M}\propto(\Delta W)^{\Gamma/
(\Gamma-1)}$, $\Delta W$ being the potential barrier between the
inner edge of the disk and the cusp, and $\Gamma$ the adiabatic index.

\begin{figure}[t]
\centerline{\epsfig{figure=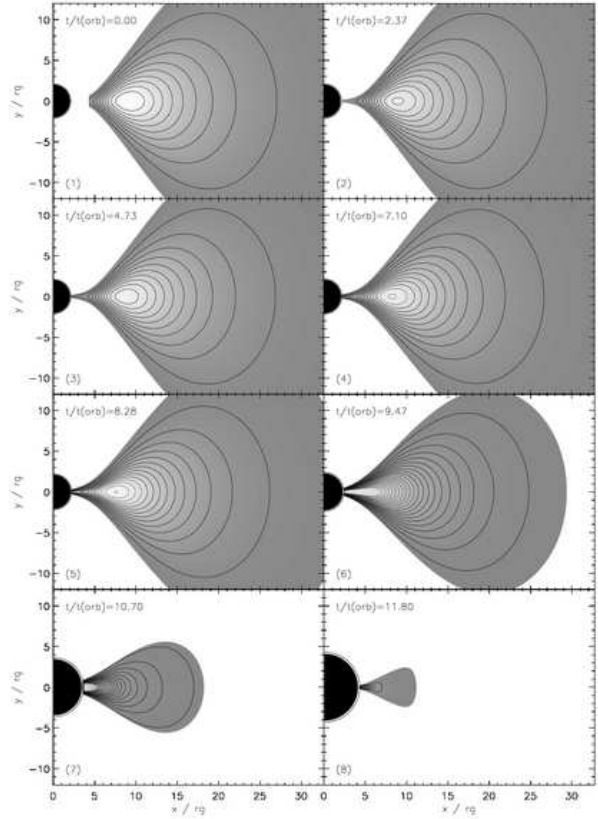,width=7.85cm}}
\vspace{-0.3cm}
\caption{Runaway instability of an unstable thick disk: contour levels of the rest-mass
density $\rho$ plotted at irregular times from $t=0$ to $t=11.80\ t_{\mathrm{orb}}$
once the disk has almost been entirely destroyed. See \cite{font02b} for details.}
\label{runaway}
\end{figure}

Thick accretion disks orbiting black holes, on the other hand, may be 
subjected to the so-called {\it runaway instability}, as first proposed by 
Abramowicz et al.~\cite{abramowicz83}. Starting from an initial disk 
filling its Roche lobe, so that mass transfer is possible through the 
cusp located at the $\mathrm{L}_{1}$ Lagrange point, two evolutions are 
feasible when the mass of the black hole increases: (i) either the cusp 
moves inwards towards the black hole, which slows down the mass transfer, 
resulting in a stable situation, or (ii) the cusp moves deeper inside 
the disk material. In this case the mass transfer speeds up, leading to 
the runaway instability. This instability, whose existence is still a 
matter of debate (see e.g.~\cite{font02b} and references therein),
is an important issue for most models of cosmic GRBs, where the 
central engine responsible for the initial energy release is such a system 
consisting of a thick disk surrounding a black hole.  

In \cite{font02b} Font and Daigne presented time-dependent simulations of 
the runaway instability of constant angular momentum thick disks around black 
holes. The study was performed using a fully relativistic hydrodynamics code 
based on HRSC schemes and the conservative formulation discussed in 
Section~\ref{valencia}. The self-gravity of the disk was neglected and the 
evolution of the central black hole was assumed to be that of a sequence of 
Schwarzschild black holes of varying mass. The black hole mass increase is 
determined by the mass accretion rate across the event horizon. In agreement 
with previous studies based on stationary models, \cite{font02b} found that 
by allowing the mass of the black hole to grow the disk becomes unstable. 
For all disk-to-hole mass ratios considered (between 1 and 0.05), the runaway 
instability appears very fast on a dynamical timescale of a few orbital periods 
(1 $\to$ 100), typically a few 10 ms and never exceeding 1 s, for a $2.5\ 
\mathrm{M_\odot}$ black hole and a large range of mass fluxes ($\dot{m} \ge 
10^{-3}\ \mathrm{M_{\odot}/s}$). 

An example of the simulations performed by \cite{font02b} appears in 
Fig.~\ref{runaway}. This figure shows eight snapshots of the time-evolution 
of the rest-mass density, from $t=0$ to $t=11.8\ t_{\rm orb}$. The contour 
levels are linearly spaced with $\Delta\rho=0.1 \rho_{\mathrm{c}}^{0}$, where 
$\rho_{\mathrm{c}}^{0}$ is the maximum value of the density at the center 
of the initial disk. In Fig.~\ref{runaway} one can clearly follow the 
transition from a quasi-stationary accretion regime (panels (1) to (5)) to 
the rapid development of the runaway instability in about two orbital periods
(panels (6) to (8)). At $t=11.80\ t_{\mathrm{orb}}$, the disk has almost 
entirely disappeared inside the black hole whose size has, in turn, noticeably 
grown.

Extensions of this work to marginally stable (or even stable) constant 
angular momentum disks are reported in Zanotti et al.~\cite{zanotti02} 
(animations can be visualized at the web site listed in Ref.~\cite{tnsweb}). 
Furthermore, recent simulations with non-constant angular momentum disks
and rotating black holes \cite{font02d}, show that the instability is 
strongly suppressed when adding a small increase outwards of the specific 
angular momentum of the disk (much smaller than the Keplerian value).

\subsubsection{Jet formation}
\label{jetformation}

Numerical simulations of relativistic jets propagating through progenitor stellar
models of collapsars have been presented in \cite{aloy00a}. The collapsar scenario,
proposed by~\cite{woosley93}, is currently the most favoured model for 
explaining long duration GRBs. The simulations performed by~\cite{aloy00a}
employ the three-dimensional code GENESIS \cite{aloy99}, with a 2D spherical 
grid and equatorial plane symmetry. The gravitational field of the black hole 
is described by the Schwarzschild metric, and the relativistic hydrodynamic 
equations are solved in the test fluid approximation using a Godunov-type 
scheme. Aloy et al.~\cite{aloy00a} showed that the jet, initially formed by 
an {\it ad hoc} energy deposition of a few $10^{50}$ erg s$^{-1}$ within a 
$30^{\circ}$ cone around the rotation axis, reaches the surface of the collapsar 
progenitor intact, with a maximum Lorentz factor of $\sim 33$.

The most promising processes for producing relativistic jets as those observed 
in AGNs, microquasars and GRBs, involve the hydromagnetic centrifugal acceleration 
of material from the accretion disk~\cite{blandford82}, or the extraction of 
rotational energy from the ergosphere of a Kerr black hole 
\cite{penrose69,blandford77}. Koide and coworkers have performed the first MHD 
simulations of jet formation in general relativity \cite{koide98,koide00,koide02a}.
Their code uses the 3+1 formalism of general relativistic conservation laws of 
particle number, momentum and energy, and Maxwell equations with infinite 
electric conductivity. The MHD equations are numerically solved in the test-fluid 
approximation (in the background geometry of Kerr spacetime) using a finite 
difference symmetric scheme~\cite{davis84}. The Kerr metric is described in 
Boyer-Lindquist coordinates, with a radial tortoise coordinate to enhance the 
resolution near the horizon.

\begin{figure}[t]
\centerline{\psfig{figure=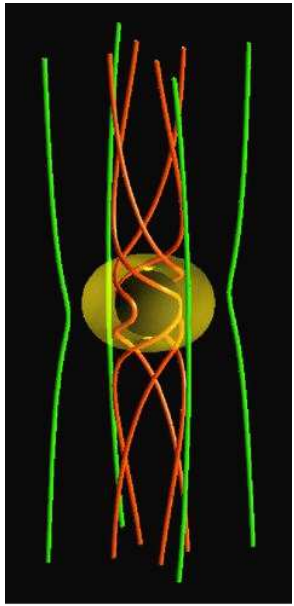,width=1.5in}}
\caption{{Jet formation: twisting of magnetic field lines around a
Kerr black hole (black sphere) The yellow surface is the ergosphere. 
The red tubes show the magnetic field lines that cross into the 
ergosphere. Figure taken from~\cite{koide02a} (used with permission).
}}
\label{fig:koide}
\end{figure}

In \cite{koide02a}, the general relativistic magneto-hydrodynamic behaviour of a 
plasma flowing into a rapidly rotating black hole ($a=0.99995$) in a large-scale 
magnetic field is investigated numerically. The initial magnetic field
is uniform and strong, $B_0=10\rho_0 c^2$, $\rho_0$ being the mass density. The
initial Alfv\'en speed is $v_A=0.953c$. The simulation shows how the rotating black 
hole drags the inertial frames around in the ergosphere. The azimuthal component of 
the magnetic field increases because of the azimuthal twisting of the magnetic field 
lines, as is depicted in Fig.~\ref{fig:koide}. This frame-dragging dynamo amplifies 
the magnetic field to a value which, by the end of the simulation, is three 
times larger than the initial one. The twist of the magnetic field lines 
propagates outwards as a torsional Alfv\'en wave. The magnetic tension torques 
the plasma inside the ergosphere in a direction opposite to that of the black 
hole rotation. Therefore, the angular momentum of the plasma outside receives 
a net increase. Even though the plasma falls into the black hole, electromagnetic 
energy is ejected along the magnetic field lines from the ergosphere, due to 
the propagation of the Alfv\'en wave. By total energy conservation arguments, 
Koide et al.~\cite{koide02a} conclude that the ultimate result of the generation 
of an outward Alfv\'en wave is the magnetic extraction of rotational energy of 
the Kerr black hole, a process the authors call MHD Penrose process. Koide and 
coworkers argue that such process can be applicable to jet formation, both in 
AGNs and GRBs.

We note that, recently, van Putten and Levinson \cite{vanputten02a} have 
considered, theoretically, the evolution of an accretion torus in suspended 
accretion, in connection with GRBs. These authors claim that the formation of 
baryon-poor outflows may be associated with a dissipative gap in a 
differentially rotating magnetic flux tube supported by an equilibrium 
magnetic moment of the black hole. Numerical simulations of non-ideal
MHD, incorporating radiative processes, are necessary to validate this picture.

\subsubsection{Wind accretion}

The term ``wind" or hydrodynamic accretion refers to the capture of matter 
by a moving object under the effects of the underlying gravitational field.
The canonical astrophysical scenario in which matter is accreted in such a
non-spherical way was suggested originally by Bondi, Hoyle and 
Lyttleton~\cite{hoyle39,bondi44}, who studied, using Newtonian gravity, the 
accretion on to a gravitating point mass moving with constant velocity 
through a non-relativistic gas of uniform density. The matter flow
inside the accretion radius, after being decelerated by a conical shock, is 
ultimately captured by the central object.  Such a process applies to describe
mass transfer and accretion in compact X-ray binaries, in particular
in the case in which the donor (giant) star lies inside its Roche lobe and
loses mass via a stellar wind. This wind impacts on the orbiting
compact star forming a bow-shaped shock front around it. This 
process is also believed to occur during the common envelope phase in
the evolution of a binary system.

Numerical simulations have extended the simplified analytic models and have 
helped to develop a thorough understanding of the hydrodynamic accretion
scenario, in its fully three-dimensional character (see, 
e.g.,~\cite{ruffert94,benensohn97} and references therein). The numerical
investigations have revealed the formation of accretion disks and the 
appearance of non-trivial phenomena such as shock waves and tangential
({\it flip-flop}) instabilities. 

Most of the existing numerical work has used Newtonian hydrodynamics
to study the accretion onto non-relativistic stars \cite{ruffert94}.
For compact accretors such as neutron stars or black holes, the correct 
numerical modeling requires a general relativistic hydrodynamical description.
Within the relativistic, frozen star framework, wind accretion onto ``moving" 
black holes was first studied in axisymmetry by Petrich et al.~\cite{petrich89}. 
In this work Wilson's formulation of the hydrodynamic equations was adopted. 
The integration algorithm was borrowed from~\cite{stark87} with the transport 
terms finite-differenced following the prescription given in~\cite{hawley84b}. 
An artificial viscosity term of the form $Q=a\rho(\Delta v)^2$, with $a$ a 
constant, was added to the pressure terms of the equations in order to stabilize 
the numerical scheme in regions of sharp pressure gradients.

An extensive survey of the morphology and dynamics of relativistic wind 
accretion past a Schwarzschild black hole was later performed 
by~\cite{font98a,font98b}. This investigation differs from~\cite{petrich89} 
in both, the use of a conservative formulation for the hydrodynamic 
equations (see Section 2.1.3) and the use of advanced HRSC schemes.  
Axisymmetric computations were compared to~\cite{petrich89} finding 
major differences in the shock location, opening angle and accretion 
rates of mass and momentum. The reasons for the discrepancies are
related to the use of different formulations, numerical schemes and
grid resolution, much higher in \cite{font98a,font98b}. Non-axisymmetric 
two-dimensional studies, restricted to the equatorial plane of the black 
hole, were discussed in~\cite{font98b}, motivated by the non-stationary 
patterns found in Newtonian simulations (see, e.g.,~\cite{benensohn97}). 
The relativistic computations revealed that initially asymptotic uniform 
flows always accrete on to the hole in a stationary way which closely 
resembles the previous axisymmetric patterns.


\begin{figure}[t]
\centerline{\psfig{figure=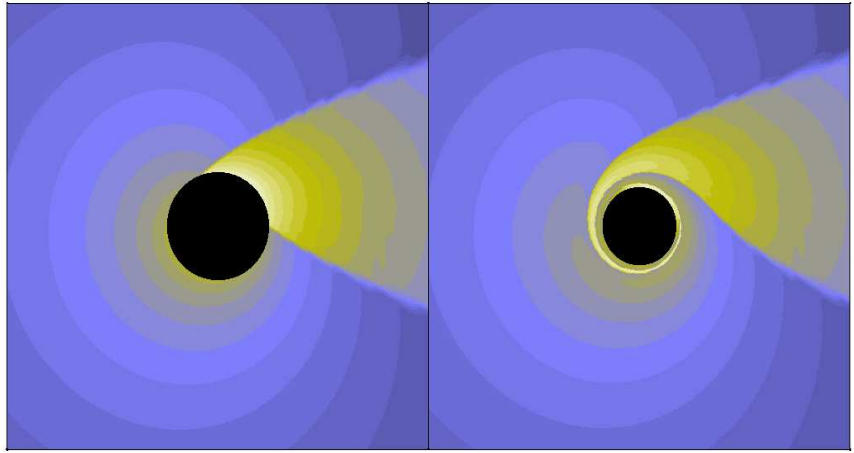,width=5.5in,height=2.9in}}
\caption{{Relativistic wind accretion onto a rapidly rotating
Kerr black hole ($a=0.999M$, the hole spin is counter-clock wise)
in Kerr-Schild coordinates (left panel). Isocontours of the
logarithm of the density are plotted at the final stationary
time $t=500M$. Brighter colors (yellow-white) indicate high density
regions while darker colors (blue) correspond to low density zones.
The right panel shows how the flow solution
looks like when transformed to Boyer-Lindquist coordinates.
The shock appears here totally wrapped around the horizon of
the black hole. The box is $12M$ units long. The simulation
employed a $(r,\phi)$-grid of $200\times 160$ zones. Further
details are given in~\cite{font98c}.
}}
\label{fig4}
\end{figure}

In~\cite{papadopoulos98a} Papadopoulos and Font presented a procedure 
which simplifies the numerical integration of the general relativistic 
hydrodynamic equations near black holes. This procedure is based on 
identifying classes of coordinates in which the black hole metric is 
free of coordinate singularities at the horizon (unlike the commonly 
adopted Boyer-Lindquist coordinates), independent of time, and admits 
a spacelike decomposition. With those coordinates the innermost radial 
boundary can be placed {\it inside} the horizon, allowing for an 
unambiguous treatment of the entire (exterior) physical domain. 
In~\cite{font98c,font99b} this approach was applied to the study of 
relativistic wind accretion onto rapidly rotating black holes. The 
effects of the black hole spin on the flow morphology were found to 
be confined to the inner regions of the black hole potential well. 
Within this region, the black hole angular momentum drags the flow, 
wrapping the shock structure around. An illustrative example is 
depicted in Fig.~\ref{fig4}. The left panel of this figure corresponds 
to a simulation employing the Kerr-Schild form of the Kerr metric, 
regular at the horizon. The right panel shows how the accretion pattern 
would look like were the computation performed using the more common
Boyer-Lindquist coordinates. The transformation induces a noticeable
wrapping of the shock around the central hole. The shock would wrap
infinitely many times before reaching the horizon. As a result, the 
computation in these coordinates would be much more challenging than 
in Kerr-Schild coordinates.

\subsubsection{Gravitational radiation}

Semi-analytical studies of finite-sized collections of dust, shaped in the
form of stars or shells, falling isotropically onto a black hole are
available in the literature~\cite{naka81b,haugan82,shapiro82,
oohara83,petrich85}. These investigations approximate gravitational 
collapse by a dust shell of mass $m$ falling into a Schwarzschild black 
hole of mass $M\gg m$. These studies have shown that 
for a fixed amount of infalling mass the gravitational radiation efficiency
is reduced compared to the point particle limit ($\Delta E\sim 0.0104 m^2/M$), 
never exceeding that of a particle with the same mass, the reason being 
cancellations of the emission from distinct parts of the extended object.


\begin{figure}[t]
\centerline{\psfig{figure=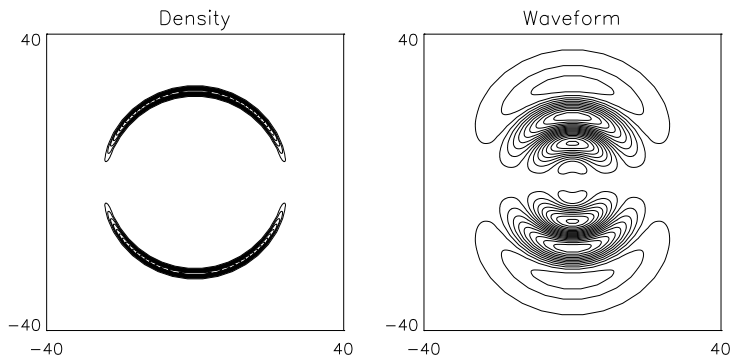,width=4.2in,height=2.5in}}
\caption{{
Stills from a movie showing the time evolution of the accretion/collapse 
of a quadrupolar shell onto a Schwarzshild black hole. The left panel shows 
isodensity contours and the right panel the associated gravitational waveform.
The shell, initially centered at $r_*=35M$, is gradually accreted by the
black hole, a process which perturbs the black hole and triggers the
emission of gravitational radiation. After the burst, the remaining
evolution shows the decay of the black hole quasi-normal mode ringing.
By the end of the simulation a spherical accretion solution is reached.
Further details are given in Ref.~\cite{papadopoulos99a}. (To see the movie, 
please go to the electronic version of this review article at 
{\tt http://www.livingreviews.org/Articles/Volume3/2000-2font.})
}}
\label{mpeg2}
\end{figure}

In~\cite{papadopoulos99a} such conclusions were corroborated with
numerical simulations of the gravitational radiation emitted during the 
accretion process of an extended object onto a black hole. The first
order deviations from the exact black hole geometry were approximated
using curvature perturbations induced by matter sources whose non-linear
evolution was integrated using a (non-linear) hydrodynamics code (adopting 
the conservative formulation of Section~\ref{valencia} and HRSC schemes). All 
possible types of curvature perturbations are captured in the real and 
imaginary parts of the Weyl tensor scalar (see, e.g.,~\cite{chandrasekhar83}). 
In the framework of the Newman-Penrose formalism the equations for the 
perturbed Weyl tensor components decouple, and when written in the 
frequency domain even separate~\cite{teukolsky72}. Papadopoulos and 
Font~\cite{papadopoulos99a} used the limiting case for Schwarzschild 
black holes, i.e., the inhomogeneous Bardeen-Press equation~\cite{bardeen72}. 
The simulations showed the gradual excitation of the black hole quasi-normal 
mode frequency by sufficiently compact shells.

An example of these simulations appears in the MPEG movie of Fig.~\ref{mpeg2}.
This movie shows the time evolution of the shell density (left panel) 
and the associated gravitational waveform during a complete accretion/collapse 
event. The (quadrupolar) shell is parametrized according to 
$\rho=\rho_0+e^{-k(r_*-r_0)^2}\sin^2\theta$ with $k=2, \rho_0=10^{-2}$ and 
$r_0=35M$. Additionally $r_*$ denotes a logarithmic radial (Schwarzschild)
coordinate. The animation shows the gradual collapse of the shell on to
the black hole. This process triggers the emission of gravitational
radiation. In the movie it can be clearly seen how the burst of the
emission coincides with the most dynamical accretion phase, when the shell
crosses the peak of the potential and is subsequently captured by the hole.
The gravitational wave signal coincides with the quasinormal ringing
frequency of the Schwarzschild black hole, $17M$. The existence of an 
initial burst, separated in time from the {\it physical} burst, is also 
noticeable in the movie. It just reflects the gravitational radiation content
of the initial data (see~\cite{papadopoulos99a} for a detailed explanation).

One-dimensional numerical simulations of a self-gravitating perfect fluid 
accreting onto a black hole were presented in~\cite{papadopoulos01a}, where 
the effects of mass accretion during the gravitational wave emission from 
a black hole of growing mass were explored. Using the conservative formulation 
outlined in Section 2.2.2 and HRSC schemes, Papadopoulos and 
Font~\cite{papadopoulos01a} performed the simulations adopting an ingoing 
null foliation of a spherically symmetric black hole spacetime 
\cite{papadopoulos99b}. Such a foliation penetrates the black hole horizon, 
allowing for an unambiguous numerical treatment of the inner boundary. 
The essence of non-spherical gravitational perturbations was captured by 
adding to the (characteristic) Einstein-perfect fluid system, the evolution 
equation for a minimally coupled massless scalar field. The simulations 
showed the familiar damped-oscillatory radiative decay, with both decay rate 
and frequencies being modulated by the mass accretion rate. Any appreciable 
increase in the horizon mass during the emission reflects on the instantaneous 
signal frequency, $f$, which shows a prominent negative branch in the 
$\dot{f}(f)$ evolution diagram. The features of the frequency evolution 
pattern reveal key properties of the accretion event, such as the total 
accreted mass and the accretion rate.  

Recently, Zanotti et al.~\cite{zanotti02b} have performed hydrodynamical
simulations of constant angular momentum thick disks (of typical neutron 
star densities) orbiting a Schwarzschild black hole. Upon the introduction 
of perturbations, these systems either become unstable to the runaway
instability~\cite{font02b} or exhibit a regular oscillatory behaviour 
resulting in a quasi-periodic variation of the accretion rate as well as 
of the mass quadrupole (animations can be visualized at the web site listed 
in Ref.~\cite{tnsweb}). Zanotti et al.~\cite{zanotti02b} have found 
that the latter is responsible for the emission of intense gravitational 
radiation whose amplitude is comparable or larger than the one
expected in stellar core collapse. The strength of the gravitational
waves emitted and their periodicity are such that signal-to-noise ratios
$\sim {\cal O}(1)-{\cal O}(10)$ can be reached for sources at 20 Mpc or
10 Kpc, respectively, making these new sources of gravitational waves
potentially detectable.


\subsection{Hydrodynamical evolution of neutron stars}

The numerical investigation of many interesting astrophysical processes 
involving neutron stars, such as the rotational evolution of proto-neutron 
stars, which can be affected by a dynamical bar-mode instability and by the
Chandrasekhar-Friedman-Schutz instability, or the gravitational radiation 
from unstable pulsation modes or, more importantly, from the catastrophic 
coalescence and merger of neutron star compact binaries, requires the 
ability of accurate, long-term hydrodynamical evolutions employing 
relativistic gravity. These scenarios are receiving increasing attention in 
recent years~\cite{shibata98,mathews98,font99a,miller99,shibata99,shibata99b,
shibata99c,font00a,font02a,shibata02c}. 

\subsubsection{Pulsations of relativistic stars}
\label{pulsations}

The use of relativistic hydrodynamical codes to study the stability properties 
of neutron stars and to compute mode frequencies of oscillations of such objects
is increasing in recent years (see, e.g. the {\it Living Reviews} article by 
Stergioulas~\cite{stergioulas02} and references therein). An obvious advantage 
of the hydrodynamical approach is that it includes, by construction, nonlinear 
effects, which are important in situations where the linearized equations, 
commonly employed in calculations of mode-frequencies of pulsating stars, break 
down. In addition, due to the current status of hydrodynamics codes, the computation
of mode frequencies in rapidly rotating relativistic stars might be easier to 
achieve through nonlinear numerical evolutions than using perturbative 
computations (see, e.g. the results in \cite{font02a,shibata02c}). 

Hydrodynamical evolutions of polytropic models of spherical neutron stars 
can be used as test-bed computations for multidimensional codes. Representative
examples are the simulations by \cite{gourgoulhon91}, with pseudo-spectral 
methods, and by~\cite{romero96} with HRSC schemes. These investigations 
adopted radial gauge polar slicing coordinates in which the general relativistic 
equations are expressed in a simple way which resembles Newtonian hydrodynamics. 
Gourgoulhon~\cite{gourgoulhon91} used a numerical code to detect, dynamically, 
the zero value of the fundamental mode of a neutron star against radial 
oscillations. Romero et al.~\cite{romero96} highlighted the accuracy of HRSC 
schemes by finding, numerically, a change in the stability behavior of two 
slightly different initial neutron star models: for a given EOS,
a model with mass $1.94532M_{\odot}$ 
is stable and a model of $1.94518M_{\odot}$ is unstable. More recently, in 
\cite{sperhake01a} a method based on the non-linear evolution of deviations 
from a background stationary equilibrium star was applied to study nonlinear 
radial oscillations of a neutron star. The accuracy of the approach permitted 
a detailed investigation of nonlinear features such as quadratic and higher 
order mode coupling and nonlinear transfer of energy.

Axisymmetric pulsations of rotating neutron stars can be excited in
several scenarios, such as core-collapse, crust and core-quakes and
binary mergers, and could become detectable either in gravitational
waves or high-energy radiation. An observational detection of such
pulsations would yield valuable information about the EOS of relativistic
stars.  As a first step towards the study of pulsations of rapidly-rotating 
relativistic stars, Font, Stergioulas and Kokkotas \cite{font00a} 
developed an axisymmetric numerical code which integrates the equations 
of general relativistic hydrodynamics in a fixed background spacetime. 
The finite-difference code is based on a state-of-the-art approximate 
Riemann solver \cite{donat96} and incorporates different second- and 
third-order TVD and ENO numerical schemes. This code is capable of 
accurately evolving rapidly rotating stars for many rotational periods,
even for stars at the mass-shedding limit. The test simulations reported 
in \cite{font00a} showed that, for non-rotating stars, small amplitude 
oscillations have frequencies that agree below the 1\% level with linear, 
radial and non-radial, normal mode frequencies in the so-called Cowling 
approximation (i.e. when the evolution of the spacetime variables is
neglected). Axisymmetric modes of pulsating non rotating stars are 
computed in \cite{siebel02b} both in Cowling and fully coupled evolutions. 
Contrary to the 2+1 approach followed by \cite{font00a}, the code used in 
Ref.~\cite{siebel02b} evolves the relativistic stars on null spacetime 
foliations (see Section \ref{papafont}).

Until very recently (see below), the quasi-radial modes of rotating 
relativistic stars had been studied only under simplifying assumptions, 
such as in the slow-rotation approximation or in the relativistic Cowling 
approximation. An example of the latter is presented in \cite{font01a}, 
where using the code developed by \cite{font00a}, a comprehensive study 
of all low-order axisymmetric modes of uniformly and rapidly rotating 
relativistic stars was presented. The frequencies of quasi-radial and 
non-radial modes with spherical harmonic indices $\ell=0,1,2$ and $3$ 
were computed through Fourier transforms of the time evolution of selected 
fluid variables. This was done for a sequence of appropriately perturbed 
stationary rotating stars, from the non-rotating limit to the mass-shedding 
limit. The frequencies of the axisymmetric modes are affected significantly 
by rotation only when the rotation rate exceeds about 50\% of the maximum 
allowed. As expected, at large rotation rates, apparent mode crossings 
between different modes appear. 

In \cite{font02a}, the first mode frequencies of uniformly rotating stars in 
full general relativity and rapid rotation were obtained, using the 
three-dimensional code \verb+GR_ASTRO+ described in Section \ref{grastro}. 
Such frequencies were computed both in fixed spacetime evolutions (Cowling 
approximation) and in coupled hydrodynamical and spacetime evolutions. The 
simulations used a sequence of (perturbed) $N=1$, $K=100$ ($G=c=M_{\odot}=1$) 
polytropes of central density $\rho_{\mbox{c}}=1.28 \times 10^{-3}$, in which 
the rotation rate $\Omega$ varies from zero to 97\% of the maximum allowed 
rotational frequency, $\Omega_K = 0.5363 \times 10^4$ s$^{-1}$. The Cowling 
runs allowed a comparison with earlier results reported by \cite{font01a}, 
obtaining agreement at the 0.5\% level. The fundamental mode-frequencies and 
their first overtones obtained in fully coupled evolutions show a dependence 
on the increased rotation which is similar to the one observed for the 
corresponding frequencies in the Cowling approximation~\cite{font01a}.

Relativistic hydrodynamical simulations of nonlinear $r$-modes are presented
in \cite{stergioulas01a} (see also \cite{lindblom01a} for Newtonian simulations). 
The gravitational radiation reaction driven instability of the $r$-modes might
have important astrophysical implications, provided that the instability 
does not saturate at low amplitudes by nonlinear effects or by dissipative 
mechanisms. Using a version of the \verb+GR_ASTRO+ code, Stergioulas and 
Font~\cite{stergioulas01a} found evidence that the maximum $r$-mode amplitude 
in isentropic stars is of order unity. The dissipative mechanisms proposed by 
different authors to limit the mode amplitude include shear and bulk viscosity, 
energy loss to a magnetic field driven by differential rotation, shock waves, 
or the slow leak of the $r$-mode energy into some short wavelength oscillation 
modes (see \cite{arras02a} and references therein). The latter mechanism would 
dramatically limit the $r$-mode amplitude to a small value, much smaller than 
those found in the simulations of~\cite{stergioulas01a,lindblom01a} (see 
\cite{stergioulas02} for a complete list of references on the subject). Energy 
leak of the $r$-mode into other fluid modes has been recently considered by 
\cite{gressman02a} through Newtonian hydrodynamical simulations, finding a 
catastrophic decay of the amplitude only once it has grown to a value larger 
than that reported by \cite{arras02a}.

The bar-mode instability in differentially rotating stars in general relativity
has been analyzed by \cite{shibata00a} by means of 3+1 hydrodynamical simulations.
Using the code discussed in Section \ref{shibata}, Shibata et al.~\cite{shibata00a} 
found that the critical ratio of rotational kinetic energy to gravitational binding 
energy for compact stars with $M/R\ge 0.1$ is $\sim 0.24 - 0.25$, slightly below
the Newtonian value $\sim 0.27$ for incompressible Maclaurin spheroids. 
All unstable stars are found to form bars on a dynamical timescale.

\subsubsection{Binary neutron star coalescence}

Many of the current efforts in code development in relativistic astrophysics 
aim at the simulation of the coalescence of compact binaries. Neutron 
star binaries are among the most promising sources of gravitational 
radiation to be detected by the various ground-based interferometers  
worldwide. The computation of the gravitational waveform during the most 
dynamical phase of the coalescence and plunge depends crucially on 
hydrodynamical, finite-size effects. This phase begins once the stars, 
initially in quasi-equilibrium orbits of gradually smaller orbital radius, 
due to the emission of gravitational waves, reach the so-called innermost 
stable circular orbit. About $\sim 10^8$ years after formation of the binary 
system the gravitational wave frequency enters the LIGO/VIRGO high frequency 
band. The final plunge of the two objects takes place on a dynamical timescale 
of a few ms. During the last 15 minutes or so until the stars finally merge, 
the frequency is inside the LIGO/VIRGO sensitivity range. About 16000 cycles 
of waveform oscillation can be monitored, while the frequency gradually 
shifts from $\sim 10$ Hz to $\sim 1$ kHz. A perturbative treatment of the 
gravitational radiation in the quadrupole approximation is valid as long 
as $M/R\ll1$ and $M/r\ll1$ simultaneously, $M$ being the total mass 
of the binary, $R$ the neutron star radius and $r$ the separation of the 
two stars. As the stars approach each other and merge, both inequalities 
are less valid and fully relativistic hydrodynamical calculations become 
necessary.

The accurate simulation of a binary neutron star coalescence is, however,
one of the most challenging tasks in numerical relativity. These scenarios
involve strong gravitational fields, matter motion with (ultra-) relativistic 
speeds, and relativistic shock waves. The numerical difficulties are exacerbated 
by the intrinsic multidimensional character of the problem and by the 
inherent complexities in Einstein's theory of gravity, such as coordinate 
degrees of freedom and the possible formation of curvature singularities 
(e.g. collapse of matter configurations to black holes). It is thus not 
surprising that most of the (few) available simulations have been attempted 
in the Newtonian (and post-Newtonian) framework (see \cite{rasio99} for a 
review). Many of these studies employ Lagrangian particle methods such as 
SPH, and only a few have considered (less viscous) high-order finite 
difference methods such as PPM~\cite{ruffert98}. 

Concerning relativistic simulations Wilson's formulation of the hydrodynamic
equations (see Section \ref{wilson}) was used in 
Refs.~\cite{wilson95,wilson96,mathews99}. Such investigations assumed a 
conformally flat 3-metric, which reduces the (hyperbolic) gravitational 
field equations to a coupled set of elliptic (Poisson-like) equations for 
the lapse function, the shift vector, and the conformal factor. These early 
simulations revealed the unexpected appearance of a ``binary-induced collapse 
instability" of the neutron stars, prior to the eventual collapse of the 
final merged object. This effect was reduced, but not eliminated fully, in 
revised simulations \cite{mathews99}, after Flanagan~\cite{flanagan98} 
pointed out an error in the momentum constraint equation as implemented 
by Wilson and coworkers~\cite{wilson95,wilson96}. A summary of this controversy 
can be found in~\cite{rasio99}. Subsequent numerical simulations with the 
full set of Einstein equations (see below) did not find this effect.

Nakamura and co-workers have been pursueing a programme to simulate
neutron star binary coalescence in general relativity since the late
1980's (see, e.g.,~\cite{naka98}). The group developed a three-dimensional code
which solves the full set of Einstein equations and self-gravitating matter 
fields~\cite{oohara96}. The equations are finite-differenced in a uniform 
Cartesian grid using van Leer's scheme~\cite{vanleer79} with TVD flux 
limiters. Shock waves are spread out using a tensor artificial viscosity 
algorithm. The hydrodynamic equations follow Wilson's Eulerian formulation 
and the ADM formalism is adopted for the Einstein equations. This code has 
been tested by the study of the gravitational collapse of a rotating polytrope 
to a black hole (comparing to the axisymmetric computation of Stark and 
Piran~\cite{stark85}). Further work to achieve long term stability in 
simulations of neutron star binary coalescence is under way~\cite{oohara96}. 
We note that the hydrodynamics part of this code is at the basis of Shibata's 
code (Section \ref{shibata}) which has successfully been applied to simulate 
the binary coalescence problem (see below).


\begin{figure}[t]
\centerline{\psfig{figure=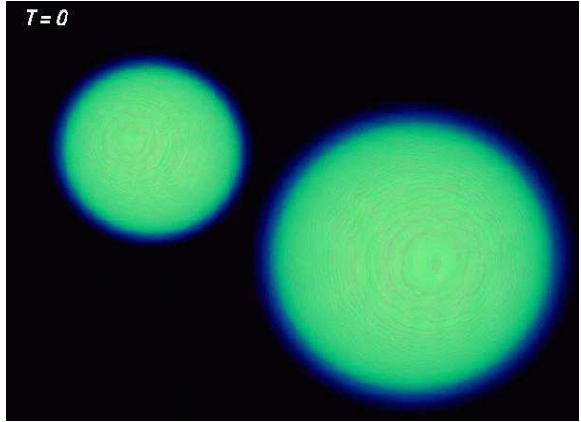,width=3.0in,height=2.2in}}
\caption{{Still from a movie showing the animation of a head-on collision 
simulation of two $1.4M_{\odot}$ neutron stars obtained with a relativistic 
code~\cite{font99a,miller99}. The movie shows the evolution of the density 
and internal energy. The formation of the black hole in prompt timescales is 
demonstrated by the sudden appearance of the apparent horizon at $t=0.16$ ms 
($t=63.194$ in code units), which is indicated by violet dotted circles 
representing the trapped photons. 
(To see the movie, please go to the electronic version of
this review article at 
{\tt http://www.livingreviews.org/Articles/Volume3/2000-2font.})
See~\cite{headon} for download options of higher quality versions of this movie.
}}
\label{movie4}
\end{figure}

The head-on collision of two neutron stars (a limiting case of a coalescence
event) was considered by Miller et al.~\cite{miller99}, who performed 
time-dependent relativistic simulations using the code described in Section 
\ref{grastro}. These simulations analyzed whether the collapse of the final 
object occurs in prompt timescales (a few milliseconds) or delayed (after 
neutrino cooling) timescales (a few seconds). In~\cite{shapiro98} it was argued 
that in a head-on collision event, sufficient thermal pressure is generated 
to support the remnant in quasi-static equilibrium against (prompt) collapse 
prior to slow cooling via neutrino emission (delayed collapse). 
In~\cite{miller99}, prompt collapse to a black hole was found in the head-on 
collision of two $1.4M_{\odot}$ neutron stars modeled by a polytropic EOS 
with $\Gamma=2$ and $K=1.16\times 10^5$ cm$^5$/g\,s$^2$. The stars, initially 
separated by a proper distance of $d=44$ km, were boosted towards one another 
at a speed of $\sqrt{GM/d}$ (the Newtonian infall velocity). The simulation 
employed a Cartesian grid of $192^3$ points. The time evolution of this simulation 
can be followed in the Quicktime movie in Fig.~\ref{movie4}. This animation 
simultaneously shows the rest-mass density and the internal energy evolution 
during the on-axis collision. The formation of the black hole in prompt 
timescales is signalled by the sudden appearance of the apparent horizon at 
$t=0.16$ ms ($t=63.194$ in code units). The violet dotted circles indicate the 
trapped photons. The animation also shows a moderately relativistic shock wave 
(Lorentz factor $\sim 1.2$) appearing at $t\sim 36$ (code units; yellow-white 
colors), which eventually is followed by two opposite moving shocks (along 
the infalling $z$ direction) which propagate along the atmosphere surrounding 
the black hole.


The most advanced simulations of neutron star coalescence in full general 
relativity are those performed by Shibata and 
Uryu \cite{shibata99,shibata99c,shibata02a}. Their numerical code, briefly 
described in Section~\ref{shibata}, allows the long-term simulation of the 
coalescences of both, irrotational and corotational binaries, from the 
innermost stable circular orbit up to the formation and ringdown of the final 
collapsed object (either a black hole or a stable neutron star), includes an 
apparent horizon finder, and can extract the gravitational waveforms emitted 
in the collisions. Shibata and Uryu have performed simulations for a large 
sample of parameters of the binary system, such as the compactness of the 
(equal mass) neutron stars ($0.12 \le M/R \le 0.16$), the adiabatic index of 
the $\Gamma$-law EOS ($1.8 \le \Gamma \le 2.5$), and the maximum density, 
rest mass, gravitational mass or total angular momentum. The initial data 
correspond to quasi-equilibrium states, either corotational or irrotational, 
the latter being more realistic from considerations of viscous versus 
gravitational radiation timescales. These initial data are obtained by solving 
the Einstein constraint equations and the equations for the gauge variables 
under the assumption of a conformally flat 3-metric and the existence of a 
helicoidal Killing vector (see \cite{shibata02a} for a detailed explanation). 
The binaries are chosen at the innermost orbits for which the Lagrange points 
appear at the inner edge of the neutron stars, and the plunge is induced by 
reducing the initial angular momentum by $\sim 2-3\%$.

\begin{figure}[t]
\begin{center}
\epsfxsize=1.40in
\leavevmode
\epsffile{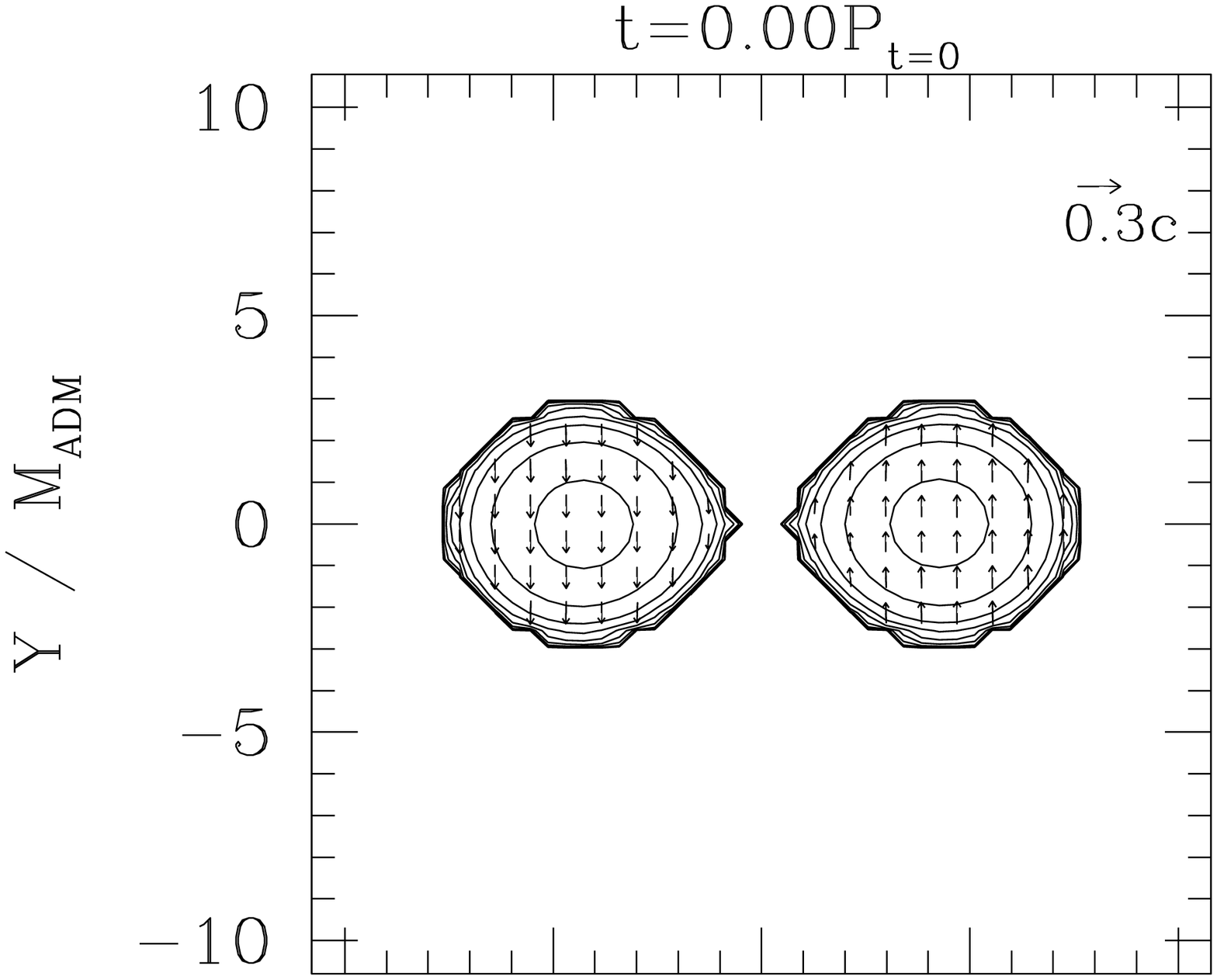}
\epsfxsize=1.40in
\leavevmode
\hspace{-1.0cm}\epsffile{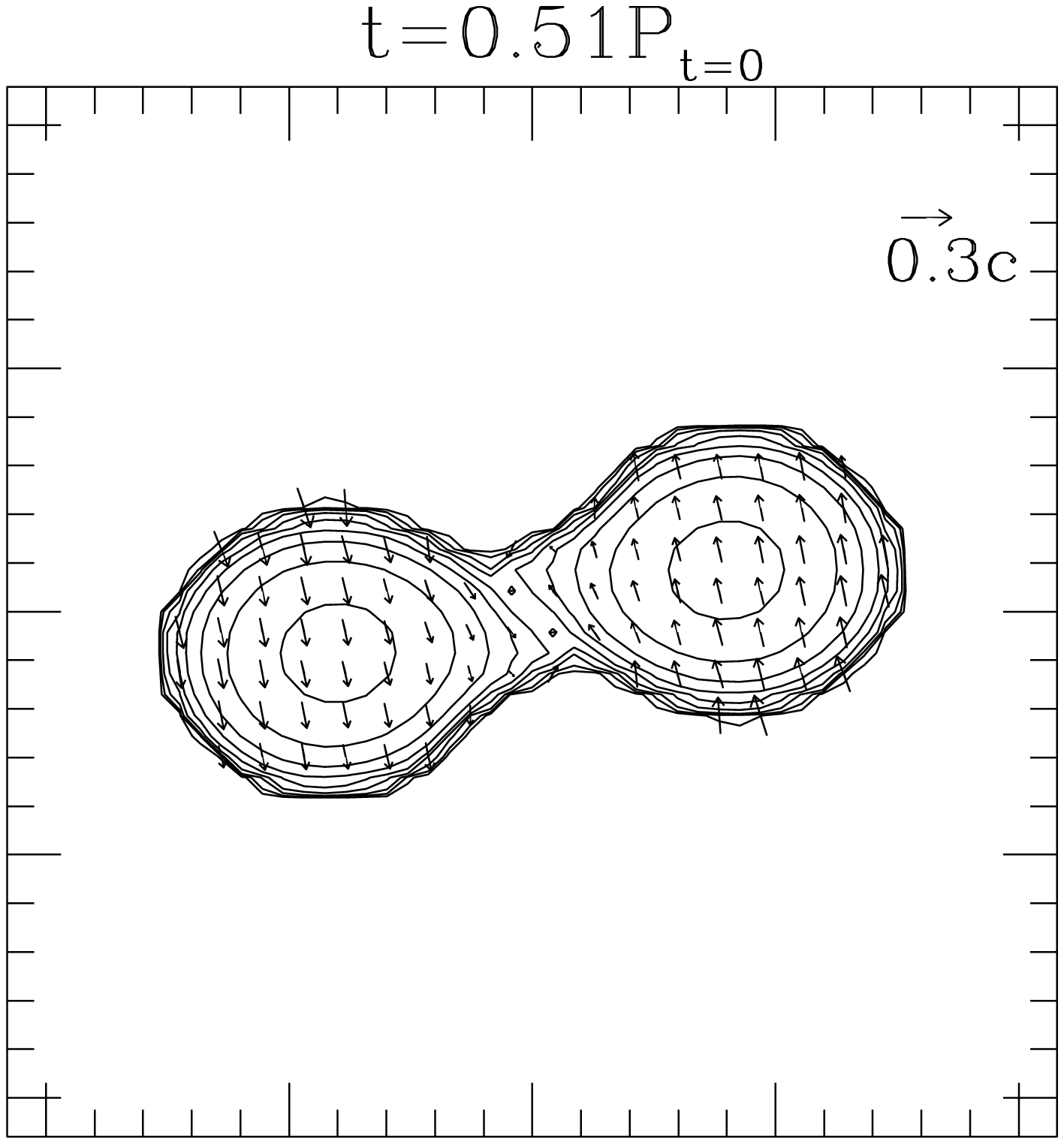} 
\epsfxsize=1.40in
\leavevmode
\hspace{-1.0cm}\epsffile{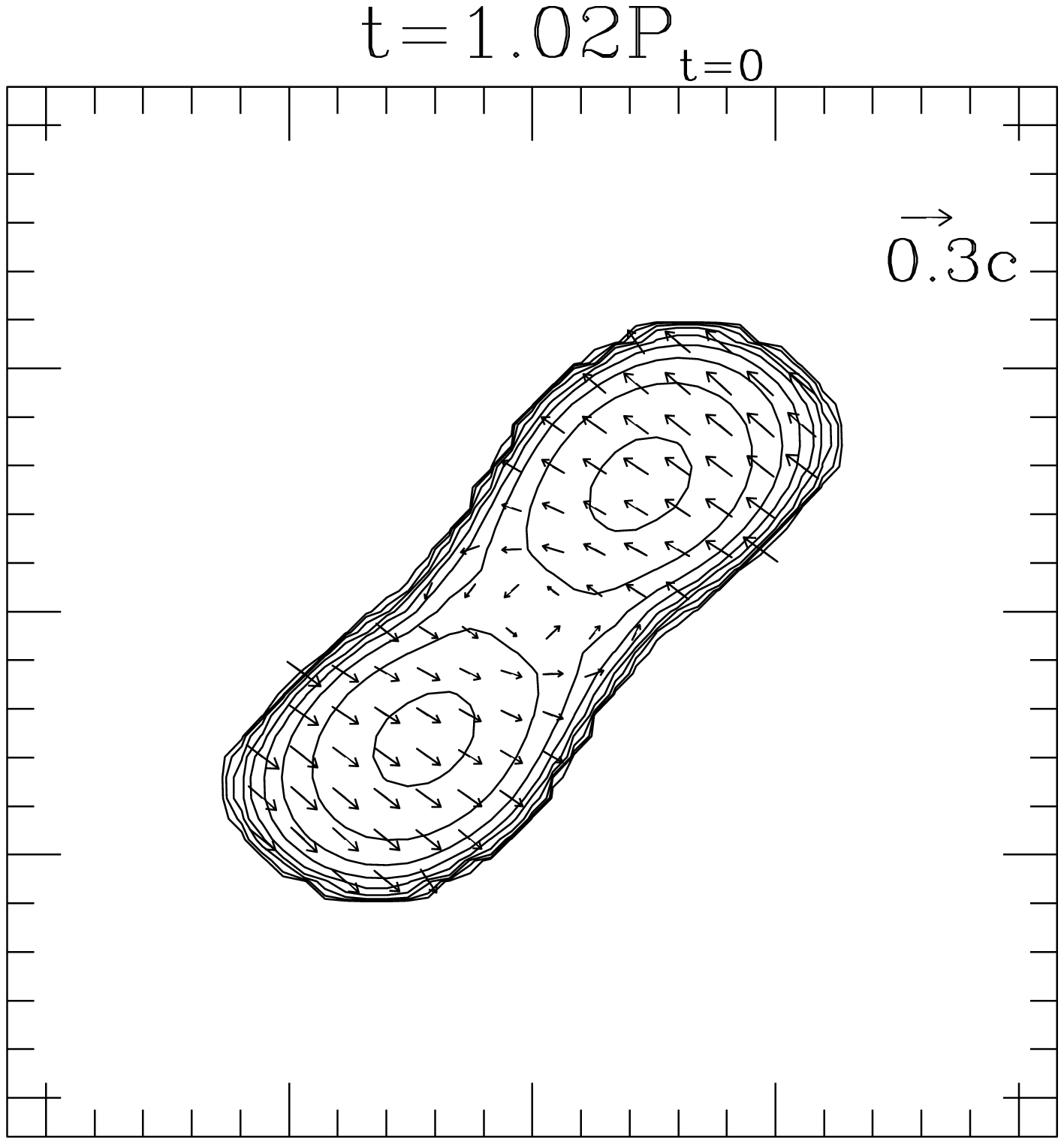} \\
\vspace{-0.7cm}
\epsfxsize=1.40in
\leavevmode
\epsffile{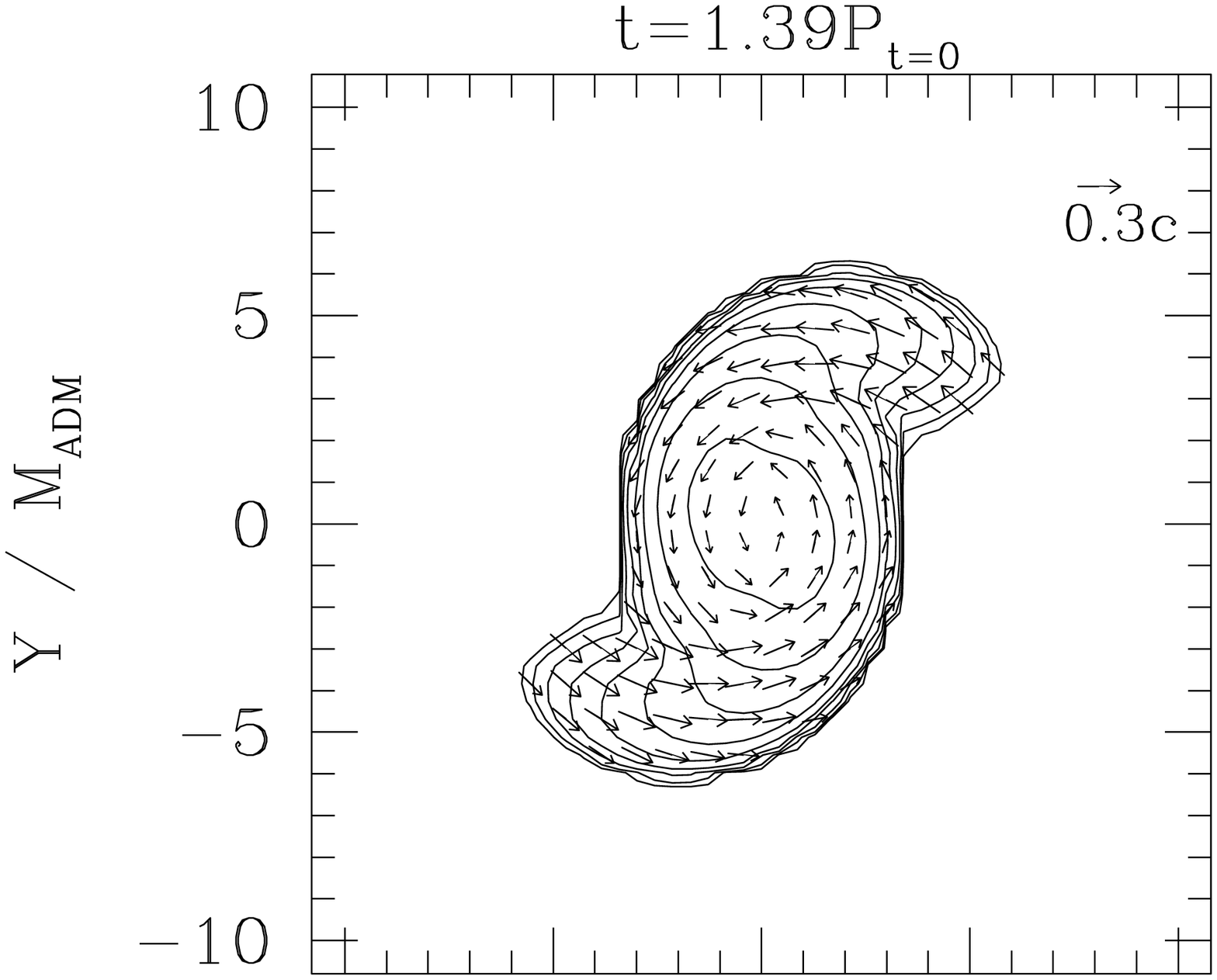} 
\epsfxsize=1.40in
\leavevmode
\hspace{-1.0cm}\epsffile{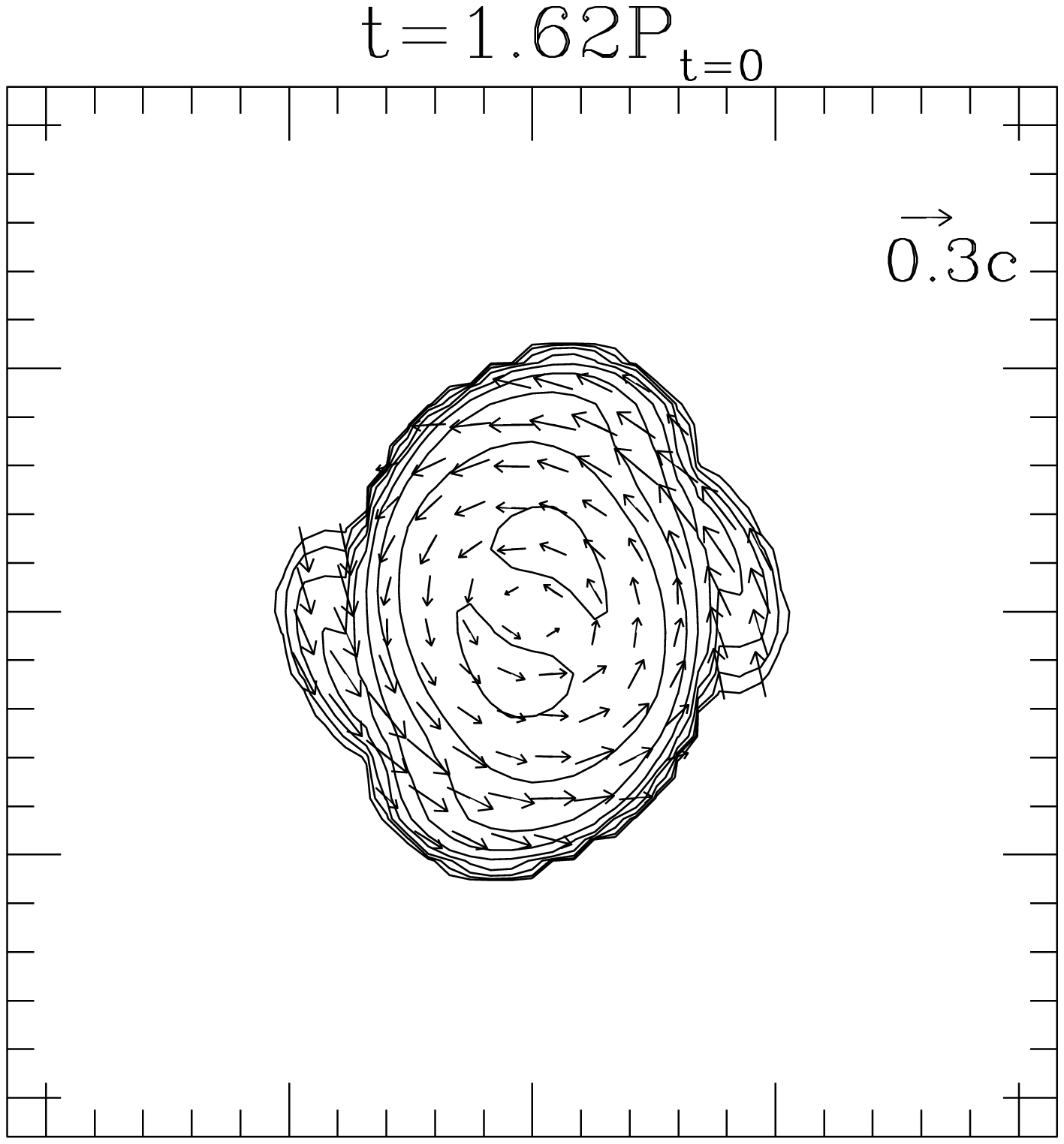}
\epsfxsize=1.40in
\leavevmode
\hspace{-1.0cm}\epsffile{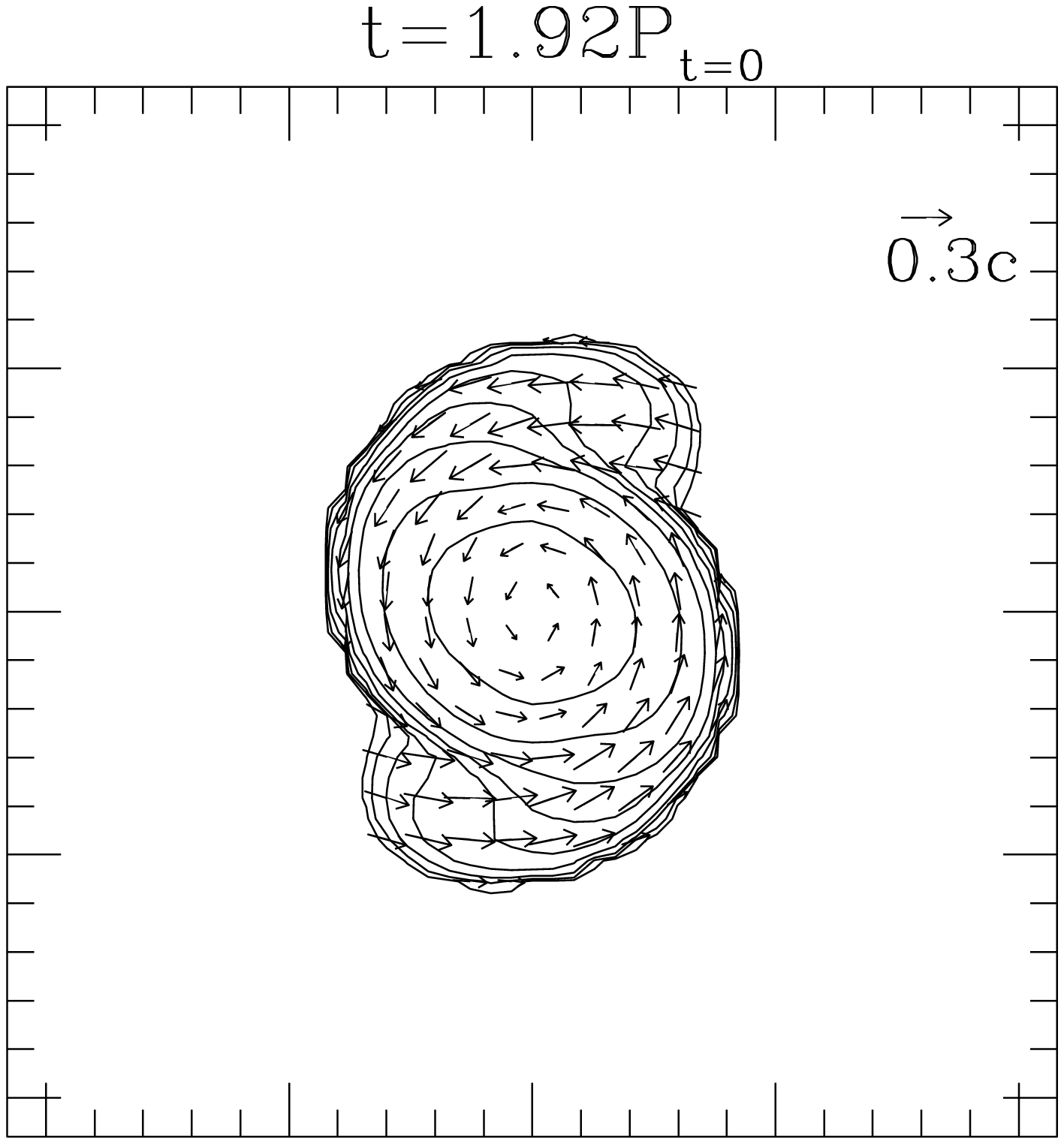}\\ 
\vspace{-0.7cm}
\epsfxsize=1.40in
\leavevmode
\epsffile{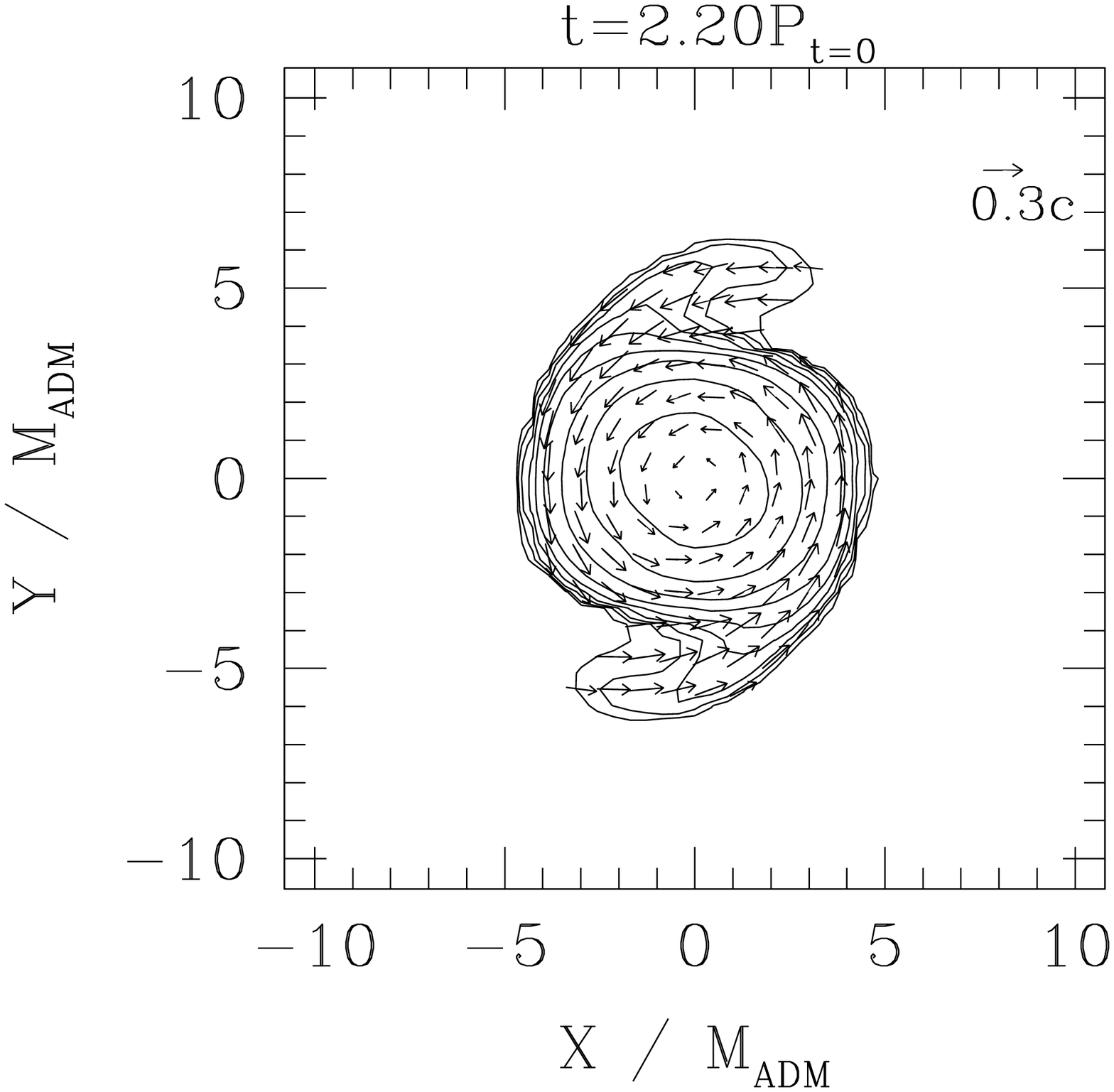} 
\epsfxsize=1.40in
\leavevmode
\hspace{-1.0cm}\epsffile{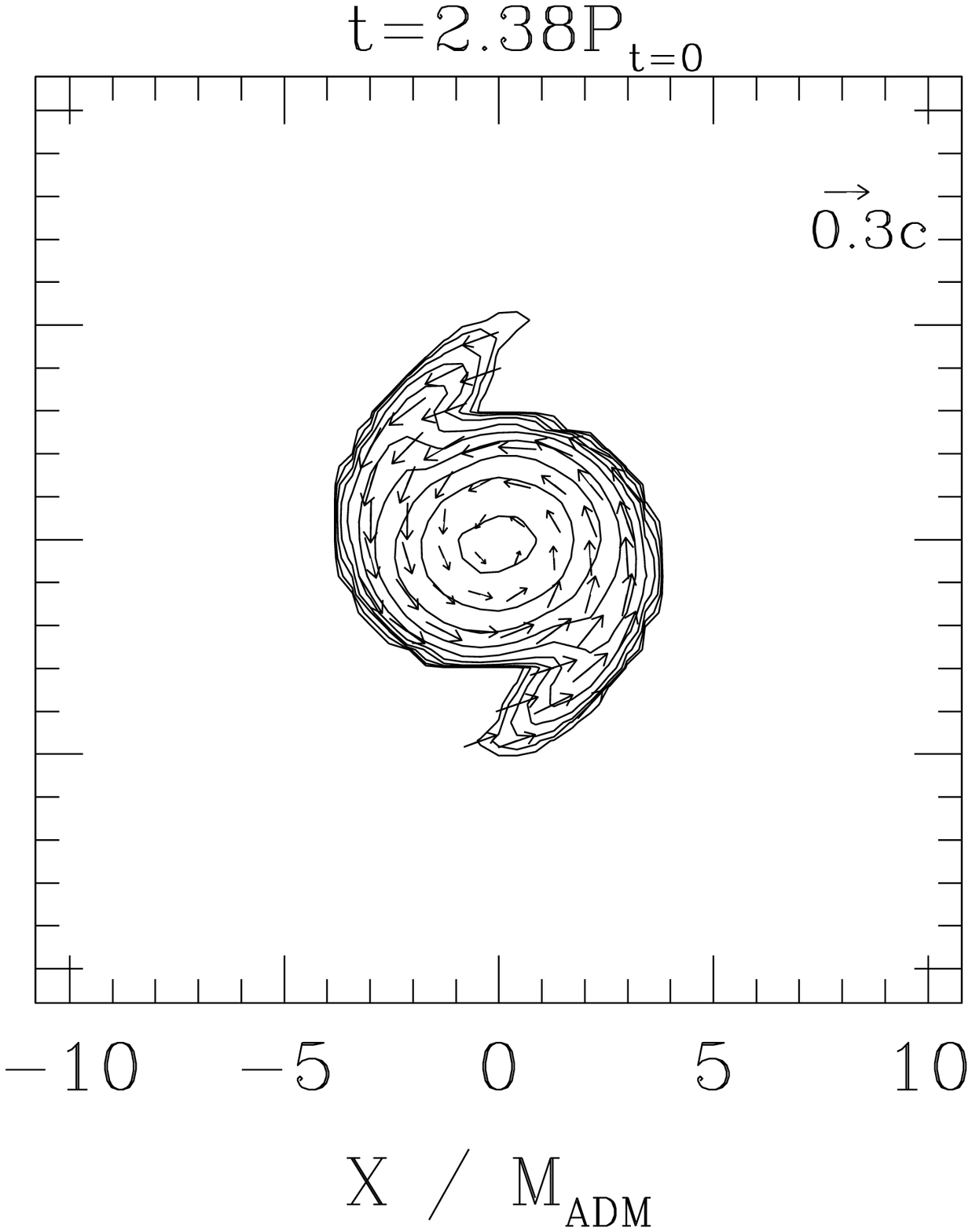}
\epsfxsize=1.40in
\leavevmode
\hspace{-1.0cm}\epsffile{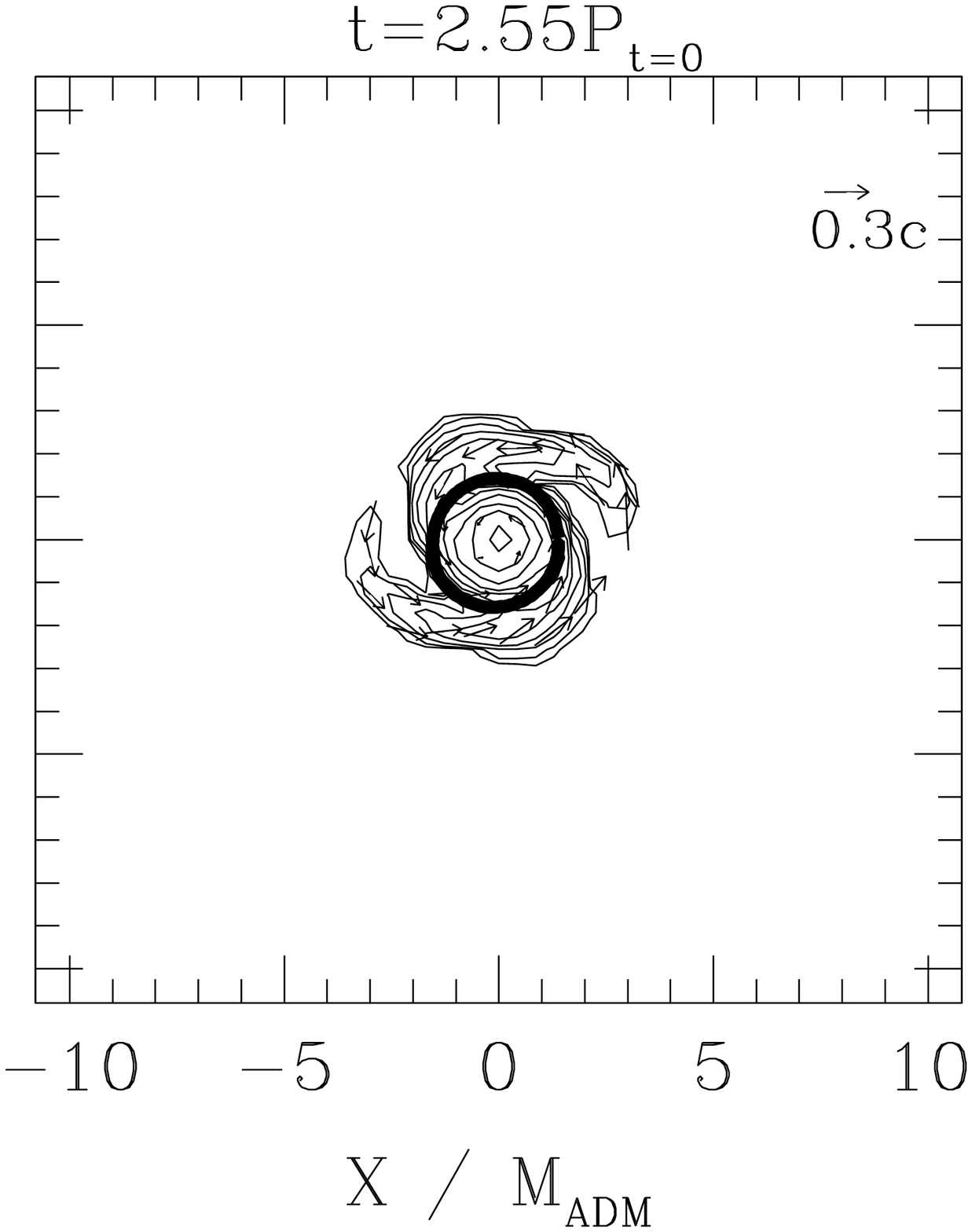} 
\end{center}
\vspace{-0.1cm}
\caption{Snapshots of the density contours in the equatorial plane for
a binary neutron star coalescence which leads to a rotating black hole
(see \cite{shibata02a} for the characteristics of the model). 
Vectors indicate the local velocity field $(v^x,v^y)$. P$_{t=0}$ denotes
the orbital period of the initial configuration.
The thick solid circle in the final panel indicates the apparent horizon. 
The figure is taken from \cite{shibata02a} (used with permission).
\label{fig:nsns}}
\end{figure}

The comprehensive parameter space survey carried out by 
\cite{shibata99,shibata99c,shibata02a} shows that the final outcome of 
the merger depends sensitively on the initial compactness of the neutron 
stars before plunge. Hence, depending on the stiffness of the EOS, 
controlled through the value of $\Gamma$, if the total rest mass of the 
system is $\sim 1.3-1.7$ times larger than the maximum rest mass of a 
spherical star in isolation, the end product is a black hole. Otherwise, a 
marginally-stable massive neutron star forms. In the latter case the star 
is supported against self-gravity by rapid differential rotation. The star 
could eventually collapse to a black hole once sufficient angular momentum 
has dissipated via neutrino emission or gravitational radiation. 
The different outcome of the merger is reflected in the gravitational 
waveforms \cite{shibata02a}. Therefore, future detection of high-frequency 
gravitational waves could help to constraint the maximum allowed mass of 
neutron stars. In addition, for prompt black hole formation, a disk 
orbiting around the black hole forms, with a mass less than 1\% the total 
rest mass of the system. Disk formation during binary neutron star 
coalescence, a fundamental issue for cosmological models of short duration 
GRBs, is enhanced for unequal mass neutron stars, in which the less massive 
one is tidally disrupted during the evolution (Shibata, private communication).

A representative example of one of the models simulated by Shibata and 
Uryu is shown in Fig.~\ref{fig:nsns}. This figure is taken from 
Ref.~\cite{shibata02a}. The compactness of each star in isolation is 
$M/R=0.14$ and $\Gamma=2.25$. Additional properties of the initial model 
can be found in Table 1 of \cite{shibata02a}. The figure shows nine snapshots 
of density isocontours and the velocity field in the equatorial plane 
($z=0$) of the computational domain. At the end of the simulation a 
black hole has formed, as indicated by the thick solid circle in the final 
snapshot, representing the apparent horizon. The formation timescale of 
the black hole is larger the smaller the initial compactness of each star. 
The snapshots depicted in Fig.~\ref{fig:nsns} show that once the stars 
have merged the object starts oscillating quasi-radially before the complete 
collapse takes place, the lapse function approaching zero not monotonically 
\cite{shibata02a}. The collapse toward a black hole sets in after the 
angular momentum of the merged object is dissipated through gravitational 
radiation. Animations of various simulations (including this example) can 
be found at Shibata's website \cite{shibataweb}.

To close this section we mention the work of Duez et al.~\cite{duez01a} where,
through analytic modelling of the inspiral of corotational and irrotational
relativistic binary neutron stars as a sequence of quasi-equilibrium
configurations, the gravitational wave-train from the last phase (a few
hundred cycles) of the inspiral is computed. These authors further show a
practical procedure to construct the entire wave-train through coalescence by 
matching the late inspiral waveform to the one obtained by fully relativistic 
hydrodynamical simulations as those discussed in the above paragraphs 
\cite{shibata99c}. Detailed theoretical waveforms of the inspiral and plunge 
as those reported by \cite{duez01a} are crucial to enhance the chances of 
experimental detection in conjunction with matched-filtering techniques.

\section{Additional information}

This last section of the review contains technical information concerning
recent developments for the implementation of Riemann solver based numerical 
schemes in general relativistic hydrodynamics.

\subsection{Riemann problems in locally Minkowskian coordinates}

In~\cite{pons98} a procedure to integrate the general relativistic
hydrodynamic equations (as formulated in Section 2.1.3),
taking advantage of the multitude of Riemann
solvers developed in special relativity, was presented. The approach
relies on a local change of coordinates
in terms of which the spacetime metric is locally Minkowskian.
This procedure allows, for 1D problems, the use of the exact solution 
of the special relativistic Riemann problem~\cite{marti96}. 

Such a coordinate transformation to locally Minkowskian coordinates at 
each numerical interface, assumes that the solution of the Riemann 
problem is the one in special relativity and planar symmetry. This last
assumption is equivalent to the approach followed in classical fluid
dynamics, when using the solution of Riemann problems in slab symmetry
for problems in cylindrical or spherical coordinates, which breaks
down near the singular points (e.g., the polar axis in cylindrical
coordinates). In analogy to classical fluid dynamics, the numerical
error depends on the magnitude of the Christoffel symbols, which might
be large whenever huge gradients or large temporal variations 
of the gravitational field are present. Finer grids and improved
time advancing methods will be required in those circumstances.

Following~\cite{pons98} we illustrate the procedure for computing
the second flux integral in Eq.~(\ref{integ}), which we call ${\cal I}$. 
We begin by expressing the 
integral on a basis ${\bf e}_{\hat{\alpha}}$ with ${\bf e}_{\hat{0}} \equiv 
n^{\mu}$ and ${\bf e}_{\hat{i}}$ forming an orthonormal basis in the plane
orthogonal to $n^{\mu}$ with ${\bf e}_{\hat{1}}$ normal to the surface
$\Sigma_{x^1}$ and ${\bf e}_{\hat{2}}$ and ${\bf e}_{\hat{3}}$ tangent
to that surface. The vectors of this basis verify
${\bf e}_{\hat{\alpha}}\cdot {\bf e}_{\hat{\beta}} = 
\eta_{\hat{\alpha}\hat{\beta}}$ with $\eta_{\hat{\alpha}\hat{\beta}}$
the Minkowski metric (in the following, caret subscripts will refer
to vector components in this basis).

Denoting by $x^\alpha_0$ the coordinates at the center of the interface
at time $t$, we introduce the following locally Minkowskian coordinate
system $x^{\hat{\alpha}} = \left.M\right._\alpha^{\hat{\alpha}}
(x^\alpha-x^\alpha_0)$, 
where the matrix $M_\alpha^{\hat{\alpha}}$ is given by
$\partial_{\alpha}=M_{\alpha}^{\hat{\alpha}} {\bf e}_{\hat{\alpha}}$, 
calculated at $x^\alpha_0$.
In this system of coordinates the equations of general relativistic 
hydrodynamics transform into the equations of special relativistic 
hydrodynamics, in Cartesian coordinates, but with non-zero sources, 
and the flux integral reads
\begin{eqnarray}
\label{intg2}
{\cal I}\equiv\displaystyle{\int_{\Sigma_{x^1}}}\sqrt{-g} F^{1}
dx^0 dx^2 dx^3 =
\displaystyle{\int_{\Sigma_{x^1}}}
\left(F^{\hat{1}} - \frac{\beta^{\hat{1}}} {\alpha}F^{\hat{0}}\right)
\sqrt{-\hat{g}}\,dx^{\hat{0}}dx^{\hat{2}}dx^{\hat{3}}, 
\end{eqnarray}
(the caret symbol representing the numerical flux in Eq.~(\ref{integ}) is
now removed to avoid confusion) with $\sqrt{-\hat{g}} = 1+ {\cal O} 
(x^{\hat{\alpha}})$, where we have taken into account that, in the 
coordinates $x^{\hat{\alpha}}$, 
$\Sigma_{x^1}$ is described by the equation $x^{\hat{1}} - 
\frac{\beta^{\hat{1}}}
{\alpha} x^{\hat{0}} = 0$ (with $\beta^{\hat{i}} = M^{\hat{i}}_i \beta^i$),
where the metric elements ${\beta^{1}}$ and ${\alpha}$ are calculated 
at $x^{\alpha}_0$.
Therefore, this surface is not at rest but moves with {\it speed} 
$\,\,{\beta^{\hat{1}}}/{\alpha}$.

At this point all the theoretical work on special relativistic
Riemann solvers developed in recent years, can be exploited. 
The quantity in parenthesis in Eq.~(\ref{intg2}) represents
the numerical flux across $\Sigma_{x^1}$, which can,
now, be calculated by solving the special relativistic Riemann problem
defined with the values at the two sides of $\Sigma_{x^1}$
of two independent thermodynamical 
variables (namely, the rest mass density $\rho$ and the specific internal 
energy $\varepsilon$) and the components of the velocity in the orthonormal 
spatial basis $v^{\hat{i}}$ ($v^{\hat{i}}=M_i^{\hat{i}} v^i$).

Once the Riemann problem has been solved, we can take advantage
of the self-similar character of the solution of the Riemann problem,
which makes it constant on the surface $\Sigma_{x^1}$ simplifying
the calculation of the above integral enormously:

\begin{equation}
\label{intg3}
{\cal I}=
\left(F^{\hat{1}} - \frac{\beta^{\hat{1}}} {\alpha}F^{\hat{0}}\right)^{*}
\int_{\Sigma_{x^1}} \sqrt{-\hat{g}}\,dx^{\hat{0}}dx^{\hat{2}}dx^{\hat{3}}, 
\end{equation}
\noindent
where the superscript (*) stands for the value 
on $\Sigma_{x^1}$ obtained from the solution of the Riemann problem.
Notice that the numerical fluxes correspond to the vector
fields $F^{1} = \{ {\bf J}$, ${\bf T}\cdot {\bf n}$,
${\bf T}\cdot e_{\hat {1}}$, ${\bf T}\cdot e_{\hat {2}}$, ${\bf T}
\cdot e_{\hat {3}} \}$ and linearized Riemann solvers provide
the numerical fluxes as defined in Eq.~(\ref{intg2}). Thus the
additional relation
${\bf T}\cdot \partial_i=M_{i}^{\hat{j}}({\bf T}\cdot {\bf e}_{\hat{j}})$
has to be used for the momentum equations.
The integral in the right hand side of Eq.~(\ref{intg3}) is the 
area of the surface $\Sigma_{x^1}$ and can be expressed in terms
of the original coordinates as
\begin{equation}
\label{intg4}
\int_{\Sigma_{x^1}} \sqrt{-\hat{g}}\,dx^{\hat{0}}dx^{\hat{2}}dx^{\hat{3}}=
\int_{\Sigma_{x^1}}\sqrt{\gamma^{11}}
\sqrt{-{g}}\,dx^{{0}}dx^{{2}}dx^{{3}}, 
\end{equation}
which can be evaluated for a given metric.
The interested reader is addressed to~\cite{pons98} for
details on the testing and calibration of this procedure.

\subsection{Characteristic fields in the 3+1 conservative Eulerian 
formulation of Section~\ref{valencia}}

This section collects all information concerning the characteristic
structure of the general relativistic hydrodynamic equations in the Eulerian
formulation of Section~\ref{valencia}.  As explained in Section 3.1.2 this 
information is necessary in order to implement approximate Riemann solvers in HRSC 
finite difference schemes.

We only present the characteristic speeds and fields corresponding to the 
$x$-direction. Equivalent expressions for the two other directions can be 
easily obtained by symmetry considerations. The characteristic speeds 
(eigenvalues) of the system are given by:
\begin{eqnarray}
\lambda_0 &=& \alpha v^x - \beta^x \,\,\,\,\,\,\,\,\,\,\mbox{(triple)},
\\
\lambda_{\pm} &=& \frac{\alpha}{1-{v}^{2}c_{s}^{2}}
\left\{ v^x(1-c_{s}^{2}) \pm c_s
\sqrt{(1-{v}^{2}) [\gamma^{xx}(1-{v}^{2}c_{s}^{2})
-v^x v^x (1-c_{s}^{2})]} \right\} 
\nonumber \\
&-&  \beta^x,
\end{eqnarray}
\noindent
where $c_s$ denotes the local sound speed, which can be obtained from
$h c_s^2 = \chi + \frac{p}{\rho^2}\kappa$, with 
$\chi\equiv\frac{\partial p}{\partial \rho}$ and $\kappa\equiv
\frac{\partial p}{\partial \varepsilon}$. We note that the Minkowskian limit of 
these expressions is recovered properly (see~\cite{donat98}) as well as
the Newtonian one ($\lambda_0=v^x$, $\lambda_{\pm}=v^x\pm c_s$).

A complete set of right-eigenvectors is given by (superscript $T$ denotes 
transpose):
\begin{eqnarray}
{\bf r_{0,1}} = \left(\frac{{\cal K}}{h W},v_x,v_y,v_z,1-\frac{{\cal K}}{h W}
\right)^T,
\end{eqnarray}
\begin{eqnarray}
{\bf r_{0,2}} = \left(W v_y, h (\gamma_{xy} + 2 W^2 v_x v_y),
                        h (\gamma_{yy} + 2 W^2 v_y v_y),
                        h (\gamma_{zy} + 2 W^2 v_z v_y),
\right. \nonumber \\
\left.
W v_y (2hW - 1)\right)^T,
\end{eqnarray}
\begin{eqnarray}
{\bf r_{0,3}} = \left(W v_z, h (\gamma_{xz} + 2 W^2 v_x v_z),
                        h (\gamma_{yz} + 2 W^2 v_y v_z),
                        h (\gamma_{zz} + 2 W^2 v_z v_z),
\right. \nonumber \\
\left.
W v_z (2hW - 1)\right)^T,
\end{eqnarray}
\begin{eqnarray}
{\bf r}_{\pm} = (1, h W {\cal C}^x_{\pm}, h W v_y, h W v_z, 
h W \tilde{\cal A}^x_{\pm} -1)^T,
\end{eqnarray}
\noindent
where the following auxiliary quantities are used:
\begin{eqnarray}
{\cal K} \equiv {\displaystyle{\frac{\tilde{\kappa}}
{\tilde{\kappa}-c_s^2}}},
\hspace{0.8cm}
\tilde{\kappa}\equiv \kappa/\rho,
\hspace{0.8cm}
{\cal C}^x_{\pm} \equiv v_x - {\cal V}^x_{\pm},
\end{eqnarray}

\begin{eqnarray}
{\cal V}^x_{\pm} \equiv {\displaystyle{\frac{v^x - \Lambda_{\pm}^x}
{\gamma^{xx} - v^x \Lambda_{\pm}^x}}}
\hspace{0.8cm}
{\tilde {\cal A}}^x_{\pm} \equiv
{\displaystyle{\frac{\gamma^{xx} - v^x v^x}
{\gamma^{xx} - v^x {\Lambda}^x_{\pm}}}},
\end{eqnarray}

\begin{eqnarray}
\Lambda_{\pm}^i \equiv {\tilde{\lambda}}_{\pm} + \tilde{\beta}^i,
\hspace{0.8cm}
\tilde{\lambda} \equiv \lambda/\alpha,
\hspace{0.8cm}
\tilde{\beta}^i \equiv \beta^i/\alpha.
\end{eqnarray}

Finally, a complete set of left-eigenvectors is given by:

\begin{eqnarray}
{\bf l}_{0,1} = \frac{W}{{\cal K} - 1}\left(h-W,W v^x,W v^y,W v^z,- W\right)^T,
\end{eqnarray}

\begin{eqnarray}
{\bf l}_{0,2} &= &
{\displaystyle{\frac{1}{h \xi}}}
\left[ \begin{array}{c}
- \gamma_{zz} v_y + \gamma_{yz} v_z
\\  \\
v^x (\gamma_{zz} v_y - \gamma_{yz} v_z)
\\  \\
\gamma_{zz} (1 - v_x v^x) + \gamma_{xz} v_z v^x
\\  \\
- \gamma_{yz} (1 - v_x v^x) - \gamma_{xz} v_y v^x
\\  \\
- \gamma_{zz} v_y + \gamma_{yz} v_z
\end{array} \right],
\\
\nonumber \\
\nonumber \\
{\bf l}_{0,3} &=&
{\displaystyle{\frac{1}{h \xi}}}
\left[ \begin{array}{c}
- \gamma_{yy} v_z + \gamma_{zy} v_y
\\  \\
v^x (\gamma_{yy} v_z - \gamma_{zy} v_y)
\\  \\
- \gamma_{zy} (1 - v_x v^x) - \gamma_{xy} v_z v^x
\\  \\
\gamma_{yy} (1 - v_x v^x) + \gamma_{xy} v_y v^x
\\  \\
- \gamma_{yy} v_z + \gamma_{zy} v_y
\end{array} \right],
\\
\nonumber \\
\nonumber \\
{\bf l}_{\mp} &=&
{\pm}{\displaystyle{\frac{h^2}{\Delta}}}
\left[ \begin{array}{c}
h W {\cal V}^x_{\pm} \xi + \frac{\Delta}{h^2}l_{\mp}^{(5)}
\\  \\
\Gamma_{xx} (1 - {\cal K} {\tilde {\cal A}}^x_{\pm}) +
(2 {\cal K} - 1){\cal V}^x_{\pm} (W^2 v^x \xi - \Gamma_{xx} v^x)
\\  \\
\Gamma_{xy} (1 - {\cal K} {\tilde {\cal A}}^x_{\pm}) +
(2 {\cal K} - 1) {\cal V}^x_{\pm} (W^2 v^y \xi - \Gamma_{xy} v^x)
\\  \\
\Gamma_{xz} (1 - {\cal K} {\tilde {\cal A}}^x_{\pm}) +
(2 {\cal K} - 1) {\cal V}^x_{\pm} (W^2 v^z \xi - \Gamma_{xz} v^x)
\\  \\
(1 - {\cal K})[- \gamma v^x + {\cal V}^x_{\pm} (W^2 \xi - \Gamma_{xx})]
- {\cal K} W^2 {\cal V}^x_{\pm} \xi
\end{array} \right],
\end{eqnarray}

where the following relations and auxiliary quantities have been used:
\begin{eqnarray}
1 - {\tilde {\cal A}}^x_{\pm} = v^x {\cal V}^x_{\pm},
\hspace{0.8cm}
{\tilde {\cal A}}^x_{\pm} - {\tilde {\cal A}}^x_{\mp} =
v^x ({\cal C}^x_{\pm} - {\cal C}^x_{\mp}),
\end{eqnarray}
\begin{eqnarray}
({\cal C}^x_{\pm} - {\cal C}^x_{\pm}) +
({\tilde {\cal A}}^x_{\mp} {\cal V}^x_{\pm} -
{\tilde {\cal A}}^x_{\pm} {\cal V}^x_{\mp}) = 0,
\end{eqnarray}
\begin{eqnarray}
\Delta \equiv h^3 W ({\cal K} - 1) ({\cal C}^x_{+} - {\cal C}^x_{-} ) \xi
\hspace{0.8cm}
\xi \equiv \Gamma_{xx}  - \gamma v^x v^x,
\end{eqnarray}
\begin{eqnarray}
\gamma \equiv \det \gamma_{ij} = \gamma_{xx} \Gamma_{xx} +
\gamma_{xy} \Gamma_{xy} + \gamma_{xz} \Gamma_{xz},
\end{eqnarray}
\begin{eqnarray}
\Gamma_{xx} = \gamma_{yy} \gamma_{zz} - \gamma_{yz}^2,
\hspace{0.2cm}
\Gamma_{xy} = \gamma_{yz} \gamma_{xz} - \gamma_{xy} \gamma_{zz},
\hspace{0.2cm}
\Gamma_{xz} = \gamma_{xy} \gamma_{yz} - \gamma_{xz} \gamma_{yy},
\end{eqnarray}
\begin{eqnarray}
\Gamma_{xx} v_x + \Gamma_{xy} v_y + \Gamma_{xz} v_z= \gamma v^x.
\end{eqnarray}
\noindent
These two sets of eigenfields reduce to the
corresponding ones in the Minkowskian limit~\cite{donat98}.

\section*{Acknowledgments}

It is a pleasure to acknowledge Jos\'e M. Ib\'a\~nez and Jos\'e M. Mart\'{\i} for 
valuable suggestions and a careful reading of the manuscript. The many colleagues 
who kindly helped me to update the previous version of the review by providing 
relevant information, are warmly acknowledged. I am particularly grateful to 
Harald Dimmelmeier, Konstantinos Kifonidis, Shin Koide, Jos\'e V. Romero, Masaru 
Shibata and Wai-Mo Suen for providing some of the figures and animations contained 
in the article. This research has been supported in part by the Spanish Ministerio 
de Ciencia y Tecnolog\'{\i}a (AYA2001-3490-C02-C01) and by a Marie Curie fellowship 
from the European Union (HPMF-CT-2001-01172). The author has made use of NASA's 
Astrophysics Data System Abstract Service, which is gratefully acknowledged.


\begin{thebibliography}{100}

\bibitem{whisky}\comment{misc}
Information about the hydrodynamics code \verb+whisky+ of the European Union
  Network on Sources of Gravitational Radiation is available at
  \verb+http://www.aei-potsdam.mpg.de/~hawke/Whisky.html+. \keywords{}

\bibitem{GR_ASTRO}\comment{misc}
Washington University Gravity Group, ``Neutron Star Grand Challenge",
  \verb+http://wugrav.wustl.edu/Relativ/nsgc.html+. \keywords{}

\bibitem{headon}\comment{misc}
Max Planck Institut for Gravitational Physics, ``Jean-Luc's Movies: Neutron
  Stars", quicktime and mpeg movies for download:\\
  \verb+http://jean-luc.ncsa.uiuc.edu/NCSA1999/NeutronStars/Headon/+.
  \keywords{}

\bibitem{abramowicz78}\comment{article}
\comment{author}Abramowicz, M., Jaroszynski, M., and Sikora, M.,
  \comment{title}``Relativistic, accreting disks'', \comment{journal}{\em
  Astron. Astrophys.}, \comment{volume}{\bf 63}, \comment{pages}221--224,
  (\comment{year}1978). \keywords{accretion disks, relativistic
  hydrodynamics,black holes}

\bibitem{abramowicz83}\comment{article}
\comment{author}{Abramowicz}, M.~A., {Calvani}, M., and {Nobili}, L.,
  \comment{title}``Runaway instability in accretion disks orbiting black
  holes'', \comment{journal}{\em Nature}, \comment{volume}{\bf 302},
  \comment{pages}597--599, (\comment{year}1983). \keywords{accretion disks,
  relativistic hydrodynamics,black holes}

\bibitem{abramowicz88}\comment{article}
\comment{author}Abramowicz, M.~A., Czerny, B., Lasota, J.~P., and Szuszkiewicz,
  E., \comment{title}``Slim accretion disks'', \comment{journal}{\em Astrophys.
  J.}, \comment{volume}{\bf 332}, \comment{pages}646--658,
  (\comment{year}1988). \keywords{accretion disks, black holes, astrophysics}

\bibitem{AEIgroup}\comment{misc}
Albert Einstein Institute, Numerical Relativity Group,\\
  \verb+http://jean-luc.aei.mpg.de+. \keywords{}

\bibitem{alcubierre99}\comment{article}
\comment{author}Alcubierre, M., Allen, G., Br\"ugmann, B., Dramlitsch, Th.,
  Font, J.~A., Papadopoulos, P., Seidel, E., Stergioulas, N., Suen, W.-M., and
  Takahashi, R., \comment{title}``Towards a Stable Numerical Evolution of
  Strongly Gravitating Systems in General Relativity: The Conformal
  Treatments'', \comment{journal}{\em Phys. Rev. D}, \comment{volume}{\bf 62},
  \comment{pages}044034, (\comment{year}2000). For a related online version
  see: \comment{author}M.~Alcubierre, et al., \comment{onlinetitle}``Towards a
  Stable Numerical Evolution of Strongly Gravitating Systems in General
  Relativity: The Conformal Treatments'', (\comment{onlinemonth}March,
  \comment{onlineyear}2000), \comment{fileformat}[Online Los Alamos Archive
  Preprint]: cited on \comment{cited}28 March 2000,
  \comment{onlineaddress}http://xxx.lanl.gov/abs/gr-qc/0003071. \keywords{ADM
  formalsim, black holes, constraint equations, Einstein equations, numerical
  relativity, relativistic hydrodynamics, relativistic stars, numerical
  methods}

\bibitem{alcubierre99b}\comment{article}
\comment{author}Alcubierre, M., Allen, G., Br\"{u}gmann, B., Seidel, E., and
  Suen, W.-M., \comment{title}``Towards an understanding of the stability
  properties of the 3+1 evolution equations in general relativity'',
  \comment{journal}{\em Phys. Rev. D}, \comment{volume}{\bf 62},
  \comment{pages}124011, (\comment{year}2000). For a related online version
  see: \comment{author}M.~Alcubierre, et al., \comment{onlinetitle}``Towards an
  understanding of the stability properties of the 3+1 evolution equations in
  general relativity'', (\comment{onlinemonth}August,
  \comment{onlineyear}1999), \comment{fileformat}[Online Los Alamos Archive
  Preprint]: cited on \comment{cited}5 July 2002,
  \comment{onlineaddress}http://xxx.lanl.gov/abs/gr-qc/9908079. \keywords{ADM
  formalsim, constraint equations, Einstein equations, numerical relativity}

\bibitem{alcubierre00b}\comment{article}
\comment{author}Alcubierre, M., Brandt, B., Br\"ugmann, B., Holz, D., Seidel,
  E., Takahashi, R., and Thornburg, J., \comment{title}``Symmetry without
  symmetry: Numerical simulations of axisymmetric systems using Cartesian
  grids'', \comment{journal}{\em Int. J. Mod. Phys.}, \comment{volume}{\bf
  D10}, \comment{pages}273--290, (\comment{year}2001). For a related online
  version see: \comment{author}M.~Alcubierre, et al.,
  \comment{onlinetitle}``Symmetry without symmetry: Numerical simulations of
  axisymmetric systems using Cartesian grids'', (\comment{onlinemonth}August,
  \comment{onlineyear}1999), \comment{fileformat}[Online Los Alamos Archive
  Preprint]: cited on \comment{cited}5 July 2002,
  \comment{onlineaddress}http://xxx.lanl.gov/abs/gr-qc/9908012.
  \keywords{Numerical relativity, numerical methods}

\bibitem{alcubierre02a}\comment{article}
\comment{author}Alcubierre, M., Br\"ugmann, B., Diener, P., Koppitz, M.,
  Pollney, D., Seidel, E., and Takahashi, R., \comment{title}``Gauge conditions
  for long-term numerical black hole evolutions without excision'',
  \comment{journal}{\em Phys.\ Rev.~D}, (\comment{year}2002). For a related
  online version see: \comment{author}M.~Alcubierre, et al.,
  \comment{onlinetitle}``Gauge conditions for long-term numerical black hole
  evolutions without excision'', (\comment{onlinemonth}June,
  \comment{onlineyear}2002), \comment{fileformat}[Online Los Alamos Archive
  Preprint]: cited on \comment{cited}26 June 2002,
  \comment{onlineaddress}http://xxx.lanl.gov/abs/gr-qc/0206072. submitted.
  \keywords{ADM formalsim, black holes, constraint equations, Einstein
  equations, numerical relativity}

\bibitem{aloy99}\comment{article}
\comment{author}Aloy, M.~A., Ib{\'a}\~nez, J.~M., Mart\'{\i}, and M\"uller, E.,
  \comment{title}``GENESIS: A high-resolution code for three-dimensional
  relativistic hydrodynamics'', \comment{journal}{\em Astrophys. J. Suppl.
  Ser.}, \comment{volume}{\bf 122}, \comment{pages}151--166,
  (\comment{year}1999). For a related online version see: \comment{author}M.~A.
  Aloy, et al., \comment{onlinetitle}``GENESIS: A high-resolution code for
  three-dimensional relativistic hydrodynamics'', (\comment{onlinemonth}March,
  \comment{onlineyear}1999), \comment{fileformat}[Online Los Alamos Archive
  Preprint]: cited on \comment{cited}1 April 1999,
  \comment{onlineaddress}http://xxx.lanl.gov/abs/astro-ph/9903352.
  \keywords{computational fluid dynamics, relativistic hydrodynamics, finite
  difference and finite volume methods, hyperbolic equations}

\bibitem{aloy00a}\comment{article}
\comment{author}Aloy, M.~A., M\"uller, E., Ib\'a\~nez, J.~M$^{\underline{\rm
  a}}$, Mart\'{\i}, J.~M$^{\underline{\rm a}}$, and MacFadyen, A.,
  \comment{title}``Relativistic jets from collapsars'', \comment{journal}{\em
  Astrophys.~J.\ Lett.}, \comment{volume}{\bf 531}, \comment{pages}L119--L122,
  (\comment{year}2000). For a related online version see: \comment{author}M.~A.
  Aloy, et al., \comment{onlinetitle}``Relativistic jets from collapsars'',
  (\comment{onlinemonth}November, \comment{onlineyear}1999),
  \comment{fileformat}[Online Los Alamos Archive Preprint]: cited on
  \comment{cited}15 July 2002,
  \comment{onlineaddress}http://xxx.lanl.gov/abs/astro-ph/9911098.
  \keywords{numerical relativistic hydrodynamics, gamma ray bursts,
  extragalactic jets}

\bibitem{anile89}\comment{book}
\comment{author}Anile, A.~M., \comment{title}{\em Relativistic fluids and
  magneto-fluids}, (\comment{publisher}Cambridge University Press,
  \comment{address}Cambridge, England, \comment{year}1989).
  \keywords{relativistic hydrodynamics and magnetohydrodynamics, hyperbolic
  equations, relativistic astrophysics}

\bibitem{anninos97}\comment{article}
\comment{author}Anninos, P., \comment{title}``Computational Cosmology: from the
  Early Universe to the Large Scale Structure'', \comment{journal}{\em Living
  Reviews in Relativity}, \comment{volume}{\bf 1}, (\comment{year}1998). For a
  related online version see: \comment{author}P.~Anninos,
  \comment{onlinetitle}``Computational Cosmology: from the Early Universe to
  the Large Scale Structure'', (\comment{onlinemonth}August,
  \comment{onlineyear}2001), \comment{fileformat}[Online Los Alamos Archive
  Preprint]: cited on \comment{cited}5 July 2002,
  \comment{onlineaddress}http://xxx.lanl.gov/abs/gr-qc/0108083.
  \keywords{Cosmology, relativistic hydrodynamics, Big Bang, cosmic microwave
  background, cosmological spacetimes, inflation, large scale structure,
  Newtonian cosmology, singularities, physical cosmology}

\bibitem{anninos98}\comment{article}
\comment{author}Anninos, P., \comment{title}``Plane-symmetric cosmology with
  relativistic hydrodynamics'', \comment{journal}{\em Phys. Rev. D},
  \comment{volume}{\bf 58}, \comment{pages}064010, (\comment{year}1998).
  \keywords{ADM formalism, cosmology, Einstein equations, relativistic
  hydrodynamics, numerical methods, cosmological spacetimes}

\bibitem{anninos02}\comment{article}
\comment{author}Anninos, P., and Fragile, P.~C.,
  \comment{title}``Non-oscillatory central difference and artificial viscosity
  schemes for relativistic hydrodynamics'', \comment{journal}{\em Astrophys. J.
  Supp. Ser.}, \comment{volume}{\bf 144}, \comment{pages}243--257,
  (\comment{year}2002). For a related online version see:
  \comment{author}P.~Anninos, et al., \comment{onlinetitle}``Non-oscillatory
  central difference and artificial viscosity schemes for relativistic
  hydrodynamics'', (\comment{onlinemonth}June, \comment{onlineyear}2002),
  \comment{fileformat}[Online Los Alamos Archive Preprint]: cited on
  \comment{cited}18 June 2002,
  \comment{onlineaddress}http://xxx.lanl.gov/abs/astro-ph/0206265.
  \keywords{numerical methods, numerical relativistic hydrodynamics,
  computational fluid dynamics, finite difference and finite volume methods}

\bibitem{arnett66}\comment{article}
\comment{author}Arnett, W.~D., \comment{title}``Gravitational collapse and weak
  interactions'', \comment{journal}{\em Can. J. Phys.}, \comment{volume}{\bf
  44}, \comment{pages}2553--2594, (\comment{year}1966). \keywords{Gravitational
  collapse, supernovae, nuclear physics, relativistic hydrodynamics}

\bibitem{arnowitt62}\comment{inbook}
\comment{author}Arnowitt, R., Deser, S., and Misner, C.~W.,
  \comment{title}``The Dynamics of General Relativity'', in
  \comment{editor}Witten, L., ed., \comment{booktitle}{\em Gravitation: An
  Introduction to Current Research}, \comment{pages} 227--265,
  (\comment{publisher}John Wiley, \comment{address}New York,
  \comment{year}1962). \keywords{ADM formalism, Einstein equations, initial
  value problem, numerical relativity, numerical methods}

\bibitem{arras02a}\comment{article}
\comment{author}Arras, P., Flanagan, E.~E., Morsink, S.~M., Schenk, A.~K.,
  Teukolsky, S.~A., and Wasserman, I., \comment{title}``Saturation of the
  $r$-mode instability'', \comment{journal}{\em Astrophys. J.},
  (\comment{year}2002). For a related online version see:
  \comment{author}P.~Arras, et al., \comment{onlinetitle}``Saturation of the
  $r$-mode instability'', (\comment{onlinemonth}February,
  \comment{onlineyear}2002), \comment{fileformat}[Online Los Alamos Archive
  Preprint]: cited on \comment{cited}5 July 2002,
  \comment{onlineaddress}http://xxx.lanl.gov/abs/astro-ph/0202345. submitted.
  \keywords{}

\bibitem{balbus01}\comment{article}
\comment{author}{Balbus}, S.~A., \comment{title}``Convective and Rotational
  Stability of a Dilute Plasma'', \comment{journal}{\em Astrophys. J.},
  \comment{volume}{\bf 562}, \comment{pages}909--917, (\comment{month}December,
  \comment{year}2001). For a related online version see: \comment{author}S.~A.
  {Balbus}, \comment{onlinetitle}``Convective and Rotational Stability of a
  Dilute Plasma'', (\comment{onlinemonth}June, \comment{onlineyear}2001),
  \comment{fileformat}[Online Los Alamos Archive Preprint]: cited on
  \comment{cited}15 July 2002,
  \comment{onlineaddress}http://xxx.lanl.gov/abs/astro-ph/0106283.
  \keywords{Accretion disks}

\bibitem{balbus98}\comment{article}
\comment{author}Balbus, S.~A., and Hawley, J.~A., \comment{title}``Instability,
  turbulence, and enhanced transport in accretion disks'',
  \comment{journal}{\em Rev. of Mod. Phys.}, \comment{volume}{\bf 70},
  \comment{pages}1--53, (\comment{year}1998). \keywords{Accretion disks}

\bibitem{balsara94}\comment{article}
\comment{author}Balsara, D., \comment{title}``Riemann solver for relativistic
  hydrodynamics'', \comment{journal}{\em J. Comput. Phys.},
  \comment{volume}{\bf 114}, \comment{pages}284--297, (\comment{year}1994).
  \keywords{computational fluid dynamics, relativistic hydrodynamics, finite
  difference and finite volume methods, hyperbolic equations}

\bibitem{balsara01}\comment{article}
\comment{author}Balsara, D., \comment{title}``Total variation diminishing
  scheme for relativistic magnetohydrodynamics'', \comment{journal}{\em
  Astrophys. J. Suppl. Ser.}, \comment{volume}{\bf 132},
  \comment{pages}83--101, (\comment{year}2001). \keywords{relativistic
  hydrodynamics and magnetohydrodynamics, numerical methods, finite difference
  and finite volume methods}

\bibitem{banyuls97}\comment{article}
\comment{author}Banyuls, F., Font, J.~A., Ib{\'a}\~nez, J.~M., Mart\'{\i},
  J.~M., and Miralles, J.~A., \comment{title}``Numerical 3+1 General
  Relativistic Hydrodynamics: A Local Characteristic Approach'',
  \comment{journal}{\em Astrophys. J.}, \comment{volume}{\bf 476},
  \comment{pages}221--231, (\comment{year}1997). \keywords{computational fluid
  dynamics, relativistic hydrodynamics, numerical relativity, finite difference
  and finite volume methods, hyperbolic equations}

\bibitem{bardeen83}\comment{article}
\comment{author}Bardeen, J.~M., and Piran, T., \comment{title}``General
  relativistic axisymmetric rotating systems: coordinates and equations'',
  \comment{journal}{\em Phys. Rep.}, \comment{volume}{\bf 96},
  \comment{pages}No. 4, 205--250, (\comment{year}1983). \keywords{ADM
  formalism, Einstein equations, numerical relativity, relativistic
  hydrodynamics, gravitational collapse, numerical methods, gauge conditions}

\bibitem{bardeen72}\comment{article}
\comment{author}Bardeen, J.~M., and Press, W.~H., \comment{title}``Radiation
  fields in the Schwarzschild background'', \comment{journal}{\em J. Math.
  Phys.}, \comment{volume}{\bf 14}, \comment{pages}7--19, (\comment{year}1972).
  \keywords{Approximation methods, Schwarzschild solution, perturbation theory}

\bibitem{baron85}\comment{article}
\comment{author}Baron, E., Cooperstein, J., and Kahana, S.,
  \comment{title}``Type-II Supernovae in $12M_{\odot}$ and $15M_{\odot}$ stars:
  the equation of state and general relativity'', \comment{journal}{\em Phys.
  Rev. Lett.}, \comment{volume}{\bf 55}, \comment{pages}126--129,
  (\comment{year}1985). \keywords{supernovae, gravitational collapse, nuclear
  physics, computational fluid dynamics, relativistic astrophysics}

\bibitem{baumgarte98}\comment{article}
\comment{author}Baumgarte, T.~W., and Shapiro, S.~L., \comment{title}``On the
  numerical integration of Einstein's field equations'', \comment{journal}{\em
  Phys. Rev. D}, \comment{volume}{\bf 59}, \comment{pages}024007,
  (\comment{year}1999). For a related online version see: \comment{author}T.~W.
  Baumgarte, et al., \comment{onlinetitle}``On the numerical integration of
  Einstein's field equations'', (\comment{onlinemonth}October,
  \comment{onlineyear}1998), \comment{fileformat}[Online Los Alamos Archive
  Preprint]: cited on \comment{cited}1 November 1998,
  \comment{onlineaddress}http://xxx.lanl.gov/abs/gr-qc/9810065. \keywords{ADM
  formliam, constraint equations, gauge conditions, numerical relativity,
  Einstein equations, numerical methods, stability theory}

\bibitem{baumgarte95}\comment{article}
\comment{author}Baumgarte, T.~W., Shapiro, S.~L., and Teukolsky, S.~A.,
  \comment{title}``Computing supernova collapse to neutron stars and black
  holes'', \comment{journal}{\em Astrophys. J.}, \comment{volume}{\bf 443},
  \comment{pages}717--734, (\comment{year}1995). \keywords{Gravitational
  collapse, supernovae, neutron stars, black holes, numerical methods, null
  surfaces, relativistic hydrodynamics}

\bibitem{benensohn97}\comment{article}
\comment{author}Benensohn, J.~S., Lamb, D.~Q., and Taam, R.~E.,
  \comment{title}``Hydrodynamical studies of wind accretion onto compact
  objects: Two-dimensional calculations'', \comment{journal}{\em Astrophys.
  J.}, \comment{volume}{\bf 478}, \comment{pages}723--733,
  (\comment{year}1997). For a related online version see: \comment{author}J.~S.
  Benensohn, et al., \comment{onlinetitle}``Hydrodynamical studies of wind
  accretion onto compact objects: Two-dimensional calculations'',
  (\comment{onlinemonth}October, \comment{onlineyear}1996),
  \comment{fileformat}[Online Los Alamos Archive Preprint]: cited on
  \comment{cited}15 February 2000,
  \comment{onlineaddress}http://xxx.lanl.gov/abs/astro-ph/9610245.
  \keywords{hydrodynamics, accretion, computational fluid dynamics, numerical
  methods}

\bibitem{bethe90}\comment{article}
\comment{author}Bethe, H.~A., \comment{title}``Supernova mechanisms'',
  \comment{journal}{\em Rev. Mod. Phys.}, \comment{volume}{\bf 62},
  \comment{pages}801--866, (\comment{year}1990). \keywords{Gravitational
  collapse, supernovae, nuclear physics, neutrinos, neutron stars, shock waves,
  hydrodynamics, relativistic astrophysics}

\bibitem{bethe85}\comment{article}
\comment{author}Bethe, H.~A., and Wilson, J.~R., \comment{title}``Revival of a
  stalled supernova shock by neutrino heating'', \comment{journal}{\em
  Astrophys. J.}, \comment{volume}{\bf 295}, \comment{pages}14--23,
  (\comment{year}1985). \keywords{Gravitational collapse, supernovae, nuclear
  physics, neutrinos, neutron stars, shock waves, hydrodynamics, relativistic
  astrophysics}

\bibitem{bishop99}\comment{article}
\comment{author}Bishop, N.~T., G\'omez, R., Lehner, L., Maharaj, M., and
  Winicour, J., \comment{title}``The incorporation of matter into
  characteristic numerical relativity'', \comment{journal}{\em Phys. Rev. D},
  \comment{volume}{\bf 60}, \comment{pages}024005, (\comment{year}1999). For a
  related online version see: \comment{author}N.~T. Bishop, et al.,
  \comment{onlinetitle}``The incorporation of matter into characteristic
  numerical relativity'', (\comment{onlinemonth}January,
  \comment{onlineyear}1999), \comment{fileformat}[Online Los Alamos Archive
  Preprint]: cited on \comment{cited}1 February 1999,
  \comment{onlineaddress}http://xxx.lanl.gov/abs/gr-qc/9901056.
  \keywords{Einstein equations, numerical relativity, numerical methods, null
  surfaces, relativistic hydrodynamics, black holes, energy and momentum,
  initial value problem}

\bibitem{blandford99}\comment{inproceedings}
\comment{author}Blandford, R.~D., \comment{title}``Relativistic accretion'', in
  \comment{editor}Sellwood, J.~A., and Goodman, J., eds.,
  \comment{booktitle}{\em ASP Conf. Ser. 160: Astrophysical Discs - an EC
  Summer School}, \comment{pages} 265, (\comment{year}1999). For a related
  online version see: \comment{author}R.~D. Blandford,
  \comment{onlinetitle}``Relativistic accretion'',
  (\comment{onlinemonth}February, \comment{onlineyear}1999),
  \comment{fileformat}[Online Los Alamos Archive Preprint]: cited on
  \comment{cited}1 February 1999,
  \comment{onlineaddress}http://xxx.lanl.gov/abs/astro-ph/9902001. in press.
  \keywords{Accretion, Accretion Disks, Black holes}

\bibitem{blandford98}\comment{article}
\comment{author}Blandford, R.~D., and Begelman, M.~C., \comment{title}``On the
  fate of gas accreting at a low rate on to a black hole'',
  \comment{journal}{\em Mon. Not. R. Astron. Soc.}, \comment{volume}{\bf 303},
  \comment{pages}L1--L5, (\comment{year}1999). For a related online version
  see: \comment{author}R.~D. Blandford, et al., \comment{onlinetitle}``On the
  fate of gas accreting at a low rate on to a black hole'',
  (\comment{onlinemonth}September, \comment{onlineyear}1998),
  \comment{fileformat}[Online Los Alamos Archive Preprint]: cited on
  \comment{cited}15 February 2000,
  \comment{onlineaddress}http://xxx.lanl.gov/abs/astro-ph/9809083.
  \keywords{Accretion, accretion disks, black holes, hydrodynamics}

\bibitem{blandford82}\comment{article}
\comment{author}Blandford, R.~D., and Payne, D.~G.,
  \comment{title}``Hydromagnetic flows from accretion discs and the production
  of radio jets'', \comment{journal}{\em Mon. Not. R. Astron. Soc.},
  \comment{volume}{\bf 199}, \comment{pages}883--903, (\comment{year}1982).
  \keywords{active galactic nuclei, magnetohydrodynamics, relativistic
  astrophysics, accretion disks, extragalactic jets}

\bibitem{blandford74}\comment{article}
\comment{author}Blandford, R.~D., and Rees, M., \comment{title}``A
  'twin-exhaust' model for double radio sources'', \comment{journal}{\em Mon.
  Not. R. Astron. Soc.}, \comment{volume}{\bf 169}, \comment{pages}395--415,
  (\comment{year}1974). \keywords{active galactic nuclei, magnetohydrodynamics,
  relativistic astrophysics, accretion disks, extragalactic jets}

\bibitem{blandford77}\comment{article}
\comment{author}Blandford, R.~D., and Znajek, R.~L.,
  \comment{title}``Electromagnetic extraction of energy from Kerr black
  holes'', \comment{journal}{\em Mon. Not. R. Astron. Soc.},
  \comment{volume}{\bf 179}, \comment{pages}433--456, (\comment{year}1977).
  \keywords{active galactic nuclei, magnetohydrodynamics, relativistic
  astrophysics, accretion disks, extragalactic jets, Kerr metric}

\bibitem{bodenheimer83a}\comment{article}
\comment{author}Bodenheimer, P., and Woosley, S.~E., \comment{title}``A
  two-dimensional supernova model with rotation and nuclear burning'',
  \comment{journal}{\em Astrophys.~J.}, \comment{volume}{\bf 269},
  \comment{pages}281--291, (\comment{year}1983). \keywords{}

\bibitem{bona93}\comment{inproceedings}
\comment{author}Bona, C., Ib\'a\~nez, J.~M., Mart\'{\i}, J.~M., and Mass\'o,
  J., \comment{title}``Gravitation and General Relativity: rotating bodies and
  other topics'', volume 423 of {\em Lecture Notes in Physics},
  (\comment{publisher}Springer-Verlag, \comment{address}New York,
  \comment{year}1993). \keywords{Numerical relativity, numerical relativistic
  hydrodynamics, hyperbolic equations, numerical methods, gravitational
  collapse}

\bibitem{bona89}\comment{article}
\comment{author}Bona, C., and Mass\'o, J., \comment{title}``Einstein's
  evolution equations as a system of balance laws'', \comment{journal}{\em
  Phys. Rev. D}, \comment{volume}{\bf 40}, \comment{pages}1022--1026,
  (\comment{year}1989). \keywords{ADM formalism, Einstein equations, hyperbolic
  equations, numerical relativity, numerical methods, gauge conditions,
  constraint equations}

\bibitem{bona95}\comment{article}
\comment{author}Bona, C., Mass\'o, J., Seidel, E., and Stela, J.,
  \comment{title}``A new formalism for numerical relativity'',
  \comment{journal}{\em Phys. Rev. Lett.}, \comment{volume}{\bf 75},
  \comment{pages}600--603, (\comment{year}1995). For a related online version
  see: \comment{author}C.~Bona, et al., \comment{onlinetitle}``New formalism
  for numerical relativity'', (\comment{onlinemonth}December,
  \comment{onlineyear}1994), \comment{fileformat}[Online Los Alamos Archive
  Preprint]: cited on \comment{cited}15 September 1996,
  \comment{onlineaddress}http://xxx.lanl.gov/abs/gr-qc/9412071. \keywords{ADM
  formalism, Einstein equations, hyperbolic equations, numerical relativity,
  numerical methods, gauge conditions, constraint equations}

\bibitem{bonazzola99}\comment{article}
\comment{author}Bonazzola, S., Gourgoulhon, and Marck, J.-A.,
  \comment{title}``Spectral methods in general relativistic astrophysics'',
  \comment{journal}{\em J. Comput. Appl. Math.}, \comment{volume}{\bf 109},
  \comment{pages}433, (\comment{year}1999). For a related online version see:
  \comment{author}S.~Bonazzola, et al., \comment{onlinetitle}``Spectral methods
  in general relativistic astrophysics'', (\comment{onlinemonth}November,
  \comment{onlineyear}1998), \comment{fileformat}[Online Los Alamos Archive
  Preprint]: cited on \comment{cited}15 February 2000,
  \comment{onlineaddress}http://xxx.lanl.gov/abs/gr-qc/9811089.
  \keywords{Spectral methods, Relativistic Astrophysics}

\bibitem{bonazzola93}\comment{article}
\comment{author}Bonazzola, S., and Marck, J.-A., \comment{title}``Efficiency of
  gravitational radiation from axisymmetric and 3D stellar collapse. I.
  Polytropic case'', \comment{journal}{\em Astron. Astrophys.},
  \comment{volume}{\bf 267}, \comment{pages}623--633, (\comment{year}1993).
  \keywords{Spectral methods, gravitational radiation, hydrodynamics,
  computational fluid dynamics, gravitational collapse}

\bibitem{bondi52}\comment{article}
\comment{author}Bondi, H., \comment{title}``On spherically symmetric
  accretion'', \comment{journal}{\em Mon. Not. R. Astron. Soc.},
  \comment{volume}{\bf 112}, \comment{pages}195--204, (\comment{year}1952).
  \keywords{accretion, astrophysics}

\bibitem{bondi44}\comment{article}
\comment{author}Bondi, H., and Hoyle, F., \comment{title}``On the mechanism of
  accretion by stars'', \comment{journal}{\em Mon. Not. R. Astron. Soc.},
  \comment{volume}{\bf 104}, \comment{pages}273--282, (\comment{year}1944).
  \keywords{accretion, stars, astrophysics}

\bibitem{bondi62}\comment{article}
\comment{author}Bondi, H., van~der Burg, M.~J.~G., and Metzner, A.~W.~K.,
  \comment{title}``Gravitational waves in general relativity. VII. Waves from
  axi-symmetric isolated systems'', \comment{journal}{\em Proc. R. Soc.
  London}, \comment{volume}{\bf Sect. A 269}, \comment{pages}21--52,
  (\comment{year}1962). \keywords{}

\bibitem{boris73}\comment{article}
\comment{author}Boris, J.~P., and Book, D.~L., \comment{title}``Flux corrected
  transport I, SHASTA, a fluid transport algorithm that works'',
  \comment{journal}{\em J. Comput. Phys.}, \comment{volume}{\bf 11},
  \comment{pages}38--69, (\comment{year}1973). \keywords{computational fluid
  dynamics, finite difference and finite volume methods, numerical methods}

\bibitem{bromley98}\comment{article}
\comment{author}Bromley, B.~C., Miller, W.A., and Pariev, V.~I.,
  \comment{title}``The inner edge of the accretion disk around a supermassive
  black hole'', \comment{journal}{\em Nature}, \comment{volume}{\bf 391},
  \comment{pages}54--56, (\comment{year}1998). \keywords{Relativistic
  astrophysics, black holes, accretion disks, active galactic nuclei}

\bibitem{brown01a}\comment{inproceedings}
\comment{author}Brown, J.~D., \comment{title}``Rotational instabilities in
  post-collapse stellar cores'', in \comment{editor}Centrella, J.~M., ed.,
  \comment{booktitle}{\em AIP Conference Proc., Vol.~575, ``Astrophysical
  sources for ground-based gravitational wave detectors''}, \comment{pages}
  234--245, (\comment{publisher}American Institute of Physics, Melville,
  \comment{address}New York, USA, \comment{year}2001). For a related online
  version see: \comment{author}J.~D. Brown, \comment{onlinetitle}``Rotational
  instabilities in post-collapse stellar cores'',
  (\comment{onlinemonth}December, \comment{onlineyear}2000),
  \comment{fileformat}[Online Los Alamos Archive Preprint]: cited on
  \comment{cited}24 June 2002,
  \comment{onlineaddress}http://xxx.lanl.gov/abs/gr-qc/0012084.
  \keywords{Hydrodynamics, supernovae, neutron stars}

\bibitem{bruenn85}\comment{article}
\comment{author}Bruenn, S.~W., \comment{title}``Stellar core collapse:
  numerical model and infall epoch'', \comment{journal}{\em Astrophys. J.
  Suppl. Ser.}, \comment{volume}{\bf 58}, \comment{pages}771--841,
  (\comment{year}1985). \keywords{Gravitational collapse, neutrinos, Stars,
  Supernovae}

\bibitem{bruenn89}\comment{article}
\comment{author}Bruenn, S.~W., \comment{title}``The prompt-shock supernova
  mechanism. I - The effect of the free-proton mass fraction and the neutrino
  transport algorithm'', \comment{journal}{\em Astrophys. J},
  \comment{volume}{\bf 340}, \comment{pages}955--965, (\comment{year}1989).
  \keywords{Gravitational collapse, supernovae, nuclear physics, neutrinos,
  computational fluid dynamics, shock waves}

\bibitem{bruenn93}\comment{inproceedings}
\comment{author}Bruenn, S.~W., in \comment{editor}Guidry, M.~W., and Strayer,
  M.~R., eds., \comment{booktitle}{\em Nuclear physics in the universe},
  (\comment{publisher}IOP, \comment{address}Bristol, \comment{year}1993).
  \keywords{Gravitational collapse, supernovae, nuclear physics, neutrinos,
  computational fluid dynamics, shock waves}

\bibitem{Cactusweb}\comment{misc}
\verb+CACTUS+ code, \verb+http://www.cactuscode.org+. \keywords{}

\bibitem{canuto88}\comment{book}
\comment{author}Canuto, C., Hussaini, M.~Y., Quarteroni, A., and Zang, T.~A.,
  \comment{title}{\em Spectral methods in fluid dynamics},
  (\comment{publisher}Springer-Verlag, \comment{address}Berlin,
  \comment{year}1988). \keywords{computational fluid dynamics, spectral
  methods, numerical methods}

\bibitem{centrella83}\comment{article}
\comment{author}Centrella, J., and Wilson, J.~R., \comment{title}``Planar
  numerical cosmology. I. The differential equations'', \comment{journal}{\em
  Astrophys. J.}, \comment{volume}{\bf 273}, \comment{pages}428--435,
  (\comment{year}1983). \keywords{cosmology, differential equations,
  relativistic hydrodynamics, numerical methods}

\bibitem{centrella84}\comment{article}
\comment{author}Centrella, J., and Wilson, J.~R., \comment{title}``Planar
  numerical cosmology. II. The difference equations and numerical tests'',
  \comment{journal}{\em Astrophys. J. Suppl. Ser.}, \comment{volume}{\bf 54},
  \comment{pages}229--249, (\comment{year}1984). \keywords{cosmology,
  computational fluid dynamics, relativistic hydrodynamics, numerical
  relativity, finite difference and finite volume methods, numerical methods}

\bibitem{chandrasekhar83}\comment{book}
\comment{author}Chandrasekhar, S., \comment{title}{\em The mathematical theory
  of black holes}, (\comment{publisher}Oxford University Press,
  \comment{address}New York, \comment{year}1983). \keywords{Black holes, Kerr
  metric, Schwarzshild solution, mathematical physics, perturbation theory,
  singularities, Kerr-Newman metric, black hole thermodynamics}

\bibitem{choptuik93}\comment{article}
\comment{author}Choptuik, M.~W., \comment{title}``Universality and scaling in
  gravitational collapse of a massless scalar field'', \comment{journal}{\em
  Phys. Rev. Lett.}, \comment{volume}{\bf 70}, \comment{pages}9--12,
  (\comment{year}1993). \keywords{Scalar field, gravitational collapse, black
  holes, critical phenomena}

\bibitem{chow97}\comment{article}
\comment{author}Chow, E., and Monaghan, J.~J.,
  \comment{title}``Ultrarelativistic SPH'', \comment{journal}{\em J. Comput.
  Phys.}, \comment{volume}{\bf 134}, \comment{pages}296--305,
  (\comment{year}1997). \keywords{computational fluid dynamics, relativistic
  hydrodynamics, hyperbolic equations, spectral methods, numerical methods}

\bibitem{colella84}\comment{article}
\comment{author}Colella, P., and Woodward, P.~R., \comment{title}``The
  piecewise parabolic method (PPM) for gas-dynamical simulations'',
  \comment{journal}{\em J. Comput. Phys.}, \comment{volume}{\bf 54},
  \comment{pages}174--201, (\comment{year}1984). \keywords{computational fluid
  dynamics, finite difference and finite volume methods, hyperbolic equations,
  numerical methods}

\bibitem{colgate89}\comment{article}
\comment{author}Colgate, S.~A., \comment{title}``Hot bubbles drive
  explosions'', \comment{journal}{\em Nature}, \comment{volume}{\bf 341},
  \comment{pages}489--490, (\comment{year}1989). \keywords{Gravitational
  collapse, supernovae, shock waves, hydrodynamics, neutrinos}

\bibitem{colgate66}\comment{article}
\comment{author}Colgate, S.~A., and White, R.~H., \comment{title}``The
  hydrodynamic behaviour of supernovae explosions'', \comment{journal}{\em
  Astrophys. J.}, \comment{volume}{\bf 143}, \comment{pages}626--681,
  (\comment{year}1966). \keywords{Gravitational collapse, supernovae, shock
  waves, hydrodynamics, computational fluid dynamics}

\bibitem{davis84}\comment{article}
\comment{author}Davis, S.~F., \comment{title}``A simplified TVD finite
  difference scheme via artificial viscosity'', \comment{journal}{\em ICASE
  Report}, \comment{volume}{\bf 84}, \comment{pages}20, (\comment{year}1984).
  \keywords{numerical methods, finite difference and finite volume methods,
  computational fluid dynamics}

\bibitem{devilliers02a}\comment{article}
\comment{author}De~Villiers, J.-P., and Hawley, J.~F.,
  \comment{title}``Three-dimensional hydrodynamic simulations of accretion tori
  in Kerr spacetimes'', \comment{journal}{\em Astrophys. J.},
  \comment{volume}{\bf 577}, \comment{pages}866--879, (\comment{year}2002). For
  a related online version see: \comment{author}J.-P. De~Villiers, et al.,
  \comment{onlinetitle}``Three-dimensional hydrodynamic simulations of
  accretion tori in Kerr spacetimes'', (\comment{onlinemonth}April,
  \comment{onlineyear}2002), \comment{fileformat}[Online Los Alamos Archive
  Preprint]: cited on \comment{cited}14 June 2002,
  \comment{onlineaddress}http://xxx.lanl.gov/abs/astro-ph/0204163.
  \keywords{Numerical relativistic hydrodynamics, accretion, black holes}

\bibitem{delzanna02a}\comment{article}
\comment{author}Del~Zanna, L., and Bucciantini, N., \comment{title}``An
  efficient shock-capturing central-type scheme for multidimensional
  relativistic flows. I. Hydrodynamics'', \comment{journal}{\em Astron.
  Astrophys.}, \comment{volume}{\bf 390}, \comment{pages}1177--1186,
  (\comment{year}2002). For a related online version see:
  \comment{author}L.~Del~Zanna, et al., \comment{onlinetitle}``An efficient
  shock-capturing central-type scheme for multidimensional relativistic flows.
  I. Hydrodynamics'', (\comment{onlinemonth}May, \comment{onlineyear}2002),
  \comment{fileformat}[Online Los Alamos Archive Preprint]: cited on
  \comment{cited}13 June 2002,
  \comment{onlineaddress}http://xxx.lanl.gov/abs/astro-ph/0205290.
  \keywords{Numerical relativistic hydrodynamics, numerical methods,
  computational fluid dynamics, finite difference and finite volume methods}

\bibitem{dimmelmeier01a}\comment{article}
\comment{author}Dimmelmeier, H., Font, J.~A., and M\"uller, E.,
  \comment{title}``Gravitational waves from relativistic rotational core
  collapse'', \comment{journal}{\em Astrophys. J.}, \comment{volume}{\bf 560},
  \comment{pages}L163--L166, (\comment{year}2001). For a related online version
  see: \comment{author}H.~Dimmelmeier, et al.,
  \comment{onlinetitle}``Gravitational waves from relativistic rotational core
  collapse'', (\comment{onlinemonth}March, \comment{onlineyear}2001),
  \comment{fileformat}[Online Los Alamos Archive Preprint]: cited on
  \comment{cited}13 June 2002,
  \comment{onlineaddress}http://xxx.lanl.gov/abs/astro-ph/0103088.
  \keywords{Numerical relativistic hydrodynamics, gravitational collapse,
  supernovae, gravitational radiation, Einstein equations}

\bibitem{dimmelmeier02a}\comment{article}
\comment{author}Dimmelmeier, H., Font, J.~A., and M\"uller, E.,
  \comment{title}``Gravitational waves from relativistic rotational core
  collapse in axisymmetry'', \comment{journal}{\em Class. Quant. Grav.},
  \comment{volume}{\bf 19}, \comment{pages}1291--1296, (\comment{year}2002).
  \keywords{Numerical relativistic hydrodynamics, gravitational collapse,
  supernovae, gravitational radiation, Einstein equations}

\bibitem{dimmelmeier02b}\comment{article}
\comment{author}Dimmelmeier, H., Font, J.~A., and M\"uller, E.,
  \comment{title}``Relativistic simulations of rotational core collapse. I.
  Methods, initial models and code tests'', \comment{journal}{\em Astron.
  Astrophys.}, \comment{volume}{\bf 388}, \comment{pages}917--935,
  (\comment{year}2002). For a related online version see:
  \comment{author}H.~Dimmelmeier, et al., \comment{onlinetitle}``Relativistic
  simulations of rotational core collapse. I. Methods, initial models and code
  tests'', (\comment{onlinemonth}April, \comment{onlineyear}2002),
  \comment{fileformat}[Online Los Alamos Archive Preprint]: cited on
  \comment{cited}13 June 2002,
  \comment{onlineaddress}http://xxx.lanl.gov/abs/astro-ph/0204288.
  \keywords{Numerical relativistic hydrodynamics, gravitational collapse,
  supernovae, gravitational radiation, Einstein equations, numerical methods}

\bibitem{dimmelmeier02c}\comment{article}
\comment{author}Dimmelmeier, H., Font, J.~A., and M\"uller, E.,
  \comment{title}``Relativistic simulations of rotational core collapse. II.
  Collapse dynamics and gravitational radiation'', \comment{journal}{\em
  Astron. Astrophys.}, \comment{volume}{\bf 393}, \comment{pages}523--542,
  (\comment{year}2002). For a related online version see:
  \comment{author}H.~Dimmelmeier, et al., \comment{onlinetitle}``Relativistic
  simulations of rotational core collapse. II. Collapse dynamics and
  gravitational radiation'', (\comment{onlinemonth}April,
  \comment{onlineyear}2002), \comment{fileformat}[Online Los Alamos Archive
  Preprint]: cited on \comment{cited}13 June 2002,
  \comment{onlineaddress}http://xxx.lanl.gov/abs/astro-ph/0204289. in press.
  \keywords{Numerical relativistic hydrodynamics, gravitational collapse,
  supernovae, gravitational radiation, Einstein equations}

\bibitem{dolezal95}\comment{article}
\comment{author}Dolezal, A., and Wong, S.~S.~M., \comment{title}``Relativistic
  hydrodynamics and Essentially Non-Oscillatory shock capturing schemes'',
  \comment{journal}{\em J. Comput. Phys.}, \comment{volume}{\bf 120},
  \comment{pages}266--277, (\comment{year}1995). \keywords{computational fluid
  dynamics, relativistic hydrodynamics, finite difference and finite volume
  methods, hyperbolic equations}

\bibitem{donat98}\comment{article}
\comment{author}Donat, R., Font, J.~A., Ib{\'a}\~nez, J.~M., and Marquina, A.,
  \comment{title}``A Flux-Split Algorithm applied to Relativistic Flows'',
  \comment{journal}{\em J. Comput. Phys.}, \comment{volume}{\bf 146},
  \comment{pages}58--81, (\comment{year}1998). \keywords{computational fluid
  dynamics, relativistic hydrodynamics, finite difference and finite volume
  methods, hyperbolic equations}

\bibitem{donat96}\comment{article}
\comment{author}Donat, R., and Marquina, A., \comment{title}``Capturing shock
  reflections: an improved flux formula'', \comment{journal}{\em J. Comput.
  Phys.}, \comment{volume}{\bf 125}, \comment{pages}42--58,
  (\comment{year}1996). \keywords{computational fluid dynamics, finite
  difference and finite volume methods, hyperbolic equations}

\bibitem{dubal98}\comment{article}
\comment{author}Dubal, M.~R., d'Inverno, R.~A., and Vickers, J.~A.,
  \comment{title}``Combining Cauchy and characteristic codes. V. CCM for a
  spherical spacetime containing a perfect fluid'', \comment{journal}{\em Phys.
  Rev. D}, \comment{volume}{\bf 58}, \comment{pages}044019,
  (\comment{year}1998). \keywords{Numerical relativity, numerical relativistic
  hydrodynamics, initial value problem, Cauchy problem, null surfaces,
  numerical methods}

\bibitem{duez01a}\comment{article}
\comment{author}Duez, M.~D., Baumgarte, T.~B., Shapiro, S.~L., Shibata, M., and
  Uryu, K., \comment{title}``Comparing the inspiral of irrotational and
  corotational binary neutron stars'', \comment{journal}{\em Phys. Rev. D},
  \comment{volume}{\bf 65}, \comment{pages}024016, (\comment{year}2002). For a
  related online version see: \comment{author}M.~D. Duez, et al.,
  \comment{onlinetitle}``Comparing the inspiral of irrotational and
  corotational binary neutron stars'', (\comment{onlinemonth}October,
  \comment{onlineyear}2001), \comment{fileformat}[Online Los Alamos Archive
  Preprint]: cited on \comment{cited}13 June 2002,
  \comment{onlineaddress}http://xxx.lanl.gov/abs/gr-qc/0110006.
  \keywords{Numerical relativity, numerical relativistic hydrodynamics,
  Einstein equations, neutron stars, binary systems, gravitational radiation}

\bibitem{duez02}\comment{article}
\comment{author}Duez, M.~D., Marronetti, P., Shapiro, S.~L., and Baumgarte,
  T.~B., \comment{title}``Hydrodynamic simulations in 3+1 general relativity'',
  \comment{journal}{\em Phys. Rev. D}, \comment{volume}{\bf 67},
  \comment{pages}024004, (\comment{year}2003). For a related online version
  see: \comment{author}M.~D. Duez, et al., \comment{onlinetitle}``Hydrodynamic
  simulations in 3+1 general relativity'', (\comment{onlinemonth}September,
  \comment{onlineyear}2002), \comment{fileformat}[Online Los Alamos Archive
  Preprint]: cited on \comment{cited}2 October 2002,
  \comment{onlineaddress}http://xxx.lanl.gov/abs/gr-qc/0209102.
  \keywords{Numerical relativity, numerical relativistic hydrodynamics,
  Einstein equations}

\bibitem{dykema80}\comment{phdthesis}
\comment{author}Dykema, P.~G., \comment{title}{\em Numerical simulation of
  axisymmetric gravitational collapse}, PhD thesis, (\comment{school}University
  of Texas at Austin, \comment{year}1980). \keywords{Gravitational collapse,
  numerical relativity, Einstein equations, ADM formalism, relativistic
  hydrodynamics, finite difference methods, numerical methods}

\bibitem{einfeldt88}\comment{article}
\comment{author}Einfeldt, B., \comment{title}``On Godunov-type methods for gas
  dynamics'', \comment{journal}{\em SIAM J. Num. Anal.}, \comment{volume}{\bf
  25}, \comment{pages}294--318, (\comment{year}1988). \keywords{computational
  fluid dynamics, finite difference and finite volume methods, hyperbolic
  equations}

\bibitem{eulderink93}\comment{phdthesis}
\comment{author}Eulderink, F., \comment{title}{\em Numerical relativistic
  hydrodynamics}, PhD thesis, (\comment{school}Rijksuniversitet Leiden,
  \comment{year}1993). \keywords{computational fluid dynamics, numerical
  relativity, relativistic hydrodynamics, finite difference and finite volume
  methods, hyperbolic equations, numerical methods}

\bibitem{eulderink94}\comment{article}
\comment{author}Eulderink, F., and Mellema, G., \comment{title}``Special
  relativistic jet collimation by inertial confinement'', \comment{journal}{\em
  Astron. Astrophys.}, \comment{volume}{\bf 284}, \comment{pages}654--662,
  (\comment{year}1994). \keywords{special relativity, relativistic
  hydrodynamics, computational fluid dynamics, numerical methods, extragalactic
  jets}

\bibitem{eulderink95}\comment{article}
\comment{author}Eulderink, F., and Mellema, G., \comment{title}``General
  relativistic hydrodynamics with a Roe solver'', \comment{journal}{\em Astron.
  Astrophys. Suppl. Ser.}, \comment{volume}{\bf 110}, \comment{pages}587--623,
  (\comment{year}1995). For a related online version see:
  \comment{author}F.~Eulderink, et al., \comment{onlinetitle}``General
  relativistic hydrodynamics with a Roe solver'',
  (\comment{onlinemonth}November, \comment{onlineyear}1994),
  \comment{fileformat}[Online Los Alamos Archive Preprint]: cited on
  \comment{cited}15 February 2000,
  \comment{onlineaddress}http://xxx.lanl.gov/abs/astro-ph/9411056.
  \keywords{computational fluid dynamics, numerical relativity, relativistic
  hydrodynamics, finite difference and finite volume methods, hyperbolic
  equations, numerical methods}

\bibitem{evans86}\comment{inbook}
\comment{author}Evans, C., \comment{title}``An Approach for Calculating
  Axisymmetric Gravitational Collapse'', in \comment{editor}Centrella, J., ed.,
  \comment{booktitle}{\em Dynamical Spacetimes and Numerical Relativity},
  \comment{pages} 3--39, (\comment{publisher}Cambridge University Press,
  \comment{address}Cambridge, England, \comment{year}1986).
  \keywords{computational fluid dynamics, gravitational collapse, gravitational
  radiation, numerical relativity, relativistic hydrodynamics, finite
  difference and finite volume methods, hyperbolic equations, numerical
  methods}

\bibitem{evans88}\comment{article}
\comment{author}Evans, C., and Hawley, J.~F., \comment{title}``Simulation of
  magnetohydrodynamic flows: a constrained transport method'',
  \comment{journal}{\em Astrophys. J.}, \comment{volume}{\bf 332},
  \comment{pages}659--677, (\comment{year}1988). \keywords{computational fluid
  dynamics, relativistic magnetohydrodynamics, finite difference and finite
  volume methods, numerical methods}

\bibitem{evans93}\comment{article}
\comment{author}Evans, C.~R., and Coleman, J.~S., \comment{title}``Critical
  phenomena and self-similarity in the gravitational collapse of radiation
  fluid'', \comment{journal}{\em Phys. Rev. Lett.}, \comment{volume}{\bf 72},
  \comment{pages}1782--1785, (\comment{year}1994). \keywords{Gravitational
  collapse, critical phenomena}

\bibitem{evans86b}\comment{inproceedings}
\comment{author}Evans, C.~R., Smarr, L.~L., and Wilson, J.~R.,
  \comment{title}``Numerical relativistic gravitational collapse with spatial
  time slices'', in \comment{editor}Norman, M.~L., and Winkler, K-H.~A., eds.,
  \comment{booktitle}{\em Astrophysical Radiation Hydrodynamics},
  \comment{pages} 491--529. Reidel Publishing Company, (\comment{year}1986).
  \keywords{ADM formalism, Einstein equations, relativistic hydrodynamics,
  gravitational collapse, gauge conditions, constraint equations, Cauchy
  problem, differential equations, numerical relativity, numerical methods}

\bibitem{falle96}\comment{article}
\comment{author}Falle, S.~A.~E.~G., and Komissarov, S.~S., \comment{title}``An
  upwind numerical scheme for relativistic hydrodynamics with a general
  equation of state'', \comment{journal}{\em Mon. Not. Roy. Astron. Soc.},
  \comment{volume}{\bf 278}, \comment{pages}586--602, (\comment{year}1996).
  \keywords{computational fluid dynamics, relativistic hydrodynamics, finite
  difference and finite volume methods, hyperbolic equations}

\bibitem{finn90a}\comment{article}
\comment{author}Finn, L.~S., and Evans, C.~R., \comment{title}``Determining
  gravitational radiation from Newtonian self-gravitating systems'',
  \comment{journal}{\em Astrophys.~J.}, \comment{volume}{\bf 351},
  \comment{pages}588--600, (\comment{year}1990). \keywords{}

\bibitem{flanagan98}\comment{article}
\comment{author}Flanagan, \'E., \comment{title}``Possible explanation for
  star-crushing effect in binary neutron star simulations'',
  \comment{journal}{\em Phys. Rev. Lett.}, \comment{volume}{\bf 82},
  \comment{pages}1354--1357, (\comment{year}1999). For a related online version
  see: \comment{author}\'E. Flanagan, \comment{onlinetitle}``Possible
  explanation for star-crushing effect in binary neutron star simulations'',
  (\comment{onlinemonth}November, \comment{onlineyear}1998),
  \comment{fileformat}[Online Los Alamos Archive Preprint]: cited on
  \comment{cited}15 February 2000,
  \comment{onlineaddress}http://xxx.lanl.gov/abs/astro-ph/9811132.
  \keywords{Einstein equations, constraint equations, numerical relativistic
  hydrodynamics, numerical relativity, binary systems, neutron stars}

\bibitem{font02d}\comment{article}
\comment{author}Font, J.~A., and Daigne, F., \comment{title}``On the stability
  of thick accretion disks around black holes'', \comment{journal}{\em
  Astrophys. J.}, \comment{volume}{\bf 581}, \comment{pages}L23--L26,
  (\comment{year}2002). For a related online version see: \comment{author}J.~A.
  Font, et al., \comment{onlinetitle}``On the stability of thick accretion
  disks around black holes'', (\comment{onlinemonth}November,
  \comment{onlineyear}2002), \comment{fileformat}[Online Los Alamos Archive
  Preprint]: cited on \comment{cited}27 March 2003,
  \comment{onlineaddress}http://xxx.lanl.gov/abs/astro-ph/0211102.
  \keywords{Accretion, Black holes, Numerical relativistic hydrodynamics}

\bibitem{font02b}\comment{article}
\comment{author}Font, J.~A., and Daigne, F., \comment{title}``The runaway
  instability of thick discs around black holes - I. The constant angular
  momentum case'', \comment{journal}{\em Mon. Not. R. Astron. Soc.},
  \comment{volume}{\bf 334}, \comment{pages}383--400, (\comment{year}2002). For
  a related online version see: \comment{author}J.~A. Font, et al.,
  \comment{onlinetitle}``The runaway instability of thick discs around black
  holes. I. The constant angular momentum case'', (\comment{onlinemonth}March,
  \comment{onlineyear}2002), \comment{fileformat}[Online Los Alamos Archive
  Preprint]: cited on \comment{cited}13 June 2002,
  \comment{onlineaddress}http://xxx.lanl.gov/abs/astro-ph/0203403.
  \keywords{Accretion, Black holes, Numerical relativistic hydrodynamics, shock
  waves, relativistic hydrodynamics}

\bibitem{font01a}\comment{article}
\comment{author}Font, J.~A., Dimmelmeier, H., Gupta, A., and Stergioulas, N.,
  \comment{title}``Axisymmetric modes of rotating relativistic stars in the
  Cowling approximation'', \comment{journal}{\em Mon. Not. R. Astron. Soc.},
  \comment{volume}{\bf 325}, \comment{pages}1463--1470, (\comment{year}2001).
  For a related online version see: \comment{author}J.~A. Font, et al.,
  \comment{onlinetitle}``Axisymmetric modes of rotating relativistic stars in
  the Cowling approximation'', (\comment{onlinemonth}December,
  \comment{onlineyear}2000), \comment{fileformat}[Online Los Alamos Archive
  Preprint]: cited on \comment{cited}13 June 2002,
  \comment{onlineaddress}http://xxx.lanl.gov/abs/astro-ph/0012477.
  \keywords{Numerical methods, neutron stars, numerical relativistic
  hydrodynamics}

\bibitem{font02a}\comment{article}
\comment{author}Font, J.~A., Goodale, T., Iyer, S., Miller, M., Rezzolla, L.,
  Seidel, E., Stergioulas, N., Suen, W.-M., and Tobias, M.,
  \comment{title}``Three-dimensional general relativistic hydrodynamics II:
  long-term dynamics of single relativistic stars'', \comment{journal}{\em
  Phys. Rev. D}, \comment{volume}{\bf 65}, \comment{pages}084024,
  (\comment{year}2002). For a related online version see: \comment{author}J.~A.
  Font, et al., \comment{onlinetitle}``Three-dimensional general relativistic
  hydrodynamics II: long-term dynamics of single relativistic stars'',
  (\comment{onlinemonth}October, \comment{onlineyear}2001),
  \comment{fileformat}[Online Los Alamos Archive Preprint]: cited on
  \comment{cited}13 June 2002,
  \comment{onlineaddress}http://xxx.lanl.gov/abs/gr-qc/0110047. \keywords{ADM
  formalism, computational fluid dynamics, relativistic hydrodynamics,
  numerical relativity, finite difference and finite volume methods, numerical
  methods, numerical relativistic hydrodynamics}

\bibitem{font98b}\comment{article}
\comment{author}Font, J.~A., and Ib{\'a}\~nez, J.~M.,
  \comment{title}``Non-axisymmetric Relativistic Bondi-Hoyle Accretion onto a
  Schwarzschild Black Hole'', \comment{journal}{\em Mon. Not. R. Astron. Soc.},
  \comment{volume}{\bf 298}, \comment{pages}835--846, (\comment{year}1998). For
  a related online version see: \comment{author}J.~A. Font, et al.,
  \comment{onlinetitle}``Non-axisymmetric Relativistic Bondi-Hoyle Accretion
  onto a Schwarzschild Black Hole'', (\comment{onlinemonth}April,
  \comment{onlineyear}1998), \comment{fileformat}[Online Los Alamos Archive
  Preprint]: cited on \comment{cited}1 May 1998,
  \comment{onlineaddress}http://xxx.lanl.gov/abs/astro-ph/9804254.
  \keywords{relativistic hydrodynamics, accretion, numerical methods, black
  holes, shock waves, computational fluid dynamics, finite difference and
  finite volume methods, numerical relativistic hydrodynamics}

\bibitem{font98a}\comment{article}
\comment{author}Font, J.~A., and Ib{\'a}\~nez, J.~M., \comment{title}``A
  Numerical Study of Relativistic Bondi-Hoyle Accretion onto a Moving Black
  Hole: Axisymmetric Computations in a Schwarzschild Background'',
  \comment{journal}{\em Astrophys. J.}, \comment{volume}{\bf 494},
  \comment{pages}297--316, (\comment{year}1998). \keywords{relativistic
  hydrodynamics, accretion, numerical methods, black holes, shock waves,
  computational fluid dynamics, finite difference and finite volume methods,
  numerical relativistic hydrodynamics}

\bibitem{font02c}\comment{article}
\comment{author}Font, J.~A., Ib\'a\~nez, J.~M., and Mart\'{\i}, J.~M.,
  (\comment{year}2002). unpublished. \keywords{}

\bibitem{font94}\comment{article}
\comment{author}Font, J.~A., Ib{\'a}\~nez, J.~M., Mart\'{\i}, J.~M., and
  Marquina, A., \comment{title}``Multidimensional relativistic hydrodynamics:
  characteristic fields and modern high-resolution shock-capturing schemes'',
  \comment{journal}{\em Astron. Astrophys.}, \comment{volume}{\bf 282},
  \comment{pages}304--314, (\comment{year}1994). \keywords{computational fluid
  dynamics, relativistic hydrodynamics, finite difference and finite volume
  methods, hyperbolic equations}

\bibitem{font98c}\comment{article}
\comment{author}Font, J.~A., Ib{\'a}\~nez, J.~M., and Papadopoulos, P.,
  \comment{title}``A horizon-adapted approach to the study of relativistic
  accretion flows onto rotating black holes'', \comment{journal}{\em Astrophys.
  J. Lett.}, \comment{volume}{\bf 507}, \comment{pages}L67--L70,
  (\comment{year}1998). For a related online version see: \comment{author}J.~A.
  Font, et al., \comment{onlinetitle}``A horizon-adapted approach to the study
  of relativistic accretion flows onto rotating black holes'',
  (\comment{onlinemonth}May, \comment{onlineyear}1998),
  \comment{fileformat}[Online Los Alamos Archive Preprint]: cited on
  \comment{cited}1 June 1998,
  \comment{onlineaddress}http://xxx.lanl.gov/abs/astro-ph/9805269.
  \keywords{Accretion, Black holes, Numerical relativistic hydrodynamics, shock
  waves, relativistic hydrodynamics}

\bibitem{font99b}\comment{article}
\comment{author}Font, J.~A., Ib{\'a}\~nez, J.~M., and Papadopoulos, P.,
  \comment{title}``Non-axisymmetric Relativistic Bondi-Hoyle Accretion onto a
  Kerr Black Hole'', \comment{journal}{\em Mon. Not. R. Astron. Soc.},
  \comment{volume}{\bf 305}, \comment{pages}920--936, (\comment{year}1999). For
  a related online version see: \comment{author}J.~A. Font, et al.,
  \comment{onlinetitle}``Non-axisymmetric Relativistic Bondi-Hoyle Accretion
  onto a Kerr Black Hole'', (\comment{onlinemonth}October,
  \comment{onlineyear}1998), \comment{fileformat}[Online Los Alamos Archive
  Preprint]: cited on \comment{cited}1 November 1998,
  \comment{onlineaddress}http://xxx.lanl.gov/abs/astro-ph/9810344.
  \keywords{relativistic hydrodynamics, accretion, numerical methods, black
  holes, shock waves, computational fluid dynamics, finite difference and
  finite volume methods, numerical relativistic hydrodynamics}

\bibitem{font99a}\comment{article}
\comment{author}Font, J.~A., Miller, M., Suen, W.-M., and Tobias, M.,
  \comment{title}``Three Dimensional Numerical General Relativistic
  Hydrodynamics: Formulations, Methods and Code Tests'', \comment{journal}{\em
  Phys. Rev. D}, \comment{volume}{\bf 61}, \comment{pages}044011,
  (\comment{year}2000). For a related online version see: \comment{author}J.~A.
  Font, et al., \comment{onlinetitle}``Three Dimensional Numerical General
  Relativistic Hydrodynamics: Formulations, Methods and Code Tests'',
  (\comment{onlinemonth}November, \comment{onlineyear}1998),
  \comment{fileformat}[Online Los Alamos Archive Preprint]: cited on
  \comment{cited}1 December 1998,
  \comment{onlineaddress}http://xxx.lanl.gov/abs/astro-ph/9811015.
  \keywords{ADM formalism, computational fluid dynamics, relativistic
  hydrodynamics, numerical relativity, finite difference and finite volume
  methods, numerical methods, numerical relativistic hydrodynamics}

\bibitem{font00a}\comment{article}
\comment{author}Font, J.~A., Stergioulas, N., and Kokkotas, K.,
  \comment{title}``Nonlinear hydrodynamical evolution of rotating relativistic
  stars: Numerical methods and code tests'', \comment{journal}{\em Mon. Not. R.
  Astron. Soc.}, \comment{volume}{\bf 313}, \comment{pages}678--688,
  (\comment{year}2000). For a related online version see: \comment{author}J.~A.
  Font, et al., \comment{onlinetitle}``Nonlinear hydrodynamical evolution of
  rotating relativistic stars: Numerical methods and code tests'',
  (\comment{onlinemonth}August, \comment{onlineyear}1999),
  \comment{fileformat}[Online Los Alamos Archive Preprint]: cited on
  \comment{cited}15 February 2000,
  \comment{onlineaddress}http://xxx.lanl.gov/abs/gr-qc/9908010.
  \keywords{Numerical methods, finite difference and finite volume methods,
  neutron stars, numerical relativistic hydrodynamics, hyperbolic equations}

\bibitem{frank92}\comment{book}
\comment{author}Frank, J., King, A., and Raine, D., \comment{title}{\em
  Accretion power in astrophysics}, (\comment{publisher}Cambridge University
  Press, \comment{address}Cambridge, England, \comment{year}1992).
  \keywords{Accretion, accretion disks, astrophysics, stars}

\bibitem{friedrich02}\comment{inbook}
\comment{author}Friedrich, H., \comment{title}``Conformal Einstein evolution'',
  in \comment{editor}Friedrich, H., and Frauendiener, J., eds.,
  \comment{booktitle}{\em Lecture Notes in Physics. Vol. 604. The conformal
  structure of spacetime: Geometry, analysis, numerics}, \comment{pages} 1--50,
  (\comment{publisher}Springer, \comment{address}Springer Verlag, Berlin,
  Heidelberg, New York, \comment{year}2002). For a related online version see:
  \comment{author}H.~Friedrich, \comment{onlinetitle}``Conformal Einstein
  evolution'', (\comment{onlinemonth}September, \comment{onlineyear}2002),
  \comment{fileformat}[Online Los Alamos Archive Preprint]: cited on
  \comment{cited}15 April 2003,
  \comment{onlineaddress}http://xxx.lanl.gov/abs/gr-qc/0209018.
  \keywords{Einstein equations, numerical relativity, differential equations}

\bibitem{friedrichs74}\comment{article}
\comment{author}Friedrichs, K.~O., \comment{title}``On the laws of relativistic
  electromagneto-fluid dynamics'', \comment{journal}{\em Commun. Pure Appl.
  Math.}, \comment{volume}{\bf 27}, \comment{pages}749--808,
  (\comment{year}1974). \keywords{}

\bibitem{fryer00a}\comment{article}
\comment{author}Fryer, C., and Heger, A., \comment{title}``Core--collapse
  simulations of rotating stars'', \comment{journal}{\em Astrophys.~J.},
  \comment{volume}{\bf 541}, \comment{pages}1033--1050, (\comment{year}2000).
  For a related online version see: \comment{author}C.~Fryer, et al.,
  \comment{onlinetitle}``Core--collapse simulations of rotating stars'',
  (\comment{onlinemonth}July, \comment{onlineyear}1999),
  \comment{fileformat}[Online Los Alamos Archive Preprint]: cited on
  \comment{cited}15 July 2002,
  \comment{onlineaddress}http://xxx.lanl.gov/abs/astro-ph/9907433.
  \keywords{Gravitational collapse, supernovae}

\bibitem{fryer02a}\comment{article}
\comment{author}Fryer, C., Holz, D.~E., and Heger, A.,
  \comment{title}``Gravitational wave emission from core collapse of massive
  stars'', \comment{journal}{\em Astrophys.~J.}, \comment{volume}{\bf 565},
  \comment{pages}430--446, (\comment{year}2002). For a related online version
  see: \comment{author}C.~Fryer, et al., \comment{onlinetitle}``Gravitational
  wave emission from core collapse of massive stars'',
  (\comment{onlinemonth}June, \comment{onlineyear}2001),
  \comment{fileformat}[Online Los Alamos Archive Preprint]: cited on
  \comment{cited}15 July 2002,
  \comment{onlineaddress}http://xxx.lanl.gov/abs/astro-ph/0106113.
  \keywords{Gravitational collapse, supernovae, gravitational radiation}

\bibitem{fryxell89}\comment{article}
\comment{author}Fryxell, B.~A., M\"uller, E., and Arnett, W.~D.,
  \comment{journal}{\em Max-Planck-Institut f\"ur Astrophysik preprint},
  \comment{volume}{\bf 449}, (\comment{year}1989). \keywords{}

\bibitem{gingold77}\comment{article}
\comment{author}Gingold, R.~A., and Monaghan, J.~J., \comment{title}``Smoothed
  particle hydrodynamics - Theory and application to non-spherical stars'',
  \comment{journal}{\em Mon. Not. R. Astron. Soc.}, \comment{volume}{\bf 181},
  \comment{pages}375--389, (\comment{year}1977). \keywords{computational fluid
  dynamics, smoothed particle hydrodynamics, numerical methods, hydrodynamics}

\bibitem{gingold82}\comment{article}
\comment{author}Gingold, R.~A., and Monaghan, J.~J., \comment{title}``Kernel
  estimates as a basis for general particle methods in hydrodynamics'',
  \comment{journal}{\em J. Comput. Phys.}, \comment{pages}429--453,
  (\comment{year}1982). \keywords{computational fluid dynamics, smoothed
  particle hydrodynamics, numerical methods, hydrodynamics}

\bibitem{glaister88}\comment{article}
\comment{author}Glaister, P., \comment{title}``An approximate linearised
  Riemann solver for the Euler equations for real gases'',
  \comment{journal}{\em J. Comput. Phys.}, \comment{volume}{\bf 74},
  \comment{pages}382--408, (\comment{year}1988). \keywords{computational fluid
  dynamics, finite difference and finite volume methods, hyperbolic equations,
  numerical methods}

\bibitem{glendening}\comment{book}
\comment{author}Glendening, N.~K., \comment{title}{\em Compact stars. Nuclear
  physics, particle physics and general relativity}, Astronomy and astrophysics
  library, (\comment{publisher}Springer-Verlag, \comment{address}Berlin,
  \comment{year}1997). \keywords{Astrophysics, relativistic stars, nuclear
  physics, neutron stars, pulsars, black holes, relativistic astrophysics}

\bibitem{godunov59}\comment{article}
\comment{author}Godunov, S.~K., \comment{title}``A finite difference method for
  the numerical computation and discontinuous solutions of the equations of
  fluid dynamics'', \comment{journal}{\em Mat. Sb.}, \comment{volume}{\bf 47},
  \comment{pages}271--306, (\comment{year}1959). in Russian.
  \keywords{computational fluid dynamics, finite difference and finite volume
  methods, hyperbolic equations, numerical methods}

\bibitem{gomez94}\comment{article}
\comment{author}G\'omez, R., Papadopoulos, P., and Winicour, J.,
  \comment{title}``Null cone evolution of axisymmetric vacuum space-times'',
  \comment{journal}{\em J. Math. Phys.}, \comment{volume}{\bf 35},
  \comment{pages}4184--4204, (\comment{year}1994). \keywords{Numerical
  relativity, null surfaces, Einstein equations, initial value problem,
  numerical methods}

\bibitem{gottlieb77}\comment{book}
\comment{author}Gottlieb, D., and Orszag, S.~A., \comment{title}{\em Numerical
  analysis of spectral methods: theory and applications},
  (\comment{publisher}Society for Industrial and Applied Mathematics,
  \comment{address}Philadelphia, \comment{year}1977). \keywords{Numerical
  methods, spectral methods, differential equations}

\bibitem{gourgoulhon91}\comment{article}
\comment{author}Gourgoulhon, E., \comment{title}``Simple equations for general
  relativistic hydrodynamics in spherical symmetry applied to neutron star
  collapse'', \comment{journal}{\em Astron. Astrophys.}, \comment{volume}{\bf
  252}, \comment{pages}651--663, (\comment{year}1991). \keywords{Gravitational
  collapse, Numerical methods, Spectral Methods, Relativistic Hydrodynamics,
  Neutron Stars}

\bibitem{gourgoulhon01a}\comment{article}
\comment{author}Gourgoulhon, E., Grandcl\'ement, P., Taniguchi, K., Marck,
  J.-A., and Bonazzola, S., \comment{title}``Quasiequilibrium sequences of
  synchronized and irrotational binary neutron stars in general relativity:
  method and tests'', \comment{journal}{\em Phys. Rev. D}, \comment{volume}{\bf
  63}, \comment{pages}064029, (\comment{year}2001). For a related online
  version see: \comment{author}E.~Gourgoulhon, et al.,
  \comment{onlinetitle}``Quasiequilibrium sequences of synchronized and
  irrotational binary neutron stars in general relativity: method and tests'',
  (\comment{onlinemonth}July, \comment{onlineyear}2000),
  \comment{fileformat}[Online Los Alamos Archive Preprint]: cited on
  \comment{cited}24 June 2002,
  \comment{onlineaddress}http://xxx.lanl.gov/abs/gr-qc/0007028.
  \keywords{Einstein equations, relativistic hydrodynamics, initial value
  problem, hyperbolic equations, elliptic equations, binary systems, spectral
  methods, neutron stars}

\bibitem{gressman02a}\comment{article}
\comment{author}Gressman, P., Lin, L-P., Suen, W-M., Stergioulas, N., and
  Friedman, J.~L., \comment{title}``Nonlinear $r$-modes in neutron stars:
  instability of an unstable mode'', \comment{journal}{\em Phys. Rev. D},
  \comment{volume}{\bf 66}, \comment{pages}041303, (\comment{year}2002). For a
  related online version see: \comment{author}P.~Gressman, et al.,
  \comment{onlinetitle}``Nonlinear $r$-modes in neutron stars: instability of
  an unstable mode'', (\comment{onlinemonth}January, \comment{onlineyear}2003),
  \comment{fileformat}[Online Los Alamos Archive Preprint]: cited on
  \comment{cited}27 March 2003,
  \comment{onlineaddress}http://xxx.lanl.gov/abs/gr-qc/0301014.
  \keywords{neutron stars}

\bibitem{gundlach}\comment{article}
\comment{author}Gundlach, C., \comment{title}``Critical phenomena in
  gravitational collapse'', \comment{journal}{\em Liv. Rev. Relativ.},
  \comment{volume}{\bf 2}, \comment{pages}4, (\comment{year}1999). For a
  related online version see: \comment{author}C.~Gundlach,
  \comment{onlinetitle}``Critical phenomena in gravitational collapse'',
  (\comment{onlinemonth}January, \comment{onlineyear}2000),
  \comment{fileformat}[Online Los Alamos Archive Preprint]: cited on
  \comment{cited}4 July 2002,
  \comment{onlineaddress}http://xxx.lanl.gov/abs/gr-qc/0001046.
  \keywords{Gravitational collapse, critical phenomena}

\bibitem{harten84}\comment{article}
\comment{author}Harten, A., \comment{title}``On a class of high resolution
  total-variation stable finite difference schemes'', \comment{journal}{\em
  SIAM J. Numer. Anal.}, \comment{volume}{\bf 21}, \comment{pages}1--23,
  (\comment{year}1984). \keywords{computational fluid dynamics, finite
  difference and finite volume methods, hyperbolic equations, numerical
  methods}

\bibitem{harten87}\comment{article}
\comment{author}Harten, A., Engquist, B., Osher, S., and Chakrabarthy, S.~R.,
  \comment{title}``Uniformly high order accurate essentially non-oscillatory
  schemes, III'', \comment{journal}{\em J. Comp. Phys.}, \comment{volume}{\bf
  71}, \comment{pages}231--303, (\comment{year}1987). \keywords{computational
  fluid dynamics, finite difference and finite volume methods, hyperbolic
  equations, numerical methods}

\bibitem{harten83}\comment{article}
\comment{author}Harten, A., Lax, P.~D., and van Leer, B., \comment{title}``On
  upstream differencing and Godunov-type schemes for hyperbolic conservation
  laws'', \comment{journal}{\em SIAM Review}, \comment{volume}{\bf 25},
  \comment{pages}35--61, (\comment{year}1983). \keywords{computational fluid
  dynamics, finite difference and finite volume methods, hyperbolic equations,
  numerical methods}

\bibitem{haugan82}\comment{article}
\comment{author}Haugan, M.~P., Shapiro, S.~L., and Wasserman, I.,
  \comment{title}``The suppression of gravitational radiation from finite-size
  stars falling into black holes'', \comment{journal}{\em Astrophys. J.},
  \comment{volume}{\bf 257}, \comment{pages}283--290, (\comment{year}1982).
  \keywords{Gravitational radiation, accretion, black holes, approximation
  methods}

\bibitem{hawley86}\comment{inbook}
\comment{author}Hawley, J.~F., \comment{title}``General relativistic
  hydrodynamics near black holes'', in \comment{editor}Centrella, J., ed.,
  \comment{booktitle}{\em Dynamical Spacetimes and Numerical Relativity},
  \comment{pages} 101--122, (\comment{publisher}Cambridge University Press,
  \comment{address}Cambridge, England, \comment{year}1986). \keywords{black
  holes, accretion, numerical relativistic hydrodynamics, numerical methods,
  finite difference methods}

\bibitem{hawley91}\comment{article}
\comment{author}Hawley, J.~F., \comment{title}``Three-dimensional simulations
  of black hole tori'', \comment{journal}{\em Astrophys. J.},
  \comment{volume}{\bf 381}, \comment{pages}496--507, (\comment{year}1991).
  \keywords{black holes, accretion, computational fluid dynamics, relativistic
  hydrodynamics}

\bibitem{hawley84a}\comment{article}
\comment{author}Hawley, J.~F., Smarr, L.~L., and Wilson, J.~R.,
  \comment{title}``A numerical study of nonspherical black hole accretion. I.
  Equations and test problems'', \comment{journal}{\em Astrophys. J.},
  \comment{volume}{\bf 277}, \comment{pages}296--311, (\comment{year}1984).
  \keywords{black holes, computational fluid dynamics, relativistic
  hydrodynamics, accretion, finite difference and finite volume methods,
  numerical methods, differential equations}

\bibitem{hawley84b}\comment{article}
\comment{author}Hawley, J.~F., Smarr, L.~L., and Wilson, J.~R.,
  \comment{title}``A numerical study of nonspherical black hole accretion. II.
  Finite differencing and code calibration'', \comment{journal}{\em Astrophys.
  J. Suppl. Ser.}, \comment{volume}{\bf 55}, \comment{pages}211--246,
  (\comment{year}1984). \keywords{black holes, computational fluid dynamics,
  relativistic hydrodynamics, accretion, finite difference and finite volume
  methods, numerical methods}

\bibitem{hernandez66}\comment{article}
\comment{author}Hern\'andez, W.~C., and Misner, C.~W.,
  \comment{title}``Observer time as a coordinate in relativistic spherical
  hydrodynamics'', \comment{journal}{\em Astrophys. J.}, \comment{volume}{\bf
  143}, \comment{pages}452--464, (\comment{year}1966). \keywords{Relativistic
  hydrodynamics, null surfaces, gravitational collapse}

\bibitem{hoyle39}\comment{article}
\comment{author}Hoyle, F., and Lyttleton, R.~A., \comment{journal}{\em Proc.
  Cambridge Phil. Soc.}, \comment{volume}{\bf 35}, \comment{pages}405,
  (\comment{year}1939). \keywords{Accretion}

\bibitem{ibanez93}\comment{inproceedings}
\comment{author}Ib\'a\~nez, J.~M., in \comment{editor}Chinea, F.~J., and
  Gonz\'alez-Romero, L.~M., eds., \comment{booktitle}{\em Lecture Notes in
  Physics, Vol. 423, Rotating objects and relativistic physics},
  \comment{pages} 149, (\comment{publisher}Springer-Verlag,
  \comment{address}Berlin, \comment{year}1993). \keywords{Numerical
  relativistic hydrodynamics, numerical methods, hyperbolic equations, finite
  difference and finite volume methods}

\bibitem{ibanez99b}\comment{inproceedings}
\comment{author}Ib\'a\~nez, J.~M., Aloy, M.~A., Font, J.~A., Mart\'{\i}, J.~M.,
  Miralles, J.~A., and Pons, J.~A., \comment{title}``Riemann solvers in general
  relativistic hydrodynamics'', in \comment{editor}Toro, E.F., ed.,
  \comment{booktitle}{\em Godunov methods: theory and applications},
  \comment{pages} 485--496, (\comment{publisher}Kluwer Academic / Plenum
  Publishers, \comment{address}New York, \comment{year}2001). For a related
  online version see: \comment{author}J.~M. Ib\'a\~nez, et al.,
  \comment{onlinetitle}``Riemann solvers in general relativistic
  hydrodynamics'', (\comment{onlinemonth}November, \comment{onlineyear}1999),
  \comment{fileformat}[Online Los Alamos Archive Preprint]: cited on
  \comment{cited}15 July 2002,
  \comment{onlineaddress}http://xxx.lanl.gov/abs/astro-ph/9911034.
  \keywords{Relativistic hydrodynamics, numerical methods, finite difference
  and finite volume methods, hyperbolic equations}

\bibitem{ibanez99}\comment{article}
\comment{author}Ib\'a\~nez, J.~M., and Mart\'{\i}, J.~M.,
  \comment{title}``Riemann solvers in relativistic astrophysics'',
  \comment{journal}{\em J. Comput. Appl. Math.}, \comment{volume}{\bf 109},
  \comment{pages}173--211, (\comment{year}1999). \keywords{Relativistic
  astrophysics, Relativistic hydrodynamics, numerical relativistic
  hydrodynamics, computational fluid dynamics, numerical methods, finite
  difference and finite volume methods, hyperbolic equations}

\bibitem{ibanez92}\comment{inproceedings}
\comment{author}Ib\'a\~nez, J.~M., Mart\'{\i}, J.~M., Miralles, J.~A., and
  Romero, J.~V., in \comment{editor}d'Inverno, R., ed., \comment{booktitle}{\em
  Approaches to numerical relativity}, \comment{pages} 223,
  (\comment{publisher}Cambridge University Press, \comment{address}Cambridge,
  England, \comment{year}1992). \keywords{gravitational collapse, computational
  fluid dynamics, relativistic hydrodynamics, numerical relativity, finite
  difference and finite volume methods, hyperbolic equations, numerical
  methods}

\bibitem{igumenshchev00}\comment{article}
\comment{author}{Igumenshchev}, I.~V., {Abramowicz}, M.~A., and {Narayan}, R.,
  \comment{title}``Numerical Simulations of Convective Accretion Flows in Three
  Dimensions'', \comment{journal}{\em Astrophys. J}, \comment{volume}{\bf 537},
  \comment{pages}L27--L30, (\comment{year}2000). For a related online version
  see: \comment{author}I.~V. {Igumenshchev}, et al.,
  \comment{onlinetitle}``Numerical Simulations of Convective Accretion Flows in
  Three Dimensions'', (\comment{onlinemonth}April, \comment{onlineyear}2000),
  \comment{fileformat}[Online Los Alamos Archive Preprint]: cited on
  \comment{cited}15 July 2002,
  \comment{onlineaddress}http://xxx.lanl.gov/abs/astro-ph/0004006.
  \keywords{accretion}

\bibitem{igumenshchev97}\comment{article}
\comment{author}Igumenshchev, I.~V., and Belodorov, A.~M.,
  \comment{title}``Numerical simulations of thick disc accretion on to a
  rotating black hole'', \comment{journal}{\em Mon. Not. R. Astron. Soc.},
  \comment{volume}{\bf 284}, \comment{pages}767--772, (\comment{year}1997).
  \keywords{Accretion, Accretion disks, black holes, numerical relativistic
  hydrodynamics, numerical methods}

\bibitem{imshennik92a}\comment{article}
\comment{author}Imshennik, V.~S., and Nadezhin, D.~K.,
  \comment{title}``SN~1987A and rotating neutron star formation'',
  \comment{journal}{\em Sov. Astron. Lett.}, \comment{volume}{\bf 18},
  \comment{pages}79--88, (\comment{year}1992). \keywords{}

\bibitem{isaacson83}\comment{article}
\comment{author}Isaacson, R.~A., Welling, J.~S., and Winicour, J.,
  \comment{title}``Null cone computation of gravitational radiation'',
  \comment{journal}{\em J. Math. Phys.}, \comment{volume}{\bf 24},
  \comment{pages}1824--1834, (\comment{year}1983). \keywords{Numerical
  relativity, gravitational radiation, null surfaces, Einstein equations,
  energy and momentum, initial value problem, numerical methods}

\bibitem{janka89a}\comment{article}
\comment{author}Janka, H.-T., and M\"onchmeyer, R.,
  \comment{title}``Hydrostatic post bounce configurations of collapse rotating
  cores: neutrino emission'', \comment{journal}{\em Astron.\ Astrophys.},
  \comment{volume}{\bf 226}, \comment{pages}69--87, (\comment{year}1989).
  \keywords{}

\bibitem{janka01}\comment{article}
\comment{author}Janka, H.-Th., Kifonidis, K., and Rampp, M.,
  \comment{title}``Supernova explosions and neutron star formation'',
  \comment{journal}{\em Lect. Notes Phys.}, \comment{volume}{\bf 578},
  \comment{pages}333--363, (\comment{year}2001). For a related online version
  see: \comment{author}H.-Th. Janka, et al., \comment{onlinetitle}``Supernova
  explosions and neutron star formation'', (\comment{onlinemonth}March,
  \comment{onlineyear}2001), \comment{fileformat}[Online Los Alamos Archive
  Preprint]: cited on \comment{cited}21 June 2002,
  \comment{onlineaddress}http://xxx.lanl.gov/abs/astro-ph/0103015.
  \keywords{hydrodynamics, gravitational collapse, supernovae, nuclear physics,
  neutrinos, nucleosynthesis}

\bibitem{janka93}\comment{article}
\comment{author}Janka, H.-Th., Zwerger, Th., and M\"onchmeyer, R,
  \comment{title}``Does artificial viscosity destroy prompt type-II supernova
  explosions?'', \comment{journal}{\em Astron. Astrophys.},
  \comment{volume}{\bf 268}, \comment{pages}360--368, (\comment{year}1993).
  \keywords{Hydrodynamics, shock waves, numerical methods, supernovae,
  gravitational collapse}

\bibitem{kheyfets90}\comment{article}
\comment{author}Kheyfets, A., Miller, W.~A., and Zurek, W.~H.,
  \comment{title}``Covariant smoothed particle hydrodynamics on a curved
  background'', \comment{journal}{\em Phys. Rev. D}, \comment{volume}{\bf 41},
  \comment{pages}451--454, (\comment{year}1990). \keywords{computational fluid
  dynamics, relativistic hydrodynamics, smoothed particle methods, numerical
  methods}

\bibitem{kifonidis99}\comment{article}
\comment{author}Kifonidis, K., Plewa, T., Janka, H.-Th., and M\"uller, E.,
  \comment{title}``Nucleosynthesis and clump formation in a core collapse
  supernova'', \comment{journal}{\em Astrophys. J. Lett.}, \comment{volume}{\bf
  531}, \comment{pages}L123--L126, (\comment{year}2000). For a related online
  version see: \comment{author}K.~Kifonidis, et al.,
  \comment{onlinetitle}``Nucleosynthesis and clump formation in a core collapse
  supernova'', (\comment{onlinemonth}November, \comment{onlineyear}1999),
  \comment{fileformat}[Online Los Alamos Archive Preprint]: cited on
  \comment{cited}15 February 2000,
  \comment{onlineaddress}http://xxx.lanl.gov/abs/astro-ph/9911183.
  \keywords{hydrodynamics, computational fluid dynamics, gravitational
  collapse, supernovae, nuclear physics, neutrinos, nucleosynthesis}

\bibitem{kley98}\comment{article}
\comment{author}Kley, W., and Sch\"afer, G., \comment{title}``Relativistic dust
  disks and the Wilson-Mathews approach'', \comment{journal}{\em Phys. Rev. D},
  \comment{volume}{\bf 60}, \comment{pages}027501, (\comment{year}1999). For a
  related online version see: \comment{author}W.~Kley, et al.,
  \comment{onlinetitle}``Relativistic dust disks and the Wilson-Mathews
  approach'', (\comment{onlinemonth}December, \comment{onlineyear}1998),
  \comment{fileformat}[Online Los Alamos Archive Preprint]: cited on
  \comment{cited}15 February 2000,
  \comment{onlineaddress}http://xxx.lanl.gov/abs/gr-qc/9812068.
  \keywords{Einstein equations, numerical relativity, accretion disks,
  post-Newtonian approximations}

\bibitem{koide00}\comment{article}
\comment{author}Koide, S., Meier, D.~L., Shibata, K., and Kudoh, T.,
  \comment{title}``General relativistic simulations of early jet formation in a
  rapidly rotating black hole magnetosphere'', \comment{journal}{\em Astrophys.
  J.}, \comment{volume}{\bf 536}, \comment{pages}668--674,
  (\comment{year}2000). For a related online version see:
  \comment{author}S.~Koide, et al., \comment{onlinetitle}``General relativistic
  simulations of jet formation in a rapidly rotating black hole
  magnetosphere'', (\comment{onlinemonth}July, \comment{onlineyear}1999),
  \comment{fileformat}[Online Los Alamos Archive Preprint]: cited on
  \comment{cited}15 July 2002,
  \comment{onlineaddress}http://xxx.lanl.gov/abs/astro-ph/9907435.
  \keywords{black holes, relativistic astrophysics, computational fluid
  dynamics, relativistic hydrodynamics and magnetohydrodynamics, AGNs,
  extragalactic jets}

\bibitem{koide98}\comment{article}
\comment{author}Koide, S., Shibata, K., and Kudoh, T., \comment{title}``General
  relativistic magnetohydrodynamic simulations of jets from black hole
  accretion disks: Two-component jets driven by nonsteady accretion of
  magnetized disks'', \comment{journal}{\em Astrophys. J.},
  \comment{volume}{\bf 495}, \comment{pages}L63--L66, (\comment{year}1998).
  \keywords{black holes, relativistic astrophysics, computational fluid
  dynamics, relativistic hydrodynamics and magnetohydrodynamics, finite
  difference and finite volume methods, hyperbolic equations, AGNs,
  extragalactic jets}

\bibitem{koide02a}\comment{article}
\comment{author}Koide, S., Shibata, K., Kudoh, T., and Meier, D.~L.,
  \comment{title}``Extraction of black hole rotational energy by a magnetic
  field and the formation of relativistic jets'', \comment{journal}{\em
  Science}, \comment{volume}{\bf 295}, \comment{pages}1688--1691,
  (\comment{year}2002). \keywords{black holes, relativistic astrophysics,
  computational fluid dynamics, relativistic hydrodynamics and
  magnetohydrodynamics, finite difference and finite volume methods, AGNs,
  extragalactic jets}

\bibitem{komissarov99}\comment{article}
\comment{author}Komissarov, S.~S., \comment{title}``A Godunov-type scheme for
  relativistic magnetohydrodynamics'', \comment{journal}{\em Mon. Not. Roy.
  Astron. Soc.}, \comment{volume}{\bf 303}, \comment{pages}343--366,
  (\comment{year}1999). \keywords{computational fluid dynamics, relativistic
  hydrodynamics and magnetohydrodynamics, finite difference and finite volume
  methods, hyperbolic equations, numerical methods}

\bibitem{kormendy95}\comment{article}
\comment{author}Kormendy, J., and Richstone, D., \comment{title}``Inward
  Bound---The Search For Supermassive Black Holes In Galactic Nuclei'',
  \comment{journal}{\em Ann. Rev. Astron. Astrophys.}, \comment{volume}{\bf
  33}, \comment{pages}581--624, (\comment{year}1995). \keywords{Relativistic
  astrophysics, active galactic nuclei, black holes}

\bibitem{kurganov00}\comment{article}
\comment{author}Kurganov, A., and Tadmor, E., \comment{title}``Solution of
  two-dimensional Riemann problems for gas dynamics without Riemann problem
  solvers'', \comment{journal}{\em J.~Comput.\ Phys.}, \comment{volume}{\bf
  160}, \comment{pages}214, (\comment{year}2000). \keywords{computational fluid
  dynamics, numerical methods, finite difference and finite volume methods}

\bibitem{laguna93a}\comment{article}
\comment{author}Laguna, P., Miller, W.~A., and Zurek, W.~H.,
  \comment{title}``Smoothed particle hydrodynamics near a black hole'',
  \comment{journal}{\em Astrophys. J.}, \comment{volume}{\bf 404},
  \comment{pages}678--685, (\comment{year}1993). \keywords{computational fluid
  dynamics, relativistic hydrodynamics, smoothed particle methods, numerical
  methods, black holes}

\bibitem{laguna93b}\comment{article}
\comment{author}Laguna, P., Miller, W.~A., Zurek, W.~H., and Davies, M.~B.,
  \comment{title}``Tidal disruptions by supermassive black holes: Hydrodynamic
  evolution of stars on a Schwarzschild background'', \comment{journal}{\em
  Astrophys. J.}, \comment{volume}{\bf 410}, \comment{pages}L83--L86,
  (\comment{year}1993). \keywords{computational fluid dynamics, relativistic
  hydrodynamics, smoothed particle methods, numerical methods, black holes}

\bibitem{lattimer91}\comment{article}
\comment{author}Lattimer, J.~M., and Swesty, F.~D., \comment{title}``A
  generalized equation of state for hot, dense matter'', \comment{journal}{\em
  Nucl. Phys. A}, \comment{volume}{\bf 535}, \comment{pages}331--376,
  (\comment{year}1991). \keywords{}

\bibitem{lax72}\comment{article}
\comment{author}Lax, P.~D., \comment{title}``Hyperbolic systems of conservation
  laws and the mathematical theory of shock waves'', \comment{journal}{\em SIAM
  Reginal Conference Series in Applied Mathematics}, \comment{volume}{\bf 11},
  (\comment{year}1972). \keywords{Numerical methods, hyperbolic equations,
  finite difference and finite volume methods, shock waves}

\bibitem{lax60}\comment{article}
\comment{author}Lax, P.~D., and Wendroff, B., \comment{title}``Systems of
  conservation laws'', \comment{journal}{\em Comm. Pure Appl. Math.},
  \comment{volume}{\bf 13}, \comment{pages}217--237, (\comment{year}1960).
  \keywords{Numerical methods, hyperbolic equations, finite difference and
  finite volume methods}

\bibitem{lehner}\comment{article}
\comment{author}Lehner, L., \comment{title}``Numerical relativity: a review'',
  \comment{journal}{\em Class. Quantum Grav.}, \comment{volume}{\bf 18},
  \comment{pages}25--86, (\comment{year}2001). For a related online version
  see: \comment{author}L.~Lehner, \comment{onlinetitle}``Numerical relativity:
  a review'', (\comment{onlinemonth}June, \comment{onlineyear}2001),
  \comment{fileformat}[Online Los Alamos Archive Preprint]: cited on
  \comment{cited}11 September 2002,
  \comment{onlineaddress}http://xxx.lanl.gov/abs/gr-qc/0106072. \keywords{}

\bibitem{leveque92}\comment{book}
\comment{author}LeVeque, R.~J., \comment{title}{\em Numerical Methods for
  Conservation Laws}, (\comment{publisher}Birkhauser Verlag,
  \comment{address}Basel, \comment{year}1992). \keywords{computational fluid
  dynamics, finite difference and finite volume methods, hyperbolic equations,
  numerical methods}

\bibitem{leveque98}\comment{inbook}
\comment{author}LeVeque, R.~J., \comment{title}``Nonlinear conservation laws
  and finite volume methods for astrophysical fluid flow'', in
  \comment{editor}Steiner, O., and Gautschy, A., eds., \comment{booktitle}{\em
  Computational methods for astrophysical fluid flow}, \comment{pages} 1--159,
  (\comment{publisher}Springer-Verlag, \comment{address}Berlin,
  \comment{year}1998). \keywords{computational fluid dynamics, finite
  difference and finite volume methods, hyperbolic equations, numerical
  methods}

\bibitem{liebendoerfer01}\comment{article}
\comment{author}Liebend\"orfer, M., Mezzacappa, A., Tielemann, F.-K., Messer,
  O.~E.~B., Hix, W.~R., and Bruenn, S.~W., \comment{title}``Probing the
  gravitational well: no supernova explosion in spherical symmetry with general
  relativistic Boltzmann neutrino transport'', \comment{journal}{\em Phys. Rev.
  D}, \comment{volume}{\bf 63}, \comment{pages}103004, (\comment{year}2001).
  For a related online version see: \comment{author}M.~Liebend\"orfer, et al.,
  \comment{onlinetitle}``Probing the gravitational well: no supernova explosion
  in spherical symmetry with general relativistic Boltzmann neutrino
  transport'', (\comment{onlinemonth}June, \comment{onlineyear}2000),
  \comment{fileformat}[Online Los Alamos Archive Preprint]: cited on
  \comment{cited}21 June 2002,
  \comment{onlineaddress}http://xxx.lanl.gov/abs/astro-ph/0006418.
  \keywords{gravitational collapse, supernovae, neutrinos, nuclear physics}

\bibitem{lindblom01a}\comment{article}
\comment{author}Lindblom, L., Tohline, J.~E., and Vallisneri, M.,
  \comment{title}``Nonlinear Evolution of the r-Modes in Neutron Stars'',
  \comment{journal}{\em Phys. Rev. Lett.}, \comment{volume}{\bf 86},
  \comment{pages}1152--1155, (\comment{year}2001). For a related online version
  see: \comment{author}L.~Lindblom, et al., \comment{onlinetitle}``Nonlinear
  evolution of the r-modes in neutron stars'', (\comment{onlinemonth}October,
  \comment{onlineyear}2000), \comment{fileformat}[Online Los Alamos Archive
  Preprint]: cited on \comment{cited}15 July 2002,
  \comment{onlineaddress}http://xxx.lanl.gov/abs/astro-ph/0010653.
  \keywords{neutron stars}

\bibitem{linke01a}\comment{article}
\comment{author}Linke, F., Font, J.~A., Janka, H.-Th., M\"uller, E., and
  Papadopoulos, P., \comment{title}``Spherical collapse of supermassive stars:
  neutrino emission and gamma ray bursts'', \comment{journal}{\em Astron.
  Astrophys.}, \comment{volume}{\bf 376}, \comment{pages}568--579,
  (\comment{year}2001). For a related online version see:
  \comment{author}F.~Linke, et al., \comment{onlinetitle}``Spherical collapse
  of supermassive stars: neutrino emission and gamma ray bursts'',
  (\comment{onlinemonth}March, \comment{onlineyear}2001),
  \comment{fileformat}[Online Los Alamos Archive Preprint]: cited on
  \comment{cited}13 June 2002,
  \comment{onlineaddress}http://xxx.lanl.gov/abs/astro-ph/0103144.
  \keywords{Numerical relativistic hydrodynamics, gravitational collapse,
  Einstein equations, numerical relativity, null surfaces, initial value
  problem, black holes, gamma-ray bursts}

\bibitem{lorene}\comment{misc}
\verb+LORENE+ code, \verb+http://www.lorene.obspm.fr+. \keywords{}

\bibitem{lucy77}\comment{article}
\comment{author}Lucy, L.~B., \comment{title}``A numerical approach to the
  testing of the fission hypothesis'', \comment{journal}{\em Astronom. J.},
  \comment{volume}{\bf 82}, \comment{pages}1013--1024, (\comment{year}1977).
  \keywords{computational fluid dynamics, smoothed particle hydrodynamics,
  numerical methods, dynamical systems}

\bibitem{maison96}\comment{article}
\comment{author}Maison, D., \comment{title}``Non-universality of critical
  behaviour in spherically symmetric gravitational collapse'',
  \comment{journal}{\em Phys. Lett.}, \comment{volume}{\bf B366},
  \comment{pages}82--84, (\comment{year}1996). For a related online version
  see: \comment{author}D.~Maison, \comment{onlinetitle}``Non-universality of
  critical behaviour in spherically symmetric gravitational collapse'',
  (\comment{onlinemonth}April, \comment{onlineyear}1995),
  \comment{fileformat}[Online Los Alamos Archive Preprint]: cited on
  \comment{cited}15 July 2002,
  \comment{onlineaddress}http://xxx.lanl.gov/abs/gr-qc/9504008.
  \keywords{gravitational collapse, critical phenomena}

\bibitem{mann91}\comment{article}
\comment{author}Mann, P.~J., \comment{title}``A relativistic smoothed particle
  hydrodynamics method tested with the shock tube'', \comment{journal}{\em
  Comput. Phys. Comm.}, \comment{volume}{\bf 67}, \comment{pages}245--260,
  (\comment{year}1991). \keywords{computational fluid dynamics, relativistic
  hydrodynamics, smoothed particle methods, numerical methods}

\bibitem{marti91b}\comment{phdthesis}
\comment{author}Mart\'{\i}, J.~M., \comment{title}{\em Hidrodin\'amica
  relativista num\'erica: aplicaciones al colapso estelar}, PhD thesis,
  (\comment{school}Universidad de Valencia, \comment{year}1991). In Spanish.
  \keywords{Numerical relativistic hydrodynamics, numerical relativity,
  numerical methods, finite difference and finite volume methods, Einstein
  equations, differential equations, hyperbolic equations, gravitational
  collapse, supernovae}

\bibitem{marti90}\comment{article}
\comment{author}Mart\'{\i}, J.~M., Ib{\'a}\~nez, J.~M., and Miralles, J.~A.,
  \comment{title}``Godunov-type methods for stellar collapse'',
  \comment{journal}{\em Astron. Astrophys.}, \comment{volume}{\bf 235},
  \comment{pages}535--542, (\comment{year}1990). \keywords{Hydrodynamics, shock
  waves, numerical methods, supernovae, gravitational collapse}

\bibitem{marti91}\comment{article}
\comment{author}Mart\'{\i}, J.~M., Ib{\'a}\~nez, J.~M., and Miralles, J.~A.,
  \comment{title}``Numerical relativistic hydrodynamics: local characteristic
  approach'', \comment{journal}{\em Phys. Rev. D}, \comment{volume}{\bf 43},
  \comment{pages}3794--3801, (\comment{year}1991). \keywords{computational
  fluid dynamics, relativistic hydrodynamics, numerical relativity, finite
  difference and finite volume methods, hyperbolic equations, numerical
  methods}

\bibitem{marti94}\comment{article}
\comment{author}Mart\'{\i}, J.~M., and M\"uller, E., \comment{title}``The
  analytical solution of the Riemann problem in relativistic hydrodynamics'',
  \comment{journal}{\em J. Fluid. Mech.}, \comment{volume}{\bf 258},
  \comment{pages}317--333, (\comment{year}1994). \keywords{relativistic
  hydrodynamics, initial value problem, shock waves}

\bibitem{marti96}\comment{article}
\comment{author}Mart\'{\i}, J.~M., and M\"uller, E., \comment{title}``Extension
  of the piecewise parabolic method to one-dimensional relativistic
  hydrodynamics'', \comment{journal}{\em J. Comput. Phys.},
  \comment{volume}{\bf 123}, \comment{pages}1--14, (\comment{year}1996).
  \keywords{computational fluid dynamics, relativistic hydrodynamics, finite
  difference and finite volume methods, hyperbolic equations, numerical
  methods}

\bibitem{marti99}\comment{article}
\comment{author}Mart\'{\i}, J.~M., and M\"uller, E., \comment{title}``Numerical
  hydrodynamics in special relativity'', \comment{journal}{\em Living Reviews
  in Relativity}, \comment{volume}{\bf 1}, (\comment{year}1999). For a related
  online version see: \comment{author}J.~M. Mart\'{\i}, et al.,
  \comment{onlinetitle}``Numerical hydrodynamics in special relativity'',
  (\comment{onlinemonth}June, \comment{onlineyear}1999),
  \comment{fileformat}[Online Los Alamos Archive Preprint]: cited on
  \comment{cited}1 July 1999,
  \comment{onlineaddress}http://xxx.lanl.gov/abs/gr-qc/9906333.
  \keywords{Relativistic hydrodynamics, numerical relativistic hydrodynamics,
  numerical methods, finite difference and finite volume methods, smoothed
  particle hydrodynamics, gamma-ray bursts, extragalactic jets}

\bibitem{marti97}\comment{article}
\comment{author}Mart\'{\i}, J.~M., M\"uller, E., Font, J.~A., Ib{\'a}\~nez,
  J.~M., and Marquina, A., \comment{title}``Morphology and dynamics of
  relativistic jets'', \comment{journal}{\em Astrophys. J.},
  \comment{volume}{\bf 479}, \comment{pages}151--163, (\comment{year}1997).
  \keywords{relativistic astrophysics, computational fluid dynamics,
  relativistic hydrodynamics, finite difference and finite volume methods,
  hyperbolic equations, AGNs, extragalactic jets}

\bibitem{mathews98}\comment{article}
\comment{author}Mathews, G.~J., Marronetti, P., and Wilson, J.~R.,
  \comment{title}``Relativistic Hydrodynamics in Close Binary Systems: Analysis
  of Neutron-Star Collapse'', \comment{journal}{\em Phys. Rev. D},
  \comment{volume}{\bf 58}, \comment{pages}043003, (\comment{year}1998). For a
  related online version see: \comment{author}G.~J. Mathews, et al.,
  \comment{onlinetitle}``Relativistic Hydrodynamics in Close Binary Systems:
  Analysis of Neutron-Star Collapse'', (\comment{onlinemonth}October,
  \comment{onlineyear}1997), \comment{fileformat}[Online Los Alamos Archive
  Preprint]: cited on \comment{cited}1 November 1997,
  \comment{onlineaddress}http://xxx.lanl.gov/abs/gr-qc/9710140.
  \keywords{Relativistic hydrodynamics, Numerical relativistic hydrodynamics,
  computational fluid dynamics, Einstein equations, post-Newtonian
  approximations, binary systems, neutron stars, nuclear physics, gravitational
  collapse}

\bibitem{mathews99}\comment{article}
\comment{author}Mathews, G.~J., and Wilson, J.~R., \comment{title}``Revised
  Relativistic Hydrodynamical Model for Neutron-Star Binaries'',
  \comment{journal}{\em Phys. Rev .D}, \comment{volume}{\bf 61},
  \comment{pages}127304, (\comment{year}2000). For a related online version
  see: \comment{author}G.~J. Mathews, et al., \comment{onlinetitle}``Revised
  Relativistic Hydrodynamical Model for Neutron-Star Binaries'',
  (\comment{onlinemonth}November, \comment{onlineyear}1999),
  \comment{fileformat}[Online Los Alamos Archive Preprint]: cited on
  \comment{cited}15 February 2000,
  \comment{onlineaddress}http://xxx.lanl.gov/abs/gr-qc/9911047.
  \keywords{Relativistic hydrodynamics, Numerical relativistic hydrodynamics,
  computational fluid dynamics, Einstein equations, post-Newtonian
  approximations, binary systems, neutron stars, nuclear physics}

\bibitem{may66}\comment{article}
\comment{author}May, M.~M., and White, R.~H., \comment{title}``Hydrodynamic
  calculations of general relativistic collapse'', \comment{journal}{\em Phys.
  Rev. D}, \comment{volume}{\bf 141}, \comment{pages}1232--1241,
  (\comment{year}1966). \keywords{Gravitational collapse, Differential
  equations, Einstein equations, Relativistic hydrodynamics, numerical methods}

\bibitem{may67}\comment{article}
\comment{author}May, M.~M., and White, R.~H., \comment{title}``Stellar dynamics
  and gravitational collapse'', \comment{journal}{\em Meth. Computat. Phys.},
  \comment{volume}{\bf 7}, \comment{pages}219--258, (\comment{year}1967).
  \keywords{Gravitational collapse, Differential equations, Einstein equations,
  Relativistic hydrodynamics, numerical methods}

\bibitem{mayle87}\comment{article}
\comment{author}Mayle, R., Wilson, J.~R., and Schramm, D.~N.,
  \comment{title}``Neutrinos from gravitational collapse'',
  \comment{journal}{\em Astrophys. J.}, \comment{volume}{\bf 318},
  \comment{pages}288--306, (\comment{year}1987). \keywords{neutrinos,
  gravitational collapse, supernovae, nuclear physics, numerical relativistic
  hydrodynamics}

\bibitem{mcabee94}\comment{article}
\comment{author}McAbee, T.~L., and Wilson, J.~R., \comment{title}``Mean-field
  pion calculations of heavy-ion collisions at Bevalac energies'',
  \comment{journal}{\em Nucl. Phys. A}, \comment{volume}{\bf 576},
  \comment{pages}626--638, (\comment{year}1994). \keywords{Nuclear physics,
  numerical relativistic hydrodynamics}

\bibitem{meier99a}\comment{article}
\comment{author}Meier, D.~L., \comment{title}``Multidimensional astrophysical
  structural and dynamical analysis. I. Development of a nonlinear finite
  element approach'', \comment{journal}{\em Astrophys. J.},
  \comment{volume}{\bf 518}, \comment{pages}788--813, (\comment{year}1999).
  \keywords{numerical methods, numerical relativistic hydrodynamics}

\bibitem{mezzacappa01}\comment{article}
\comment{author}Mezzacappa, A., Liebend\"orfer, M., Messer, O.~E.~B., Hix,
  W.~R., Tielemann, F.-K., and Bruenn, S.~W., \comment{title}``Simulation of
  the spherically symmetric stellar core collapse, bounce and postbounce
  evolution of a 13 solar mass star with Boltzmann neutrino transport and its
  implications for the supernova mechanism'', \comment{journal}{\em Phys. Rev.
  Lett.}, \comment{volume}{\bf 86}, \comment{pages}1935--1938,
  (\comment{year}2001). For a related online version see:
  \comment{author}A.~Mezzacappa, et al., \comment{onlinetitle}``Simulation of
  the spherically symmetric stellar core collapse, bounce and postbounce
  evolution of a 13 solar mass star with Boltzmann neutrino transport and its
  implications for the supernova mechanism'', (\comment{onlinemonth}May,
  \comment{onlineyear}2000), \comment{fileformat}[Online Los Alamos Archive
  Preprint]: cited on \comment{cited}21 June 2002,
  \comment{onlineaddress}http://xxx.lanl.gov/abs/astro-ph/0005366.
  \keywords{Hydrodynamics, neutrinos, gravitaional collapse, supernovae}

\bibitem{mezzacappa89}\comment{article}
\comment{author}Mezzacappa, A., and Matzner, R.~A., \comment{title}``Computer
  simulation of time-dependent, spherically symmetric spacetimes containing
  radiating fluids - Formalism and code tests'', \comment{journal}{\em
  Astrophys. J.}, \comment{volume}{\bf 343}, \comment{pages}853--873,
  (\comment{year}1989). \keywords{Hydrodynamics, neutrinos, numerical methods,
  Einstein equations}

\bibitem{michel72}\comment{article}
\comment{author}Michel, F.~C., \comment{title}``Accretion of matter by
  condensed objects'', \comment{journal}{\em Astrophys. Spa. Sci.},
  \comment{volume}{\bf 15}, \comment{pages}153--160, (\comment{year}1972).
  \keywords{Accretion, relativistic hydrodynamics, Schwarzschild solution}

\bibitem{mihalas84}\comment{book}
\comment{author}Mihalas, D., and Mihalas, B., \comment{title}{\em Foundations
  of radiation hydrodynamics}, (\comment{publisher}Oxford University Press,
  \comment{address}Oxford, England, \comment{year}1984). \keywords{}

\bibitem{miller89}\comment{article}
\comment{author}Miller, J.~C., and Motta, S., \comment{title}``Computations of
  spherical gravitational collapse using null slicing'', \comment{journal}{\em
  Class. Quant. Grav.}, \comment{volume}{\bf 6}, \comment{pages}185--193,
  (\comment{year}1989). \keywords{Gravitational collapse, black holes,
  numerical relativity, null surfaces}

\bibitem{misci}\comment{inbook}
\comment{author}Miller, J.~C., and Sciama, D.~W.,
  \comment{title}``Gravitational collapse to the black hole state'', in
  \comment{editor}Held, A., ed., \comment{booktitle}{\em General relativity and
  gravitation, II}, \comment{pages} 359--391, (\comment{publisher}Plenum Press,
  \comment{address}New York and London, \comment{year}1980). \keywords{}

\bibitem{miller99}\comment{article}
\comment{author}Miller, M., Suen, W.-M., and Tobias, M.,
  \comment{title}``Shapiro conjecture: Prompt or delayed collapse in the
  head-on collision of neutron stars?'', \comment{journal}{\em Phys. Rev. D},
  \comment{volume}{\bf 63}, \comment{pages}121501, (\comment{year}2001). For a
  related online version see: \comment{author}M.~Miller, et al.,
  \comment{onlinetitle}``The Shapiro Conjecture: Prompt or Delayed Collapse in
  the head-on collision of neutron stars?'', (\comment{onlinemonth}April,
  \comment{onlineyear}1999), \comment{fileformat}[Online Los Alamos Archive
  Preprint]: cited on \comment{cited}15 February 2000,
  \comment{onlineaddress}http://xxx.lanl.gov/abs/gr-qc/9904041.
  \keywords{Gravitational collapse, numerical relativity, numerical
  relativistic hydrodynamics, neutron stars}

\bibitem{miralles91}\comment{article}
\comment{author}Miralles, J.~A., Ib\'a\~nez, J.~M., Mart\'{\i}, J.~M., and
  P\'erez, A., \comment{title}``Incompressibility of hot nuclear matter,
  general relativistic stellar collapse and shock propagation'',
  \comment{journal}{\em Astron. Astrophys. Suppl. Ser.}, \comment{volume}{\bf
  90}, \comment{pages}283--299, (\comment{year}1991). \keywords{Hydrodynamics,
  supernovae, gravitational collapse, neutron stars, shock waves, nuclear
  physics}

\bibitem{misner64}\comment{article}
\comment{author}Misner, C.~W., and Sharp, D.~H., \comment{title}``Relativistic
  equations for adiabatic, spherically symmetric, gravitational collapse'',
  \comment{journal}{\em Phys. Rev.}, \comment{volume}{\bf 136},
  \comment{pages}571--576, (\comment{year}1964). \keywords{Numerical
  relativity, gravitational collapse, supernovae, Einstein equations,
  relativistic hydrodynamics}

\bibitem{MTW}\comment{book}
\comment{author}Misner, C.~W., Thorne, K.~S., and Wheeler, J.~A.,
  \comment{title}{\em Gravitation}, (\comment{publisher}W. H. Freeman,
  \comment{address}San Francisco, \comment{year}1973). \keywords{}

\bibitem{monaghan92}\comment{article}
\comment{author}Monaghan, J.~J., \comment{title}``Smoothed particle
  hydrodynamics'', \comment{journal}{\em Ann. Rev. Astron. Astrophys.},
  \comment{volume}{\bf 30}, \comment{pages}543--574, (\comment{year}1992).
  \keywords{computational fluid dynamics, smoothed particle hydrodynamics,
  numerical methods, astrophysics}

\bibitem{moenchmeyer91a}\comment{article}
\comment{author}M\"onchmeyer, R., Sch\"afer, G., M\"uller, E., and Kates,
  R.~E., \comment{title}``Gravitational waves from the collapse of rotating
  stellar cores'', \comment{journal}{\em Astron.\ Astrophys.},
  \comment{volume}{\bf 246}, \comment{pages}417--440, (\comment{year}1991).
  \keywords{}

\bibitem{MPAweb}\comment{misc}
Max Planck Institute for Astrophysics, ``MPA Hydro Gang Home Page",
  \verb+http://www.mpa-garching.mpg.de/Hydro/hydro.html+. \keywords{}

\bibitem{mueller82}\comment{article}
\comment{author}M\"uller, E, \comment{title}``Gravitational radiation from
  collapsing rotating stellar cores'', \comment{journal}{\em Astron.
  Astrophys.}, \comment{volume}{\bf 114}, \comment{pages}53--59,
  (\comment{year}1982). \keywords{Gravitational collapse, supernovae,
  gravitational radiation, computational fluid dynamics}

\bibitem{mueller98}\comment{inbook}
\comment{author}M\"uller, E, \comment{title}``Simulation of astrophysical fluid
  flow'', in \comment{editor}Steiner, O., and Gautschy, A., eds.,
  \comment{booktitle}{\em Computational methods for astrophysical fluid flow},
  \comment{pages} 343--494, (\comment{publisher}Springer-Verlag,
  \comment{address}Berlin, \comment{year}1998). \keywords{computational fluid
  dynamics, numerical methods, astrophysics, supernovae}

\bibitem{mueller99}\comment{article}
\comment{author}M\"uller, I, \comment{title}``Speeds of propagation in
  classical and relativistic extended thermodynamics'', \comment{journal}{\em
  Living Reviews in Relativity}, \comment{volume}{\bf 2}, (\comment{year}1999).
  \keywords{extended thermodynamics}

\bibitem{naka81}\comment{article}
\comment{author}Nakamura, T., \comment{title}``General relativistic collapse of
  axially symmetric stars leading to the formation of rotating black holes'',
  \comment{journal}{\em Prog. Theor. Phys.}, \comment{volume}{\bf 65},
  \comment{pages}1876--1890, (\comment{year}1981). \keywords{gravitational
  collapse, black holes, computational fluid dynamics, relativistic
  hydrodynamics, numerical relativity, finite difference and finite volume
  methods, numerical methods}

\bibitem{naka83a}\comment{article}
\comment{author}Nakamura, T., \comment{title}``General relativistic collapse of
  accreting neutron stars with rotation'', \comment{journal}{\em Prog.\ Theor.\
  Phys.}, \comment{volume}{\bf 70}, \comment{pages}1144--1147,
  (\comment{year}1983). \keywords{gravitational collapse, numerical
  relativistic hydrodynamics, numerical relativity}

\bibitem{naka80}\comment{article}
\comment{author}Nakamura, T., Maeda, K., Miyama, S., and Sasaki, M.,
  \comment{title}``General relativistic collapse of an axially symmetric
  star'', \comment{journal}{\em Prog. Theor. Phys.}, \comment{volume}{\bf 63},
  \comment{pages}1229--1244, (\comment{year}1980). \keywords{gravitational
  collapse, computational fluid dynamics, relativistic hydrodynamics, numerical
  relativity, finite difference and finite volume methods, numerical methods}

\bibitem{naka98}\comment{inproceedings}
\comment{author}Nakamura, T., and Oohara, K., \comment{title}``A Way to 3D
  Numerical Relativity'', in \comment{editor}Miyama, S.~M., Tomisaka, K., and
  Hanawa, T., eds., \comment{booktitle}{\em ASSL Vol. 240: Numerical
  Astrophysics}, \comment{pages} 247, (\comment{year}1999). For a related
  online version see: \comment{author}T.~Nakamura, et al.,
  \comment{onlinetitle}``A Way to 3D Numerical Relativity - Coalescing Binary
  Neutron Stars'', (\comment{onlinemonth}December, \comment{onlineyear}1998),
  \comment{fileformat}[Online Los Alamos Archive Preprint]: cited on
  \comment{cited}1 February 1999,
  \comment{onlineaddress}http://xxx.lanl.gov/abs/gr-qc/9812054.
  \keywords{Numerical relativity, binary systems, neutron stars, numerical
  relativistic hydrodynamics}

\bibitem{naka87a}\comment{article}
\comment{author}Nakamura, T., Oohara, K., and Kojima, Y.,
  \comment{title}``General relativistic collapse to black holes and
  gravitational waves from black holes'', \comment{journal}{\em Prog.\ Theor.\
  Phys.}, \comment{volume}{\bf 90}, \comment{pages}1--218,
  (\comment{year}1987). \keywords{gravitational collapse, numerical
  relativistic hydrodynamics, numerical relativity, gravitational radiation}

\bibitem{naka81b}\comment{article}
\comment{author}Nakamura, T., and Sasaki, M., \comment{title}``Is collapse of a
  deformed star always effectual for gravitational radiation?'',
  \comment{journal}{\em Phys. Letters}, \comment{volume}{\bf 106 B},
  \comment{pages}69--72, (\comment{year}1981). \keywords{Gravitational
  collapse, gravitational radiation, Black holes}

\bibitem{naka82}\comment{article}
\comment{author}Nakamura, T., and Sato, H., \comment{title}``General
  relativistic collapse of non-rotating, axisymmetric stars'',
  \comment{journal}{\em Prog. Theor. Phys.}, \comment{volume}{\bf 67},
  \comment{pages}1396--1405, (\comment{year}1982). \keywords{gravitational
  collapse, computational fluid dynamics, relativistic hydrodynamics, numerical
  relativity, finite difference and finite volume methods, numerical methods}

\bibitem{narayan98}\comment{inbook}
\comment{author}Narayan, R., Mahadevan, R., and Quataert, E.,
  \comment{title}``Advection-Dominated Accretion around Black Holes'', in
  \comment{editor}M.~A.~Abramowicz, G.~Bjornsson, and Pringle, J.~E., eds.,
  \comment{booktitle}{\em The theory of black hole accretion disks},
  \comment{pages} 148, (\comment{publisher}Cambridge University Press,
  \comment{year}1998). For a related online version see:
  \comment{author}R.~Narayan, et al.,
  \comment{onlinetitle}``Advection-Dominated Accretion around Black Holes'',
  (\comment{onlinemonth}March, \comment{onlineyear}1998),
  \comment{fileformat}[Online Los Alamos Archive Preprint]: cited on
  \comment{cited}15 February 2000,
  \comment{onlineaddress}http://xxx.lanl.gov/abs/astro-ph/9803141.
  \keywords{accretion, black holes, accretion disks, hydrodynamics}

\bibitem{narayan92}\comment{article}
\comment{author}Narayan, R., Paczy\'nski, B., and Piran, T.,
  \comment{title}``Gamma-ray bursts as the death throes of massive binary
  stars'', \comment{journal}{\em Astrophys. J.}, \comment{volume}{\bf 395},
  \comment{pages}L83--L86, (\comment{year}1992). \keywords{Astrophysics,
  gamma-ray bursts, stars}

\bibitem{narayan94}\comment{article}
\comment{author}Narayan, R., and Yi, I., \comment{title}``Advection-dominated
  accretion: A self-similar solution'', \comment{journal}{\em Astrophys. J.},
  \comment{volume}{\bf 428}, \comment{pages}L13--L16, (\comment{year}1994). For
  a related online version see: \comment{author}R.~Narayan, et al.,
  \comment{onlinetitle}``Advection-dominated accretion: A self-similar
  solution'', (\comment{onlinemonth}March, \comment{onlineyear}1994),
  \comment{fileformat}[Online Los Alamos Archive Preprint]: cited on
  \comment{cited}15 February 2000,
  \comment{onlineaddress}http://xxx.lanl.gov/abs/astro-ph/9403052.
  \keywords{accretion, black holes, accretion disks, hydrodynamics}

\bibitem{neilsen99b}\comment{article}
\comment{author}Neilsen, D.~W., and Choptuik, M.~W., \comment{title}``Critical
  phenomena in perfect fluids'', \comment{journal}{\em Class. Quant. Grav.},
  \comment{volume}{\bf 17}, \comment{pages}761--782, (\comment{year}2000). For
  a related online version see: \comment{author}D.~W. Neilsen, et al.,
  \comment{onlinetitle}``Critical phenomena in perfect fluids'',
  (\comment{onlinemonth}December, \comment{onlineyear}1998),
  \comment{fileformat}[Online Los Alamos Archive Preprint]: cited on
  \comment{cited}1 February 1999,
  \comment{onlineaddress}http://xxx.lanl.gov/abs/gr-qc/9812053.
  \keywords{Relativistic hydrodynamics, critical phenomena}

\bibitem{neilsen99a}\comment{article}
\comment{author}Neilsen, D.~W., and Choptuik, M.~W.,
  \comment{title}``Ultrarelativistic fluid dynamics'', \comment{journal}{\em
  Class. Quant. Grav.}, \comment{volume}{\bf 17}, \comment{pages}733--759,
  (\comment{year}2000). For a related online version see: \comment{author}D.~W.
  Neilsen, et al., \comment{onlinetitle}``Ultrarelativistic fluid dynamics'',
  (\comment{onlinemonth}April, \comment{onlineyear}1999),
  \comment{fileformat}[Online Los Alamos Archive Preprint]: cited on
  \comment{cited}1 May 1999,
  \comment{onlineaddress}http://xxx.lanl.gov/abs/gr-qc/9804052.
  \keywords{Relativistic hydrodynamics, numerical methods, critical phenomena}

\bibitem{tadmor90}\comment{article}
\comment{author}Nessyahu, H., and Tadmor, E., \comment{title}``Non-oscillatory
  central differencing for hyperbolic conservation laws'',
  \comment{journal}{\em J. Comput. Phys.}, \comment{volume}{\bf 87},
  \comment{pages}408--463, (\comment{year}1990). \keywords{numerical methods,
  computational fluid dynamics, hyperbolic systems, finite difference and
  finite volume methods}

\bibitem{new02}\comment{article}
\comment{author}New, K.~C.~B., \comment{title}``Gravitational waves from
  gravitational collapse'', \comment{journal}{\em Living Reviews in
  Relativity}, \comment{volume}{\bf 6}, \comment{pages}2, (\comment{year}2003).
  For a related online version see: \comment{author}K.~C.~B. New,
  \comment{onlinetitle}``Gravitational waves from gravitational collapse'',
  (\comment{onlinemonth}June, \comment{onlineyear}2002),
  \comment{fileformat}[Online Los Alamos Archive Preprint]: cited on
  \comment{cited}18 June 2002,
  \comment{onlineaddress}http://xxx.lanl.gov/abs/gr-qc/0206041.
  \keywords{gravitational collapse, gravitational radiation}

\bibitem{noh87}\comment{article}
\comment{author}Noh, W.~F., \comment{title}``Errors for calculations of strong
  shocks using an artificial viscosity and an artificial heat flux'',
  \comment{journal}{\em J. Comput. Phys.}, \comment{volume}{\bf 72},
  \comment{pages}78--120, (\comment{year}1987). \keywords{computational fluid
  dynamics, finite difference and finite volume methods, hyperbolic equations,
  numerical methods}

\bibitem{norman86}\comment{inproceedings}
\comment{author}Norman, M.~L., and Winkler, K-H.~A., \comment{title}``Why
  ultrarelativistic numerical hydrodynamics is difficult?'', in
  \comment{editor}Norman, M.~L., and Winkler, K-H.~A., eds.,
  \comment{booktitle}{\em Astrophysical Radiation Hydrodynamics},
  \comment{pages} 449--475. Reidel Publishing Company, (\comment{year}1986).
  \keywords{computational fluid dynamics, relativistic hydrodynamics, energy
  and momentum, finite difference and finite volume methods, numerical methods}

\bibitem{novak98}\comment{article}
\comment{author}Novak, J., \comment{title}``Spherical neutron star collapse in
  tensor-scalar theory of gravity'', \comment{journal}{\em Phys. Rev. D},
  \comment{volume}{\bf 57}, \comment{pages}4789--4801, (\comment{year}1998).
  For a related online version see: \comment{author}J.~Novak,
  \comment{onlinetitle}``Spherical neutron star collapse toward a black hole in
  tensor-scalar theory of gravity'', (\comment{onlinemonth}July,
  \comment{onlineyear}1997), \comment{fileformat}[Online Los Alamos Archive
  Preprint]: cited on \comment{cited}15 February 2000,
  \comment{onlineaddress}http://xxx.lanl.gov/abs/gr-qc/9707041.
  \keywords{Gravitational collapse, spectral methods, theories of gravity}

\bibitem{novak01}\comment{article}
\comment{author}Novak, J., \comment{title}``Velocity-induced collapses of
  stable neutron stars'', \comment{journal}{\em Astron. Astrophys.},
  \comment{volume}{\bf 376}, \comment{pages}606--613, (\comment{year}2001). For
  a related online version see: \comment{author}J.~Novak,
  \comment{onlinetitle}``Velocity-induced collapses of stable neutron stars'',
  (\comment{onlinemonth}July, \comment{onlineyear}2001),
  \comment{fileformat}[Online Los Alamos Archive Preprint]: cited on
  \comment{cited}24 June 2002,
  \comment{onlineaddress}http://xxx.lanl.gov/abs/gr-qc/0107045.
  \keywords{Gravitational collapse, spectral methods, numerical methods}

\bibitem{novak99}\comment{article}
\comment{author}Novak, J., and Ib\'a\~nez, J.~M.,
  \comment{title}``Gravitational waves from the collapse and bounce of a
  stellar core in tensor-scalar gravity'', \comment{journal}{\em Astrophys, J},
  \comment{volume}{\bf 533}, \comment{pages}392--405, (\comment{year}2000). For
  a related online version see: \comment{author}J.~Novak, et al.,
  \comment{onlinetitle}``Gravitational waves from the collapse and bounce of a
  stellar core in tensor-scalar gravity'', (\comment{onlinemonth}November,
  \comment{onlineyear}1999), \comment{fileformat}[Online Los Alamos Archive
  Preprint]: cited on \comment{cited}15 February 2000,
  \comment{onlineaddress}http://xxx.lanl.gov/abs/astro-ph/9911298.
  \keywords{Gravitational collapse, spectral methods, theories of gravity,
  finite difference and finite volume methods, numerical methods}

\bibitem{oleinik57}\comment{article}
\comment{author}Oleinik, O., \comment{title}``Discontinuous solutions and
  non-linear differential equations'', \comment{journal}{\em Amer. Math. Soc.
  Transl. Ser.}, \comment{volume}{\bf 26}, \comment{pages}95--172,
  (\comment{year}1957). \keywords{numerical methods, hyperbolic equations,
  finite difference and finite volume methods}

\bibitem{oohara83}\comment{article}
\comment{author}Oohara, K., and Nakamura, T., \comment{title}``Gravitational
  radiation from a particle scattered by a non-rotating black hole'',
  \comment{journal}{\em Phys. Lett. A}, \comment{volume}{\bf 98},
  \comment{pages}407--410, (\comment{year}1983). \keywords{gravitational
  radiation, approximation methods, black holes}

\bibitem{oohara96}\comment{inproceedings}
\comment{author}Oohara, K.-I., and Nakamura, T., \comment{title}``Coalescence
  of Binary Neutron Stars'', in \comment{editor}Lasota, Jean-Pierre, and Marck,
  Jean-Alain, eds., \comment{booktitle}{\em Relativistic Gravitation and
  Gravitational Radiation}, (\comment{publisher}Cambridge University Press,
  \comment{address}Cambridge, England, \comment{year}1997). For a related
  online version see: \comment{author}K.-I. Oohara, et al.,
  \comment{onlinetitle}``Coalescence of Binary Neutron Stars'',
  (\comment{onlinemonth}June, \comment{onlineyear}1996),
  \comment{fileformat}[Online Los Alamos Archive Preprint]: cited on
  \comment{cited}1 September 1996,
  \comment{onlineaddress}http://xxx.lanl.gov/abs/astro-ph/9606179.
  \keywords{Numerical relativity, numerical relativistic hydrodynamics,
  Einstein equations, binary systems, neutron stars, gravitational radiation,
  numerical methods}

\bibitem{paczynski86}\comment{article}
\comment{author}Paczy\'nski, B., \comment{title}``Gamma-ray bursters at
  cosmological distances'', \comment{journal}{\em Astrophys. J.},
  \comment{volume}{\bf 308}, \comment{pages}L43--L46, (\comment{year}1986).
  \keywords{relativistic astrophysics, extragalactic astronomy, GRBs}

\bibitem{paczynski98}\comment{article}
\comment{author}Paczy\'nski, B., \comment{title}``Are gamma-ray bursts in
  star-forming regions?'', \comment{journal}{\em Astrophys. J.},
  \comment{volume}{\bf 494}, \comment{pages}L45--L48, (\comment{year}1998).
  \keywords{relativistic astrophysics, GRBs, supernovae}

\bibitem{paczynski80}\comment{article}
\comment{author}Paczy\'nski, B., and Wiita, P.~J., \comment{title}``Thick
  accretion disks and supercritical luminosities'', \comment{journal}{\em
  Astron. Astrophys.}, \comment{volume}{\bf 88}, \comment{pages}23--31,
  (\comment{year}1980). \keywords{Accretion disks, black holes}

\bibitem{papadopoulos98a}\comment{article}
\comment{author}Papadopoulos, P., and Font, J.~A.,
  \comment{title}``Relativistic Hydrodynamics around Black Holes and Horizon
  Adapted Coordinate Systems'', \comment{journal}{\em Phys. Rev. D},
  \comment{volume}{\bf 58}, \comment{pages}024005,1--9, (\comment{year}1998).
  For a related online version see: \comment{author}P.~Papadopoulos, et al.,
  \comment{onlinetitle}``Relativistic Hydrodynamics around Black Holes and
  Horizon Adapted Coordinate Systems'', (\comment{onlinemonth}March,
  \comment{onlineyear}1998), \comment{fileformat}[Online Los Alamos Archive
  Preprint]: cited on \comment{cited}1 April 1998,
  \comment{onlineaddress}http://xxx.lanl.gov/abs/gr-qc/9803087.
  \keywords{Relativistic hydrodynamics, numerical relativistic hydrodynamics,
  computational fluid dynamics, accretion, black holes, numerical methods,
  shock waves}

\bibitem{papadopoulos99c}\comment{article}
\comment{author}Papadopoulos, P., and Font, J.~A., \comment{title}``Analysis of
  relativistic hydrodynamics in conservation form'', (\comment{year}1999). For
  a related online version see: \comment{author}P.~Papadopoulos, et al.,
  \comment{onlinetitle}``Analysis of relativistic hydrodynamics in conservation
  form'', (\comment{onlinemonth}December, \comment{onlineyear}1999),
  \comment{fileformat}[Online Los Alamos Archive Preprint]: cited on
  \comment{cited}1 February 2000,
  \comment{onlineaddress}http://xxx.lanl.gov/abs/gr-qc/9912054. MPA preprint.
  \keywords{Relativistic hydrodynamics, hyperbolic partial differential
  equations, energy and momentum}

\bibitem{papadopoulos99a}\comment{article}
\comment{author}Papadopoulos, P., and Font, J.~A., \comment{title}``Matter
  Flows around Black Holes and Gravitational Radiation'', \comment{journal}{\em
  Phys. Rev. D}, \comment{volume}{\bf 59}, \comment{pages}044014,1--17,
  (\comment{year}1999). For a related online version see:
  \comment{author}P.~Papadopoulos, et al., \comment{onlinetitle}``Matter Flows
  around Black Holes and Gravitational Radiation'',
  (\comment{onlinemonth}August, \comment{onlineyear}1998),
  \comment{fileformat}[Online Los Alamos Archive Preprint]: cited on
  \comment{cited}1 September 1998,
  \comment{onlineaddress}http://xxx.lanl.gov/abs/gr-qc/9808054.
  \keywords{Gravitational radiation, numerical relativistic hydrodynamics,
  black holes, perturbation theory, numerical methods}

\bibitem{papadopoulos99b}\comment{article}
\comment{author}Papadopoulos, P., and Font, J.~A.,
  \comment{title}``Relativistic hydrodynamics on spacelike and null surfaces:
  Formalism and computations of spherically symmetric spacetimes'',
  \comment{journal}{\em Phys. Rev. D}, \comment{volume}{\bf 61},
  \comment{pages}024015, (\comment{year}1999). For a related online version
  see: \comment{author}P.~Papadopoulos, et al.,
  \comment{onlinetitle}``Relativistic hydrodynamics on spacelike and null
  surfaces: Formalism and computations of spherically symmetric spacetimes'',
  (\comment{onlinemonth}February, \comment{onlineyear}1999),
  \comment{fileformat}[Online Los Alamos Archive Preprint]: cited on
  \comment{cited}1 March 1999,
  \comment{onlineaddress}http://xxx.lanl.gov/abs/gr-qc/9902018.
  \keywords{computational fluid dynamics, relativistic hydrodynamics, initial
  value problem, null surfaces, finite difference and finite volume methods,
  hyperbolic equations, numerical methods}

\bibitem{papadopoulos01a}\comment{article}
\comment{author}Papadopoulos, P., and Font, J.~A., \comment{title}``Imprints of
  accretion on gravitational waves from black holes'', \comment{journal}{\em
  Phys. Rev. D}, \comment{volume}{\bf 63}, \comment{pages}044016,
  (\comment{year}2001). For a related online version see:
  \comment{author}P.~Papadopoulos, et al., \comment{onlinetitle}``Imprints of
  accretion on gravitational waves from black holes'',
  (\comment{onlinemonth}September, \comment{onlineyear}2000),
  \comment{fileformat}[Online Los Alamos Archive Preprint]: cited on
  \comment{cited}13 June 2002,
  \comment{onlineaddress}http://xxx.lanl.gov/abs/gr-qc/0009024.
  \keywords{computational fluid dynamics, relativistic hydrodynamics, initial
  value problem, null surfaces, finite difference and finite volume methods,
  numerical methods}

\bibitem{papaloizou84}\comment{article}
\comment{author}Papaloizou, J.~C.~B., and Pringle, J.~E., \comment{title}``The
  dynamical stability of differentially rotating discs with constant specific
  angular momentum'', \comment{journal}{\em Mon. Not. R. Astron. Soc.},
  \comment{volume}{\bf 208}, \comment{pages}721--750, (\comment{year}1984).
  \keywords{Accretion disks}

\bibitem{peitz99}\comment{article}
\comment{author}Peitz, J., and Appl, S., \comment{title}``Dissipative fluid
  dynamics in the 3+1 formalism'', \comment{journal}{\em Class. Quantum Grav.},
  \comment{volume}{\bf 16}, \comment{pages}979--989, (\comment{year}1999).
  \keywords{ADM formalism, relativistic hydrodynamics, energy and momentum}

\bibitem{penrose69}\comment{article}
\comment{author}Penrose, R., \comment{title}``Gravitational collapse: The role
  of general relativity'', \comment{journal}{\em Riv. Nuovo Cimento},
  \comment{volume}{\bf 1}, \comment{pages}252--276, (\comment{year}1969).
  \keywords{}

\bibitem{petrich89}\comment{article}
\comment{author}Petrich, L.~I., Shapiro, S.~L., Stark, R.~F., and Teukolsky,
  S.~A., \comment{title}``Accretion onto a moving black hole: a fully
  relativistic treatment'', \comment{journal}{\em Astrophys. J.},
  \comment{volume}{\bf 336}, \comment{pages}313--349, (\comment{year}1989).
  \keywords{Relativistic hydrodynamics, numerical relativistic hydrodynamics,
  computational fluid dynamics, accretion, black holes, shock waves, numerical
  methods}

\bibitem{petrich85}\comment{article}
\comment{author}Petrich, L.~I., Shapiro, S.~L., and Wasserman, I.,
  \comment{title}``Gravitational radiation from nonspherical infall into black
  holes. II. A catalog of ``exact" waveforms'', \comment{journal}{\em
  Astrophys. J. Suppl. Ser.}, \comment{volume}{\bf 58},
  \comment{pages}297--320, (\comment{year}1985). \keywords{Gravitational
  radiation, black holes, approximation methods}

\bibitem{piran86}\comment{inbook}
\comment{author}Piran, T., and Stark, R.~F., \comment{title}``Numerical
  relativity, rotating gravitational collapse and gravitational radiation'', in
  \comment{editor}Centrella, J., ed., \comment{booktitle}{\em Dynamical
  Spacetimes and Numerical Relativity}, \comment{pages} 40--73,
  (\comment{publisher}Cambridge University Press, \comment{address}Cambridge,
  England, \comment{year}1986). \keywords{ADM formalism, Einstein equations,
  Gravitational collapse, Numerical methods, Relativistic hydrodynamics,
  Gravitational radiation, Numerical relativity}

\bibitem{pons98}\comment{article}
\comment{author}Pons, J.~A., Font, J.~A., Ib\'a\~nez, J.~M., Mart\'{\i}, J.~M.,
  and Miralles, J.~A., \comment{title}``General Relativistic Hydrodynamics with
  Special Relativistic Riemann Solvers'', \comment{journal}{\em Astron.
  Astrophys.}, \comment{volume}{\bf 339}, \comment{pages}629--637,
  (\comment{year}1998). For a related online version see: \comment{author}J.~A.
  Pons, et al., \comment{onlinetitle}``General Relativistic Hydrodynamics with
  Special Relativistic Riemann Solvers'', (\comment{onlinemonth}July,
  \comment{onlineyear}1998), \comment{fileformat}[Online Los Alamos Archive
  Preprint]: cited on \comment{cited}1 August 1998,
  \comment{onlineaddress}http://xxx.lanl.gov/abs/astro-ph/9807215.
  \keywords{computational fluid dynamics, relativistic hydrodynamics, numerical
  relativity, finite difference and finite volume methods, hyperbolic
  equations, numerical methods}

\bibitem{pons00b}\comment{article}
\comment{author}Pons, J.~A., Ib{\' a}{\~ n}ez, J.~M., and Miralles, J.~A.,
  \comment{title}``Hyperbolic character of the angular moment equations of
  radiative transfer and numerical methods'', \comment{journal}{\em Mon. Not.
  R. Astron. Soc.}, \comment{volume}{\bf 317}, \comment{pages}550--562,
  (\comment{year}2000). For a related online version see: \comment{author}J.~A.
  Pons, et al., \comment{onlinetitle}``Hyperbolic character of the angular
  moment equations of radiative transfer and numerical methods'',
  (\comment{onlinemonth}May, \comment{onlineyear}2000),
  \comment{fileformat}[Online Los Alamos Archive Preprint]: cited on
  \comment{cited}28 October 2002,
  \comment{onlineaddress}http://xxx.lanl.gov/abs/astro-ph/0005310. \keywords{}

\bibitem{pons00}\comment{article}
\comment{author}Pons, J.~A., Mart\'{\i}, J.~M$^{\underline{\rm a}}$., and
  M\"uller, E., \comment{title}``The exact solution of the Riemann problem with
  non-zero tangential velocities in relativistic hydrodynamics'',
  \comment{journal}{\em J.~Fluid Mech.}, \comment{volume}{\bf 422},
  \comment{pages}125--139, (\comment{year}2000). For a related online version
  see: \comment{author}J.~A. Pons, et al., \comment{onlinetitle}``The exact
  solution of the Riemann problem with non-zero tangential velocities in
  relativistic hydrodynamics'', (\comment{onlinemonth}May,
  \comment{onlineyear}2000), \comment{fileformat}[Online Los Alamos Archive
  Preprint]: cited on \comment{cited}15 July 2002,
  \comment{onlineaddress}http://xxx.lanl.gov/abs/astro-ph/0005038.
  \keywords{relativistic hydrodynamics, initial value problem, shock waves,
  numerical methods}

\bibitem{rampp00a}\comment{article}
\comment{author}Rampp, M., and Janka, H.-Th., \comment{title}``Spherically
  symmetric simulation with Boltzmann neutrino transport of core-collapse and
  post-bounce evolution of a 15 solar mass star'', \comment{journal}{\em
  Astrophys. J.}, \comment{volume}{\bf 539}, \comment{pages}L33--L36,
  (\comment{year}2000). For a related online version see:
  \comment{author}M.~Rampp, et al., \comment{onlinetitle}``Spherically
  symmetric simulation with Boltzmann neutrino transport of core-collapse and
  post-bounce evolution of a 15 solar mass star'', (\comment{onlinemonth}May,
  \comment{onlineyear}2000), \comment{fileformat}[online Los Alamos Archive
  Preprint]: cited on \comment{cited}14 June 2002,
  \comment{onlineaddress}http://xxx.lanl.gov/abs/astro-ph/0005438.
  \keywords{gravitational collapse, supernovae}

\bibitem{rampp02a}\comment{article}
\comment{author}Rampp, M., and Janka, H.-Th., \comment{title}``Radiation
  hydrodynamics with neutrinos: Variable Eddington factor method for
  core-collapse supernova simulations'', \comment{journal}{\em Astron.
  Astrophys.}, \comment{volume}{\bf 396}, \comment{pages}361--392,
  (\comment{year}2002). For a related online version see:
  \comment{author}M.~Rampp, et al., \comment{onlinetitle}``Radiation
  hydrodynamics with neutrinos: Variable Eddington factor method for
  core-collapse supernova simulations'', (\comment{onlinemonth}March,
  \comment{onlineyear}2002), \comment{fileformat}[online Los Alamos Archive
  Preprint]: cited on \comment{cited}14 June 2002,
  \comment{onlineaddress}http://xxx.lanl.gov/abs/astro-ph/0203101.
  \keywords{gravitational collapse, supernovae}

\bibitem{rampp98a}\comment{article}
\comment{author}Rampp, M., M\"uller, E., and Ruffert, M.,
  \comment{title}``Simulations of non-axisymmetric rotational core collapse'',
  \comment{journal}{\em Astron.\ Astrophys.}, \comment{volume}{\bf 332},
  \comment{pages}969--983, (\comment{year}1998). For a related online version
  see: \comment{author}M.~Rampp, et al., \comment{onlinetitle}``Simulations of
  non-axisymmetric rotational core collapse'', (\comment{onlinemonth}November,
  \comment{onlineyear}1997), \comment{fileformat}[online Los Alamos Archive
  Preprint]: cited on \comment{cited}15 July 2002,
  \comment{onlineaddress}http://xxx.lanl.gov/abs/astro-ph/9711122.
  \keywords{gravitational collapse, supernovae}

\bibitem{rasio99}\comment{article}
\comment{author}Rasio, F.~A., and Shapiro, S.~L., \comment{title}``Coalescing
  binary neutron stars'', \comment{journal}{\em Class. Quantum Grav.},
  \comment{volume}{\bf 16}, \comment{pages}R1--R29, (\comment{year}1999). For a
  related online version see: \comment{author}F.~A. Rasio, et al.,
  \comment{onlinetitle}``Coalescing binary neutron stars'',
  (\comment{onlinemonth}February, \comment{onlineyear}1999),
  \comment{fileformat}[online Los Alamos Archive Preprint]: cited on
  \comment{cited}1 March 1999,
  \comment{onlineaddress}http://xxx.lanl.gov/abs/gr-qc/9902019.
  \keywords{Neutron stars, Astrophysics, Gravitational radiation,
  hydrodynamics}

\bibitem{rezzolla94}\comment{article}
\comment{author}Rezzolla, L., and Miller, J.~C., \comment{title}``Relativistic
  radiative transfer for spherical flows'', \comment{journal}{\em Class.
  Quantum Grav.}, \comment{volume}{\bf 11}, \comment{pages}1815--1832,
  (\comment{year}1994). \keywords{}

\bibitem{rezzolla00}\comment{article}
\comment{author}Rezzolla, L., and Zanotti, O., \comment{title}``An improved
  exact Riemann solver for relativistic hydrodynamics'', \comment{journal}{\em
  J. Fluid Mech.}, \comment{volume}{\bf 449}, \comment{pages}395,
  (\comment{year}2001). For a related online version see:
  \comment{author}L.~Rezzolla, et al., \comment{onlinetitle}``An improved exact
  Riemann solver for relativistic hydrodynamics'', (\comment{onlinemonth}March,
  \comment{onlineyear}2001), \comment{fileformat}[online Los Alamos Archive
  Preprint]: cited on \comment{cited}15 July 2002,
  \comment{onlineaddress}http://xxx.lanl.gov/abs/gr-qc/0103005.
  \keywords{Relativistic hydrodynamics}

\bibitem{zanotti02}\comment{article}
\comment{author}Rezzolla, L., Zanotti, O., and Pons, J.~A., \comment{title}``An
  improved exact Riemann solver for multidimensional relativistic flows'',
  \comment{journal}{\em J. Fluid Mech.}, (\comment{year}2002). For a related
  online version see: \comment{author}L.~Rezzolla, et al.,
  \comment{onlinetitle}``An improved exact Riemann solver for multidimensional
  relativistic flows'', (\comment{onlinemonth}May, \comment{onlineyear}2002),
  \comment{fileformat}[Online Los Alamos Archive Preprint]: cited on
  \comment{cited}15 July 2002,
  \comment{onlineaddress}http://xxx.lanl.gov/abs/astro-ph/0205034. submitted.
  \keywords{Relativistic hydrodynamics}

\bibitem{richardson02}\comment{article}
\comment{author}Richardson, G.~A., and Chung, T.~J.,
  \comment{title}``Computational relativistic astrophysics using the flow
  field-dependent variation theory'', \comment{journal}{\em Astrophys. J.
  Suppl. Ser.}, \comment{volume}{\bf 139}, \comment{pages}539--563,
  (\comment{year}2002). \keywords{}

\bibitem{richtmyer67}\comment{book}
\comment{author}Richtmyer, R.~D., and Morton, K.~W., \comment{title}{\em
  Difference methods for initial value problems},
  (\comment{publisher}Wiley-Interscience, \comment{address}New York,
  \comment{year}1967). \keywords{differential equations, finite difference and
  finite volume methods, numerical methods}

\bibitem{roe81}\comment{article}
\comment{author}Roe, P.~L., \comment{title}``Approximate Riemann solvers,
  parameter vectors and difference schemes'', \comment{journal}{\em J. Comput.
  Phys.}, \comment{volume}{\bf 43}, \comment{pages}357--372,
  (\comment{year}1981). \keywords{computational fluid dynamics, finite
  difference and finite volume methods, hyperbolic equations, numerical
  methods}

\bibitem{roe84}\comment{article}
\comment{author}Roe, P.~L., \comment{title}``Generalized formulation of TVD
  Lax-Wendroff schemes'', \comment{journal}{\em ICASE Report},
  \comment{volume}{\bf 84}, \comment{pages}53, (\comment{year}1984).
  \keywords{computational fluid dynamics, finite difference and finite volume
  methods, hyperbolic equations, numerical methods}

\bibitem{romero96}\comment{article}
\comment{author}Romero, J.~V., Ib{\'a}nez, J.~M., Mart\'{\i}, J.~M., and
  Miralles, J.~A., \comment{title}``A new spherically symmetric general
  relativistic hydrodynamical code'', \comment{journal}{\em Astrophys. J.},
  \comment{volume}{\bf 462}, \comment{pages}839--854, (\comment{year}1996). For
  a related online version see: \comment{author}J.~V. Romero, et al.,
  \comment{onlinetitle}``A new spherically symmetric general relativistic
  hydrodynamical code'', (\comment{onlinemonth}September,
  \comment{onlineyear}1995), \comment{fileformat}[Online Los Alamos Archive
  Preprint]: cited on \comment{cited}1 October 1995,
  \comment{onlineaddress}http://xxx.lanl.gov/abs/astro-ph/9509121.
  \keywords{gravitational collapse, computational fluid dynamics, relativistic
  hydrodynamics, numerical relativity, finite difference and finite volume
  methods, hyperbolic equations, numerical methods}

\bibitem{ruffert94}\comment{article}
\comment{author}Ruffert, M., and Arnett, D., \comment{title}``Three-dimensional
  hydrodynamic Bondi-Hoyle accretion. 2: Homogeneous medium at Mach 3 with
  gamma = 5/3'', \comment{journal}{\em Astrophys. J.}, \comment{volume}{\bf
  427}, \comment{pages}351--376, (\comment{year}1994). \keywords{Hydrodynamics,
  computational fluid dynamics, shock waves, accretion}

\bibitem{ruffert98}\comment{article}
\comment{author}Ruffert, M., and Janka, H.-T., \comment{title}``Colliding
  neutron stars. Gravitational waves, neutrino emission, and gamma-ray
  bursts'', \comment{journal}{\em Astron. Astrophys.}, \comment{volume}{\bf
  338}, \comment{pages}535--555, (\comment{year}1998). For a related online
  version see: \comment{author}M.~Ruffert, et al.,
  \comment{onlinetitle}``Colliding neutron stars --- Gravitational waves,
  neutrino emission, and gamma-ray bursts'', (\comment{onlinemonth}April,
  \comment{onlineyear}1998), \comment{fileformat}[Online Los Alamos Archive
  Preprint]: cited on \comment{cited}1 May 1998,
  \comment{onlineaddress}http://xxx.lanl.gov/abs/astro-ph/9804132.
  \keywords{Hydrodynamics, computational fluid dynamics, astrophysics, binary
  systems, gamma-ray bursts, gravitational radiation, neutrinos, neutron stars,
  nuclear physics}

\bibitem{sachs62}\comment{article}
\comment{author}Sachs, R.~K., \comment{title}``Gravitational waves in general
  relativity. VIII. Waves in asymptotically flat space-time'',
  \comment{journal}{\em Proc. R. Soc. London}, \comment{volume}{\bf Sect. A
  270}, \comment{pages}103--126, (\comment{year}1962). \keywords{}

\bibitem{schinder89}\comment{inbook}
\comment{author}Schinder, P.~J., \comment{title}``General relativistic implicit
  radiation hydrodynamics in polar sliced spacetime'', in
  \comment{editor}C.~R.~Evans, L.~S.~Finn, and Hobill, D.~W., eds.,
  \comment{booktitle}{\em Frontiers in numerical relativity}, \comment{pages}
  163--170, (\comment{publisher}Cambridge University Press,
  \comment{address}Cambridge, England, \comment{year}1989). \keywords{Numerical
  relativity, numerical relativistic hydrodynamics, gauge conditions,
  gravitational collapse}

\bibitem{schinder88}\comment{article}
\comment{author}Schinder, P.~J., Bludmann, S.~A., and Piran, T.,
  \comment{title}``General relativistic implicit hydrodynamics in polar sliced
  spacetime'', \comment{journal}{\em Phys. Rev. D}, \comment{volume}{\bf 37},
  \comment{pages}2722--2731, (\comment{year}1988). \keywords{Numerical
  relativity, numerical relativistic hydrodynamics, gauge conditions,
  gravitational collapse}

\bibitem{schneider93}\comment{article}
\comment{author}Schneider, V., Katscher, V., Rischke, D.~H., Waldhauser, B.,
  Marhun, J.~A., and Munz, C.-D., \comment{title}``New algorithms for
  ultra-relativistic numerical hydrodynamics'', \comment{journal}{\em J.
  Comput. Phys.}, \comment{volume}{\bf 105}, \comment{pages}92--107,
  (\comment{year}1993). \keywords{computational fluid dynamics, relativistic
  hydrodynamics, finite difference and finite volume methods, hyperbolic
  equations, numerical methods}

\bibitem{seidel87}\comment{article}
\comment{author}Seidel, E., and Moore, T., \comment{title}``Gravitational
  radiation from realistic relativistic stars: Odd-parity fluid
  perturbations'', \comment{journal}{\em Phys. Rev. D}, \comment{volume}{\bf
  35}, \comment{pages}2287--2296, (\comment{year}1987). \keywords{Gravitational
  radiation, relativistic stars, perturbation theory, gravitational collapse}

\bibitem{seidel88}\comment{article}
\comment{author}Seidel, E., Myra, E.~S., and Moore, T.,
  \comment{title}``Gravitational radiation from type II supernovae - The effect
  of the high-density equation of state'', \comment{journal}{\em Phys. Rev. D},
  \comment{volume}{\bf 38}, \comment{pages}2349--2356, (\comment{year}1988).
  \keywords{Gravitational radiation, relativistic stars, perturbation theory,
  gravitational collapse, supernovae}

\bibitem{shakura73}\comment{article}
\comment{author}Shakura, N.~I., and Sunyaev, R.~A., \comment{title}``Black
  holes in binary systems. Observational appearance'', \comment{journal}{\em
  Astron. Astrophys.}, \comment{volume}{\bf 24}, \comment{pages}337--355,
  (\comment{year}1973). \keywords{Accretion, accretion disks, astrophysics,
  black holes, binary systems}

\bibitem{shapiro98}\comment{article}
\comment{author}Shapiro, S.~L., \comment{title}``Head-On collision of neutron
  stars as a thought experiment'', \comment{journal}{\em Phys. Rev. D},
  \comment{volume}{\bf 58}, \comment{pages}103002, (\comment{year}1998). For a
  related online version see: \comment{author}S.~L. Shapiro,
  \comment{onlinetitle}``Head-On collision of neutron stars as a thought
  experiment'', (\comment{onlinemonth}September, \comment{onlineyear}1998),
  \comment{fileformat}[Online Los Alamos Archive Preprint]: cited on
  \comment{cited}15 February 2000,
  \comment{onlineaddress}http://xxx.lanl.gov/abs/gr-qc/9809060.
  \keywords{Gravitational collapse, neutron stars, relativistic astrophysics}

\bibitem{shapiro80}\comment{article}
\comment{author}Shapiro, S.~L., and Teukolsky, S.~A.,
  \comment{title}``Gravitational collapse to neutron stars and black holes:
  computer generation of spherical spacetimes'', \comment{journal}{\em
  Astrophys. J.}, \comment{volume}{\bf 235}, \comment{pages}199--215,
  (\comment{year}1980). \keywords{Black holes, relativistic hydrodynamics,
  neutron stars, gravitational collapse}

\bibitem{shapiro82}\comment{article}
\comment{author}Shapiro, S.~L., and Wasserman, I.,
  \comment{title}``Gravitational radiation from nonspherical infall into black
  holes'', \comment{journal}{\em Astrophys. J.}, \comment{volume}{\bf 260},
  \comment{pages}838--848, (\comment{year}1982). \keywords{Gravitational
  radiation, black holes, approximation methods}

\bibitem{shibata99}\comment{article}
\comment{author}Shibata, M., \comment{title}``Fully general relativistic
  simulation of coalescing binary neutron stars: Preparatory tests'',
  \comment{journal}{\em Phys. Rev. D}, \comment{volume}{\bf 60},
  \comment{pages}104052, (\comment{year}1999). For a related online version
  see: \comment{author}M.~Shibata, \comment{onlinetitle}``Fully general
  relativistic simulation of coalescing binary neutron stars: Preparatory
  tests'', (\comment{onlinemonth}August, \comment{onlineyear}1999),
  \comment{fileformat}[Online Los Alamos Archive Preprint]: cited on
  \comment{cited}15 February 2000,
  \comment{onlineaddress}http://xxx.lanl.gov/abs/gr-qc/9908027.
  \keywords{Numerical relativity, numerical relativistic hydrodynamics,
  Einstein equations, binary systems, neutron stars, black holes, numerical
  methods}

\bibitem{shibata00b}\comment{article}
\comment{author}Shibata, M., \comment{title}``Axisymmetric simulations of
  rotating stellar collapse in full general relativity: Criteria for prompt
  collapse to black holes'', \comment{journal}{\em Prog. Theor. Phys.},
  \comment{volume}{\bf 104}, \comment{pages}325--358, (\comment{year}2000). For
  a related online version see: \comment{author}M.~Shibata,
  \comment{onlinetitle}``Axisymmetric simulations of rotating stellar collapse
  in full general relativity: Criteria for prompt collapse to black holes'',
  (\comment{onlinemonth}July, \comment{onlineyear}2000),
  \comment{fileformat}[Online Los Alamos Archive Preprint]: cited on
  \comment{cited}13 June 2002,
  \comment{onlineaddress}http://xxx.lanl.gov/abs/gr-qc/0007049.
  \keywords{Numerical relativity, numerical relativistic hydrodynamics, neutron
  stars, relativistic stars, black holes, gravitational collapse}

\bibitem{shibata02c}\comment{article}
\comment{author}Shibata, M., \comment{title}``Axisymmetric general relativistic
  hydrodynamics: Long-term evolution of neutron stars and stellar collapse to
  neutron stars and black holes'', \comment{journal}{\em Phys. Rev. D},
  (\comment{year}2002). For a related online version see:
  \comment{author}M.~Shibata, \comment{onlinetitle}``Axisymmetric general
  relativistic hydrodynamics: Long-term evolution of neutron stars and stellar
  collapse to neutron stars and black holes'', (\comment{onlineyear}2002),
  \comment{onlineaddress}http://xxx.lanl.gov/abs/gr-qc/0301103. submitted.
  \keywords{Numerical relativity, numerical relativistic hydrodynamics,
  gravitational collapse, black holes,neutron stars}

\bibitem{shibata00a}\comment{article}
\comment{author}Shibata, M., Baumgarte, T.~B., and Shapiro, S.~L.,
  \comment{title}``The bar-mode instability in differentially rotating neutron
  stars: Simulations in full general relativity'', \comment{journal}{\em
  Astrophys. J.}, \comment{volume}{\bf 542}, \comment{pages}453--463,
  (\comment{year}2000). For a related online version see:
  \comment{author}M.~Shibata, et al., \comment{onlinetitle}``The bar-mode
  instability in differentially rotating neutron stars: Simulations in full
  general relativity'', (\comment{onlinemonth}May, \comment{onlineyear}2000),
  \comment{fileformat}[Online Los Alamos Archive Preprint]: cited on
  \comment{cited}13 June 2002,
  \comment{onlineaddress}http://xxx.lanl.gov/abs/astro-ph/0005378.
  \keywords{Numerical relativity, numerical relativistic hydrodynamics,
  Einstein equations, neutron stars, relativistic stars}

\bibitem{shibata99b}\comment{article}
\comment{author}Shibata, M., Baumgarte, T.~W., and Shapiro, S.~L,
  \comment{title}``Stability and collapse of rapidly rotating, supramassive
  neutron stars: 3D simulations in general relativity'', \comment{journal}{\em
  Phys. Rev. D}, \comment{volume}{\bf 61}, \comment{pages}044012,
  (\comment{year}2000). For a related online version see:
  \comment{author}M.~Shibata, et al., \comment{onlinetitle}``Stability and
  collapse of rapidly rotating, supramassive neutron stars: 3D simulations in
  general relativity'', (\comment{onlinemonth}November,
  \comment{onlineyear}1999), \comment{fileformat}[Online Los Alamos Archive
  Preprint]: cited on \comment{cited}15 February 2000,
  \comment{onlineaddress}http://xxx.lanl.gov/abs/astro-ph/9911308.
  \keywords{Relativistic hydrodynamics, numerical relativistic hydrodynamics,
  computational fluid dynamics, gravitational collapse, neutron stars, black
  holes, numerical relativity, Einstein equations}

\bibitem{shibata95}\comment{article}
\comment{author}Shibata, M., and Nakamura, T., \comment{title}``Evolution of
  three-dimensional gravitational waves: Harmonic slicing case'',
  \comment{journal}{\em Phys. Rev. D}, \comment{volume}{\bf 52},
  \comment{pages}5428--5444, (\comment{year}1995). \keywords{Numerical
  relativity, Einstein equations, constraint equations, gauge conditions,
  gravitational radiation}

\bibitem{shibata02b}\comment{article}
\comment{author}Shibata, M., and Shapiro, S.~L., \comment{title}``Collapse of a
  rotating supermassive star to a supermassive black hole: Fully relativistic
  simulations'', \comment{journal}{\em Astrophys. J.}, \comment{volume}{\bf
  572}, \comment{pages}L39--L43, (\comment{year}2002). For a related online
  version see: \comment{author}M.~Shibata, et al.,
  \comment{onlinetitle}``Collapse of a rotating supermassive star to a
  supermassive black hole: Fully relativistic simulations'',
  (\comment{onlinemonth}May, \comment{onlineyear}2002),
  \comment{fileformat}[Online Los Alamos Archive Preprint]: cited on
  \comment{cited}13 June 2002,
  \comment{onlineaddress}http://xxx.lanl.gov/abs/astro-ph/0205091.
  \keywords{Numerical relativity, numerical relativistic hydrodynamics,
  gravitational collapse, black holes}

\bibitem{shibata99c}\comment{article}
\comment{author}Shibata, M., and Uryu, K., \comment{title}``Simulation of
  merging binary neutron stars in full general relativity: $\Gamma=2$ case'',
  \comment{journal}{\em Phys. Rev. D}, \comment{volume}{\bf 61},
  \comment{pages}064001, (\comment{year}2000). For a related online version
  see: \comment{author}M.~Shibata, et al., \comment{onlinetitle}``Simulation of
  merging binary neutron stars in full general relativity: $\Gamma=2$ case'',
  (\comment{onlinemonth}November, \comment{onlineyear}1999),
  \comment{fileformat}[Online Los Alamos Archive Preprint]: cited on
  \comment{cited}15 February 2000,
  \comment{onlineaddress}http://xxx.lanl.gov/abs/gr-qc/9911058.
  \keywords{Numerical relativity, numerical relativistic hydrodynamics,
  Einstein equations, binary systems, neutron stars, black holes, numerical
  methods}

\bibitem{shibata02a}\comment{article}
\comment{author}Shibata, M., and Uryu, K., \comment{title}``Gravitational waves
  from the merger of binary neutron stars in a fully general relativistic
  simulation'', \comment{journal}{\em Prog. Theor. Phys.}, \comment{volume}{\bf
  107}, \comment{pages}265--303, (\comment{year}2002). For a related online
  version see: \comment{author}M.~Shibata, et al.,
  \comment{onlinetitle}``Gravitational waves from the merger of binary neutron
  stars in a fully general relativistic simulation'',
  (\comment{onlinemonth}March, \comment{onlineyear}2002),
  \comment{fileformat}[Online Los Alamos Archive Preprint]: cited on
  \comment{cited}13 June 2002,
  \comment{onlineaddress}http://xxx.lanl.gov/abs/gr-qc/0203037.
  \keywords{Numerical relativity, numerical relativistic hydrodynamics, neutron
  stars, relativistic stars, gravitational radiation, binary systems}

\bibitem{shibata98}\comment{article}
\comment{author}Shibata, M., W., Baumgarte.~T., and Shapiro, S.~L.,
  \comment{title}``Stability of coalescing binary stars against gravitational
  collapse: hydrodynamical simulations'', \comment{journal}{\em Phys. Rev. D},
  \comment{volume}{\bf 58}, \comment{pages}023002, (\comment{year}1998). For a
  related online version see: \comment{author}M.~Shibata, et al.,
  \comment{onlinetitle}``Stability of coalescing binary stars against
  gravitational collapse: hydrodynamical simulations'',
  (\comment{onlinemonth}May, \comment{onlineyear}1998),
  \comment{fileformat}[Online Los Alamos Archive Preprint]: cited on
  \comment{cited}15 February 2000,
  \comment{onlineaddress}http://xxx.lanl.gov/abs/gr-qc/9805026.
  \keywords{Gravitational collapse, numerical relativistic hydrodynamics, ADM
  formalism, Einstein equations, constraint equations, numerical relativity,
  neutron stars}

\bibitem{shibataweb}\comment{misc}
M.~Shibata Web Page,\\ \verb+http://esa.c.u-tokyo.ac.jp/~shibata/anim.html+.
  \keywords{}

\bibitem{shu87}\comment{article}
\comment{author}Shu, C.~W., \comment{title}``TVB uniformly high-order schemes
  for conservation laws'', \comment{journal}{\em Math. Comp.},
  \comment{volume}{\bf 49}, \comment{pages}105--121, (\comment{year}1987).
  \keywords{computational fluid dynamics, finite difference and finite volume
  methods, hyperbolic equations, numerical methods}

\bibitem{siebel02b}\comment{article}
\comment{author}Siebel, F., Font, J.~A., M\"uller, E., and Papadopoulos, P.,
  \comment{title}``Simulating the dynamics of relativistic stars via a
  light-cone approach'', \comment{journal}{\em Phys. Rev. D},
  \comment{volume}{\bf 65}, \comment{pages}064038, (\comment{year}2002). For a
  related online version see: \comment{author}F.~Siebel, et al.,
  \comment{onlinetitle}``Simulating the dynamics of relativistic stars via a
  light-cone approach'', (\comment{onlinemonth}November,
  \comment{onlineyear}2001), \comment{fileformat}[Online Los Alamos Archive
  Preprint]: cited on \comment{cited}13 June 2002,
  \comment{onlineaddress}http://xxx.lanl.gov/abs/gr-qc/0111093.
  \keywords{Numerical relativistic hydrodynamics, numerical methods, Einstein
  equations, numerical relativity, null surfaces, initial value problem,
  relativistic stars}

\bibitem{siebel02a}\comment{article}
\comment{author}Siebel, F., Font, J.~A., and Papadopoulos, P.,
  \comment{title}``Scalar field induced oscillations of relativistic stars and
  gravitational collapse'', \comment{journal}{\em Phys. Rev. D},
  \comment{volume}{\bf 65}, \comment{pages}024021, (\comment{year}2002). For a
  related online version see: \comment{author}F.~Siebel, et al.,
  \comment{onlinetitle}``Scalar field induced oscillations of neutron stars and
  gravitational collapse'', (\comment{onlinemonth}August,
  \comment{onlineyear}2001), \comment{fileformat}[Online Los Alamos Archive
  Preprint]: cited on \comment{cited}13 June 2002,
  \comment{onlineaddress}http://xxx.lanl.gov/abs/gr-qc/0108006.
  \keywords{Numerical relativistic hydrodynamics, numerical methods,
  gravitational collapse, Einstein equations, numerical relativity, null
  surfaces, initial value problem, relativistic stars}

\bibitem{siegler99}\comment{article}
\comment{author}Siegler, S., and Riffert, H., \comment{title}``Smoothed
  particle hydrodynamics simulations of ultra-relativistic shocks with
  artificial viscosity'', \comment{journal}{\em Astrophys. J.},
  \comment{volume}{\bf 531}, \comment{pages}1053--1066, (\comment{year}2000).
  For a related online version see: \comment{author}S.~Siegler, et al.,
  \comment{onlinetitle}``Smoothed particle hydrodynamics simulations of
  ultra-relativistic shocks with artificial viscosity'',
  (\comment{onlinemonth}April, \comment{onlineyear}1999),
  \comment{fileformat}[Online Los Alamos Archive Preprint]: cited on
  \comment{cited}1 May 1999,
  \comment{onlineaddress}http://xxx.lanl.gov/abs/astro-ph/9904070.
  \keywords{computational fluid dynamics, relativistic hydrodynamics, smoothed
  particle methods, numerical methods}

\bibitem{sloan85}\comment{inbook}
\comment{author}Sloan, J., and Smarr, L.~L., \comment{title}``General
  relativistic magnetohydrodynamics'', in \comment{editor}J.~Centrella,
  J.~LeBlanc, and Bowers, R., eds., \comment{booktitle}{\em Numerical
  Astrophysics}, \comment{pages} 52--68, (\comment{publisher}Jones and
  Bartlett, \comment{address}Boston, \comment{year}1985).
  \keywords{Relativistic magnetohydrodynamics, partial differential equations,
  numerical methods}

\bibitem{smarr75}\comment{phdthesis}
\comment{author}Smarr, L.~L., \comment{title}{\em The structure of general
  relativity with a numerical illustration: the collision of two black holes},
  PhD thesis, (\comment{school}University of Texas at Austin,
  \comment{year}1975). \keywords{numerical relativity, numerical methods,
  Einstein equations, Cauchy problem, initial value problem, constraint
  equations, ADM formalism, black holes, gauge conditions}

\bibitem{sperhake01a}\comment{article}
\comment{author}Sperhake, U., Papadopoulos, P., and Andersson, N.,
  \comment{title}``Nonlinear radial oscillations of neutron stars:
  Mode-coupling results'', \comment{journal}{\em Mon. Not. R. Astron. Soc.},
  (\comment{year}2001). For a related online version see:
  \comment{author}U.~Sperhake, et al., \comment{onlinetitle}``Nonlinear radial
  oscillations of neutron stars: Mode-coupling results'',
  (\comment{onlinemonth}October, \comment{onlineyear}2001),
  \comment{fileformat}[Online Los Alamos Archive Preprint]: cited on
  \comment{cited}5 July 2002,
  \comment{onlineaddress}http://xxx.lanl.gov/abs/astro-ph/0110487. submitted.
  \keywords{}

\bibitem{stark89}\comment{inbook}
\comment{author}Stark, R.~F., \comment{title}``Non-axisymmetric rotating
  gravitational collapse and gravitational radiation'', in
  \comment{editor}C.~R.~Evans, L.~S.~Finn, and Hobill, D.~W., eds.,
  \comment{booktitle}{\em Frontiers in numerical relativity}, \comment{pages}
  281--296, (\comment{publisher}Cambridge University Press,
  \comment{address}Cambridge, England, \comment{year}1989). \keywords{Numerical
  relativity, numerical methods, gravitational collapse, gravitational
  radiation, numerical relativistic hydrodynamics}

\bibitem{stark85}\comment{article}
\comment{author}Stark, R.~F., and Piran, T.,
  \comment{title}``Gravitational-Wave Emission from Rotating Gravitational
  Collapse'', \comment{journal}{\em Phys. Rev. Lett.}, \comment{volume}{\bf
  55}, \comment{pages}891--894, (\comment{year}1985). \keywords{Numerical
  relativity, numerical methods, gravitational collapse, gravitational
  radiation, numerical relativistic hydrodynamics}

\bibitem{stark87}\comment{article}
\comment{author}Stark, R.~F., and Piran, T., \comment{title}``A general
  relativistic code for rotating axisymmetric configurations and gravitational
  radiation: numerical methods and tests'', \comment{journal}{\em Comp. Phys.
  Rep.}, \comment{volume}{\bf 5}, \comment{pages}221--264,
  (\comment{year}1987). \keywords{computational fluid dynamics, relativistic
  hydrodynamics, numerical relativity, finite difference and finite volume
  methods, gravitational collapse, gravitational radiation, numerical methods}

\bibitem{stergioulas02}\comment{article}
\comment{author}Stergioulas, N., \comment{title}``Rotating stars in
  relativity'', \comment{journal}{\em Living Reviews in Relativity},
  \comment{volume}{\bf 1}, \comment{pages}8, (\comment{year}1998). \keywords{}

\bibitem{stergioulas01a}\comment{article}
\comment{author}Stergioulas, N., and Font, J.~A., \comment{title}``Nonlinear
  $r$-modes in rapidly rotating relativistic stars'', \comment{journal}{\em
  Phys. Rev. Lett.}, \comment{volume}{\bf 86}, \comment{pages}1148--1151,
  (\comment{year}2001). For a related online version see:
  \comment{author}N.~Stergioulas, et al., \comment{onlinetitle}``Nonlinear
  $r$-modes in rapidly rotating relativistic stars'',
  (\comment{onlinemonth}July, \comment{onlineyear}2000),
  \comment{fileformat}[Online Los Alamos Archive Preprint]: cited on
  \comment{cited}13 June 2002,
  \comment{onlineaddress}http://xxx.lanl.gov/abs/gr-qc/0007086.
  \keywords{numerical relativistic hydrodynamics, numerical relativity,
  relativistic stars}

\bibitem{swesty94}\comment{article}
\comment{author}Swesty, D., Lattimer, J.~M., and Myra, E.~S.,
  \comment{title}``The role of the equation of state in the `prompt' phase of
  type II supernovae'', \comment{journal}{\em Astrophys. J.},
  \comment{volume}{\bf 425}, \comment{pages}195--204, (\comment{year}1994).
  \keywords{Nuclear physics, shock waves, neutron stars, gravitational
  collapse, supernovae, hydrodynamics}

\bibitem{tadmorweb}\comment{misc}
E. Tadmor Web Page,\\ \verb+http://www.math.ucla.edu/~tadmor+. \keywords{}

\bibitem{tanaka95}\comment{article}
\comment{author}Tanaka, Y., and et~al, \comment{title}``Gravitationally
  redshifted emission implying an accretion disk and massive black-hole in the
  active galaxy MCG:-6-30-15'', \comment{journal}{\em Nature},
  \comment{volume}{\bf 375}, \comment{pages}659--661, (\comment{year}1995).
  \keywords{relativistic astrophysics, active galactic nuclei, black holes,
  accretion disks}

\bibitem{taniguchi02a}\comment{article}
\comment{author}Taniguchi, K., and Gourgoulhon, E.,
  \comment{title}``Equilibrium sequences of synchronized and irrotational
  binary systems composed of different mass stars in Newtonian gravity'',
  \comment{journal}{\em Phys. Rev. D}, \comment{volume}{\bf 65},
  \comment{pages}044027, (\comment{year}2002). For a related online version
  see: \comment{author}K.~Taniguchi, et al., \comment{onlinetitle}``Equilibrium
  sequences of synchronized and irrotational binary systems composed of
  different mass stars in Newtonian gravity'', (\comment{onlinemonth}August,
  \comment{onlineyear}2001), \comment{fileformat}[Online Los Alamos Archive
  Preprint]: cited on \comment{cited}24 June 2002,
  \comment{onlineaddress}http://xxx.lanl.gov/abs/astro-ph/0108086.
  \keywords{Einstein equations, relativistic hydrodynamics, initial value
  problem, hyperbolic equations, elliptic equations, binary systems, spectral
  methods, neutron stars}

\bibitem{gourgoulhon01b}\comment{article}
\comment{author}Taniguchi, K., Gourgoulhon, E., and Bonazzola, S.,
  \comment{title}``Quasiequilibrium sequences of synchronized and irrotational
  binary neutron stars in general relativity. II. Newtonian limits'',
  \comment{journal}{\em Phys. Rev. D}, \comment{volume}{\bf 64},
  \comment{pages}064012, (\comment{year}2001). For a related online version
  see: \comment{author}K.~Taniguchi, et al.,
  \comment{onlinetitle}``Quasiequilibrium sequences of synchronized and
  irrotational binary neutron stars in general relativity. II. Newtonian
  limits'', (\comment{onlinemonth}March, \comment{onlineyear}2001),
  \comment{fileformat}[Online Los Alamos Archive Preprint]: cited on
  \comment{cited}24 June 2002,
  \comment{onlineaddress}http://xxx.lanl.gov/abs/gr-qc/0103041.
  \keywords{Einstein equations, relativistic hydrodynamics, initial value
  problem, hyperbolic equations, elliptic equations, binary systems, spectral
  methods, neutron stars}

\bibitem{teukolsky72}\comment{article}
\comment{author}Teukolsky, S.~A., \comment{title}``Rotating black holes:
  separable wave equations for gravitational and electromagnetic
  perturbations'', \comment{journal}{\em Phys. Rev. Lett.},
  \comment{volume}{\bf 29}, \comment{pages}1114--1118, (\comment{year}1972).
  \keywords{Black holes, Kerr metric, perturbation theory}

\bibitem{thorne96}\comment{inproceedings}
\comment{author}Thorne, K., \comment{title}``Gravitational waves'', in
  \comment{editor}Kolb, E.~W., and Peccei, R., eds., \comment{booktitle}{\em
  Proceedings of the Snowmass 95 Summer Study on Particle and Nuclear
  Astrophysics and Cosmology}, (\comment{publisher}World Scientific,
  \comment{address}Singapore, \comment{year}1996). \keywords{}

\bibitem{tnsweb}\comment{misc}
For animations of oscillating toroidal neutron stars see: \\
  \verb+http://www.sissa.it/~rezolla/movies.html+. \keywords{}

\bibitem{toro97}\comment{book}
\comment{author}Toro, E.~F., \comment{title}{\em Riemann solvers and numerical
  methods for fluid dynamics -- a practical introduction},
  (\comment{publisher}Springer Verlag, \comment{address}Berlin, Germany,
  \comment{year}1997). \keywords{numerical methods, computational fluid
  dynamics, hyperbolic systems, finite difference and finite volume methods}

\bibitem{vanderklis98}\comment{inproceedings}
\comment{author}van~der Klis, M., \comment{title}``Kilohertz Quasi-Periodic
  Oscillations in Low-Mass X-Ray Binaries'', in \comment{editor}R.~Buccheri, J.
  van~Paradijs, and Alpar, M.A., eds., \comment{booktitle}{\em Proceedings of
  the NATO Advanced Study Institute}, \comment{pages} 337,
  (\comment{publisher}Kluwer Academic Publishers, \comment{address}Boston,
  \comment{year}1998). For a related online version see:
  \comment{author}M.~van~der Klis, \comment{onlinetitle}``Kilohertz
  Quasi-Periodic Oscillations in Low-Mass X-Ray Binaries - a Review'',
  (\comment{onlinemonth}October, \comment{onlineyear}1997),
  \comment{fileformat}[Online Los Alamos Archive Preprint]: cited on
  \comment{cited}15 February 2000,
  \comment{onlineaddress}http://xxx.lanl.gov/abs/astro-ph/9710016.
  \keywords{relativistic stars, relativistic astrophysics, accretion disks}

\bibitem{vanLeer77}\comment{article}
\comment{author}van Leer, B.~J., \comment{title}``Towards the ultimate
  conservative difference scheme. III. Upstream centered finite difference
  schemes for ideal compressible flows'', \comment{journal}{\em J. Comput.
  Phys.}, \comment{volume}{\bf 23}, \comment{pages}263--275,
  (\comment{year}1977). \keywords{computational fluid dynamics, finite
  difference and finite volume methods, hyperbolic equations, numerical
  methods}

\bibitem{vanleer79}\comment{article}
\comment{author}van Leer, B.~J., \comment{title}``Towards the ultimate
  conservative difference scheme. V. A second order sequel to Godunov's
  method'', \comment{journal}{\em J. Comput. Phys.}, \comment{volume}{\bf 32},
  \comment{pages}101--136, (\comment{year}1979). \keywords{computational fluid
  dynamics, finite difference and finite volume methods, hyperbolic equations,
  numerical methods}

\bibitem{vanputten98}\comment{article}
\comment{author}van Putten, M.~H.~P.~M. For a related online version see:
  \comment{author}M.~H.~P.~M. van Putten, \comment{onlinetitle}``Uniqueness in
  MHD in divergence form: right nullvectors and well-posedness'',
  (\comment{onlinemonth}April, \comment{onlineyear}1998),
  \comment{fileformat}[Online Los Alamos Archive Preprint]: cited on
  \comment{cited}1 May 1998,
  \comment{onlineaddress}http://xxx.lanl.gov/abs/astro-ph/9804139.
  \keywords{relativistic hydrodynamics and magnetohydrodynamics, initial value
  problem, hyperbolic equations}

\bibitem{vanputten02a}\comment{article}
\comment{author}van Putten, M.~H.~P.~M., and Levinson, A.,
  \comment{title}``Detecting Energy Emissions from a Rotating Black Hole'',
  \comment{journal}{\em Science}, \comment{volume}{\bf 295},
  \comment{pages}1874--1877, (\comment{month}March, \comment{year}2002).
  \keywords{}

\bibitem{vanriper79}\comment{article}
\comment{author}van Riper, K.~A., \comment{title}``General relativistic
  hydrodynamics and the adiabatic collapse of stellar cores'',
  \comment{journal}{\em Astrophys. J.}, \comment{volume}{\bf 232},
  \comment{pages}558--571, (\comment{year}1979). \keywords{Relativistic
  hydrodynamics, gravitational collapse, nuclear physics, neutron stars,
  supernovae, black holes, numerical methods}

\bibitem{vanriper88}\comment{article}
\comment{author}van Riper, K.~A., \comment{title}``Effects of nuclear equation
  of state on general relativistic stellar core collapse models'',
  \comment{journal}{\em Astrophys. J.}, \comment{volume}{\bf 326},
  \comment{pages}235--240, (\comment{year}1988). \keywords{Relativistic
  hydrodynamics, gravitational collapse, nuclear physics, neutron stars,
  supernovae, black holes}

\bibitem{vonneumann50}\comment{article}
\comment{author}von Neumann, J., and Richtmyer, R.~D, \comment{title}``A method
  for the numerical calculation of hydrodynamic shocks'', \comment{journal}{\em
  J. Appl. Phys.}, \comment{volume}{\bf 21}, \comment{pages}232--247,
  (\comment{year}1950). \keywords{computational fluid dynamics, shock waves,
  finite difference and finite volume methods, numerical methods}

\bibitem{wen97}\comment{article}
\comment{author}Wen, L., Panaitescu, A., and Laguna, P., \comment{title}``A
  shock-patching code for ultrarelativistic fluid flows'',
  \comment{journal}{\em Astrophys. J.}, \comment{volume}{\bf 486},
  \comment{pages}919--927, (\comment{year}1997). For a related online version
  see: \comment{author}L.~Wen, et al., \comment{onlinetitle}``A shock-patching
  code for ultrarelativistic fluid flows'', (\comment{onlinemonth}December,
  \comment{onlineyear}1996), \comment{fileformat}[Online Los Alamos Archive
  Preprint]: cited on \comment{cited}15 February 2000,
  \comment{onlineaddress}http://xxx.lanl.gov/abs/astro-ph/9612045.
  \keywords{computational fluid dynamics, relativistic hydrodynamics, finite
  difference and finite volume methods, hyperbolic equations, numerical
  methods}

\bibitem{wilson71}\comment{article}
\comment{author}Wilson, J.~R., \comment{title}``A numerical study of
  gravitational stellar collapse'', \comment{journal}{\em Astrophys. J.},
  \comment{volume}{\bf 163}, \comment{pages}209--219, (\comment{year}1971).
  \keywords{Gravitational collapse, neutrinos, supernovae, relativistic
  hydrodynamics, nuclear physics}

\bibitem{wilson72}\comment{article}
\comment{author}Wilson, J.~R., \comment{title}``Numerical study of fluid flow
  in a Kerr space'', \comment{journal}{\em Astrophys. J.}, \comment{volume}{\bf
  173}, \comment{pages}431--438, (\comment{year}1972). \keywords{Black holes,
  Accretion, Numerical relativistic hydrodynamics, Numerical methods}

\bibitem{wilson79}\comment{inbook}
\comment{author}Wilson, J.~R., \comment{title}``A numerical method for
  relativistic hydrodynamics'', in \comment{editor}Smarr, L., ed.,
  \comment{booktitle}{\em Sources of Gravitational Radiation}, \comment{pages}
  423--445, (\comment{publisher}Cambridge University Press,
  \comment{address}Cambridge, England, \comment{year}1979).
  \keywords{Relativistic hydrodynamics, numerical relativistic hydrodynamics,
  computational fluid dynamics, numerical methods}

\bibitem{wilson85}\comment{inbook}
\comment{author}Wilson, J.~R, \comment{title}``Supernovae and post-collapse
  behaviour'', in \comment{editor}J.~M.~Centrella, J.~M.~LeBlanc, and Wilson,
  J.~R., eds., \comment{booktitle}{\em Numerical astrophysics}, \comment{pages}
  422--434, (\comment{publisher}Jones and Bartlett, \comment{address}Boston,
  \comment{year}1985). \keywords{Supernovae, shock waves, gravitational
  collapse, neutrinos, numerical relativistic hydrodynamics}

\bibitem{wilson89}\comment{inbook}
\comment{author}Wilson, J.~R., and Mathews, G.~J.,
  \comment{title}``Relativistic hydrodynamics'', in
  \comment{editor}C.~R.~Evans, L.~S.~Finn, and Hobill, D.~W., eds.,
  \comment{booktitle}{\em Frontiers in numerical relativity}, \comment{pages}
  306--314, (\comment{publisher}Cambridge University Press,
  \comment{address}Cambridge, England, \comment{year}1989).
  \keywords{relativistic hydrodynamics, numerical relativistic hydrodynamics,
  numerical methods}

\bibitem{wilson95}\comment{article}
\comment{author}Wilson, J.~R., and Mathews, G.~J.,
  \comment{title}``Instabilities in Close Neutron Star Binaries'',
  \comment{journal}{\em Phys. Rev. Lett.}, \comment{volume}{\bf 75},
  \comment{pages}4161--4164, (\comment{year}1995). \keywords{Relativistic
  hydrodynamics, Numerical relativistic hydrodynamics, computational fluid
  dynamics, Einstein equations, post-Newtonian approximations, binary systems,
  neutron stars, nuclear physics, gravitational collapse}

\bibitem{wilson96}\comment{article}
\comment{author}Wilson, J.~R., Mathews, G.~J., and Marronetti, P.,
  \comment{title}``Relativisitic Numerical Model for Close Neutron Star
  Binaries'', \comment{journal}{\em Phys. Rev. D}, \comment{volume}{\bf 54},
  \comment{pages}1317--1331, (\comment{year}1996). For a related online version
  see: \comment{author}J.~R. Wilson, et al.,
  \comment{onlinetitle}``Relativisitic Numerical Model for Close Neutron Star
  Binaries'', (\comment{onlinemonth}January, \comment{onlineyear}1996),
  \comment{fileformat}[Online Los Alamos Archive Preprint]: cited on
  \comment{cited}1 September 1996,
  \comment{onlineaddress}http://xxx.lanl.gov/abs/gr-qc/9601017.
  \keywords{Relativistic hydrodynamics, Numerical relativistic hydrodynamics,
  computational fluid dynamics, Einstein equations, post-Newtonian
  approximations, binary systems, neutron stars, nuclear physics, gravitational
  collapse}

\bibitem{winicour98}\comment{article}
\comment{author}Winicour, J., \comment{title}``Characteristic evolution and
  matching'', \comment{journal}{\em Living Reviews in Relativity},
  \comment{volume}{\bf 1}, \comment{pages}5, (\comment{year}1998).
  \keywords{Einstein equations, numerical relativity, null surfaces, initial
  value problem, gauge conditions, numerical methods}

\bibitem{woodward84}\comment{article}
\comment{author}Woodward, P., and Colella, P., \comment{title}``The numerical
  simulation of two-dimensional fluid flow with strong shocks'',
  \comment{journal}{\em J. Comput. Phys.}, \comment{volume}{\bf 54},
  \comment{pages}115--173, (\comment{year}1984). \keywords{Hydrodynamics,
  computational fluid dynamics, shock waves, numerical methods}

\bibitem{woosley93}\comment{article}
\comment{author}Woosley, S.~E., \comment{title}``Gamma-ray bursts from stellar
  mass accretion disks around black holes'', \comment{journal}{\em Astrophys.
  J.}, \comment{volume}{\bf 405}, \comment{pages}273--277,
  (\comment{year}1993). \keywords{relativistic astrophysics, gamma-ray bursts,
  accretion disks, black holes}

\bibitem{woosley88}\comment{article}
\comment{author}Woosley, S.~E., Pinto, P.~A., and Ensman, L.,
  \comment{title}``Supernova 1987 A - Six weeks later'', \comment{journal}{\em
  Astrophys. J.}, \comment{volume}{\bf 324}, \comment{pages}466--489,
  (\comment{year}1988). \keywords{astrophysics, gravitational collapse,
  supernovae}

\bibitem{yamada96}\comment{article}
\comment{author}Yamada, S., \comment{title}``An implicit Lagrangian code for
  spherically symmetric general relativistic hydrodynamics with an approximate
  Riemann solver'', \comment{journal}{\em Astrophys. J.}, \comment{volume}{\bf
  475}, \comment{pages}720--739, (\comment{year}1997). For a related online
  version see: \comment{author}S.~Yamada, \comment{onlinetitle}``An implicit
  Lagrangian code for spherically symmetric general relativistic hydrodynamics
  with an approximate Riemann solver'', (\comment{onlinemonth}January,
  \comment{onlineyear}1996), \comment{fileformat}[Online Los Alamos Archive
  Preprint]: cited on \comment{cited}15 February 2000,
  \comment{onlineaddress}http://xxx.lanl.gov/abs/astro-ph/9601042.
  \keywords{Relativistic hydrodynamics, supernovae, numerical methods}

\bibitem{yamada94a}\comment{article}
\comment{author}Yamada, S., and Sato, K., \comment{title}``Numerical study of
  rotating core collapse in supernova explosions'', \comment{journal}{\em
  Astrophys.~J.}, \comment{volume}{\bf 434}, \comment{pages}268--276,
  (\comment{year}1994). \keywords{}

\bibitem{yee87}\comment{article}
\comment{author}Yee, H.~C., \comment{title}``Construction of explicit and
  implicit symmetric TVD schemes and their applications'',
  \comment{journal}{\em J. Comput. Phys.}, \comment{volume}{\bf 68},
  \comment{pages}151--179, (\comment{year}1987). \keywords{numerical methods,
  computational fluid dynamics, hyperbolic equations, finite difference and
  finite volume methods}

\bibitem{yee89}\comment{inproceedings}
\comment{author}Yee, H.~C., \comment{title}``A class of high-resolution
  explicit and implicit shock-capturing methods'', in \comment{booktitle}{\em
  Computational fluid dynamics}. Von Karman Institute for Fluid Dynamics,
  Lecture Series 1989-04., (\comment{year}1989). \keywords{numerical methods,
  computational fluid dynamics, hyperbolic equations, finite difference and
  finite volume methods}

\bibitem{yokosawa93}\comment{article}
\comment{author}Yokosawa, M., \comment{title}``Energy and angular momentum
  transport in magnetohydrodynamical accretion onto a rotating black hole'',
  \comment{journal}{\em Publ. Astron. Soc. Japan}, \comment{volume}{\bf 45},
  \comment{pages}207--218, (\comment{year}1993). \keywords{Accretion disks,
  Active galactic nuclei, black holes, relativistic hydrodynamics and
  magnetohydrodynamics}

\bibitem{yokosawa95}\comment{article}
\comment{author}Yokosawa, M., \comment{title}``Structure and dynamics of an
  accretion disk around a black hole'', \comment{journal}{\em Publ. Astron.
  Soc. Japan}, \comment{volume}{\bf 47}, \comment{pages}605--615,
  (\comment{year}1995). \keywords{Accretion disks, Active galactic nuclei,
  black holes, numerical relativistic hydrodynamics}

\bibitem{york79}\comment{inproceedings}
\comment{author}York, J., \comment{title}``Kinematics and Dynamics of General
  Relativity'', in \comment{editor}Smarr, L., ed., \comment{booktitle}{\em
  Sources of Gravitational Radiation}, (\comment{publisher}Cambridge University
  Press, \comment{address}Cambridge, England, \comment{year}1979).
  \keywords{ADM formalism, Cauchy problem, Constraint equations, Einstein
  equations, Energy and momentum, Gauge conditions, Initial value problem}

\bibitem{zampieri96}\comment{article}
\comment{author}Zampieri, L., Miller, J.~C., and Turolla, R.,
  \comment{title}``Time-dependent analysis of spherical accretion on to black
  holes'', \comment{journal}{\em Mon. Not. R. Astron. Soc.},
  \comment{volume}{\bf 281}, \comment{pages}1183--1196, (\comment{year}1996).
  For a related online version see: \comment{author}L.~Zampieri, et al.,
  \comment{onlinetitle}``Time-dependent analysis of spherical accretion onto
  black holes'', (\comment{onlinemonth}July, \comment{onlineyear}1996),
  \comment{fileformat}[Online Los Alamos Archive Preprint]: cited on
  \comment{cited}15 February 2000,
  \comment{onlineaddress}http://xxx.lanl.gov/abs/astro-ph/9607030.
  \keywords{Accretion, Accretion disks, black holes, Hydrodynamics, numerical
  methods}

\bibitem{zanotti02b}\comment{article}
\comment{author}Zanotti, O., Rezzolla, and Font, J.~A.,
  \comment{title}``Quasi-periodic accretion and gravitational waves from
  oscillating ``toroidal neutron stars'' around a Schwarzschild black hole'',
  \comment{journal}{\em Mon. Not. R. Astron. Soc.}, (\comment{year}2002). For a
  related online version see: \comment{author}O.~Zanotti, et al.,
  (\comment{onlinemonth}October, \comment{onlineyear}2002),
  \comment{fileformat}[Online Los Alamos Archive Preprint]: cited on
  \comment{cited}15 October 2002,
  \comment{onlineaddress}http://xxx.lanl.gov/abs/gr-qc/0210018. submitted.
  \keywords{Accretion, gravitational waves}

\bibitem{fem}\comment{book}
\comment{author}Zienkiewicz, O.~C., \comment{title}{\em The finite element
  method}, (\comment{publisher}McGraw-Hill, \comment{address}London,
  \comment{year}1977). \keywords{Numerical methods}

\bibitem{zwerger97}\comment{article}
\comment{author}Zwerger, T., and M\"uller, E., \comment{title}``Dynamics and
  gravitational wave signature of axisymmetric rotational core collapse'',
  \comment{journal}{\em Astron. Astrophys.}, \comment{volume}{\bf 320},
  \comment{pages}209--227, (\comment{year}1997). \keywords{gravitational
  collapse, gravitational radiation}

\end{thebibliography}
\end{document}